\documentclass[11pt,twoside,notitlepage]{article}
\usepackage{amsmath}
\usepackage{amssymb}
\usepackage{amsthm}
\usepackage[title]{appendix}
\usepackage{array}
\usepackage{bm}
\usepackage[font={small},margin=20pt]{caption}
\usepackage{cite}
\usepackage{comment}
\usepackage{dcolumn}
\usepackage{enumerate}
\usepackage{float}
\usepackage[T1]{fontenc}
\usepackage{graphicx}
\usepackage[
colorlinks=true,
citecolor=MidnightBlue,
linkcolor=BrickRed,
urlcolor=blue,
breaklinks=true,
linktoc=all,
]{hyperref}
\usepackage{listings}
\usepackage{mathrsfs}
\usepackage{mathtools}
\usepackage{multicol}
\usepackage{stmaryrd}
\usepackage{subcaption}
\usepackage{textgreek}
\usepackage{titlesec}
\usepackage{tocloft}
\usepackage[normalem]{ulem}
\usepackage{wrapfig}
\usepackage[usenames,dvipsnames]{xcolor}
\usepackage{xy}

\newcommand{\scp}{\mathscr{P}}

\newcommand{\bscp}{{\partial \! \scp}}
\newcommand{\scpo}{{\scp_{\! 0}}}

\newcommand{\bnscp}{\partial^{\text{t}} \! \scp}
\newcommand{\bdscp}{\partial^{\text{v}} \! \scp}
\newcommand{\Gammad}{\Gamma^{\text{v}}}
\newcommand{\Gamman}{\Gamma^{\text{t}}}
\newcommand{\Gammane}{\Gamma^{\text{t}}_e}

\newcommand{\bmx}{\bm{x}}
\newcommand{\bmv}{\bm{v}}
\newcommand{\bmu}{\bm{u}}
\newcommand{\bmxh}{\bm{x}_h}

\newcommand{\bmvh}{\bm{v}_h}
\newcommand{\bmvmh}{\bmvm_h}
\newcommand{\lambdah}{\lambda_h}
\newcommand{\Lambdah}{\Lambda_h}

\newcommand{\bmvm}{\bm{v}^{\mathrm{m}}}

\newcommand{\Jm}{J^{\mathrm{m}}}
\newcommand{\JGammam}{J^{\mathrm{m}}_\Gamma}

\newcommand{\sz}{{i}}
\newcommand{\so}{{i+1}}

\newcommand{\tn}{{t_n}}
\newcommand{\tno}{{t_{n+1}}}

\newcommand{\mN}{[ \mathbf{N} ]}

\newcommand{\mhN}{[ \bar{\mathbf{N}} ]}

\newcommand{\mv}{[ \mathbf{v} ]}
\newcommand{\mbu}{[ \mathbf{u} ]}
\newcommand{\mvm}{[ \mathbf{v}^{\mathrm{m}} ]}
\newcommand{\mlambda}{[ \bar{\bm{\lambda}} ]}

\newcommand{\mvt}{[ \mathbf{v} (t) ]}
\newcommand{\mbut}{[ \mathbf{u} (t) ]}
\newcommand{\mvmt}{[ \mathbf{v}^{\mathrm{m}} (t) ]}
\newcommand{\mlambdat}{[ \bar{\bm{\lambda}} (t) ]}

\newcommand{\mbutn}{[ \mathbf{u} (t_n) ]}
\newcommand{\mbutno}{[ \mathbf{u} (t_{n+1}) ]}
\newcommand{\mDeltautno}{[ \Delta \mathbf{u} (t_{n+1}) ]}

\newcommand{\mvtzero}{[ \mathbf{v} (t_0) ]}
\newcommand{\mvmtzero}{[ \mathbf{v}^{\mathrm{m}} (t_0) ]}

\newcommand{\mxn}{[ \mathbf{x}_n ]}

\newcommand{\mvn}{[ \mathbf{v}_n ]}
\newcommand{\mbun}{[ \mathbf{u}_n ]}
\newcommand{\mvmn}{[ \mathbf{v}^{\mathrm{m}}_n ]}
\newcommand{\mlambdan}{[ \bar{\bm{\lambda}}_n ]}

\newcommand{\mxno}{[ \mathbf{x}_{n+1} ]}

\newcommand{\mvno}{[ \mathbf{v}_{n+1} ]}
\newcommand{\mbuno}{[ \mathbf{u}_{n+1} ]}
\newcommand{\mvmno}{[ \mathbf{v}^{\mathrm{m}}_{n+1} ]}
\newcommand{\mlambdano}{[ \bar{\bm{\lambda}}_{n+1} ]}

\newcommand{\mNe}{[ \mathbf{N}^{{e}} ]}
\newcommand{\mNeo}{[ \mathbf{N}^{{e}}_{(1)} ]}
\newcommand{\mhNe}{[ \bar{\mathbf{N}}^{{e}} ]}
\newcommand{\mbNe}{[ \breve{\mathbf{N}}^{{e}} ]}

\newcommand{\mbue}{[ \mathbf{u}^{{e}} ]}
\newcommand{\mve}{[ \mathbf{v}^{{e}} ]}
\newcommand{\mvme}{[ \mathbf{v}^{\mathrm{m}, {e}} ]}
\newcommand{\mlambdae}{[ \bar{\bm{\lambda}}^{e} ]}
\newcommand{\mblambdae}{[ \breve{\bm{\lambda}}^{e} ]}

\newcommand{\mlambdaet}{[ \bar{\bm{\lambda}}^{e} (t) ]}

\newcommand{\mven}{[ \mathbf{v}^{{e}}_n ]}
\newcommand{\mvmen}{[ \mathbf{v}^{\mathrm{m}, {e}}_n ]}
\newcommand{\mlambdaen}{[ \bar{\bm{\lambda}}^{e}_n ]}
\newcommand{\mblambdaen}{[ \breve{\bm{\lambda}}^{e}_n ]}

\newcommand{\mlambdaeno}{[ \bar{\bm{\lambda}}^{e}_{n+1} ]}
\newcommand{\mblambdaeno}{[ \breve{\bm{\lambda}}^{e}_{n+1} ]}
\newcommand{\mpeno}{[ \mathbf{p}^{e}_{n+1} ]}

\newcommand{\mdeltau}{[ \delta \mathbf{u} ]}
\newcommand{\mdeltav}{[ \delta \mathbf{v} ]}
\newcommand{\mdeltavm}{[ \delta \mathbf{v}^{\mathrm{m}} ]}
\newcommand{\mdeltalambda}{[ \delta \bar{\bm{\lambda}} ]}

\newcommand{\mdeltave}{[ \delta \mathbf{v}^{e} ]}

\newcommand{\mdeltalambdae}{[ \delta \bar{\bm{\lambda}}^{e} ]}
\newcommand{\mdeltablambdae}{[ \delta \breve{\bm{\lambda}}^{e} ]}
\newcommand{\mdeltape}{[ \delta \mathbf{p}^{e} ]}

\newcommand{\mr}{[ \mathbf{r} ]}
\newcommand{\mrv}{[ \mathbf{r}^{\mathrm{v}} ]}
\newcommand{\mrm}{[ \mathbf{r}^{\mathrm{m}} ]}
\newcommand{\mrlambda}{[ \mathbf{r}^{\bar{\lambda}} ]}

\newcommand{\mrt}{[ \mathbf{r} (t) ]}
\newcommand{\mrvt}{[ \mathbf{r}^{\mathrm{v}} (t) ]}
\newcommand{\mrmt}{[ \mathbf{r}^{\mathrm{m}} (t) ]}
\newcommand{\mrlambdat}{[ \mathbf{r}^{\bar{\lambda}} (t) ]}

\newcommand{\mrno}{[ \mathbf{r}_{n+1} ]}
\newcommand{\mrvno}{[ \mathbf{r}^{\mathrm{v}}_{n+1} ]}
\newcommand{\mrmno}{[ \mathbf{r}^{\mathrm{m}}_{n+1} ]}
\newcommand{\mrlambdano}{[ \mathbf{r}^{\bar{\lambda}}_{n+1} ]}

\newcommand{\mre}{[ \mathbf{r}^{e} ]}

\newcommand{\mrveno}{[ \mathbf{r}^{\mathrm{v}, {e}}_{n+1} ]}

\newcommand{\mrlambdaeno}{[ \mathbf{r}^{\bar{\lambda}, {e}}_{n+1} ]}

\newcommand{\mrpeno}{[ \mathbf{r}^{\mathrm{p}, {e}}_{n+1} ]}
\newcommand{\mrvrhoeno}{[ \mathbf{r}^{\mathrm{v}, {e}}_{\rho, n+1} ]}

\newcommand{\mDeltave}{[ \Delta \mathbf{v}^{e} ]}
\newcommand{\mDeltavme}{[ \Delta \mathbf{v}^{\mathrm{m}, {e}} ]}
\newcommand{\mDeltalambdae}{[ \Delta \bar{\bm{\lambda}}^{{e}} ]}

\newcommand{\mDeltav}{[ \Delta \mathbf{v} ]}
\newcommand{\mDeltavm}{[ \Delta \mathbf{v}^{\mathrm{m}} ]}
\newcommand{\mDeltalambda}{[ \Delta \bar{\bm{\lambda}} ]}

\newcommand{\mDeltauno}{[ \Delta \mathbf{u}_{n+1} ]}
\newcommand{\mDeltavno}{[ \Delta \mathbf{v}_{n+1} ]}
\newcommand{\mDeltavmno}{[ \Delta \mathbf{v}^{\mathrm{m}}_{n+1} ]}
\newcommand{\mDeltalambdano}{[ \Delta \bar{\bm{\lambda}}_{n+1} ]}

\newcommand{\mzero}{[ \bm{0} ]}
\newcommand{\mone}{[  \bm{1} ]}
\newcommand{\mT}{{\mathrm{T}}}

\newcommand{\mKno}{[   \mathbf{K}_{n+1} ]}

\newcommand{\mK}{[   \mathbf{K} ]}
\newcommand{\mKvv}{[ \mathbf{K}^{\mathrm{vv}} ]}
\newcommand{\mKvm}{[ \mathbf{K}^{\mathrm{vm}} ]}
\newcommand{\mKvl}{[ \mathbf{K}^{\mathrm{v} \bar{\lambda}} ]}

\newcommand{\mKmv}{[ \mathbf{K}^{\mathrm{mv}} ]}
\newcommand{\mKmm}{[ \mathbf{K}^{\mathrm{mm}} ]}
\newcommand{\mKml}{[ \mathbf{K}^{\mathrm{m} \bar{\lambda}} ]}
\newcommand{\mKlv}{[ \mathbf{K}^{\bar{\lambda} \mathrm{v}} ]}
\newcommand{\mKlm}{[ \mathbf{K}^{\bar{\lambda} \mathrm{m}} ]}
\newcommand{\mKll}{[ \mathbf{K}^{\bar{\lambda} \bar{\lambda}} ]}

\newcommand{\mKe}{[ \mathbf{K}^{e} ]}
\newcommand{\mKvve}{[ \mathbf{K}^{\mathrm{vv}, {e}} ]}
\newcommand{\mKvme}{[ \mathbf{K}^{\mathrm{vm}, {e}} ]}
\newcommand{\mKvle}{[ \mathbf{K}^{\mathrm{v} \bar{\lambda}, {e}} ]}

\newcommand{\mKmve}{[ \mathbf{K}^{\mathrm{mv}, {e}} ]}
\newcommand{\mKmme}{[ \mathbf{K}^{\mathrm{mm}, {e}} ]}
\newcommand{\mKmle}{[ \mathbf{K}^{\mathrm{m} \bar{\lambda}, {e}} ]}
\newcommand{\mKlve}{[ \mathbf{K}^{\bar{\lambda} \mathrm{v}, {e}} ]}
\newcommand{\mKlme}{[ \mathbf{K}^{\bar{\lambda} \mathrm{m}, {e}} ]}
\newcommand{\mKlle}{[ \mathbf{K}^{\bar{\lambda} \bar{\lambda}, {e}} ]}

\newcommand{\mKvveno}{[ \mathbf{K}^{\mathrm{vv}, {e}}_{n+1} ]}
\newcommand{\mKvmeno}{[ \mathbf{K}^{\mathrm{vm}, {e}}_{n+1} ]}
\newcommand{\mKvleno}{[ \mathbf{K}^{\mathrm{v} \bar{\lambda}, {e}}_{n+1} ]}
\newcommand{\mKvpeno}{[ \mathbf{K}^{\mathrm{vp}, {e}}_{n+1} ]}

\newcommand{\mKlveno}{[ \mathbf{K}^{\bar{\lambda} \mathrm{v}, {e}}_{n+1} ]}

\newcommand{\mKlleno}{[ \mathbf{K}^{\bar{\lambda} \bar{\lambda}, {e}}_{n+1} ]}
\newcommand{\mKlpeno}{[ \mathbf{K}^{\bar{\lambda} \mathrm{p}, {e}}_{n+1} ]}
\newcommand{\mKpveno}{[ \mathbf{K}^{\mathrm{pv}, {e}}_{n+1} ]}
\newcommand{\mKpleno}{[ \mathbf{K}^{\mathrm{p} \bar{\lambda}, {e}}_{n+1} ]}
\newcommand{\mKppeno}{[ \mathbf{K}^{\mathrm{pp}, {e}}_{n+1} ]}
\newcommand{\mKvmpeno}{[ \mathbf{K}^{\mathrm{vm}, {e}}_{\mathrm{p}, n+1} ]}
\newcommand{\mKvvrhoeno}{[ \mathbf{K}^{\mathrm{vv}, {e}}_{\rho, n+1} ]}

\newcommand{\mGdb}{[ \mathbf{G}^{\breve{\lambda} \bar{\lambda}, {e}} ]}
\newcommand{\mHdb}{[ \mathbf{H}^{\breve{\lambda} \breve{\lambda}, {e}} ]}

\newcommand{\alphaDB}{\alpha^{\mathrm{DB}}}
\newcommand{\alpham}{\alpha^{\mathrm{m}}}

\newcommand{\mcg}{\mathcal{G}}

\newcommand{\mcgDB}{\mbox{$\mcg$}_{\raisebox{-1pt}{\text{\tiny{DB}}}}}
\newcommand{\mcu}{\mathcal{U}}
\newcommand{\mcuh}{\mathcal{U}_h}
\newcommand{\mcv}{\mathcal{V}}

\newcommand{\mcvh}{\mathcal{V}_h}
\newcommand{\mcvho}{\mathcal{V}_{0, h}}
\newcommand{\mcp}{\mathcal{P}}

\newcommand{\cone}{{\check{1}}}
\newcommand{\ctwo}{{\check{2}}}
\newcommand{\calpha}{{\check{\alpha}}}
\newcommand{\cbeta}{{\check{\beta}}}
\newcommand{\cgamma}{{\check{\gamma}}}

\newcommand{\clambda}{{\check{\lambda}}}
\newcommand{\cmu}{{\check{\mu}}}
\newcommand{\cnu}{{\check{\nu}}}

\newcommand{\blambda}{\breve{\lambda}}
\newcommand{\bLambda}{\breve{\Lambda}}

\newcommand{\Rey}{\mbox{\textit{Re}}}	% Reynolds number
\newcommand{\nume}{\texttt{ne}}
\newcommand{\nn}{\texttt{nn}}
\newcommand{\nln}{\texttt{nln}}
\newcommand{\nen}{\texttt{nen}}
\newcommand{\neln}{\texttt{neln}}

\newcommand{\Omegae}{{\Omega^{e}}}

\graphicspath{{figs/}}

\definecolor{dkgreen}{rgb}{0,0.6,0}
\definecolor{gray}{rgb}{0.5,0.5,0.5}
\definecolor{mauve}{rgb}{0.58,0,0.82}

\makeatletter
\def\@fnsymbol#1{
	\ensuremath{\ifcase#1\or
	*\or				% 1
	\mathsection\or		% 2
	\mathflat\or		% 3
	\dagger\or			% 4
	\ddagger\or			% 5
	\|\or				% 6
	\ddagger\ddagger\or	% 7
	\dagger\dagger\or	% 8
	**					% 9
	\else\@ctrerr\fi}}
\makeatother

\makeatletter
\newcommand{\linkdest}[1]{\Hy@raisedlink{\hypertarget{#1}{}}}
\makeatother

\begin{document}

	\begin{center}
		{\textbf{
			\Large{Arbitrary Lagrangian--Eulerian finite element method for curved and deforming surfaces}
		}} \\
		\vspace{0.11in}
		{\textbf{
			\large{I. General theory and application to fluid interfaces}
		}} \\
		\vspace{0.21in}

		{\small
			Amaresh Sahu,$^{1,}$\hyperlink{email1}{$^{\ddag}$}
			Yannick A. D. Omar,$^{1,}$\hyperlink{email2}{$^\S$}
			Roger A. Sauer,$^{2,}$\hyperlink{email3}{$^\flat$}
			and Kranthi K. Mandadapu$^{1,3,}$\hyperlink{email4}{$^\dag$} \\
		}
		\vspace{0.25in}

		\footnotesize{
			{
				$^1$
				Department of Chemical \& Biomolecular Engineering,
				University of California, Berkeley, CA 94720, USA
				\\[3pt]
				$^2$
				Aachen Institute for Advanced Study in Computational Engineering Sciences (AICES),\\
				RWTH Aachen University, Templergraben 55, 52056 Aachen, Germany
				\\[3pt]
				$^3$
				Chemical Sciences Division, Lawrence Berkeley National Laboratory, CA 94720, USA
				\\
			}
		}
	\end{center}

	\vspace{12pt}
	%
	% *** Abstract
	%

%
% *** Abstract
%

\begin{abstract}
	An arbitrary Lagrangian--Eulerian (ALE) finite element method for arbitrarily
	curved and deforming two-dimensional materials and interfaces is presented here.
	An ALE theory is developed by endowing the surface with a mesh whose in-plane velocity
	need not depend on the in-plane material velocity, and can be specified arbitrarily.
	A finite element implementation of the theory is formulated and applied to curved and
	deforming surfaces with in-plane incompressible flows.
	Numerical inf--sup instabilities associated with in-plane incompressibility are removed
	by locally projecting the surface tension onto a discontinuous space of piecewise linear
	functions.
	The general isoparametric finite element method, based on an arbitrary surface
	parametrization with curvilinear coordinates, is tested and validated against several
	numerical benchmarks.
	A new physical insight is obtained by applying the ALE developments to cylindrical fluid
	films, which are computationally and analytically found to be stable to non-axisymmetric
	perturbations, and unstable with respect to long-wavelength axisymmetric perturbations
	when their length exceeds their circumference.
	A Lagrangian scheme is attained as a special case of the ALE formulation.
	Though unable to model fluid films with sustained shear flows,
	the Lagrangian scheme is validated by reproducing the cylindrical instability.
	However, relative to the ALE results, the Lagrangian simulations are found to have
	spatially unresolved regions with few nodes, and thus larger errors.
\end{abstract}

	\vspace{14pt}

	% authors

	\noindent\rule{5.0cm}{0.4pt}

	\footnotesize
	\noindent{\linkdest{email1}{\hspace{10pt}$^\ddag \,$}amaresh.sahu\textit{@}berkeley.edu \\
		\linkdest{email2}{\hspace{10pt}$^\S \,$}yannick.omar\textit{@}berkeley.edu \\
		\linkdest{email3}{\hspace{10pt}$^\flat \,$}sauer\textit{@}aices.rwth-aachen.de \\
		\linkdest{email4}{\hspace{10pt}$^\dag \,$}kranthi\textit{@}berkeley.edu
	}
	\small

	\vspace{15pt}

	%
	% *** CONTENT 
	%

%
% *** Introduction
%

\section{Introduction} \label{sec:sec_introduction}

In this paper, and the subsequent manuscript in the series\!\footnote{From now on, we refer to the present paper as ``Part I'' and the following one \cite{sahu-mandadapu-ale-ii} as ``Part II.''}
\cite{sahu-mandadapu-ale-ii}, we develop an arbitrary Lagrangian--Eulerian (ALE) theory for arbitrarily curved and deforming two-dimensional interfaces with in-plane fluidity.
The theory is based on a surface discretization which is independent of the in-plane material flow, such that the surface mesh need not convect with the material.
Consequently, two-dimensional materials with large in-plane flows on arbitrarily deforming surfaces can be modeled.
In Part I, we develop our theory and use standard numerical techniques to devise an isoparametric ALE finite element method for incompressible fluid films.
We then implement the finite element formulation, model the deformations and flows of such materials over time, and provide several numerical results for both flat and cylindrical geometries.
In Part II, we hope to extend the finite element formulation to lipid membranes and study membrane behavior in several biologically relevant situations.
As the equations governing single- and multi-component lipid membranes reduce to the fluid film equations in the limit where no elastic energy is stored in the membrane, such a separation is natural and allows us to present our results in a more accessible manner.

Two-dimensional fluids have played an increasingly important role in many engineering applications, in which they often arise at phase boundaries in multiphase systems \cite{prudhomme}.
For example, under the influence of gravity and capillary forces, foams will drain over time until the constituent bubbles burst \cite{stone-jpcm-2002}.
Foam lifetime plays a key role in their viability for engineering applications, and there have consequently been many efforts to improve foam stability \cite{prudhomme}.
Similar efforts have been made to stabilize emulsions and colloidal dispersions, which again are of much industrial value \cite{smith-2012, russel-1989}.
Surfactants are often used to stabilize vapor--liquid and liquid--liquid interfaces by lowering the local surface tension \cite{edwards-brenner}; surface tension gradients can drive Marangoni flows \cite{marangoni-1865, gibbs, levich} and in some cases have been shown to significantly affect material properties \cite{mysels}.

Two-dimensional materials with in-plane fluidity also play a fundamental role in biology.
Biological membranes, which are interfaces composed of lipids and proteins, are in-plane fluid and out-of-plane elastic materials \cite{evans-skalak}.
They make up the boundary of the cell as well as many of its internal organelles, including the nucleus, endoplasmic reticulum, and Golgi complex.
Lipid membranes thus add structure and organization to the cell, and furthermore play an important role in many cellular processes.
Endocytosis, for example, begins when proteins in the surrounding bulk fluid bind to the cell membrane's constituent lipids and proteins at a specific location.
The membrane forms an initially shallow invagination, which then develops into a mature bud and eventually pinches off into a membrane-bound vesicle that enters the cell \cite{mcmahon-cell-2002}.
The vesicular membrane contains lipids and proteins which were previously on the cell boundary, and furthermore the vesicle may enclose nutrients or other cargo.
Endocytosis is thus a key process in transferring nutrients to the cell, regulating the expression of proteins on the cell surface, and cell homeostasis \cite{mcmahon-nrmcb-2011}.
It involves nontrivial coupling between protein binding and unbinding reactions, in-plane lipid flow, and out-of-plane membrane shape changes.
In particular, the in-plane flow and out-of-plane bending are coupled because lipid membranes are nearly area-incompressible \cite{evans-skalak} and therefore lipids are required to flow in-plane to accommodate any shape changes.
In another example, lipid membranes can phase separate into liquid--ordered $(\mathrm{L}_\mathrm{o})$ and liquid--disordered $(\mathrm{L}_\mathrm{d})$ domains under physiological conditions \cite{veatch-acscb-2008}; the energetic penalty of the $\mathrm{L}_\mathrm{o}$--$\mathrm{L}_\mathrm{d}$ interface plays a major role in the fusion of HIV-containing vesicles with target immune cells \cite{yang-tamm-nat-comm-2016}.
This phenomenon demonstrates the value in understanding the coupling between elastic membrane shape changes and the thermodynamically irreversible processes of in-plane lipid flow and in-plane species diffusion.

The interfacial materials discussed thus far are of fundamental importance to engineering and biology.
Consequently, there have been significant theoretical efforts to better understand their physics.
The pioneering work of L.E.~Scriven \cite{scriven-1960} and R.~Aris \cite{aris} was crucial to our current understanding of interfacial flows.
In particular, Scriven recognized it is prohibitively difficult to use standard Cartesian, cylindrical, and spherical coordinate systems to solve for fluid flows on arbitrarily curved surfaces, where even expressing the surface Laplacian of the velocity field at every point on the surface is nontrivial.
Consequently, Scriven used a mathematically elegant differential geometric framework to naturally represent two-dimensional flows and their gradients on arbitrarily curved surfaces \cite{scriven-1960}.
Subsequently, Aris worked to describe three-dimensional fluids using the machinery of differential geometry, and incorporated Scriven's surface flow description within his general differential geometric perspective~\cite{aris}.
The powerful formalism developed by Scriven and Aris continue to be in widespread use today.
An excellent review of the interfacial dynamics of fluid interfaces in multiphase systems is provided in Ref.~\cite{edwards-brenner}, and for a wonderful perspective on interfaces in fluid mechanics see Ref.~\cite{stone-jfm-2010}.

While the equations of motion characterizing two-dimensional interfacial flows on
surfaces are now widespread and well-understood, theoretical developments for lipid membranes are in a less mature stage.
A major complexity arises in modeling lipid membranes because they behave as in-plane fluids, out-of-plane elastic solids, and the surface on which dynamical equations are to be written is itself curved and deforming over time.
Early membrane models were modifications of P.M. Naghdi's seminal contributions to shell theory. However, while Naghdi used a balance law formulation \cite{naghdi-1973-theory}, the first membrane models used variational methods and focused only on elastic membrane behavior.
In particular, P.~Canham \cite{canham-jtb-1970} and W.~Helfrich \cite{helfrich-1973} proposed an elastic membrane bending energy in the early 1970's; Helfrich also used variational methods to determine the Euler--Lagrange equations governing axisymmetric membrane shapes in the absence of in-plane flows.
The Euler--Lagrange equations, which by construction include only thermodynamically reversible phenomena and thus do not contain viscous forces, were not extended to non-axisymmetric settings until 1999 \cite{steigmann-arma-1999}.
However, by this time various other models which restricted membrane shapes to small deviations from flat planes \cite{seifert-epl-1993, safran-prl-1995, fournier-prl-1996, seifert-long} and cylindrical or spherical shells \cite{seifert-long, seifert-pra-1991} had also emerged.
Since then, variational methods encompassing different physical phenomena have continued to be developed \cite{prost-prl-2002, capovilla-guven-jpa-2002, guven-jpa-2004, du-wang-jcp-2004, agrawal-steigmann-bmmb-2008, wang-du-jmb-2008, leitenberger-langmuir-2008, rahimi-soft-matter-2013, ramaswamy-prl-2014}.
In a parallel development, in-plane velocities were included in some models about simple geometries \cite{seifert-long, leitenberger-langmuir-2008, voth-bpj-2004a}.
It was not until 2009, however, that the general equations governing a single-component, arbitrarily curved and deforming lipid membrane with in-plane viscosity were determined \cite{arroyo-pre-2009}---using a combination of variational methods to determine elastic contributions and the so-called Rayleigh dissipation potential to determine the viscous terms.
Since then, variational methods have been extended, with viscous stresses sometimes included in an ad-hoc manner \cite{powers-rmp-2010, rahimi-arroyo-pre-2012, agrawal-cmt-2009}.
Membrane models have also recently been developed by building on the work of Naghdi \cite{naghdi-1973-theory} and using fundamental balance laws and associated constitutive equations \cite{kranthi-bmmb-2012, agrawal-steigmann-zamp-2011, kranthi-bpj-2014, walani-pnas-2015}.

While such theoretical developments have had success in modeling certain membrane phenomena, they were difficult to extend to study how elastic out-of-plane membrane bending is coupled to different irreversible phenomena, such as in-plane lipid flow, in-plane phase transitions involving multiple components, and chemical reactions between membrane components and species in the surrounding bulk.
Our recent work \cite{sahu-mandadapu-pre-2017}, inspired by the pioneering works of I.~Prigogine \cite{prigogine}, L.~Onsager \cite{onsager-pr-1931-i, onsager-pr-1931-ii}, and S.R.~de Groot \& P.~Mazur \cite{degroot-mazur}, developed the general theory of irreversible thermodynamics for arbitrarily curved lipid membranes, provided a formalism to determine the equations governing membrane dynamics, and presented comprehensive models for all of the irreversible phenomena described thus far.

Though the equations governing both fluid interfaces and lipid membranes are now determined, the equations are highly nonlinear and in general cannot be solved analytically.
Our work entails developing an ALE theory for two-dimensional materials with in-plane fluidity.
The theory involves a surface discretization whose in-plane velocity can be (i) zero, as in an Eulerian formulation, (ii) equal to the in-plane material velocity, as in a Lagrangian formulation, or (iii) specified arbitrarily.
The flexibility of our ALE theory, as well as its similarities to bulk methods of the same name \cite{donea-2004}, explain our nomenclature.
With the ALE theory, we numerically solve the equations of motion governing the aforementioned materials of interest.
We split this effort into two pieces: in Part~I we derive the general ALE theory, develop its finite element formulation, and apply it to two-dimensional fluid films.
In Part II we extend the finite element formulation to lipid membranes, which elastically resist out-of-plane bending, and present results from our numerical simulations.

The challenges in theoretically modeling deforming interfaces with in-plane flow, and lipid membranes especially, extend to their numerical modeling as well: to model a material with arbitrarily large shape deformations and in-plane flows, standard techniques from fluid mechanics and solid mechanics are insufficient.
Regarding fluid films, many previous studies simplified the problem by assuming the film was fixed in space.
The resultant fixed-surface flow equations, derived
by Scriven \cite{scriven-1960}, have been solved using various methods: by modeling the fluid interface as a level set in $\mathbb{R}^3$ \cite{ratz-voigt-cms-2006}, with projection-based finite element methods \cite{fries-ijnmf-2018}, and through the discretization of exterior calculus operators \cite{nitschke-voigt-2017}.
On the other hand, a different study using level set methods \cite{saye-sethian-science-2013} made considerable advances in numerically modeling bubble deformation and breakup in foams, however they separated the dynamics into different steps and in each step made simplifying assumptions.
In addition, interfaces have been modeled as the boundary between bulk fluid domains, using both ALE and level set techniques---however, such works are either restricted to simple geometries \cite{woodhouse-jfm-2012} or do not include in-plane interfacial flow \cite{pozrikidis-jcp-2001, pozrikidis-jem-2004, ganesan-jcp-2009, villone-cf-2014}.
ALE methods were also used to study the evolution of scalar fields on surfaces whose time evolution is known~\cite{elliott-mjm-2012}.
While the numerical methods discussed thus far have modeled different fluid film phenomena, they do not seem easily amenable to the study of general deforming fluid interfaces or lipid membranes, with the latter having their own constitutive behavior.

Just as in the case of fluid films, several lipid membrane studies assumed the membrane shape to be fixed, and under these conditions studied how surface flows are coupled to flows in the surrounding bulk in two cases: spherical surfaces with protein inclusions \cite{sigurdsson-atzberger-sm-2016} and radial surfaces in a one-to-one correspondence with a sphere \cite{gross-atzberger-jcp-2018}.
Alternatively, many of the studies modeling the deformation of lipid membranes \cite{feng-klug-jcp-2006, barrett-jcp-2008, barrett-siam-2008, ma-klug-jcp-2008, dziuk-nm-2008, elliott-stinner-jcp-2010, bonito-pauletti-2010, mercker-siam-2013, rangarajan-gao-jcp-2015} consider only the Euler--Lagrange equations, and thus predict membrane shapes without knowledge of the in-plane flow.
However, as the in-plane flow and out-of-plane deformations are coupled through the in-plane viscosity \cite{sahu-mandadapu-pre-2017}, the predictions of the previously mentioned works are only physically relevant in the limit where velocities are negligible.
Another approach has been to include in-plane fluid flow and limit the membrane to remain in one-to-one correspondence with a flat plane \cite{kranthi-bmmb-2012, kranthi-bpj-2014}.
While such an approach is theoretically sound, it is limited in its use as it cannot, for example, model the large shape deformations observed in endocytosis \cite{mcmahon-cell-2002}.
Several recent works have avoided the computational complexity of modeling the full lipid membrane equations by assuming only axisymmetric shapes \cite{walani-pnas-2015, hassinger-pnas-2017}, however this turns out to be a poor assumption which in many cases yields incorrect results \cite{omar-biorxiv-2019}.
In our previous work we modeled the full non-axisymmetric membrane equations using a Lagrangian finite element method \cite{kranthi-jcp-2017, omar-biorxiv-2019}, which is computationally valid yet can attain locally singular Jacobians and uninvertible matrices when there are moderate in-plane flows.
Lagrangian methods are thus not suitable for the study of general fluid and lipid membrane phenomena.

The limitations of our Lagrangian finite element formulation and other computational techniques in modeling fluid interfaces and lipid membranes motivate our development of an ALE finite element method for curved and deforming surfaces.
The following aspects are new in this work: we
\begin{enumerate}
	\item develop an ALE theory, within a differential geometric setting, for general arbitrarily curved and deforming two-dimensional interfaces with in-plane flow,
	\item apply the theory to two-dimensional fluid films and derive a corresponding isoparametric, fully implicit ALE finite element method,
	\item prevent numerical inf--sup instabilities associated with the in-plane areal incompressibility by adapting the method of C.R. Dohrmann and P.B. Bochev~\cite{dohrmann-bochev-ijnmf-2004} to curved surfaces,
	\item numerically simulate an arbitrarily curved and deforming fluid film, from which we find a physical instability that is confirmed analytically with a linear stability analysis, and
	\item demonstrate how our ALE formulation can be altered to yield a Lagrangian scheme as a special case.
\end{enumerate}
As mentioned earlier, we limit our numerical calculations to fluid films in this manuscript, as the extensions of the theory and numerical methods to lipid membranes will be presented in Part~II~\cite{sahu-mandadapu-ale-ii}.
We note Ref.~\cite{torres-jfm-2019} describes a concurrent effort with a similar objective.
In particular, Ref.\ \cite{torres-jfm-2019} also derives a general ALE theory.
However, their numerical implementation is based on a surface parametrization involving small membrane deformations, generalized to arbitrarily curved surfaces, with periodic updates of the reference surface as required.
Additionally, Ref.\ \cite{torres-jfm-2019} uses a Hodge decomposition of the membrane velocity field, while the present work employs isoparametric finite element methods.

Our paper is organized as follows:
In Sec.~\ref{sec:sec_ale_theory}, we present our ALE theory for general two-dimensional interfaces.
We provide the equations of motion governing arbitrarily curved and deforming
fluid films in Sec.~\ref{sec:sec_theory_fluid_films} and develop the corresponding finite element formulation in Sec.~\ref{sec:sec_finite_element_formulation_fluid_films}.
Numerical results of an ALE implementation are presented in Sec.~\ref{sec:sec_numerical_simulations}; the modifications leading to a pure Lagrangian formulation
and corresponding results are provided in Sec.~\ref{sec:sec_lagrangian}.
We end with conclusions and avenues for future work in Sec.~\ref{sec:sec_conclusions}.
Several of the more detailed calculations regarding fluid films and our finite element implementation, as well as additional numerical benchmarks, are relegated to Appendices~\ref{sec:sec_appendix_calculations}--\ref{sec:sec_appendix_numerical_benchmarks}.
Relevant movies are provided in Appendix~\ref{sec:sec_appendix_movies}, and important symbols are listed in Appendix~\ref{sec:sec_symbol_list}.

%
% *** Arbitrary Lagrangian--Eulerian Theory
%

\section{Arbitrary Lagrangian--Eulerian Theory} \label{sec:sec_ale_theory}

While the equations of motion governing arbitrarily curved and deforming fluid films and lipid membranes
were presented in Refs.~\cite{scriven-1960, arroyo-pre-2009, kranthi-bmmb-2012, sahu-mandadapu-pre-2017},
solving the resultant equations is nontrivial.
These equations are highly nonlinear and cannot be solved analytically, yet many issues arise when trying to solve them numerically as well.
We have described the shortcomings of our previous Lagrangian finite element formulation \cite{omar-biorxiv-2019, kranthi-jcp-2017}, which is not appropriate for materials with in-plane flow.
Namely, when the surface is discretized and the corresponding
mesh travels in-plane with the material, mesh elements become highly distorted and attain nearly singular Jacobians (see Fig.~\ref{fig:fig_xi_velocity}).
In such a case, a simple example of vortex flows is not possible.
We therefore require a new numerical method to solve the equations of motion.

In this section, we derive an ALE theory for arbitrarily curved and deforming surfaces, which provides equations more amenable to numerical solution.
In particular, we seek a description of the material surface which can be easily discretized, with individual elements not undergoing large distortions when material flows in-plane.
To this end, we introduce an ALE parametrization of the two-dimensional material, which endows the surface with a mesh that deforms in the normal direction with the material---yet whose in-plane motion can be specified arbitrarily and need not depend on the material flow.
We note Ref.\ \cite{elliott-mjm-2012} introduced a mesh evolving in a similar manner, for the study of scalar fields on evolving surfaces whose time evolution is prescribed.
Our ALE description, on the other hand, introduces three new unknowns corresponding to the three components of the mesh velocity.
In what follows, we discuss various parametrizations of the surface, and in the ALE case provide the additional three equations required for our problem to be mathematically well-posed.
We also describe how the equations governing fluid films and lipid membranes, which are based on an Eulerian surface parametrization, are modified when the ALE parametrization is employed.

%
% *** Surface Geometry
%

\subsection{Surface Geometry} \label{sec:sec_surface_geometry}

\begin{figure}[p]
	\centering
	\begin{subfigure}[b]{0.30\columnwidth}
		\centering
		\includegraphics[width=0.95\textwidth]{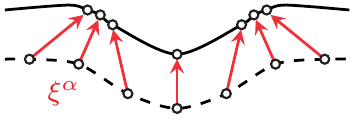}
		\caption{convected}
		\label{fig:fig_xi_velocity}
	\end{subfigure}
	~
	\begin{subfigure}[b]{0.30\columnwidth}
		\centering
		\includegraphics[width=0.95\textwidth]{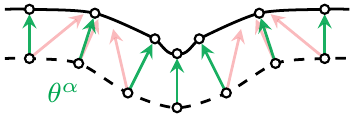}
		\caption{surface-fixed}
		\label{fig:fig_theta_velocity}
	\end{subfigure}
	~
	\begin{subfigure}[b]{0.30\columnwidth}
		\centering
		\includegraphics[width=0.95\textwidth]{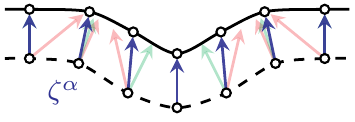}
		\caption{ALE}
		\label{fig:fig_zeta_velocity}
	\end{subfigure}
	\vspace{5pt}
	\caption{
		A schematic of how the convected (a), surface-fixed (b), and ALE (c) surface coordinates evolve in time, for a side view of the surface.
		In all cases, the initial surface is represented with a dotted line, the final surface is represented by a solid line, and the circles represent nodes of constant coordinate values.
		(a) The convected coordinates $\xi^\alpha$ move along the direction of the material velocity (red arrows), and can squeeze together or spread apart.
		Moreover, when there are in-plane flows, elements can become highly distorted---leading to complications in numerical implementations, and singularities in extreme cases~\cite{omar-biorxiv-2019, kranthi-jcp-2017}.
		(b) The surface-fixed coordinates $\theta^\alpha$ only move orthogonally to the surface (green arrows), and thus maintain regularity for arbitrarily large in-plane flows.
		However, as shown in the center of (b), in certain cases it is possible for the surface-fixed coordinates to squeeze together when the surface deforms.
		(c) The motion of the ALE coordinates $\zeta^\alpha$ can be specified arbitrarily, provided the normal components of the material velocities (red arrows) and mesh velocities (blue arrows) are equal, according to Eqs.~\eqref{eq:ale_relative_velocity} and \eqref{eq:ale_velocity_with_relative_mesh}---or equivalently Eq.~\eqref{eq:ale_mesh_velocity_normal_constraint}.
		Thus, a regular mesh can in principle always be maintained.
	}
	\label{fig:fig_velocity_schemes}
\end{figure}

\begin{figure}[p]
	\centering
		\includegraphics[width=0.95\textwidth]{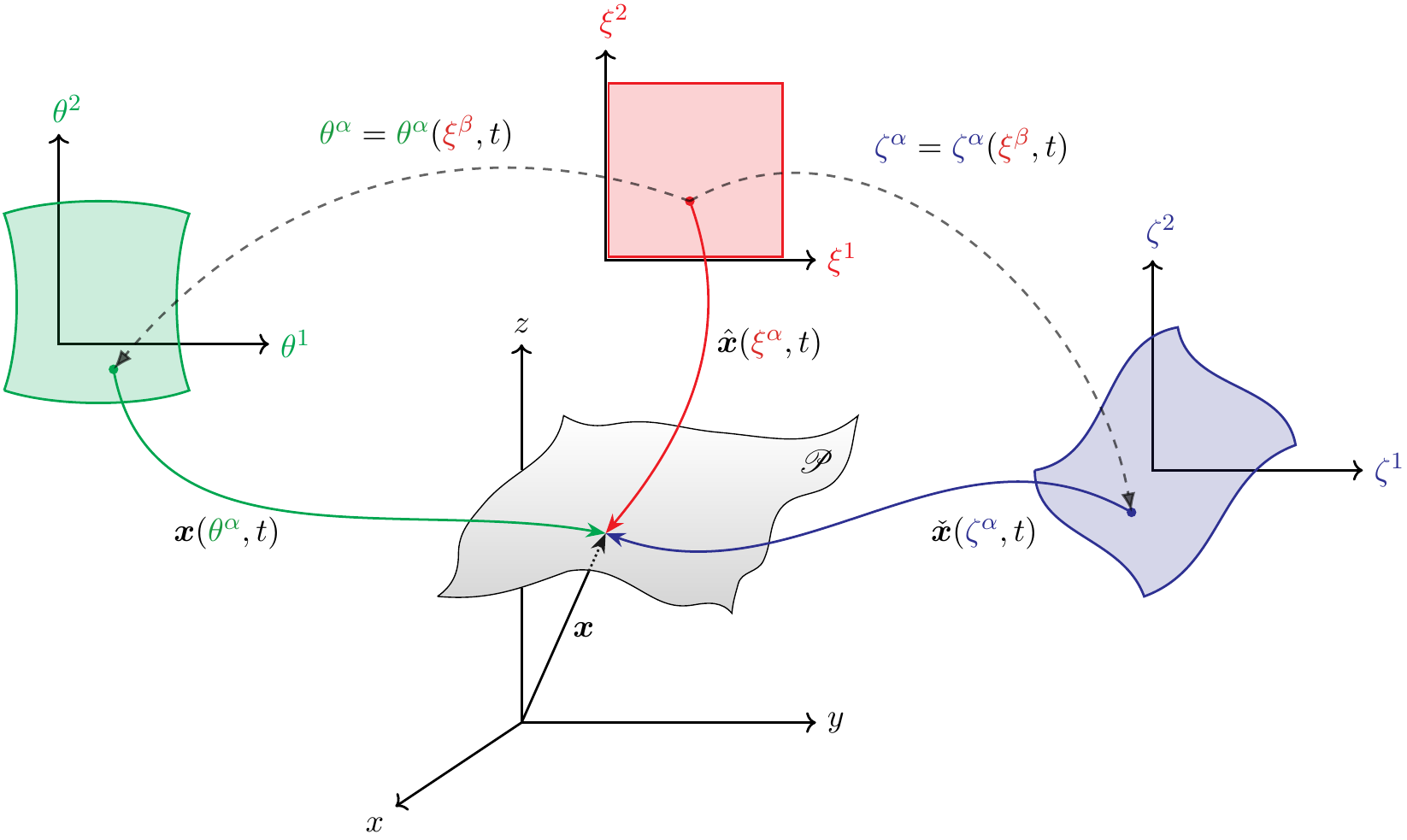}
		\vspace{10pt}
	\caption{
		A schematic of the surface parametrizations.
		The three different parametric domains, for $\theta^\alpha$, $\xi^\alpha$, and $\zeta^\alpha$, are shown respectively in \textcolor{Green}{green}, \textcolor{Red}{red}, and \textcolor{Blue}{blue}.
		The mappings from the parametric domains to the patch of surface $\scp$ are depicted at a single instant in time $t$ with solid, colored arrows.
		A single material point labeled by the convected coordinates $\xi^\beta$ has corresponding surface-fixed and mesh coordinates
		$\theta^\alpha = \theta^\alpha (\xi^\beta, t)$
		and
		$\zeta^\alpha = \zeta^\alpha (\xi^\beta, t)$,
		as shown with the dashed gray arrows.
		Accordingly, the patch position is equivalently written as (see Eq.~\eqref{eq:position_coordinate_equivalence_mesh})
		$
			\bmx
			= \hat{\bm{x}} (\xi^\alpha, \, t)
			= \bm{x} (\theta^\alpha, \, t)
			= \check{\bm{x}} (\zeta^\alpha, \, t)
		$.
	}
	\label{fig:fig_surface_parametrization}
\end{figure}

As discussed in Refs.~\cite{kranthi-bmmb-2012, sahu-mandadapu-pre-2017}, an arbitrarily curved and deforming patch of surface $\scp$ can be parametrized by either the convected coordinates $\xi^\alpha$, which are attached to material points and are convected with the material, or the surface-fixed coordinates $\theta^\alpha$, which are defined such that a point of fixed $\theta^\alpha$ moves only normal to the surface.
Schematics of the movement of points of constant $\xi^\alpha$ and $\theta^\alpha$ are provided in Figs.~\ref{fig:fig_xi_velocity} and \ref{fig:fig_theta_velocity}, respectively.
Here and from now on, we prescribe Greek indices to span the set $\{1, 2\}$, and use the Einstein summation convention in which Greek indices repeated in a subscript and superscript are summed over.
As the material occupying a point of constant $\theta^\alpha$ will in general change in time, we formally write
$\theta^\alpha = \theta^\alpha (\xi^\beta, \, t)$
and express the position in terms of either surface-fixed or convected coordinates as
\begin{equation} \label{eq:surface_position_convected_coordinates}
	\bm{x} (\theta^\alpha, \, t)
	= \bm{x} (\theta^\alpha (\xi^\beta, \, t), \, t)
	= \hat{\bm{x}} (\xi^\beta, \, t)
	~,
\end{equation}
where the `hat' accent is used to denote the position expressed in terms of the $\xi^\alpha$ parametrization (see Fig.~\ref{fig:fig_surface_parametrization}).

As discussed below, the $\theta^\alpha$ parametrization yields a surface description which is in-plane Eulerian and out-of-plane Lagrangian, and is most natural to theoretically model lipid membranes---which are in-plane fluids and out-of-plane elastic solids.
	Consequently, our theoretical developments~\cite{sahu-mandadapu-pre-2017} used the surface-fixed parametrization throughout.
	We use the same notation in this work, and review it here.
Partial and covariant differentiation with respect to $\theta^\alpha$ are respectively denoted $( \, ~ \, )_{, \alpha}$ and $( \, ~ \, )_{; \alpha}$.
The surface-fixed parametrization yields the in-plane basis vectors
$\bm{a}_\alpha := \bm{x}_{, \alpha}$
and unit normal
$\bm{n} := ( \bm{a}_1 \times \bm{a}_2 ) / \lvert \bm{a}_1 \times \bm{a}_2 \rvert$.
The metric and curvature components are respectively given by
$
	a_{\alpha \beta}
	\, := \, \bm{a}_\alpha \cdot \bm{a}_\beta
$
and
$
	b_{\alpha \beta}
	:= \bm{n} \cdot \bm{x}_{, \alpha \beta}
$,
with which the mean and Gaussian curvatures are respectively calculated as
$
	H
	:= \tfrac{1}{2} a^{\alpha \beta} b_{\alpha \beta}
$
and
$
	K
	:= \det (b_{\alpha \beta}) / \det (a_{\alpha \beta})
$.
A comprehensive geometric description of the surface can be found in Ref.~\cite{sahu-mandadapu-pre-2017} and the references provided therein.

%
% *** Surface Kinematics
%

\subsection{Surface Kinematics} \label{sec:sec_surface_kinematics}

As a point of constant $\xi^\alpha$ follows a material point over time, the velocity $\bm{v}$ of a material point is defined as
$
	\bmv
	:= [ \partial \hat{\bmx} / \partial t ] \rvert_{\xi^\beta}
$,
which upon substitution of Eq.~\eqref{eq:surface_position_convected_coordinates} and application of the chain rule yields
\begin{equation} \label{eq:velocity_surface_fixed_chain_rule}
	\bm{v}
	\, = \, \dfrac{\partial \theta^\alpha}{\partial t} \Big\rvert_{\xi^\beta} \dfrac{\partial \bmx}{\partial \theta^\alpha}
	\, + \, \dfrac{\partial \bm{x}}{\partial t} \Big\rvert_{\theta^\alpha}
	\, = \, \dfrac{\partial \theta^\alpha}{\partial t} \Big\rvert_{\xi^\beta} \bm{a}_\alpha
	\, + \, \dfrac{\partial \bm{x}}{\partial t} \Big\rvert_{\theta^\alpha}
	~.
\end{equation}
At this point, the parametrization $\theta^\alpha$ is defined such that the second term on the right-hand side of Eq.~\eqref{eq:velocity_surface_fixed_chain_rule} lies entirely in the normal direction.
In this case the normal component of the velocity, $v = \bmv \cdot \bm{n}$, satisfies
$
	v \bm{n}
	= [ \partial \bmx / \partial t ] \rvert_{\theta^\alpha}
$---and
a point of constant $\theta^\alpha$
is
unaffected by in-plane flow, as shown in Fig.~\ref{fig:fig_theta_velocity}.
We accordingly refer to $\theta^\alpha$ as a surface-fixed parametrization.
The in-plane contravariant velocity components $v^\alpha$ are defined as
$
	v^\alpha
	:= [ \partial \theta^\alpha / \partial t ] \rvert_{\xi^\beta}
$,
such that
the velocity $\bm{v}$ can be written as
$
	\bmv
	= v^\alpha \bm{a}_\alpha
	+ v \bm{n}
$---indicating
our definitions of the normal velocity $v$
and contravariant velocity components $v^\alpha$
are consistent with the geometric description of the surface.

The two different parametrizations introduced thus far offer perspectives analogous to the familiar Lagrangian and Eulerian formulations from standard continuum mechanics.
A point of constant $\xi^\alpha$ is a material point, so the convected coordinates provide a Lagrangian perspective.
A point of constant $\theta^\alpha$, on the other hand, is independent of the in-plane surface flow and so the surface-fixed coordinates provide an in-plane Eulerian perspective.
The material time derivative is calculated as
$
	\mathrm{d} (\, ~ \,) / \mathrm{d} t
	:= [\partial (\, ~ \,) / \partial t ]\rvert_{\xi^\beta}
$
in the Lagrangian perspective, and as
$
	\mathrm{d} / \mathrm{d} t (\, ~ \,)
	= v^\alpha (\, ~ \,)_{, \alpha}
	+ (\, ~ \,)_{, t}
$
for scalar quantities in the in-plane Eulerian perspective,
where the partial time derivative
is defined as
$(\, ~ \,)_{, t} := [\partial (\, ~ \,) / \partial t ]\rvert_{\theta^\alpha}$.
We often denote the material time derivative of a quantity with a dot over that quantity, as in
$\bm{v} = \dot{\bm{x}}$
or
$v^\alpha = \dot{\theta}^\alpha$
.
By applying the material time derivative to the basis vectors $\bm{a}_\alpha$ and $\bm{n}$, we find
$\dot{\bm{a}}_\alpha = \bm{v}_{, \alpha}$
and
$\dot{\bm{n}} = - (\bm{a}^\alpha \otimes \bm{n}) \, \bm{v}_{, \alpha}$,
where $\otimes$ denotes the dyadic or outer product.
We calculate the material time derivatives of the metric and curvature components as
$
	\dot{a}_{\alpha \beta}
	= \bm{v}_{, \alpha} \cdot \bm{a}_\beta
	+ \bm{v}_{, \beta} \cdot \bm{a}_\alpha
$
and
$
	\dot{b}_{\alpha \beta}
	= \bm{v}_{; \alpha \beta} \cdot \bm{n}
$.
A more detailed description of the surface geometry and kinematics is provided in our past theoretical work~\cite{sahu-mandadapu-pre-2017}.

%
% *** ALE Description and Geometry
%

\subsection{ALE Description and Geometry} \label{sec:sec_ale_mesh_description}

Thus far, the convected coordinates $\xi^\alpha$ and surface-fixed coordinates $\theta^\alpha$ were introduced to describe a surface with in-plane fluidity.
Lagrangian numerical methods based on convected coordinates are conceptually simpler to develop, yet they cannot capture in-plane flows---which highly distort mesh elements, as discussed previously.
Surface-fixed coordinates are most natural for a theoretical description of arbitrarily curved and deforming surfaces with in-plane fluidity, and their numerical implementation requires nodes to only move orthogonally to the surface (see Fig.~\ref{fig:fig_theta_velocity}).
A numerical method based on the $\theta^\alpha$ parametrization can model arbitrarily large in-plane flows, yet in rare instances the discretized nodes move towards one another as the surface deforms, also shown in Fig.~\ref{fig:fig_theta_velocity}.
In the spirit of formulating a completely general numerical method to describe evolving surfaces with in-plane fluidity, we begin by developing an ALE surface parametrization, denoted by the ALE coordinate $\zeta^\alpha$, which endows the surface with a mesh.
The main idea is to define $\zeta^\alpha$ such that the corresponding mesh deforms out-of-plane with the material, while its in-plane motion can be specified arbitrarily (see Fig.~\ref{fig:fig_zeta_velocity}).
We now describe how the geometric description introduced in Sec.~\ref{sec:sec_surface_geometry} is modified when expressing quantities in terms of the ALE
parametrization $\zeta^\alpha$.

In general, the material at any ALE coordinate
$\zeta^\alpha$ will change over time, as there is in-plane flow relative to the mesh,
and the mapping from convected to ALE
coordinates is expressed as
$\zeta^\alpha = \zeta^\alpha (\xi^\beta, \, t)$.
The surface position can be described equivalently with convected coordinates, surface-fixed coordinates, or the newly introduced ALE coordinates.
To this end, any surface position $\bmx$ can be written as
\begin{equation} \label{eq:position_coordinate_equivalence_mesh}
	\bmx
	= \hat{\bm{x}} (\xi^\alpha, \, t)
	= \bm{x} (\theta^\alpha, \, t)
	= \check{\bm{x}} (\zeta^\alpha, \, t)
	~,
\end{equation}
where in the last equality and from now on a `check' accent is used to denote the position expressed in terms of the $\zeta^\alpha$ parametrization (see Fig.~\ref{fig:fig_surface_parametrization}).

Under a change of surface parametrization from $\theta^\alpha$ to $\zeta^\alpha$, the latter of which can be specified arbitrarily, our geometric description of the surface is modified---as is any quantity with a
Greek index.
However, quantities transform in such a way that any variable without an
index is invariant to the change in parametrization.
Thus, for example, $\bmv$ and $v$ are invariant quantities while $v^\alpha$ and $a_{\alpha \beta}$ are not.
As we will see, the continuity equation and vector form of the equations of motion contain no free indices, and can be expressed in terms of any surface parametrization---which shows the utility of our differential geometric developments and notation.

In this work, quantities are expressed in terms of the $\zeta^\alpha$ parametrization by placing a `check' accent over every Greek index, where checked and unchecked indices take the same value.
For example,
$\bm{a}_\calpha := \partial \check{\bmx} / \partial \zeta^\alpha = \check{\bmx}_{, \calpha}$
are the new in-plane basis vectors,
$a_{\calpha \cbeta} := \bm{a}_\calpha \cdot \bm{a}_\cbeta$
are the new metric components, and
$b_{\calpha \cbeta} := \bm{n} \cdot \check{\bmx}_{, \calpha \cbeta}$
are the new curvature components.
In this manner, the ALE
parametrization is used throughout the rest of this manuscript.
We used a similar technique in our previous Lagrangian surface description~\cite{kranthi-jcp-2017}, where we transitioned from the $\theta^\alpha$ to the $\xi^\alpha$ parametrization.
As all quantities in Ref.~\cite{kranthi-jcp-2017} were written in the $\xi^\alpha$ representation, all Greek indices should be interpreted as having a `hat' accent to be consistent with our notation.

%
% *** ALE Kinematics
%

\subsection{ALE Kinematics} \label{sec:sec_ale_mesh_kinematics}

While the material velocity $\bmv$ is an invariant quantity, it is expressed differently for different surface parametri\-zations.
For example,
$
	\bmv
	:= [ \partial \hat{\bmx} / \partial t ] \rvert_{\xi^\beta}
$
for convected coordinates and
$
	\bmv
	= v^\alpha \bm{a}_\alpha
	+ v \bm{n}
$
for surface-fixed coordinates (see Sec.~\ref{sec:sec_surface_kinematics}).
In this section, we characterize the kinematics of an arbitrarily curved and deforming surface when the surface is parametrized by the ALE coordinates $\zeta^\alpha$.
Our developments mirror those of Sec.~\ref{sec:sec_surface_kinematics}.

Using the mapping
$\zeta^\alpha = \zeta^\alpha (\xi^\beta, \, t)$,
any surface position can be written as
\begin{equation} \label{eq:ale_position}
	\check{\bm{x}} (\zeta^\alpha, \, t)
	\, = \, \check{\bm{x}} (\zeta^\alpha(\xi^\beta, \, t), \, t)
	\, = \, \hat{\bm{x}} (\xi^\beta, \, t)
	~,
\end{equation}
which is analogous to Eq.~\eqref{eq:surface_position_convected_coordinates}.
The velocity of a point,
$
	\bmv
	= [ \partial \hat{\bmx} / \partial t ] \rvert_{\xi^\beta}
$,
can be expressed in the $\zeta^\alpha$ representation as
\begin{equation} \label{eq:ale_velocity_chain_rule}
	\bm{v}
	\, = \,
	\dfrac{\partial \zeta^\alpha}{\partial t} \Big\rvert_{\xi^\beta} \, \dfrac{\partial \check{\bm{x}}}{\partial \zeta^\alpha}
	\, + \, \dfrac{\partial \check{\bm{x}}}{\partial t} \Big\rvert_{\zeta^\alpha}
	\, = \, 
	\dfrac{\partial \zeta^\alpha}{\partial t} \Big\rvert_{\xi^\beta} \, \bm{a}_\calpha
	\, + \, \dfrac{\partial \check{\bm{x}}}{\partial t} \Big\rvert_{\zeta^\alpha}
	~,
\end{equation}
where in the second equality we substituted
$\bm{a}_\calpha = \bmx_{, \calpha}$.
The last term in Eq.~\eqref{eq:ale_velocity_chain_rule} describes how the position of a mesh point changes in time, which we denote the mesh velocity $\bmvm$:
\begin{equation} \label{eq:ale_mesh_velocity}
	\bmvm
	\, := \, \dfrac{\partial \check{\bm{x}}}{\partial t} \Big\rvert_{\zeta^\alpha}
	\, = \, \bm{x}'
	~.
\end{equation}
In Eq.~\eqref{eq:ale_mesh_velocity}, the notation
$(\, ~ \,)' := [\partial (\, ~ \,) / \partial t]\rvert_{\zeta^\alpha}$
indicates how a quantity at a mesh point evolves in time.
The partial time derivative $(\, ~ \,)'$ in the ALE parametrization is analogous to the partial time derivative $(\, ~ \,)_{, t}$ in the surface-fixed parametrization, as both describe how a quantity changes at a fixed value of the appropriate coordinates.
Note the mesh velocity $\bmvm$ need not be orthogonal to the surface, in contrast to its surface-fixed analog
$
	v \bm{n}
	= [ \partial \bmx / \partial t ] \rvert_{\theta^\alpha}
$
(see Sec.~\ref{sec:sec_surface_kinematics}).
Defining
\begin{equation} \label{eq:ale_relative_velocity}
	c^\calpha
	\, := \, \dfrac{\partial \zeta^\alpha}{\partial t} \Big\rvert_{\xi^\beta}
	\, = \, \dot{\zeta}^\alpha
	\hspace{20pt}
	\mathrm{and}
	\hspace{23pt}
	\bm{c}
	\, := \, c^\calpha \bm{a}_\calpha
	~,
\end{equation}
and using Eq.~\eqref{eq:ale_mesh_velocity}, we express Eq.~\eqref{eq:ale_velocity_chain_rule} as
\begin{equation} \label{eq:ale_velocity_with_relative_mesh}
	\bmv
	\, = \, \bm{c}
	\, + \, \bmvm
	~.
\end{equation}
Equation \eqref{eq:ale_velocity_with_relative_mesh} indicates that $\bm{c}$ is the relative velocity between the material and the mesh.
Furthermore, since
$\bm{c} \cdot \bm{n} = c^\calpha \bm{a}_\calpha \cdot \bm{n}$
and
$\bm{a}_\calpha \cdot \bm{n} = 0$,
Eq.~\eqref{eq:ale_relative_velocity}$_2$ shows that the relative velocity lies entirely in the tangent plane to the surface, as shown schematically in Fig.~\ref{fig:fig_zeta_velocity}.

The material time derivative of any invariant quantity can be expressed in the $\zeta^\alpha$ representation as
\begin{equation} \label{eq:material_derivative_expansion_ale}
	\dfrac{\mathrm{d}}{\mathrm{d} t} (\, ~ \,)
	\, = \, (\, ~ \,)'
	\, + \, c^\calpha \, (\, ~ \,)_{, \calpha}
	\, = \, (\, ~ \,)'
	\, + \, \big(
		\bmv - \bmvm
	\big) \cdot \bm{a}^\calpha \, (\, ~ \,)_{, \calpha}
	~,
\end{equation}
where the relation
$c^\calpha = (\bmv - \bmvm) \cdot \bm{a}^\calpha$,
obtained from Eqs.~\eqref{eq:ale_relative_velocity}$_2$ and \eqref{eq:ale_velocity_with_relative_mesh}, is used in the second equality.
The acceleration of a point $\bmx$ is calculated using Eq.~\eqref{eq:material_derivative_expansion_ale} as
\begin{equation} \label{eq:ale_acceleration}
	\dot{\bmv}
	\, = \, \bmv'
	\, + \, c^\calpha  \, \bmv_{, \calpha}
	\, = \, \bmv'
	\, + \, \big(
		\bmv_{, \calpha} \otimes \bm{a}^\calpha
	\big) \, \big(
		\bmv - \bmvm
	\big)
	~.
\end{equation}
Finally, in our simulations, the mesh velocity $\bmvm$  is treated as a fundamental unknown.
The mesh position is calculated by integrating the mesh velocity over time, formally written as
\begin{equation} \label{eq:ale_mesh_position}
	\check{\bm{x}} (\zeta^\alpha, \, t)
	\, = \, \check{\bm{x}} (\zeta^\alpha, \, t_0)
	\, + \, \int_{t_0}^t \bmvm (\zeta^\alpha, \, t') ~\mathrm{d}t'
	~,
\end{equation}
where $\check{\bmx} (\zeta^\alpha, \, t_0)$ is the initial mesh position at time $t_0$.

%
% *** Mesh Velocity Equations
%

\subsection{Mesh Velocity Equations} \label{sec:sec_ale_mesh_velocity_equations}

With the introduction of three new unknowns, namely the three components of the mesh velocity $\bmvm$, three additional governing equations
are required for the problem to be mathematically well-posed.
One equation is found by taking the dot product of Eq.~\eqref{eq:ale_velocity_with_relative_mesh} with the normal vector $\bm{n}$ and recognizing $\bm{c} \cdot \bm{n} = 0$, yielding
\begin{equation} \label{eq:ale_mesh_velocity_normal_constraint}
	\bmvm \cdot \bm{n}
	= \bmv \cdot \bm{n}
	~,
\end{equation}
which ensures the mesh and the surface always overlap with one another as the surface deforms.
In this sense, all schemes considered are out-of-plane Lagrangian.

The remaining two equations required to close the problem come from specifying the relative velocity $\bm{c}$, or equivalently specifying the relationship between $\bmvm \cdot \bm{a}_\calpha$ and $\bmv \cdot \bm{a}_\calpha$.
There are no restrictions on how $\bmvm \cdot \bm{a}_\calpha$ and $\bmv \cdot \bm{a}_\calpha$ are related, so one can specify their relationship arbitrarily.
If we were to choose
$\bmvm \cdot \bm{a}_\calpha = \bmv \cdot \bm{a}_\calpha$,
for example, we implicitly set $\zeta^\alpha = \xi^\alpha$ and therefore recover a Lagrangian scheme in which the mesh velocity and material velocity coincide.
If, on the other hand, we choose
$\bmvm \cdot \bm{a}_\calpha = 0$,
we implicitly set $\zeta^\alpha = \theta^\alpha$ and recover an in-plane Eulerian scheme in which the mesh moves only in the direction normal to the surface.
The theoretical developments of this section allow us to specify
$\bmvm \cdot \bm{a}^\calpha$,
or equivalently $\zeta^\alpha$, arbitrarily as is best-suited to solve the problem at hand (see Fig.~\ref{fig:fig_zeta_velocity}).
This flexibility is analogous to that of a Cartesian ALE formulation \cite{donea-2004}, and for this reason we name our scheme `arbitrary Lagrangian--Eulerian.'

For the majority of this manuscript we consider an out-of-plane Lagrangian, in-plane Eulerian scheme, for which the mesh velocity satisfies
\begin{equation} \label{eq:lagrangian_eulerian_in_plane_mesh_velocity}
	\bmvm \cdot \bm{a}_\calpha
	= 0
	~,
\end{equation}
which from now on is called a Lagrangian--Eulerian (LE) scheme.
As shown in Sec.~\ref{sec:sec_lagrangian}, the LE implementation can be easily modified to produce a pure Lagrangian scheme.
The analysis of more general mesh velocity descriptions is left to a future study, as care must be taken to avoid well-known ALE issues arising in the discretized equations---such as the violation of the geometric conservation law \cite{thomas-aiaa-1979,farhat-cmame-1996}.
In the LE case, such issues do not arise, and one can condense the three constraints on the mesh velocity into a single vector equation, given by
\begin{equation} \label{eq:ale_mesh_velocity_eulerian_equation}
	\bmvm
	= (\bm{n} \otimes \bm{n}) \, \bmv
	~.
\end{equation}
Equation \eqref{eq:ale_mesh_velocity_eulerian_equation} provides the three equations necessary to resolve the LE mesh motion, and concludes our theoretical ALE surface description.

%
% *** Fluid Film Equations
%

\section{Fluid Film Equations} \label{sec:sec_theory_fluid_films}

The equations governing an arbitrarily curved and deforming fluid film can be obtained in the form presented below by starting with the lipid membrane equations of Ref.~\cite{sahu-mandadapu-pre-2017}, obtained within the framework of irreversible thermodynamics, and setting the bending moduli to zero.
Here and from now on, all equations are written in terms of the ALE parametrization by placing `check' accents over all Greek indices, as described in the previous section.

%
% *** Strong Formulation (Fluid Films)
%

\subsection{Strong Formulation} \label{sec:sec_strong_form_fluid}

The continuity equation for an arbitrarily curved, incompressible two-dimensional material is given by
\begin{equation} \label{eq:strong_incompressibility}
	\bm{a}^\calpha \cdot \bmv_{, \calpha}
	\, = \, v^\calpha_{; \calpha}
	\, - \, 2 \, v \, H
	\, = \, 0
	~.
\end{equation}
Equation~\eqref{eq:strong_incompressibility} is also called the incompressibility constraint and is enforced by the Lagrange multiplier
$\lambda = \lambda(\zeta^\alpha, t)$---which is physically the surface tension of the material.
The local form of the linear momentum balance is given by
\begin{equation} \label{eq:strong_equations_of_motion}
	\rho \dot{\bmv}
	\, = \, \rho \bm{b}
	\, + \, \bm{T}^\calpha_{; \calpha}
	~,
\end{equation}
where $\rho$ is the density, $\dot{\bmv}$ is the acceleration, $\bm{b}$ are the body forces, and $\bm{T}^\calpha$ are the stress vectors, namely, the boundary tractions across curves of constant $\zeta^\alpha$.
For fluid films, we find
$
	\bm{T}^\calpha
	= \sigma^{\calpha \cbeta} \bm{a}_\cbeta
$,
where $\sigma^{\calpha \cbeta}$ are the in-plane stress components given by (see Appendix~\ref{sec:sec_appendix_irrev_thermo} and Ref.~\cite{sahu-mandadapu-pre-2017} for more details)
\begin{equation} \label{eq:fluid_stresses}
	\sigma^{\calpha \cbeta}
	= \lambda \, a^{\calpha \cbeta}
	+ \pi^{\calpha \cbeta}
	~.
\end{equation}
In Eq.~\eqref{eq:fluid_stresses}, $\lambda$ is the surface tension enforcing areal incompressibility \eqref{eq:strong_incompressibility} and $\pi^{\calpha \cbeta}$ are the in-plane viscous stresses.
For an isotropic and incompressible fluid film, the viscous stresses are found to be
\begin{equation} \label{eq:viscous_stresses}
	\pi^{\calpha \cbeta}
	\, = \, \zeta \, \dot{a}_{\mu \nu} \, a^{\alpha \mu} \, a^{\beta \nu}
	\, = \, \zeta \, \bm{v}_{, \cgamma} \cdot \big(
		\bm{a}^\calpha \, a^{\cgamma \cbeta}
		+ \, \bm{a}^\cbeta \, a^{\cgamma \calpha}
	\big)
	~,
\end{equation}
where $\zeta$ is the two-dimensional shear viscosity with units of force$\cdot$time/length or equivalently mass/time.

To write the equations of motion in component form, we decompose the body force $\bm{b}$ in the $\{ \bm{a}^\calpha, \bm{n} \}$ basis as
$\bm{b} = b_\calpha \, \bm{a}^\calpha + p \, \bm{n}$,
where $b_\calpha$ are the in-plane covariant components and $p$ is the pressure drop across the surface.
The in-plane equations of motion are given by
\begin{equation} \label{eq:eom_in_plane}
	\rho \dot{\bmv} \cdot \bm{a}_\calpha
	= \rho b_\calpha
	+ \lambda_{, \calpha}
	+ \zeta \, \Big(
		a^{\cbeta \cgamma} \, v_{\calpha; \cbeta \cgamma}
		+ K \, v_\calpha
		+ 2 \, v_{, \calpha} \, H
		- 2 \, v_{, \cbeta} \, b^\cbeta_\calpha
		- 2 \, v \, H_{, \calpha}
	\Big)
	~,
\end{equation}
as shown in Appendix~\ref{sec:sec_appendix_component_eqns}.
The left-hand side of Eq.~\eqref{eq:eom_in_plane} contains the inertial terms, while the right-hand side consists of the body forces, surface tension gradient, and divergence of the viscous stresses, respectively.
The viscous forces clearly show the coupling between surface curvature and the in-plane and out-of-plane velocity components.

The out-of-plane equation of motion, also called the shape equation, is found to be
(see Appendix~\ref{sec:sec_appendix_component_eqns})
\begin{equation} \label{eq:shape_eqn}
	\rho \dot{\bmv} \cdot \bm{n}
	= p
	+ 2 \, \lambda \, H
	+ \zeta \, \Big(
		2 \, b^{\calpha \cbeta} \, v_{\calpha; \cbeta}
		- 4 \, v \, \big( 2 H^2 - K\big)
	\Big)
	~.
\end{equation}
Equation \eqref{eq:shape_eqn} is an extension of the Young--Laplace equation to fluid films with nonzero velocity.
Indeed, by setting $\bmv = \bm{0}$, Eq.~\eqref{eq:shape_eqn} simplifies to
the familiar Young--Laplace equation,
$p + 2 \lambda H = 0$.
The presence of the in-plane fluid viscosity $\zeta$ and in-plane velocity components $v_\calpha$ in Eq.~\eqref{eq:shape_eqn} leads to nontrivial coupling between the in-plane and out-of-plane equations when the surface is curved, i.e.\ when
$b_{\calpha \cbeta} \ne 0$.

In our ALE formulation, the surface shape is evolved with the mesh velocity, rather than the material velocity, according to Eq.~\eqref{eq:ale_mesh_position}.
As such, there are seven unknowns to solve for: three components of the material velocity $\bmv$, three components of the mesh velocity $\bmvm$, and the surface tension $\lambda$.
The corresponding equations are the three components of the equation of motion \eqref{eq:eom_in_plane} and \eqref{eq:shape_eqn}, the three components of the mesh equation \eqref{eq:ale_mesh_velocity_eulerian_equation}, and the incompressibility constraint \eqref{eq:strong_incompressibility}.
These seven equations constitute the strong formulation of the problem.

%
% *** Boundary Conditions
%

\subsection{Boundary Conditions} \label{sec:sec_strong_form_boundary_conditions}

The decomposition of the equations of motion into in-plane and out-of-plane components allows us to determine possible boundary conditions.
The first term in parenthesis on the right-hand side of Eq.~\eqref{eq:eom_in_plane},
$ a^{\cbeta \cgamma} \, v_{\calpha; \cbeta \cgamma} $,
contains two derivatives of the in-plane velocity components $v_\calpha$.
As no higher derivative of $v_\calpha$ appears, we accordingly specify either Dirichlet or Neumann boundary conditions at every point on the patch boundary.
In analogy to the boundary conditions of a bulk fluid in three dimensions, Dirichlet boundary conditions specify the in-plane velocity component $v_\calpha$, while Neumann boundary conditions specify the in-plane boundary tractions
$\bm{T} = \sigma^{\calpha \cbeta} \, \nu_\calpha \, \bm{a}_\cbeta$,
where
$\nu_\calpha = \bm{\nu} \cdot \bm{a}_\calpha$
and $\bm{\nu}$ is the in-plane unit normal to the surface at its boundary \cite{sahu-mandadapu-pre-2017}.
Accordingly, the patch boundary $\bscp$ is separated into a Dirichlet portion $\bdscp$ and a Neumann portion $\bnscp$, such that
$\bdscp \cap \bnscp = \varnothing$
and 
$\overline{\bdscp \cup \bnscp} = \bscp$.
In this manuscript, only traction-free boundary conditions are considered, for which
$\bm{T} = \bm{0}$
on $\bnscp$.
General traction boundary conditions will be considered in Part II.

We next consider the shape equation~\eqref{eq:shape_eqn}, which describes the out-of-plane behavior of the fluid film and also provides an evolution equation for the position $\bmx$ through the normal velocity $v$.
The shape equation contains two spatial derivatives of the position through the curvature components $b^{\calpha \cbeta}$.
Consequently, at each point on the boundary we specify either the normal velocity or its in-plane gradient along the $\bm{\nu}$ direction, perpendicular to the surface boundary.
These boundary conditions are independent of the in-plane boundary conditions, and in this manuscript we always specify the normal velocity of the patch boundary.

Our boundary conditions on the equations of motion are succinctly written as
\begin{equation} \label{eq:strong_boundary_conditions}
	\bmv
	= \bar{\bmv}
	\hspace{8pt}
	\textrm{on}
	\hspace{5pt}
	\bdscp
	\hspace{30pt}
	\textrm{and}
	\hspace{40pt}
	\bm{T}
	= \bm{0}
	~,
	\hspace{5pt}
	v
	= \bar{v}
	\hspace{8pt}
	\textrm{on}
	\hspace{4pt}
	\bnscp
	~,
\end{equation}
where
$\bar{\bmv}$ is a known velocity on the boundary and $\bar{v}$ is a known normal component of the velocity on the boundary.
As the equations governing the mesh velocity \eqref{eq:ale_mesh_velocity_eulerian_equation} are algebraic equations which do not contain any derivatives, we do not specify any boundary conditions for the mesh velocity $\bmvm$.

%
% *** Finite Element Formulation for Fluid Films
%

\section{Finite Element Formulation for Fluid Films} \label{sec:sec_finite_element_formulation_fluid_films}

In this section, we determine the weak formulation of the governing equations presented in Sec.~\ref{sec:sec_strong_form_fluid}, subject to the boundary conditions in Sec.~\ref{sec:sec_strong_form_boundary_conditions}.
The weak form is modified to remove numerical inf--sup instabilities arising from the incompressibility of the fluid film, with a method inspired by Dohrmann and Bochev~\cite{dohrmann-bochev-ijnmf-2004}.
We provide the function spaces in which the solution to the weak formulation resides, and then use the standard tools of finite element analysis to find an approximate numerical solution to the weak form of the governing equations.

While inertial terms are included in the strong and weak formulations for completeness, they are not included in our simulations of arbitrarily curved and deforming fluid films---as they are negligible in many physical problems of interest.
Despite the absence of inertia, the equations of motion are nonlinear due to the many terms involving the surface geometry.
Furthermore, time derivatives still remain in the problem because the rate of change of the surface position is contained in the mesh velocity.
The only simulation including inertial terms is the lid-driven cavity benchmark problem, in which the mesh is constrained to be fixed and inertia is included purely to demonstrate the validity of our numerical method.

%
% *** Weak Formulation
%

\subsection{Weak Formulation} \label{sec:sec_fluid_weak}

Here we derive the weak formulation of the strong form equations provided in Sec.~\ref{sec:sec_strong_form_fluid}.
Let $\mcv$ be the space of functions for the material velocity $\bmv$ and mesh velocity $\bmvm$, and let $\Lambda$ be the space of functions for the surface tension $\lambda$.
We consider the arbitrary variations
$\delta \bmv \in \mcv_0$,
$\delta \bmvm \in \mcv$,
and
$\delta \lambda \in \Lambda$,
where all elements of $\mcv_0 \subset \mcv$ vanish on the Dirichlet portion of the boundary.
The variations are contracted with the appropriate strong form equations and integrated over the fluid surface to yield the weak formulation of the problem.

In our previous work \cite{sahu-mandadapu-pre-2017},
we presented surface integrals as being of the general form
$\int_\scp ( \ldots ) ~\mathrm{d}a$,
where $\mathrm{d}a$ is a differential areal element of the patch $\scp$.
While such a description is theoretically sound, it is not amenable to numerical integration.
We define
$\Omega := \{ (\zeta^1, \, \zeta^2) \}$
to be the space of all ALE coordinates $\zeta^\alpha$, shown in blue in Fig.~\ref{fig:fig_surface_parametrization}, and map areal integrals to $\Omega$ according to
\begin{equation} \label{eq:map_integral}
	\int_\scp \big( \ldots \big) ~\mathrm{d}a
	\, = \, \int_\Omega \big( \ldots \big) ~\Jm ~\mathrm{d}\Omega
	~.
\end{equation}
In Eq.~\eqref{eq:map_integral},
$\mathrm{d}\Omega := \mathrm{d}\zeta^1 \, \mathrm{d}\zeta^2$
is a differential element of $\Omega$ and
$\Jm := \sqrt{ \rule{0mm}{1.30ex} } \overline{ \det \, a_{\calpha \cbeta} } $
is the Jacobian determinant of the mapping
$\check{\bmx}(\zeta^\alpha) : \Omega \rightarrow \scp$.
In a similar way, integrals over the patch boundary $\bscp$ are mapped to integrals over the parametric domain boundary $\Gamma := \partial \Omega$ according to
\begin{equation} \label{eq:map_integral_boundary}
	\int_\bscp \big( \ldots \big) ~\mathrm{d}s
	\, = \, \int_\Gamma \big( \ldots \big) ~\JGammam ~\mathrm{d}\Gamma
	~,
\end{equation}
where $\mathrm{d}s$ is a differential line element of the patch boundary $\bscp$, $\mathrm{d}\Gamma$ is a differential line element on the parametric domain boundary, and $\JGammam$ is the Jacobian determinant of the mapping
$\check{\bmx}_{\mathrm{b}}(\zeta^\alpha) : \Gamma \rightarrow \bscp$---in which $\check{\bmx}_{\mathrm{b}}$ refers to the position of the patch boundary.
The Dirichlet and Neumann portions of $\Gamma$, denoted $\Gammad$ and $\Gamman$, map to the patch boundaries $\bdscp$ and $\bnscp$, respectively.

To obtain the weak formulation, we begin by contracting the equations of motion \eqref{eq:strong_equations_of_motion} with an arbitrary velocity variation $\delta \bmv \in \mcv_0$ and integrating over the patch $\scp$ to obtain
\begin{equation} \label{eq:fluid_contracted_eqn_motion}
	\int_\Omega
		\delta \bmv \cdot \rho \dot{\bmv}
	~\Jm
	~\mathrm{d} \Omega
	\, -\int_\Omega
		\delta \bm{v} \cdot \rho \bm{b}
	~\Jm
	~\mathrm{d} \Omega
	\, - \int_\Omega
		\delta \bm{v} \cdot \bm{T}^\calpha_{; \calpha}
	~\Jm
	~\mathrm{d} \Omega
	\, = \, 0
	\hspace{15pt}
	\forall \,\, \delta \bm{v} \in \mcv_0
	~,
\end{equation}
where all integrals are mapped to $\Omega$ through Eq.~\eqref{eq:map_integral}.
Note that in our previous Lagrangian formulation \cite{omar-biorxiv-2019, kranthi-jcp-2017}, we solved for the material position $\bmx$ and the weak form contained an arbitrary position variation $\delta \bmx$.
In this case, however, the in-plane fluidity necessitates the velocity $\bmv$ to be the fundamental unknown, such that the weak form is calculated with an arbitrary velocity variation $\delta \bmv$.
Applying the surface divergence theorem to the third term on the left-hand side in Eq. \eqref{eq:fluid_contracted_eqn_motion} yields
\begin{equation} \label{eq:fluid_contracted_manipulation}
	-\int_\Omega
		\delta \bmv \cdot \bm{T}^\calpha_{; \calpha}
	~\Jm
	~\mathrm{d} \Omega
	\, = \int_\Omega 
		\delta \bmv_{, \calpha} \cdot \bm{T}^\calpha
		~\Jm
	~\mathrm{d} \Omega
	\, -\int_{\Gamma}
		\delta \bmv \cdot \bm{T}
		~\JGammam
	~\mathrm{d} \Gamma
	~,
\end{equation}
and the boundary integral is simplified by recognizing the velocity variation $\delta \bmv = \bm{0}$ on $\Gammad$ by construction and $\bm{T} = \bm{0}$ on $\Gamman$ \eqref{eq:strong_boundary_conditions}$_2$, such that
\begin{equation} \label{eq:fluid_contracted_manipulation_final}
	-\int_\Omega
		\delta \bmv \cdot \bm{T}^\calpha_{; \calpha}
	~\Jm
	~\mathrm{d} \Omega
	\,
	= \int_\Omega 
		\delta \bmv_{, \calpha} \cdot \bm{T}^\calpha
		~\Jm
	~\mathrm{d} \Omega
	~.
\end{equation}
Substituting Eq.~\eqref{eq:fluid_contracted_manipulation_final} into Eq. \eqref{eq:fluid_contracted_eqn_motion} and recognizing
$\bm{T}^\calpha = \sigma^{\calpha \cbeta} \bm{a}_\cbeta \,$,
with $\sigma^{\calpha \cbeta}$ given by Eq.~\eqref{eq:fluid_stresses},
we obtain
\begin{equation} \label{eq:fluid_weak_eqn_motion}
	\begin{split}
		\mcg_{\mathrm{v}}
		\, := \,
		&\int_\Omega
			\delta \bmv \cdot \rho \dot{\bmv}
		~\Jm
		~\mathrm{d} \Omega
		\,
		+ \int_\Omega 
			\delta \bmv_{, \calpha} \cdot \pi^{\calpha \cbeta} \, \bm{a}_\cbeta
			~\Jm
		~\mathrm{d} \Omega
		\\
		&\hspace{24pt}
		+ \int_\Omega
			\delta \bmv_{, \calpha} \cdot \bm{a}^\calpha \, \lambda
			~\Jm
		~\mathrm{d} \Omega
		\,
		-\int_\Omega
			\delta \bm{v} \cdot \rho \bm{b}
		~\Jm
		~\mathrm{d} \Omega
		\,
		= 0 \,
		\hspace{15pt}
		\forall \,\, \delta \bm{v} \in \mcv_0
		~.
	\end{split}
\end{equation}

The weak form of the mesh equation is found by contracting Eq.~\eqref{eq:ale_mesh_velocity_eulerian_equation} with an arbitrary mesh velocity variation $\delta \bmvm$, multiplying by a constant $\alpham$, and integrating over the surface, which yields
\begin{equation} \label{eq:mesh_velocity_weak_form}
	\mcg_{\mathrm{m}}
	\, := \,
	\alpham \!
	\int_\Omega
		\delta \bmvm \cdot \Big[
			\bmvm
			- \big(
				\bm{n} \otimes \bm{n}
			\big) \bmv
		\Big]
		~\Jm
	~\mathrm{d} \Omega
	\,
	= 0
	\hspace{15pt}
	\forall \,\, \delta \bmvm \in \mcv
	~.
\end{equation}
The factor $\alpham$ is introduced such that $\mcg_{\mathrm{m}}$ has units of power, in agreement with $\mcg_{\mathrm{v}}$ \eqref{eq:fluid_weak_eqn_motion}.
In our simulations, $\alpham$ is set to unity.
While Eq.~\eqref{eq:mesh_velocity_weak_form} corresponds to the LE mesh equation~\eqref{eq:ale_mesh_velocity_eulerian_equation}, a different mesh equation can be used in our ALE formulation by setting $\mcg_{\mathrm{m}}$ to be the corresponding weak form expression.

Finally, as $\lambda$ is the Lagrange multiplier corresponding to the incompressibility constraint \eqref{eq:strong_incompressibility}, we multiply Eq.~\eqref{eq:strong_incompressibility} by an arbitrary variation $\delta \lambda \in \Lambda$ and integrate over the patch to obtain
\begin{equation} \label{eq:weak_incompressibility}
	\mcg_{\lambda}
	\, := \,
	\int_\Omega
		\delta \lambda \, \big(
			\bm{a}^\calpha \cdot \bm{v}_{, \calpha}
		\big)
		~\Jm
	~\mathrm{d}\Omega
	\,
	= 0
	\hspace{15pt}
	\forall \,\, \delta \lambda \in \Lambda
	~.
\end{equation}

Eqs.~\eqref{eq:fluid_weak_eqn_motion}--\eqref{eq:weak_incompressibility} are the weak forms corresponding to the strong forms given respectively by Eqs.~\eqref{eq:strong_equations_of_motion}, \eqref{eq:ale_mesh_velocity_eulerian_equation}, and \eqref{eq:strong_incompressibility}.
By summing them together and introducing a shorthand for the vector of unknowns, $\bm{u}$, as
$
	\bm{u}^{\mathrm{T}}
	:= (\bmv^{\mathrm{T}}, \, (\bmvm)^{\mathrm{T}}, \, \lambda)^{\mathrm{T}}
$,
its variation $\delta \bm{u}$ as
$
	\delta \bm{u}^{\mathrm{T}}
	:= (\delta \bmv^{\mathrm{T}}, \, (\delta \bmvm)^{\mathrm{T}}, \, \delta \lambda)^{\mathrm{T}}
$,
and the space of arbitrary variations $\mcu_0$ as
$\mcu_0 := \mcv_0 \times \mcv \times \Lambda$,
we obtain the overall weak formulation, given by
\begin{align*}
	&\int_\Omega
		\delta \bmv \cdot \rho \dot{\bmv}
	~\Jm
	~\mathrm{d} \Omega
	\,\,
	+ \int_\Omega 
		\delta \bmv_{, \calpha} \cdot \pi^{\calpha \cbeta} \, \bm{a}_\cbeta
		~\Jm
	~\mathrm{d} \Omega
	\,\,
	+ \int_\Omega
		\delta \bmv_{, \calpha} \cdot \bm{a}^\calpha \, \lambda
		~\Jm
	~\mathrm{d} \Omega
	\,\,
	- \int_\Omega
		\delta \bm{v} \cdot \rho \bm{b}
	~\Jm
	~\mathrm{d} \Omega
	\\[3pt]
	&\hspace{20pt}
	+ \, \alpham \! \int_\Omega
		\delta \bmvm \cdot \Big[
			\bmvm
			- \big(
				\bm{n} \otimes \bm{n}
			\big) \bmv
		\Big]
		~\Jm
	~\mathrm{d} \Omega
	\,\,
	+ \int_\Omega
		\delta \lambda \, \big(
			\bm{a}^\calpha \cdot \bm{v}_{, \calpha}
		\big)
		~\Jm
	~\mathrm{d}\Omega
	\\[8pt]
	&\hspace{40pt}
	= \,\, 0
	\hspace{10pt}
	\forall \,\, \delta \bmu \in \mcu_0
	~.
	\stepcounter{equation}
	\tag{\theequation}\label{eq:fluid_weak_form}
\end{align*}
Note the weak form is nonlinear due to the out-of-plane deformations of the fluid film, as well as the inertial terms.
Introducing $\mcg$ as the direct Galerkin expression~\cite{zienkiewicz-taylor-solids} corresponding to the left-hand side of Eq.~\eqref{eq:fluid_weak_form}, the weak form can be compactly written as
\begin{equation} \label{eq:fluid_weak_form_compact}
	\mcg \big(
		\bmu(\zeta^\alpha, \, t), \, \delta \bmu(\zeta^\alpha)
	\big)
	\, := \,
	\mcg_{\mathrm{v}} \,
	+ \, \mcg_{\mathrm{m}} \,
	+ \, \mcg_\lambda
	\, = \, 0
	\hspace{15pt}
	\forall \,\, \delta \bm{u} \in \mcu_0
	~.
\end{equation}

%
% *** Solution Spaces
%

\subsection{Solution Spaces} \label{sec:sec_strong_solution_spaces}

With the weak formulation~\eqref{eq:fluid_weak_form}, in what follows, we define the infinite-dimensional spaces in which the surface tension $\lambda$, velocity $\bmv$, and mesh velocity $\bmvm$ reside.

%
% *** Surface Tension Solution Space
%

\paragraph{Surface Tension Solution Space} \label{sec:sec_strong_lambda_space}

The surface tension $\lambda$ enters the weak form~\eqref{eq:fluid_weak_form} without any gradients, and so we require $\lambda$ only to be square-integrable.
We define the space of all possible surface tension fields, $\Lambda$, as the space of square-integrable functions on the parametric domain $\Omega$, denoted $L^2(\Omega)$.
Thus, $\Lambda$ is given by
\begin{equation} \label{eq:lambda_solution_space}
	\textstyle
	\Lambda
	= L^2 (\Omega)
	:= \Big\{~
		u (\zeta^\alpha) \! : \Omega \rightarrow \mathbb{R}
		\hspace{8pt}
		\text{such that}
		\hspace{4pt}
		\Big(
			\int_\Omega u^2 ~\mathrm{d}\Omega
		\Big)^{\!^1\!\!/_{\!2}}
		< \infty
	~\Big\}
	~.
\end{equation}

%
% *** Velocity Solution Space
%

\paragraph{Velocity Solution Space} \label{sec:sec_strong_velocity_space}

The weak formulation~\eqref{eq:fluid_weak_form} contains a gradient of both the velocity variation $\delta \bmv$ and the material velocity $\bmv$, the latter of which is contained in the viscous stresses $\pi^{\calpha \cbeta}$ \eqref{eq:viscous_stresses}.
The velocity and velocity variation are both elements of the space of functions $\mcv$, and so elements of $\mcv$ are required to be square-integrable and have square-integrable gradients in order for the weak formulation to remain bounded.
Furthermore, as the lipid membrane weak form requires the second derivatives of elements of $\mcv$ to be square-integrable, we define the space of velocities $\mcv$ as
\begin{equation} \label{eq:velocity_solution_space}
	\mcv
	:= \big( H^2(\Omega) \big)^3
	~.
\end{equation}
Each of the three Cartesian components of the velocity lies in $H^2(\Omega)$: the Sobolev space of order two on $\Omega$, defined as
\begin{equation} \label{eq:solution_space_sobolev}
	H^2(\Omega)
	:= \Big\{~
		u (\zeta^\alpha) \! : \Omega \rightarrow \mathbb{R}
		\hspace{8pt}
		\text{such that}
		\hspace{8pt}
		u \in L^2 (\Omega)
		~,
		\hspace{8pt}
		u_{, \calpha} \in L^2(\Omega)
		~,
		\hspace{8pt}
		u_{, \calpha \cbeta} \in L^2(\Omega)
	~\Big\}
	~.
\end{equation}
We also define $\mcv_0$ as the space of functions in $\mcv$ which vanish on $\Gammad$, written as
\begin{equation} \label{eq:delta_velocity_space}
	\mcv_0
	:= \Big\{~
		\bm{u} (\zeta^\alpha) \! : \Omega \rightarrow \mathbb{R}^3
		\hspace{8pt}
		\text{such that}
		\hspace{8pt}
		\bm{u} \in \mcv
		~,
		\hspace{8pt}
		\bm{u} \big\rvert_{\Gammad} = 0
	~\Big\}
	~,
	%~.
\end{equation}
such that $\delta \bmv \in \mcv_0$.

%
% *** Mesh Velocity Solution Space
%

\paragraph{Mesh Velocity Solution Space} \label{sec:sec_strong_mesh_velocity_space}

The weak formulation~\eqref{eq:fluid_weak_form} contains terms involving the gradient of both the velocity variation $\delta \bmv$ and the mesh velocity $\bmvm$. While the mesh velocity gradient is not easily recognized in Eq.~\eqref{eq:fluid_weak_form}, it is found once the weak form is linearized and discretized (see Eq.~\eqref{eq:appendix_mKvme}). As a result, we require $\bmvm \in \mcv$ in order for the weak form to remain bounded.

%
% *** Finite-Dimensional Subspaces
%

\subsection{Finite-Dimensional Subspaces} \label{sec:sec_fluid_subspaces}

We now choose the finite-dimensional subspaces in which we seek $\bmvh$, $\bmvmh$, and $\lambdah$, which are approximations of the true solutions $\bmv$, $\bmvm$, and $\lambda$, respectively.
The approximate surface position $\bmxh$ is chosen to lie in the same subspace as $\bmvh$, as is standard in isoparametric finite element methods \cite{zienkiewicz-taylor-fem}.
To this end, we discretize $\Omega$ into $\nume$ (\underline{n}umber of \underline{e}lements) non-overlapping elements
$\{ \Omega^1, \, \Omega^2, \, \ldots, \, \Omega^{\nume} \}$,
such that
$\overline{\Omega} = \overline{\cup_{e = 1}^{\,\, \nume} \, \Omegae}$
and
$\Omega^j \cap \Omega^k = \varnothing$
for $j \ne k$.
In all cases considered
the parametric domain is discretized with a rectangular grid, as required by our choice of basis functions (see Sec.~\ref{sec:sec_numerical_solution_method}), such that all elements have the same dimensions---which is denoted $h$.
The partitioning of the parametric domain naturally leads to finite-dimensional subspaces in which functions are polynomials over single elements and have certain continuity requirements across element boundaries.

The finite-dimensional subspace of velocities, $\mcvh \subset \mcv$, is defined as
\begin{equation} \label{eq:fluid_velocity_subspace}
	\mcvh
	:= \Big\{~
		\bm{u} (\zeta^\alpha) \! : \Omega \rightarrow \mathbb{R}^3
		\hspace{8pt}
		\text{such that}
		\hspace{4pt}
		\bm{u} \in \big( C^1(\Omega) \big)^3
		\cap \mcv
		~,
		\hspace{10pt}
		\bm{u} \big\rvert_{\Omegae} \in \big( \mathbb{Q}_2 (\Omegae) \big)^3
		\hspace{10pt}
		\forall
		\hspace{4pt}
		\Omegae
	~\Big\}
	~,
\end{equation}
where $C^m(\Omega)$ denotes the space of scalar functions on $\Omega$ with at least $m$ continuous derivatives, and $\mathbb{Q}_n (\Omegae)$ is the space of bi-polynomial functions of order $n$ on the parametric element $\Omegae$.
Accordingly, $\mcvh$ is the space of piecewise bi-quadratic functions with continuous first derivatives over the entire domain $\Omega$.
While in the present formulation for fluid films, first derivatives need not be continuous, they are required to be continuous when modeling more complex systems which resist bending, such as lipid membranes \cite{kranthi-jcp-2017} and viscoelastic Kirchhoff--Love shells \cite{zimmerman-arxiv-2017}.
We define the subspace $\mcvho \subset \mcv_0$ to be the space of functions in $\mcvh$ which are also zero on $\Gammad$, formally written as
\begin{equation} \label{eq:fluid_velocity_zero_subspace}
	\mcvho
	:=
	\mcvh \cap \, \mcv_0
	~.
\end{equation}

In the finite element analysis of bulk fluids, it is well-known that choosing the Lagrange multiplier space to be of the same polynomial order as the velocity leads to an unstable matrix equation, an issue resulting from the inf--sup condition, also called the Ladyzhenskaya--Babu\v{s}ka--Brezzi (LBB) condition \cite{ladyzhenskaya-1969, babuska-1973, brezzi-1974}.
We refer the reader to Ref.~\cite{brezzi-cmame-1990} for an excellent analysis of this numerical instability, which is often avoided in practice by choosing the Lagrange multiplier basis functions to be one polynomial order lower than the velocity basis functions.
As we chose our velocities to be piecewise bi-quadratic functions on $\Omega$ \eqref{eq:fluid_velocity_subspace}, the Lagrange multiplier subspace is accordingly chosen to be continuous, piecewise bi-linear functions on $\Omega$, written as
\begin{equation} \label{eq:fluid_lambda_subspace}
	\Lambdah
	:= \Big\{~
		u (\zeta^\alpha) \! : \Omega \rightarrow \mathbb{R}
		\hspace{8pt}
		\text{such that}
		\hspace{4pt}
		u \in C^0(\Omega)
		\cap \Lambda
		~,
		\hspace{10pt}
		u \big\rvert_{\Omegae} \in \mathbb{Q}_1 (\Omegae)
		\hspace{10pt}
		\forall
		\hspace{4pt}
		\Omegae
	~\Big\}
	~.
\end{equation}
Even with this choice of basis functions, our scheme is LBB-unstable.
As a result, we invoke a projection method devised by Dohrmann and Bochev~\cite{dohrmann-bochev-ijnmf-2004}, as described below, to further stabilize our numerical method.

%
% *** Inf--Sup Stabilization
%

\subsection{Inf--Sup Stabilization} \label{sec:sec_inf_sup_stabilization}

In modeling fluids with finite element methods, it is well-known that LBB errors arise when the velocity and surface tension solution spaces are identical.
We initially used bi-quadratic velocities and bi-linear surface tensions \eqref{eq:lambda_solution_space}--\eqref{eq:solution_space_sobolev}, which were successfully used in our previous work~\cite{omar-biorxiv-2019, kranthi-jcp-2017}.
However, in the present study our numerical scheme exhibited LBB instabilities (see Appendix \ref{sec:sec_flat_dohrmann_bochev_example}).
Further inspection showed our past work may have unknowingly avoided such instabilities by prescribing the surface tension along the entire boundary.
In the spirit of developing a completely general finite element formulation, we seek to modify the numerical method presented thus far to remove the LBB instability.

We note there are many techniques to overcome the LBB instability, for example
(i) using lower-order shape functions for the Lagrange multipliers \cite{larson-bengzon-2013},
(ii) reduced and selective integration of the Lagrange multiplier equations \cite{malkus-cmame-1978, zienkiewicz-cs-1984},
(iii) stabilization methods \cite{brezzi-1984}, and
(iv) macroelement approaches \cite{bressan-jna-2013, linder-ijnme-2018}.
All of these methods are valid in different cases.
In this section, we describe how our weak form is modified by
a technique developed by Dohrmann and Bochev \cite{dohrmann-bochev-ijnmf-2004} to remove LBB instabilities given general polynomial spaces for the velocity and surface tension; the analysis of other methods is left to a future study.
The main idea underlying the Dohrmann--Bochev method is to locally project the surface tension onto a space of discontinuous, piecewise linear functions, and to energetically penalize the difference between the projected and unprojected surface tensions.
The Dohrmann--Bochev method is thus based on two equations: one which projects the surface tension, and another which penalizes surface tension deviations from the projected space in a manner suitable for finite element analysis.
A thorough description of the Dohrmann--Bochev method is provided in Ref.~\cite{zienkiewicz-taylor-fem}, and we follow their notation in this manuscript.
The details of our numerical implementation can be found in Appendix~\ref{sec:sec_bochev_finite_element_implementation}.

%
% *** Theory
%

\subsubsection{Theory} \label{sec:sec_bochev_theory}

We begin by specifying the space of piecewise linear, discontinuous basis functions $\bLambda$ onto which the surface tension is projected, given by
\begin{equation} \label{eq:D_B_breve_lambda_subspace}
	\bLambda
	:= \Big\{~
		u (\zeta^\alpha) \! : \Omega \rightarrow \mathbb{R}
		\hspace{8pt}
		\text{such that}
		\hspace{4pt}
		u \big\rvert_{\Omegae} \in \mathbb{P}_1 (\Omegae)
		\hspace{10pt}
		\forall
		\hspace{4pt}
		\Omegae
	~\Big\}
	~,
\end{equation}
where $\mathbb{P}_n (\Omegae)$ is the space of polynomial functions of order $n$ on the parametric element $\Omegae$.
Note that while piecewise bi-linear functions can be continuous on quadrilateral elements \eqref{eq:fluid_lambda_subspace}, piecewise linear functions cannot be.
Accordingly, the space $\bLambda$ is discontinuous and over a single element $\Omegae$:
$\bLambda \big\rvert_\Omegae = \mathbb{P}_1 (\Omegae)$.

We next introduce the projection of the surface tension and its arbitrary variation, denoted $\blambda$ and $\delta \blambda$, respectively, such that $\blambda, \, \delta \blambda \in \bLambda$.
The $L^2$-projection of the surface tension, $\blambda$, is defined by
\begin{equation} \label{eq:D_B_projection_global}
	\int_{\Omega}
		\delta \blambda \, \big(
			\lambda
			- \blambda
		\big)
	~\mathrm{d}\Omega
	= 0
	\hspace{15pt}
	\forall \,\, \delta \blambda \in \bLambda
	~.
\end{equation}
As $\delta \blambda$ belongs to a space of linear functions which are discontinuous across elements, Eq.~\eqref{eq:D_B_projection_global} can be considered separately for individual elements $\Omegae$, and is equivalently expressed as
\begin{equation} \label{eq:D_B_projection}
	\int_{\Omegae}
		\delta \blambda \, \big(
			\lambda
			- \blambda
		\big)
	~\mathrm{d}\Omega
	= 0
	\hspace{15pt}
	\forall \,\, \delta \blambda \in \mathbb{P}_1 (\Omegae), \,\, \forall \,\, \Omegae
	~.
\end{equation}
Deviations between $\blambda$ and $\lambda$ are penalized in the weak form by subtracting the term
\begin{equation} \label{eq:D_B_weak}
	\mcgDB
	\, := \,
	\dfrac{\alphaDB}{\zeta}
	\int_{\Omega}
		\big(
			\delta \lambda
			- \delta \blambda
		\big) \, \big(
			\lambda
			- \blambda
		\big)
	~\mathrm{d}\Omega
\end{equation}
from the left-hand side of Eq.~\eqref{eq:fluid_weak_form}.
In Eq.~\eqref{eq:D_B_weak}, $\zeta$ is the two-dimensional fluid shear viscosity and $\alphaDB$ is a computational parameter having units of $\Jm$ which, as in Ref.~\cite{dohrmann-bochev-ijnmf-2004}, is chosen to be unity.
The units of $\alphaDB$ ensure that Eq.~\eqref{eq:D_B_weak} is dimensionally consistent with the other terms in the weak form~\eqref{eq:fluid_weak_form}.
The weak formulation is now given by
\begin{align*}
	&\int_\Omega
		\delta \bmv \cdot \rho \dot{\bmv}
	~\Jm
	~\mathrm{d} \Omega
	\,\,
	+ \int_\Omega 
		\delta \bmv_{, \calpha} \cdot \pi^{\calpha \cbeta} \, \bm{a}_\cbeta
		~\Jm
	~\mathrm{d} \Omega
	\,\,
	+ \int_\Omega
		\delta \bmv_{, \calpha} \cdot \bm{a}^\calpha \, \lambda
		~\Jm
	~\mathrm{d} \Omega
	\,\,
	- \int_\Omega
		\delta \bm{v} \cdot \rho \bm{b}
	~\Jm
	~\mathrm{d} \Omega
	\\[3pt]
	&\hspace{20pt}
	+ \, \alpham \! \int_\Omega
		\delta \bmvm \cdot \Big[
			\bmvm
			- \big(
				\bm{n} \otimes \bm{n}
			\big) \bmv
		\Big]
		~\Jm
	~\mathrm{d} \Omega
	\,\,
	+ \int_\Omega
		\delta \lambda \, \big(
			\bm{a}^\calpha \cdot \bm{v}_{, \calpha}
		\big)
		~\Jm
	~\mathrm{d}\Omega
	\, - \, \dfrac{\alphaDB}{\zeta}
	\int_{\Omega}
		\big(
			\delta \lambda
			- \delta \blambda
		\big) \, \big(
			\lambda
			- \blambda
		\big)
	~\mathrm{d}\Omega
	\\[8pt]
	&\hspace{40pt}
	\,\,
	= \,\, 0
	\hspace{10pt}
	\forall \,\, \delta \bmu \in \mcu_0
	~,
	\stepcounter{equation}
	\tag{\theequation}\label{eq:fluid_weak_form_with_DB}
\end{align*}
and the direct Galerkin expression $\mcg$ in Eq.~\eqref{eq:fluid_weak_form_compact} is redefined such that
\begin{equation} \label{eq:direct_galerkin_expression_including_DB}
	\mcg \big(
		\bmu(\zeta^\alpha, \, t), \, \delta \bmu(\zeta^\alpha)
	\big)
	\, := \,\,~
	\mcg_{\mathrm{v}}
	\, + \, \mcg_{\mathrm{m}}
	\, + \, \mcg_\lambda
	\, - \, \mcgDB
	\, = \, 0
	\hspace{15pt}
	\forall \,\, \delta \bm{u} \in \mcu_0
	~.
\end{equation}
We provide the details of our Dohrmann--Bochev implementation in Appendix~\ref{sec:sec_bochev_finite_element_implementation}, within which Appendix \ref{sec:sec_flat_dohrmann_bochev_example} demonstrates the success of this method.

%
% *** Summary of Numerical Solution Method
%

\subsection{Summary of Numerical Solution Method} \label{sec:sec_numerical_solution_method}

At this point, we seek an approximate solution to the weak formulation, as presented in Eqs.~\eqref{eq:fluid_weak_form_with_DB} and \eqref{eq:direct_galerkin_expression_including_DB}.
To this end, we discretize the fundamental unknowns and their arbitrary variations.
We then obtain the residual equations, discretize them temporally, and solve the resulting equations via Newton--Raphson iteration.
Our techniques are briefly summarized here, however, extensive details of our numerical implementation are provided in Appendix~\ref{sec:sec_appendix_numerical_solution_method}.

We first introduce the basis functions $\{ N_I (\zeta^\alpha) \}$ and Lagrange multiplier basis functions $\{ \bar{N}_J (\zeta^\alpha) \}$ such that
$\mcvh = (\textrm{span} \, \{ N_I (\zeta^\alpha) \} )^3$,
and
$\Lambdah = \textrm{span} \, \{ \bar{N}_J (\zeta^\alpha) \}$.
The fundamental unknowns are discretized as
\begin{equation} \label{eq:unknown_basis_expansions}
	\bmvh (\zeta^\alpha, t)
	\, = \, \mN \, \mvt
	~,
	\qquad
	\bmvmh (\zeta^\alpha, t)
	\, = \, \mN \, \mvmt
	~,
	\qquad
	\text{and}
	\qquad
	\lambdah (\zeta^\alpha, t)
	= \mhN \, \mlambdat
	~,
\end{equation}
where $\mN$ and $\mhN$ are shape function matrices and $\mvt$, $\mvmt$, and $\mlambdat$ are the velocity, mesh velocity, and surface tension degree of freedom vectors, respectively.
We introduce the shorthand
$\mbut = (\mvt^\mT, \mvmt^\mT, \mlambdat^\mT)^\mT$
as the collection of all degrees of freedom in the discretized system.
The arbitrary variations $\delta \bmv$, $\delta \bmvm$, and $\delta \lambda$ are discretized with the same basis functions as
\begin{equation} \label{eq:unknown_variation_basis_expansions}
	\delta \bmv (\zeta^\alpha)
	\, = \, \mN \, \mdeltav
	~,
	\qquad
	\delta \bmvm (\zeta^\alpha)
	\, = \, \mN \, \mdeltavm
	~,
	\qquad
	\text{and}
	\qquad
	\delta \lambda (\zeta^\alpha)
	= \mhN \, \mdeltalambda
	~,
\end{equation}
according to a Bubnov--Galerkin approximation, with $\mdeltav$, $\mdeltavm$, and $\mdeltalambda$ collectively gathered into $\mdeltau$.
The weak formulation~\eqref{eq:direct_galerkin_expression_including_DB} can then be written as
$
	\mdeltau^{\mathrm{T}} \mrt
	\, = \, 0
$
for all $\mdeltau$,
which is satisfied when the residual vector
\begin{equation} \label{eq:general_residual_equation}
	\mrt
	\, := \, \dfrac{\partial \mcg}{\partial \mdeltau}
	\, = \, \mzero
	~.
\end{equation}
To solve Eq.~\eqref{eq:general_residual_equation} at a set of $N$ discrete times $\{ t_1, \, t_2, \ldots , t_{N} \}$, we assume $\mbutn$ is known and solve for
$
	\mbutno
	:= \mbutn
	+ \mDeltautno
$
according to the Newton--Raphson method.
Again, a detailed description of our numerical procedure is provided in Appendix~\ref{sec:sec_appendix_numerical_solution_method}.

In our implementation, the spaces $\mcvh$~\eqref{eq:fluid_velocity_subspace} and $\mcvho$ \eqref{eq:fluid_velocity_zero_subspace} involve $C^1$-continuous basis functions.
We maintain basis function continuity across elements by using uniform B-spline basis functions, which have the advantage of naturally enforcing arbitrary continuity requirements yet the complication of non-interpolatory basis function coefficients, and the requirement of a rectangular parametric mesh, as well as basis functions spreading over multiple elements.
The method of using B-spline basis functions within an isoparametric finite element framework is in the spirit of Ref.~\cite{cirak-ortiz-cad-2002}, and detailed in Ref.~\cite{cottrell-iga}.
We use the algorithms described in Ref.~\cite{piegl-tiller} to efficiently calculate the basis functions and their derivatives.

We have now concluded our discussion of the LE finite element formulation, and a high-level overview of our code structure can be found in Algorithm~\ref{algo} of Appendix~\ref{sec:sec_appendix_code_structure}.

%
% *** Numerical Simulations
%

\section{Numerical Simulations} \label{sec:sec_numerical_simulations}

We now present several results from our LE finite element formulation to validate the robustness of the method and demonstrate its capabilities.
The numerical implementation of the method is tested with problems of increasing complexity, and results are compared to known analytical solutions whenever possible.
The first test cases involve fluid flows on flat planes, and once several cases are validated we move on to study fluid flows on fixed, curved surfaces.
In the scenarios mentioned thus far, the mesh is constrained to remain stationary and the mesh velocity is not solved for.
In our last example, the entire LE implementation is tested by modeling an initially cylindrical fluid film, which is allowed to deform over time.
The simulations show that fluid films are unstable with respect to long wavelength perturbations, which may explain why bubbles are often observed and long, cylindrical fluid films are not.
Namely, in any experimental system, we expect the latter to break up and form bubbles.
A linear stability analysis is performed to calculate the critical length above which the cylinder becomes unstable, as well as the time scale of the instability.
We show that the analytical predictions for the critical length and the time scale of the fastest growing unstable mode agree quantitatively with our simulation results.
Moreover, we find both theoretically and computationally that cylindrical fluid films are stable to non-axisymmetric perturbations.

%
% *** Fluid Flow on a Flat Plane
%

\subsection{Fluid Flow on a Flat Plane} \label{sec:sec_flat_numerics}

We first consider the simplest test case, fluid flowing on a flat plane, for which the surface tension can be equivalently thought of as the negative surface pressure and the governing equations simplify to the two-dimensional incompressible Navier--Stokes equations (see Appendix~\ref{sec:sec_appendix_flat_eqns}).
While the problems considered in this section are easily solved using standard Cartesian finite element methods, we solve them using our nonlinear isoparametric finite element framework, in which differential geometry is used to express the surface position and fundamental unknowns in terms of curvilinear coordinates.
Thus, even these simple problems serve as important benchmarks for our LE finite element implementation.

In our simulation of fluid flow on a flat plane, the mesh is constrained to be fixed, such that
$\bmvm = \bm{0}$, and we do not solve for the mesh velocity.
Furthermore, as the mesh is flat, there is no motion in the out-of-plane direction and we solve only for the $x$- and $y$-components of the fluid velocity as well as the surface tension.
The three corresponding strong form equations are the continuity equation~\eqref{eq:strong_incompressibility} and the two in-plane Navier--Stokes equations~\eqref{eq:appendix_flat_equations}$_3$.
We simulate five scenarios with increasingly complex solutions:
(i) a hydrostatic fluid, with zero velocity and linear surface tension;
(ii) Couette flow between parallel plates, with linear velocity and zero surface tension;
(iii) Couette flow between parallel plates under the influence of a quadratic body force, with linear velocity and cubic surface tension;
(iv) Hagen--Poiseuille flow in a channel, with quadratic velocity and linear surface tension;
and finally (v) the lid-driven cavity problem, for which no analytical solution is known.
Only the lid-driven cavity result is discussed in the main text, and the validation of the first four cases against analytical solutions is left to Appendix~\ref{sec:sec_appendix_numerical_benchmarks}.

For the numerical results presented in this manuscript, we neglect inertial terms in all cases except the lid-driven cavity, for which inertial terms are evaluated using a backwards Euler temporal discretization.
The contribution of inertial terms to the tangent matrix and residual vector, for the limited case of a fixed surface, is provided in Appendix~\ref{sec:sec_appendix_inertia_fixed_surface}.
Inertial terms are included in the lid-driven cavity problem only to further validate our numerical implementation with flows at moderate Reynolds numbers.

%
% *** Lid-Driven Cavity
%

\subsubsection{Lid-Driven Cavity} \label{sec:sec_flat_cavity}

\begin{figure}[!t]
	\centering
	\begin{subfigure}[b]{0.30\columnwidth}
		\centering
		\includegraphics[width=0.99\textwidth]{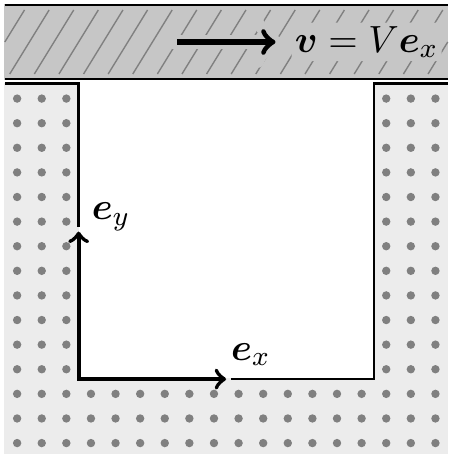}
		\caption{schematic}
		\label{fig:fig_cavity_schematic}
	\end{subfigure}
	\begin{subfigure}[b]{0.30\columnwidth}
		%\centering
		\hfill
		\includegraphics[width=0.855\textwidth]{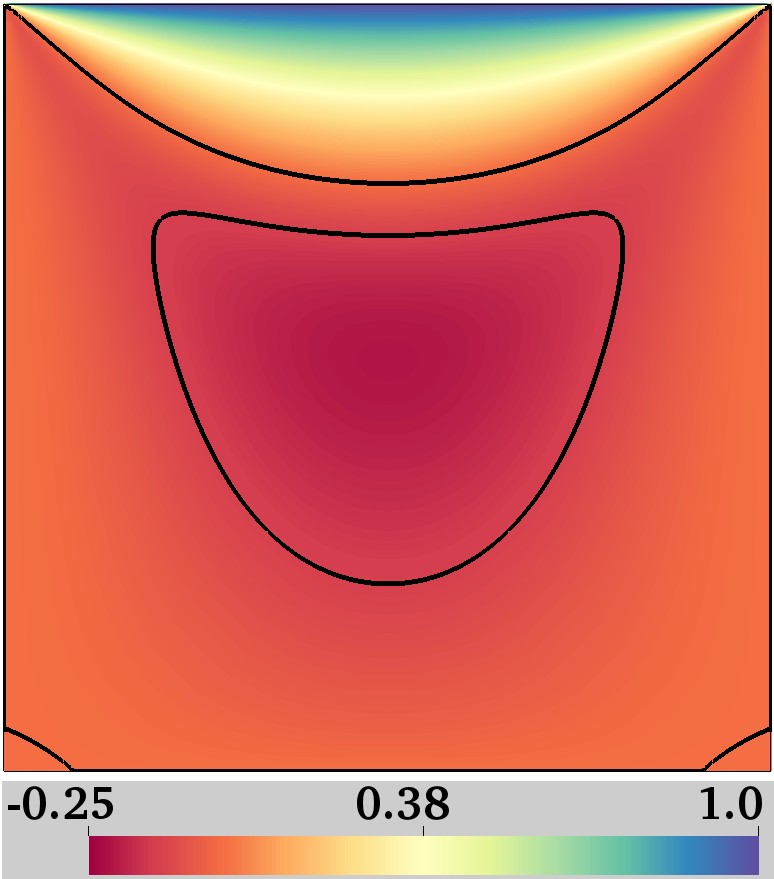}
		\caption{$v_x$, for $\Rey = 0$}
		\label{fig:fig_cavity_Re_0}
	\end{subfigure}
	\begin{subfigure}[b]{0.30\columnwidth}
		%\centering
		\hfill
		\includegraphics[width=0.855\textwidth]{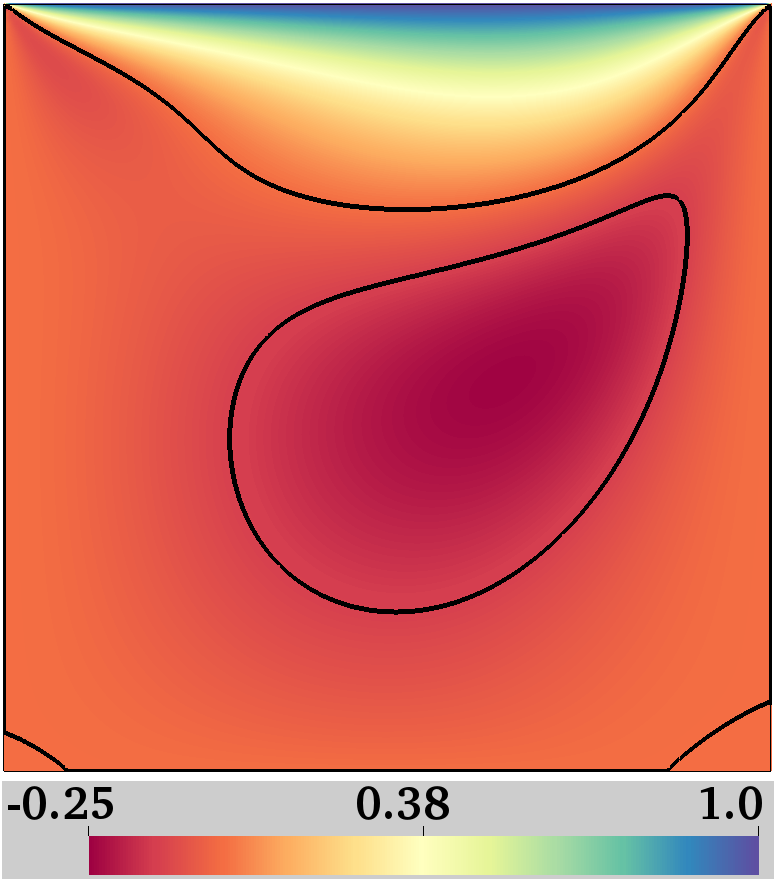}
		\caption{$v_x$, for $\Rey = 100$}
		\label{fig:fig_cavity_Re_100}
	\end{subfigure}
	\vspace{12pt}
	\\
	\begin{subfigure}[b]{0.30\columnwidth}
		%\centering
		\includegraphics[width=0.97\textwidth]{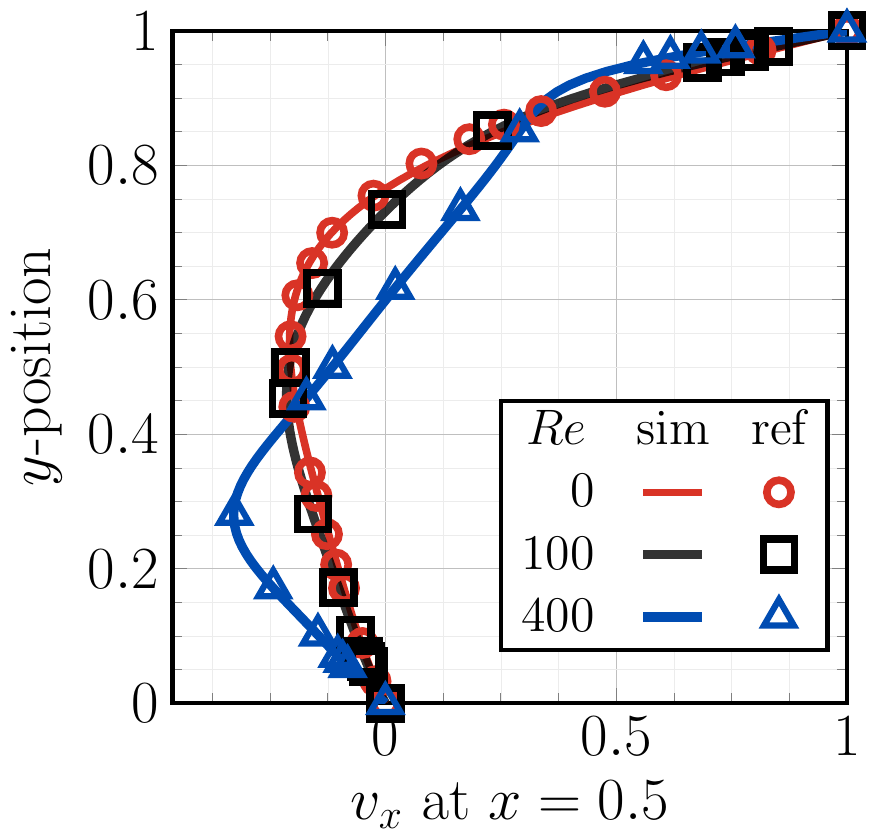}
		\caption{effect of $\Rey$ on $v_x$}
		\label{fig:fig_cavity_Re_plot_vx}
	\end{subfigure}
	\begin{subfigure}[b]{0.30\columnwidth}
		%\centering
		\includegraphics[width=0.97\textwidth]{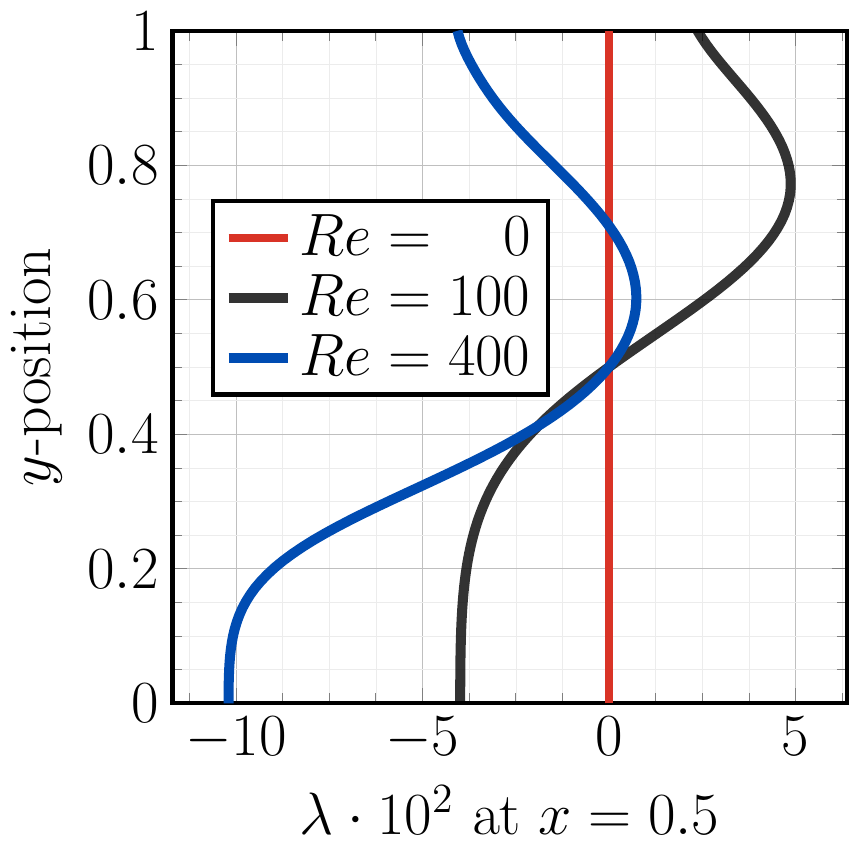}
		\caption{effect of $\Rey$ on $\lambda$}
		\label{fig:fig_cavity_Re_plot_lambda}
	\end{subfigure}
	\begin{subfigure}[b]{0.30\columnwidth}
		%\centering
		\includegraphics[width=0.99\textwidth]{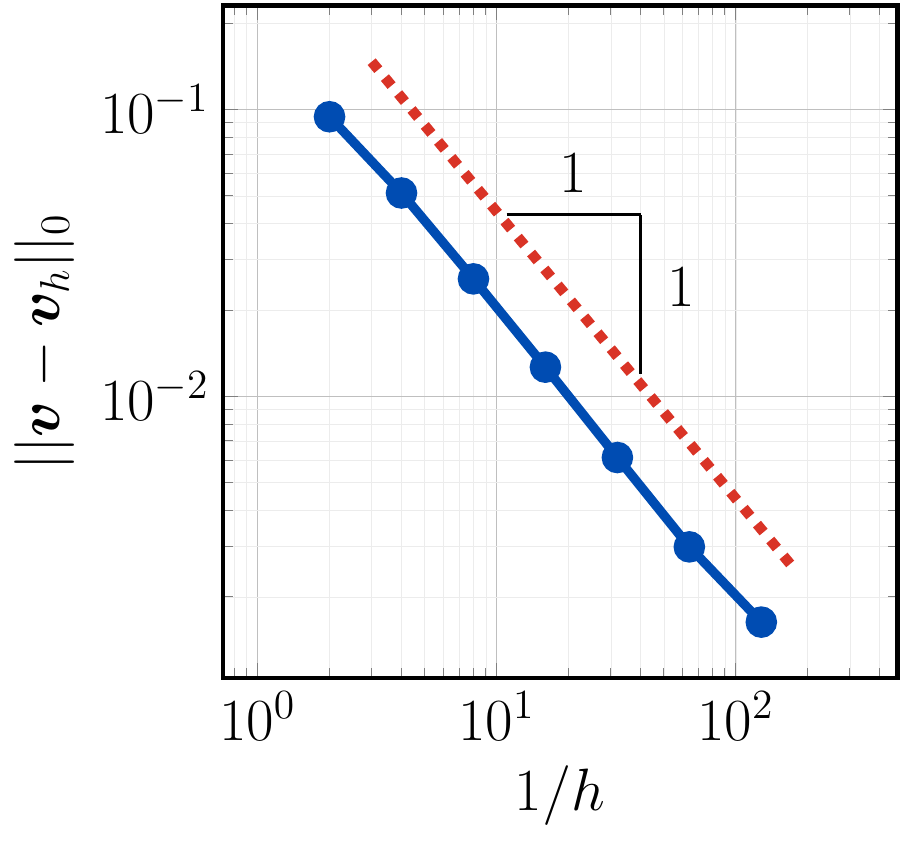}
		\caption{error in $\bmv$, $\Rey = 0$}
		\label{fig:fig_cavity_error}
	\end{subfigure}
	\vspace{5pt}
	\caption{
		Lid-driven cavity problem.
		(a) Schematic of the problem, in which fluid in a cavity (shown in white) of unit height and width is surrounded on three sides by rigid walls.
		The lid of the cavity is dragged to the right at constant speed $V$, such that at the top edge the fluid travels towards the right with the same speed.
		In all numerical calculations, we set $V = 1$.
		(b),(c) Plots of the $x$-velocity throughout the domain, on a $128 \times 128$ mesh, at Reynolds number $\Rey = 0$ (b) and $\Rey = 100$ (c).
		In both cases, fluid flows to the right at the top of the domain and then recirculates to the left in the center.
		When $\Rey = 100$, the fluid is pulled more towards the top right corner of the domain.
		The outer contour line shows where $v_x = 0$, and the inner contour depicts where $v_x = -0.12$.
		(d),(e) Plots of the $x$-velocity (d) and surface tension (e) on a $128 \times 128$ mesh at the vertical cross section $x = 0.5$.
		(d)
		The $x$-velocities in our simulations (solid lines) agree with Ref.~\cite{nallasamy-jfm-1977} (red circles) at $\Rey = 0$, and Ref.~\cite{ghia-jcp-1982} (black squares, blue triangles) at $\Rey = 100$ and $\Rey = 400$, respectively.
		(e) The surface tension has small differences for $\Rey = 0$, $100$, and $400$.
		The point where the three lines meet is the center of the domain,
		$(x, \, y) = (0.5, \, 0.5)$,
		where we set $\lambda = 0$.
		(f) Plot of the velocity $L^2$-error as a function of the element width $h$, for $\Rey = 0$.
		As no exact solution is known, we calculate the error relative to the solution on a $256 \times 256$ mesh.
		Data is shown in blue; the dashed red line is a reference with slope $-1$ to show the scaling of the error.
	}
	\vspace{-8pt}
	\label{fig:fig_cavity}
\end{figure}

A schematic of the problem is shown in Fig.~\ref{fig:fig_cavity_schematic}.
Fluid in a square cavity with stationary walls is driven by a top lid which moves to the right at constant speed $V$.
We solve for the flow field and surface tension in the cavity, which is taken to be a unit square in the $x$--$y$ plane.
The boundary condition $\bmv = \bm{0}$ is imposed on the sides and bottom of the square domain, and
$\bmv = V \bm{e}_x$
is set on the top edge.
There is a choice in what velocity to specify at the top two corners of the domain where the stationary edges meet the moving top lid.
At these locations we set
$\bmv = \bm{0}$,
as done in Ref.~\cite{zienkiewicz-taylor-fluids}.
Furthermore, as only gradients of the surface tension enter the equations governing a flat, two-dimensional fluid, the surface tension $\lambda$ is indeterminate up to a constant.
Consequently, $\lambda$ is specified to be zero at the center of the domain, located at
$(x, \, y) = (0.5, \, 0.5)$.

We first set the inertial terms to zero and solve for the Stokes flow result at Reynolds number
$\Rey := \rho \, V L / \zeta = 0$,
for which the $x$-velocity is plotted in Fig.~\ref{fig:fig_cavity_Re_0}.
The flow field is moving towards the right at the top of the domain, and the presence of the right wall requires the fluid to be recirculated towards the left side in the bulk of the domain.
We then include inertial terms and set
$\Rey = 100$
by setting $\rho = 1$ and $\zeta = 0.01$ (we already have $V = 1$ and $L = 1$).
The corresponding $x$-velocity is plotted in Fig.~\ref{fig:fig_cavity_Re_100}.
Relative to Fig.~\ref{fig:fig_cavity_Re_0}, Fig.~\ref{fig:fig_cavity_Re_100} shows that at the top of the domain, inertia pulls the fluid to the right in the direction of the moving top lid.
In Figs.~\ref{fig:fig_cavity_Re_plot_vx} and \ref{fig:fig_cavity_Re_plot_lambda}, the $x$-velocity and surface tension are plotted at different $\Rey$ across a vertical cross-section through the domain.
Our results for the lid-driven cavity problem agree with those of other numerical studies~\cite{nallasamy-jfm-1977, ghia-jcp-1982}, as plotted in Fig.~\ref{fig:fig_cavity_Re_plot_vx}.

Our final task for the lid-driven cavity problem is to analyze how our simulations converge to the true solution as we increase the number of elements.
We consider the case with no inertia, i.e. $\Rey = 0$, and calculate the error as a function of the length (and width) of a single element, denoted $h$.
As the exact solution is unknown, we treat the solution on the finest, $256 \times 256$ mesh as the true solution.
We calculate the $L^2$-error for the velocity, denoted $|| \bmv - \bmvh ||_0$ and defined as
\begin{equation} \label{eq:l2_error_def}
	\big\lvert \big\lvert
		\bmv - \bmvh
	\big\rvert \big\rvert_0
	\, := \,
	\bigg(
		\displaystyle
		\int_\Omega
			| \bmv - \bmvh |^2
		~\mathrm{d}\Omega
	\bigg)^{1/2}
	~,
\end{equation}
where $\bmv$ is the true velocity and $\bmvh$ is the approximate velocity found on coarser meshes with elements of size $h \times h$.
The $L^2$-error is denoted with a subscript `0' because $L^2(\Omega) = H^0(\Omega)$, i.e. the Sobolev space of order zero.
A plot of $|| \bmv - \bmvh ||_0$ as a function of $1/h$ is shown in Fig.~\ref{fig:fig_cavity_error}, in which we see a linear scaling of the $L^2$-error with the mesh size $h$.
The numerical treatment of the corner nodes may explain why our simulations do not converge faster:
at each mesh size, we set $\bmv = \bm{0}$ at the corner nodes and $\bmv = V \, \bm{e}_z$ at the adjacent nodes on the top edge, however, as the number of elements increases, these two nodes move closer together such that we solve a slightly different problem at each mesh size.

Due to the moving top lid being in contact with the stationary walls at the top two corners of the domain, the lid-driven cavity solution is known to have large surface tension spikes at the corners \cite{zienkiewicz-taylor-fluids}.
For this reason, numerical studies generally use meshes which are very finely discretized at the corners.
We find the surface tension spikes vary significantly with the mesh size, for example, on a $16 \times 16$ mesh $\lambda$ is $\mathcal{O}(10^2)$ at the corners while on a $128 \times 128$ mesh it is $\mathcal{O} (10^3)$.
These spikes dominate the error calculation, which even on the $128 \times 128$ mesh is $\mathcal{O}(1)$ relative to the solution on a $256 \times 256$ mesh.
The surface tension spikes are plotted in Appendix~\ref{sec:sec_flat_dohrmann_bochev_example}, in which the success of the Dohrmann--Bochev method in removing inf--sup instabilities is also shown.
As the surface tension error does not converge in this situation, in Appendix~\ref{sec:sec_appendix_couette_body} we show how it converges in the benchmark problem of Couette flow with a body force.

With the numerical results provided thus far, as well as those of Appendix~\ref{sec:sec_appendix_numerical_benchmarks}, we conclude our analysis of fluid flow on a fixed, flat plane.
Our general ALE finite element framework, based on a curvilinear coordinate description via the machinery of differential geometry, has successfully reproduced the classical benchmarks of hydrostatic flow, Couette flow, Hagen--Poiseuille flow, and lid-driven cavity flow.
We are therefore confident our finite element method can model arbitrary flows on flat surfaces.

%
% *** Fluid Flow on a Stationary, Curved Surface
%

\subsection{Fluid Flow on a Stationary, Curved Surface} \label{sec:sec_flow_curved_surface}

After testing our code on the simplest case of fluid flow on a flat plane, we turn to study fluid flows on stationary, curved surfaces.
In such problems the mesh is fixed, and thus once again the mesh velocities are not solved for.
However, a complication arises because for a fixed surface, the normal component of the material velocity
$v = \bmv \cdot \bm{n} = 0$.
In practice, there are two ways we could handle this problem.
First, we could posit that our arbitrary velocity variation
$\delta \bmv = \delta v^\calpha \bm{a}_\calpha$,
where $\bm{a}_\calpha$ is known for the fixed surface.
However, such a method cannot be easily extended to model a deforming fluid film: when the surface is deforming, the velocity $\bm{v}$ is represented as
$\bmv = v^\calpha \bm{a}_\calpha + v \bm{n}$,
and in a fully implicit numerical scheme $v^\calpha$, $\bm{a}_\calpha$, $v$, and $\bm{n}$ are all unknown.

We employ a different approach by representing the velocity and its arbitrary variation in a Cartesian basis, including both the in-plane and shape equations in our weak formulation, and enforcing
$v = \bmv \cdot \bm{n} = 0$
with a Lagrange multiplier---which
is interpreted physically as the pressure drop required to constrain the fluid to the surface.
In this section, we describe how both the strong and weak formulations are modified by the normal pressure $p$ being an unknown Lagrange multiplier field.
Our method is then tested by modeling fluid flow on a fixed, bulged cylinder.
Numerical results are compared with analytical solutions, and we calculate how our numerical error decreases on mesh refinement.

%
% *** Strong Form Modification
%

\subsubsection{Strong Form Modification} \label{sec:sec_normal_pressure_strong_form}

In satisfying the constraint
$v = 0$
with a Lagrange multiplier, we include the shape equation~\eqref{eq:shape_eqn} in our description of the fluid film.
In the shape equation~\eqref{eq:shape_eqn}, the pressure drop $p$ is an unknown Lagrange multiplier field, which at every point on the surface takes the requisite value to satisfy
$v = 0$.
There are thus five unknowns: the three components of the velocity, the surface tension, and the normal pressure drop, and five corresponding equations: the continuity equation~\eqref{eq:strong_incompressibility}, the two in-plane equations~\eqref{eq:eom_in_plane}, the shape equation~\eqref{eq:shape_eqn}, and the constraint
$v = 0$.

%
% *** Weak Form Modification
%

\subsubsection{Weak Form Modification} \label{sec:sec_normal_pressure_weak_form}

With the pressure drop $p$ being a fundamental unknown, the arbitrary pressure variation $\delta p$ is expected to enter the weak formulation.
To understand how $\delta p$ will appear, we take the variation of the virtual work associated with moving the fluid film in the normal direction and obtain
\begin{equation} \label{eq:normal_pressure_virtual_work}
	\delta \, \bigg(
		\int_\Omega
			\bmv \cdot p \, \bm{n}
			~\Jm
		~\mathrm{d} \Omega
	\bigg)
	\, =
	\, \int_\Omega
		\delta \bmv \cdot p \, \bm{n}
		~\Jm
	~\mathrm{d} \Omega
	\,\, + \int_\Omega
		\bmv \cdot \delta p \, \bm{n}
		~\Jm
	~\mathrm{d} \Omega
	~.
\end{equation}
The first term on the right-hand side of Eq.~\eqref{eq:normal_pressure_virtual_work} is already contained in Eq.~\eqref{eq:fluid_weak_form_with_DB} through the body force term.
Assuming no in-plane body forces ($b_\calpha = 0$), treating the pressure $p$ as a fundamental unknown, again recognizing inertia is negligible, and removing mesh velocity degrees of freedom yields a modified weak form (c.f.~Eq.~\eqref{eq:fluid_weak_form_with_DB})
\begin{equation} \label{eq:fluid_weak_form_fixed}
	\begin{split}
		&\int_\Omega 
			\delta \bmv_{, \calpha} \cdot \pi^{\calpha \cbeta} \, \bm{a}_\cbeta
			~\Jm
		~\mathrm{d} \Omega
		\,
		+ \int_\Omega
			\delta \bmv_{, \calpha} \cdot \bm{a}^\calpha \, \lambda
			~\Jm
		~\mathrm{d} \Omega
		\,
		+ \int_\Omega
			\delta \lambda \, \big(
				\bm{a}^\calpha \cdot \bm{v}_{, \calpha}
			\big)
			~\Jm
		~\mathrm{d}\Omega
		\\[3pt]
		&\hspace{20pt}
		- \int_\Omega
			\delta \bmv \cdot p \, \bm{n}
			~\Jm
		~\mathrm{d} \Omega
		\,
		- \int_\Omega
			\delta p \, \bmv \cdot \bm{n}
			~\Jm
		~\mathrm{d} \Omega
		\, - \, \dfrac{\alphaDB}{\zeta}
		\int_{\Omega}
			\big(
				\delta \lambda
				- \delta \blambda
			\big) \, \big(
				\lambda
				- \blambda
			\big)
		~\mathrm{d}\Omega
		\\[9pt]
		&\hspace{40pt}
		~ = ~ 0 \hspace{20pt}
		\forall \,\, \delta \bmv \in \mcv_0, \,\, \delta p \in \mcp, \,\, \delta \lambda \in \Lambda
		~,
	\end{split}
\end{equation}
where $\mathcal{P}$ is the space of pressure solutions and arbitrary pressure variations.
The weak form \eqref{eq:fluid_weak_form_fixed} contains no gradients of pressure and thus it is theoretically sound for us to choose $\mathcal{P}$ as the space of square-integrable functions on $\Omega$, i.e. $L^2 (\Omega)$.
However, we simplify our finite element analysis by using the same basis functions for the velocity and pressure.
In accordance with Eqs.~\eqref{eq:velocity_solution_space} and \eqref{eq:solution_space_sobolev}, we define
\begin{equation} \label{eq:normal_pressure_solution_space}
	\mathcal{P}
	:= H^2 (\Omega)
	~.
\end{equation}

The structure of Eq.~\eqref{eq:fluid_weak_form_fixed} indicates the pressure variation $\delta p$ enforces the normal constraint
$\bmv \cdot \bm{n} = 0$,
in the same way the surface tension variation $\delta \lambda$ enforces the incompressibility constraint
$\bm{a}^\calpha \cdot \bmv_{, \calpha} = 0$.
With the weak formulation \eqref{eq:fluid_weak_form_fixed}, an identical procedure to that of Sec.~\ref{sec:sec_finite_element_formulation_fluid_films} is followed to linearize and discretize the equation, calculate the tangent matrix and residual vector, and then iteratively solve for the unknowns via Newton--Raphson iteration.
The approximate pressure $p_h$ is an element of the finite-dimensional subspace $\mcp_h \subset \mcp$, which is chosen to be
\begin{equation} \label{eq:normal_pressure_discrete_solution_space}
	\mcp_h
	:= \Big\{~
		u (\zeta^\alpha) \! : \Omega \rightarrow \mathbb{R}
		\hspace{8pt}
		\text{such that}
		\hspace{4pt}
		u \in C^1(\Omega)
		\cap \mcp
		~,
		\hspace{10pt}
		u \big\rvert_{\Omegae} \in \mathbb{Q}_2 (\Omegae)
		\hspace{10pt}
		\forall
		\hspace{4pt}
		\Omegae
	~\Big\}
\end{equation}
in accordance with the finite-dimensional space of velocities $\mcvh$ \eqref{eq:fluid_velocity_subspace}.
The pressure can then be expressed in terms of the same set of basis functions, $\{ N_I (\zeta^\alpha) \}$, used for the velocities.
As mentioned previously, our choice of $\mcp$ \eqref{eq:normal_pressure_solution_space} and $\mcp_h$ \eqref{eq:normal_pressure_discrete_solution_space} is purely for convenience in our numerical implementation.
The details of the modifications to our finite element implementation are
provided in Appendix~\ref{sec:sec_appendix_fixed_surface}.

\begin{figure}[p]
	\centering
	\begin{subfigure}[b]{0.28\columnwidth}
		\hfill
		\includegraphics[width=0.84\textwidth]{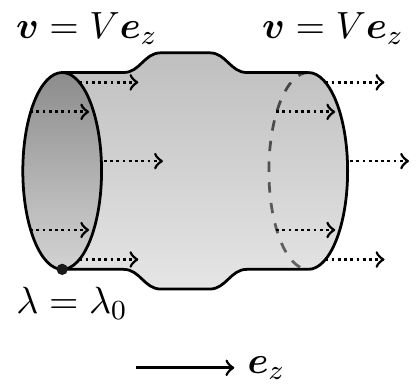}
		\vspace{10pt}
		\caption{schematic}
		\label{fig:fig_cylinder_schematic}
	\end{subfigure}
	\begin{subfigure}[b]{0.28\columnwidth}
		\centering
		\includegraphics[width=0.95\textwidth]{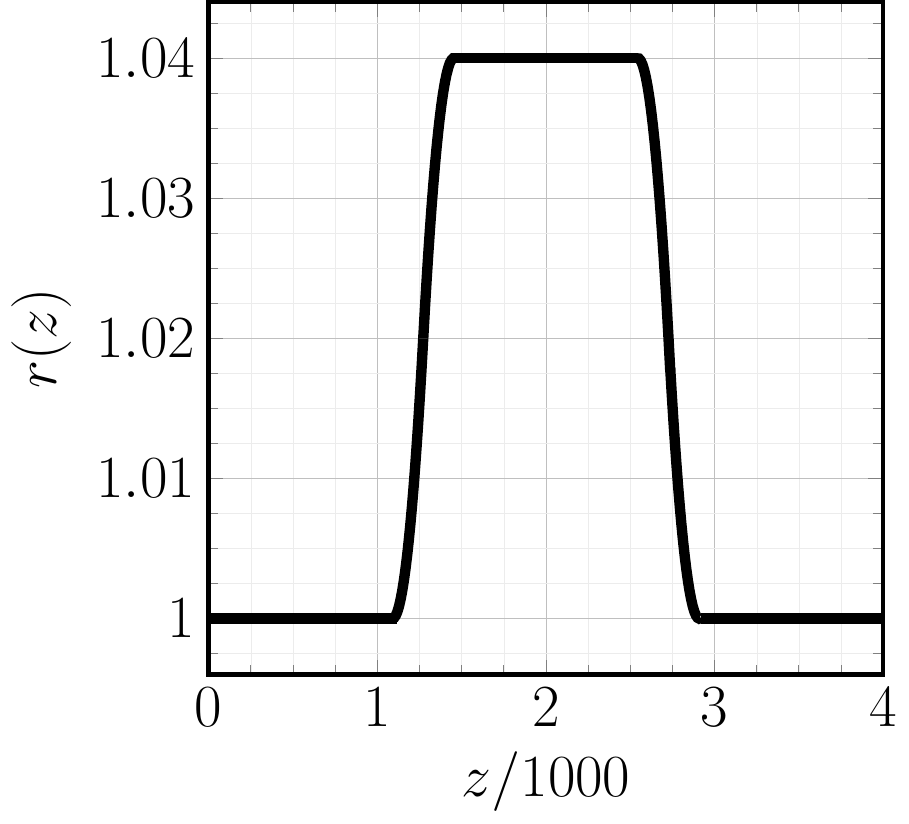}
		\caption{radius $r(z)$}
		\label{fig:fig_cylinder_radius}
	\end{subfigure}
	\begin{subfigure}[b]{0.28\columnwidth}
		\centering
		\includegraphics[width=0.90\textwidth]{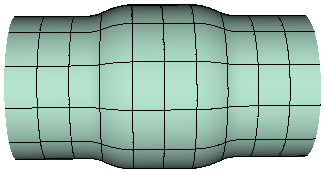}
		\vspace{33pt}
		\caption{$10 \times 10$ mesh}
		\label{fig:fig_cylinder_bulge_mesh}
	\end{subfigure}
	\\[8pt]
	\begin{subfigure}[b]{0.28\columnwidth}
		\centering
		\includegraphics[width=0.95\textwidth]{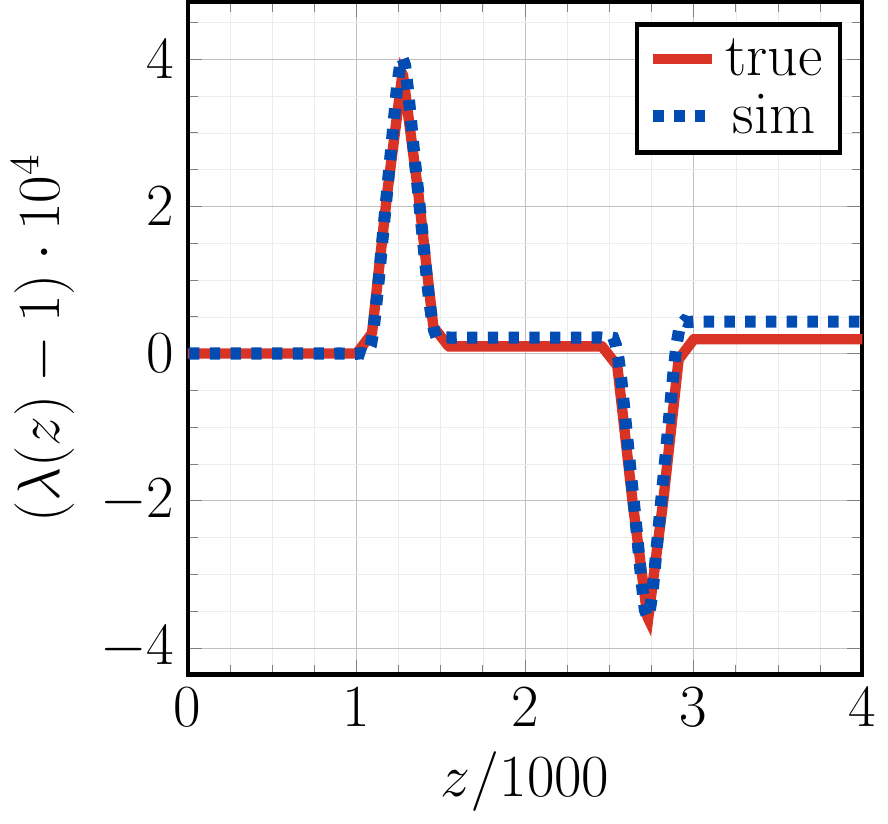}
		\caption{surface tension $\lambda(z)$}
		\label{fig:fig_cylinder_lambda}
	\end{subfigure}
	\begin{subfigure}[b]{0.28\columnwidth}
		\centering
		\includegraphics[width=0.95\textwidth]{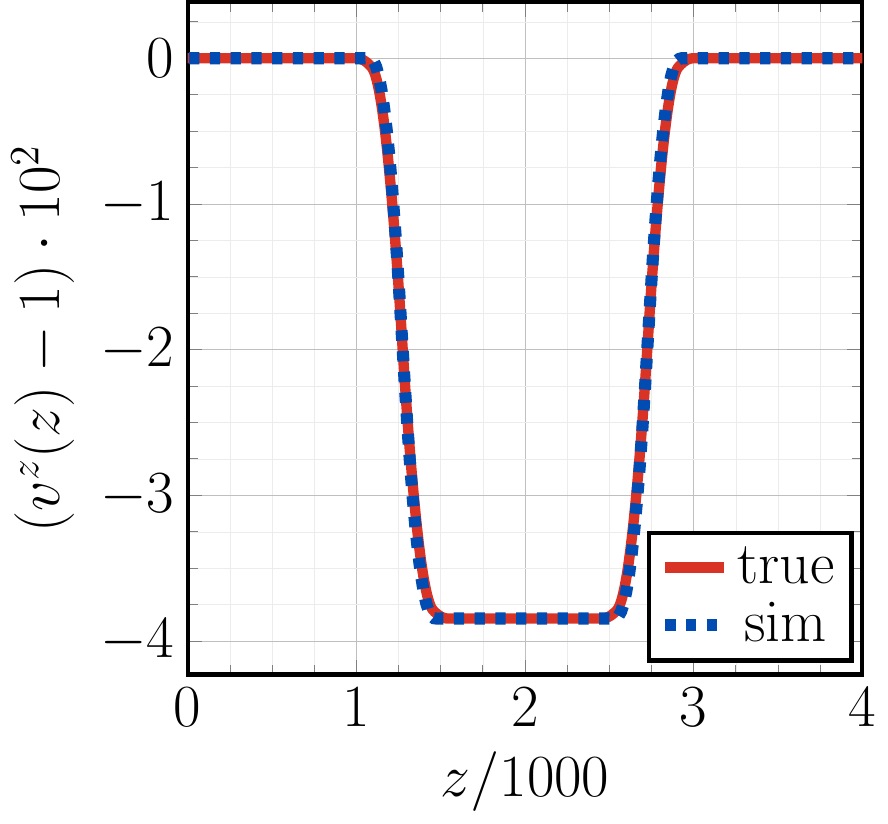}
		\caption{$z$-velocity $v_z (z)$}
		\label{fig:fig_cylinder_vz}
	\end{subfigure}
	\begin{subfigure}[b]{0.28\columnwidth}
		\centering
		\includegraphics[width=0.95\textwidth]{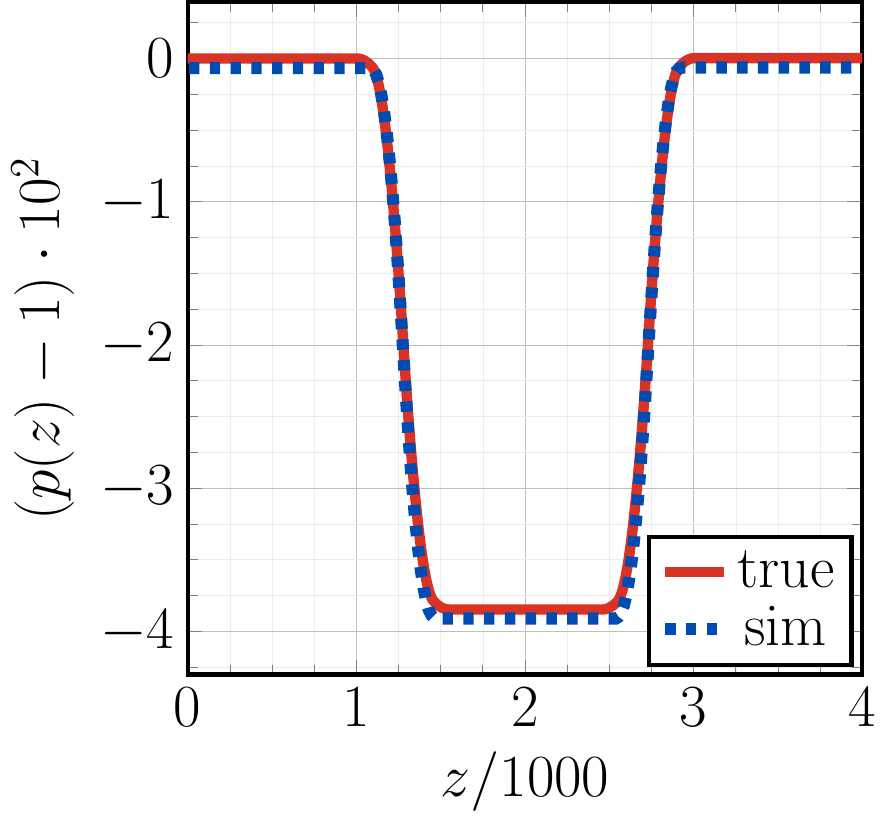}
		\caption{pressure $p(z)$}
		\label{fig:fig_cylinder_p}
	\end{subfigure}
	\\[9pt]
	\begin{subfigure}[b]{0.28\columnwidth}
		\centering
		\includegraphics[width=0.95\textwidth]{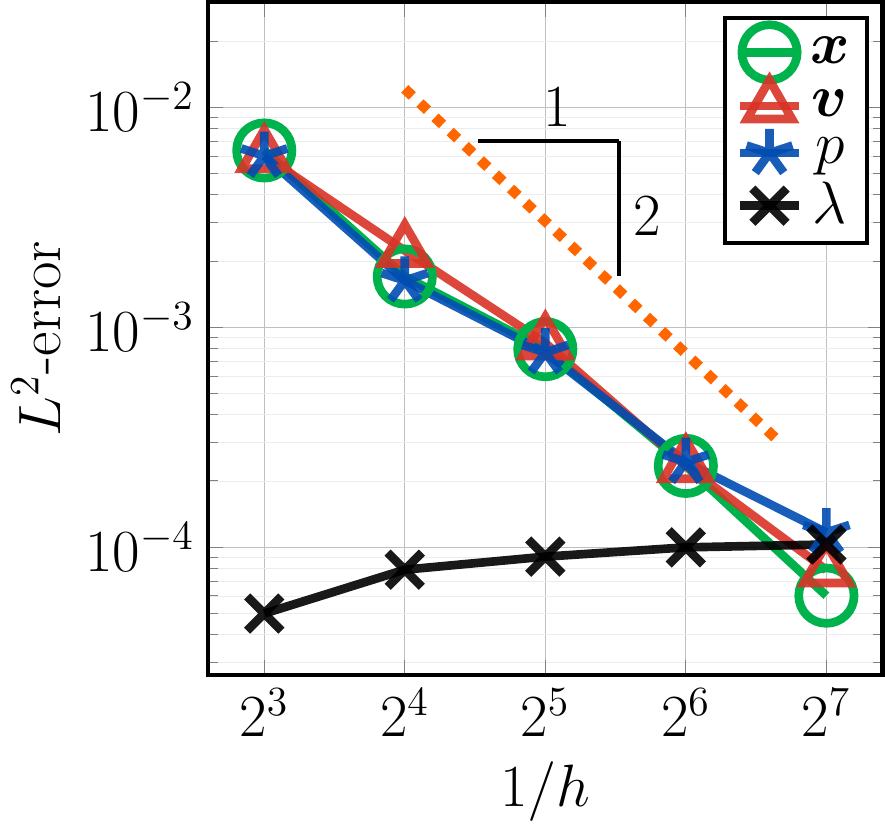}
		\caption{error, refining $z$ and $\theta$}
		\label{fig:fig_cylinder_both_true_error}
	\end{subfigure}
	\begin{subfigure}[b]{0.28\columnwidth}
		\centering
		\includegraphics[width=0.95\textwidth]{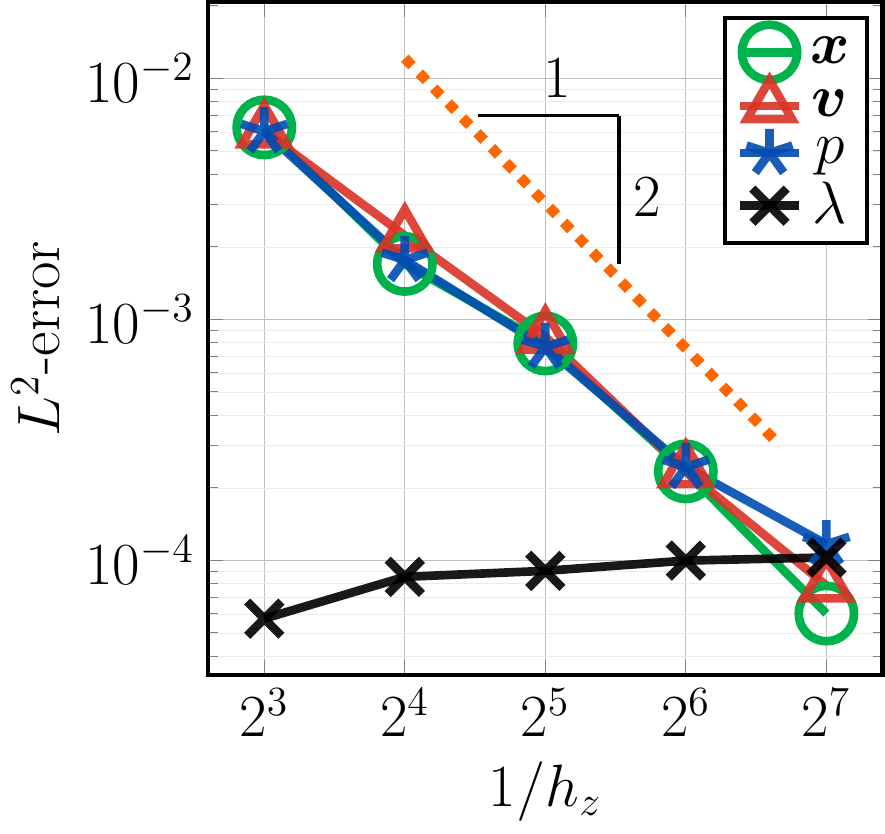}
		\caption{error, refining $z$}
		\label{fig:fig_cylinder_z_true_error}
	\end{subfigure}
	\begin{subfigure}[b]{0.28\columnwidth}
		\centering
		\includegraphics[width=0.98\textwidth]{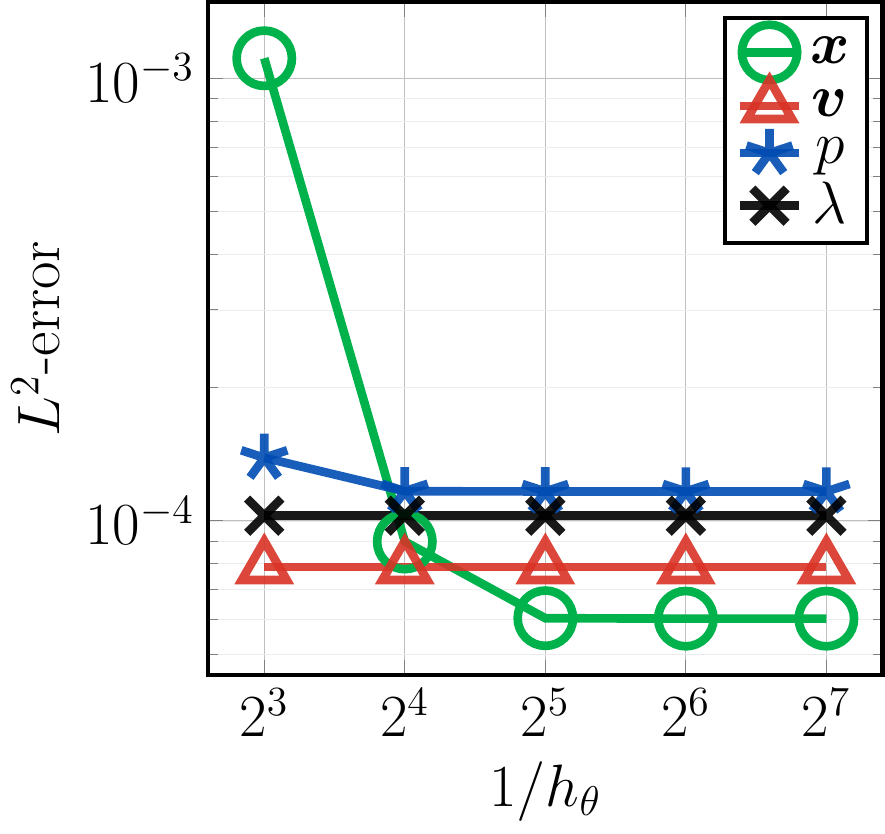}
		\caption{error, refining $\theta$}
		\label{fig:fig_cylinder_theta_true_error}
	\end{subfigure}
	\vspace{-0pt}
	\captionsetup{width=\linewidth}
	\caption{
		Fluid flow on a fixed cylinder with a bulge.
		(a) Schematic of the problem.
		Fluid enters at the left edge at velocity $V \, \bm{e}_z$ and exits at the right edge at the same velocity.
		The surface tension $\lambda$ is specified to be unity at a single point on the boundary.
		In the numerical calculation, no assumption about axisymmetry is made.
		(b) Radius as a function of axial position, with a 4\% bulge in the central region.
		(c) A coarse $10 \times 10$ mesh of the bulged cylinder.
		(d--f) Plots of the surface tension (d), $z$-velocity (e), and normal pressure (f) as a function of axial position.
		We plot the deviation in these quantities from their analytical value at the left edge of the cylinder $(z = 0)$, with axes scaled for convenience.
		The solid red lines are the analytical solutions and the dashed blue lines are our numerical results.
		Calculations were made on a $128 \times 128$ mesh.
		(g)--(i) Plots of the $L^2$-error according to Eq.~\eqref{eq:l2_error_def}, relative to the analytical solution, as we change the number of elements.
		In (g), we refine in both the $z$ and $\theta$ directions, such that there are the same number of elements in each direction.
		In (h), there are 128 $\theta$-elements and the number of $z$-elements varies from $8$ to $128$ in powers of two.
		In (i), there are 128 $z$-elements and between 8 and 128 $\theta$-elements, again in powers of two.
		In (c)--(i), the length $L = 4, \! 000$ is chosen for analytical solutions to be sufficiently smooth (see Appendix~\ref{sec:sec_appendix_cylinder_bulge_eqns}).
	}
	\label{fig:fig_cylinder_bulge}
\end{figure}

%
% *** Flow on a Fixed Cylinder with a Bulge
%

\subsubsection{Flow on a Fixed Cylinder with a Bulge} \label{sec:sec_fixed_cylinder_flow}

In our numerical implementation, we consider fluid flowing on a fixed, bulged cylinder, as shown in Figs.~\ref{fig:fig_cylinder_schematic}--\ref{fig:fig_cylinder_bulge_mesh}.
Our boundary conditions are shown schematically in Fig.~\ref{fig:fig_cylinder_schematic}, where constant inflow and outflow of the fluid is prescribed at the entrance and exit of the cylinder, respectively.
The surface tension is specified at a single point on the boundary, as only gradients of $\lambda$ enter the in-plane equations.
The bulge in the center leads to nontrivial velocity and surface tension profiles due to the coupling between curvature and fluid flow, and the symmetry of the surface shape allows us to determine the analytical solution (see Appendix~\ref{sec:sec_appendix_cylinder_bulge_eqns}).
The bulged cylinder thus serves as a useful benchmark problem for our numerical method, in the study of flows on curved yet fixed surfaces.

The position of the bulged, cylindrical surface is given by
\begin{equation} \label{eq:bulged_cylinder_position_general}
	\bmx (\theta, z)
	= r(z) \, \bm{e}_r (\theta)
	+ z \, \bm{e}_z
	~,
\end{equation}
where $\theta$ and $z$ are the polar angle and axial distance, respectively, of a standard cylindrical coordinate system.
The radius $r(z)$ is independent of $\theta$ because the surface is axisymmetric.
Our choice of radial profile is given in Appendix~\ref{sec:sec_appendix_cylinder_bulge_eqns} and shown in Fig.~\ref{fig:fig_cylinder_radius}.
We define $h_\theta$ and $h_z$ as the width of an element $\Omegae$ in the $\theta$ and $z$ directions, respectively, and denote a mesh with 16 elements in the $\theta$-direction and 32 elements in the $z$-direction, for example, as a $16 \times 32$ mesh.
A coarse $10 \times 10$ mesh of the bulged cylinder is shown in Fig.~\ref{fig:fig_cylinder_bulge_mesh}.
The surface tension, $z$-velocity, and normal pressure are calculated numerically on a $128 \times 128$ mesh, and compared to their analytical counterparts in Figs.~\ref{fig:fig_cylinder_lambda}--\ref{fig:fig_cylinder_p}, respectively.
In all cases, our numerical results show excellent agreement with the analytical calculations.

We conclude our analysis of the fixed, bulged cylinder by calculating the $L^2$-error of the velocity, surface tension, and normal pressure.
We also calculate the error between our numerical surface position $\bmxh$ and the true position $\bmx$ given in Eq.~\eqref{eq:bulged_cylinder_position_general}, which is a function of how closely our basis functions can represent the analytical geometry.
Figures~\ref{fig:fig_cylinder_both_true_error}--\ref{fig:fig_cylinder_theta_true_error} show three different plots of the error: changing $h_\theta$ and $h_z$ together (\ref{fig:fig_cylinder_both_true_error}), changing $h_z$ with fixed $h_\theta$ (\ref{fig:fig_cylinder_z_true_error}), and changing $h_\theta$ with fixed $h_z$ (\ref{fig:fig_cylinder_theta_true_error}).
Figures~\ref{fig:fig_cylinder_both_true_error} and \ref{fig:fig_cylinder_z_true_error} are nearly identical, indicating $\theta$-refinement has little effect on the errors, an observation which is confirmed with Fig.~\ref{fig:fig_cylinder_theta_true_error}.
As the analytical solution is only a function of $z$, it is unsurprising that refining in $\theta$ has no significant effect on the errors.
In particular, $\theta$-refinement only makes observable changes to our numerical surface representation when there are few elements in the $\theta$-direction.
In Figs.~\ref{fig:fig_cylinder_both_true_error} and \ref{fig:fig_cylinder_z_true_error}, the errors in the position, velocity, and normal pressure converge quadratically on mesh refinement.
However, the error in $\lambda$ is approximately constant.
As discussed in Appendix~\ref{sec:sec_appendix_cylinder_bulge_eqns}, a cylinder length $L=4,\!000$ is chosen, with radius changes over extended length scales, to avoid discontinuities in the normal pressure.
For such cylinders, $r'(z) \ll 1$ and $r''(z) \ll 1$, and therefore
$\lambda(z) \approx \lambda_0 = 1$
everywhere (see Eq.~\eqref{eq:appendix_cylinder_bulge_lambda})---irrespective of the shape of the bulge.
Analytically, the difference between $\lambda$ and $\lambda_0$ is $\mathcal{O} (10^{-4})$ and occurs when the radius changes due to nonzero $r'(z)$, $r''(z)$, and $z$-velocity \eqref{eq:appendix_cylinder_bulge_lambda}.
Numerically, errors in $\lambda$ are also $\mathcal{O}(10^{-4})$ and occur when the radius changes due to the difference between the numerical and analytical position $\bmx$; these errors persist along the length of the cylinder~(Fig.~\ref{fig:fig_cylinder_lambda}).
As can be seen from Fig.~\ref{fig:fig_cylinder_z_true_error}, the errors in position $\bmx$ are $\gtrsim 10^{-4}$ for all simulated meshes, and therefore the error in $\lambda$ remains approximately constant at $10^{-4}$.
We expect the $\lambda$ error to decrease as one further refines the mesh, thus reducing errors in the position.
However, these simulations are prohibitively expensive in our current implementation and require parallelization of our code.
Our error calculations, captured in Figs.~\ref{fig:fig_cylinder_both_true_error}--\ref{fig:fig_cylinder_theta_true_error}, conclude our analysis of fluid flows on fixed, curved surfaces.

%
% *** Curved and Deforming Fluid Films
%

\subsection{Curved and Deforming Fluid Films} \label{sec:sec_curved_deforming_films}

In this section, we demonstrate the capabilities of our full LE finite element implementation by studying the stability of cylindrical fluid films which have a constant pressure drop across their surface.
It is already known that given a surface tension and pressure drop, spherical fluid films of radius $R$ satisfy the Young--Laplace equation
$p = 2 \lambda / R$
and are stable \cite{degennes-wetting}.
The cylindrical fluid films under consideration, however, are found to be unstable to axisymmetric perturbations;
the large, nontrivial shape changes resulting from the instability are studied here.
Our numerical results are compared with analytical results from a linear stability analysis, and the two are found to be in excellent agreement.
Moreover, we find the fluid film instability is mediated by the in-plane flow resulting from the initial shape perturbation, thus demonstrating the importance of our ALE framework in studying two-dimensional surfaces with in-plane flow.
Our general LE implementation is also employed to show cylinders of any length are stable to non-axisymmetric perturbations, as confirmed analytically.
We end by showing our results are independent of our choice of time step and mesh size, thus demonstrating the necessary convergence of numerical solutions.

%
% *** Stability of Axisymmetric Perturbations
%

\subsubsection{Stability of Axisymmetric Perturbations} \label{sec:sec_axi_perturbations}

To begin, the position of an unperturbed cylinder of radius $R$ and length $L$ is given by
\begin{equation} \label{eq:unperturbed_cylinder_position}
	\bmx (\theta, z)
	\, = \, R \, \bm{e}_r (\theta)
	\, + \, z \, \bm{e}_z
	~,
\end{equation}
with the polar angle
$\theta \in [0, 2\pi)$
and the axial length
$z \in [0, L]$.
For the boundary conditions
$\bmv = \bm{0}$
on both edges of the cylinder, a valid base state solution is given by
$\bmv = \bm{0}$
and
$\lambda = \lambda_0$
everywhere, as shown in Appendix~\ref{sec:sec_appendix_cylinder_stability}.
In this case the shape equation \eqref{eq:shape_eqn} simplifies to the Young--Laplace equation
$\lambda_0 = p \, R$,
with $p$ being the constant pressure drop imposed across the fluid surface.

\begin{figure}[!b]
	\centering
	\begin{subfigure}[b]{0.32\columnwidth}
		\centering
		\includegraphics[width=0.99\textwidth]{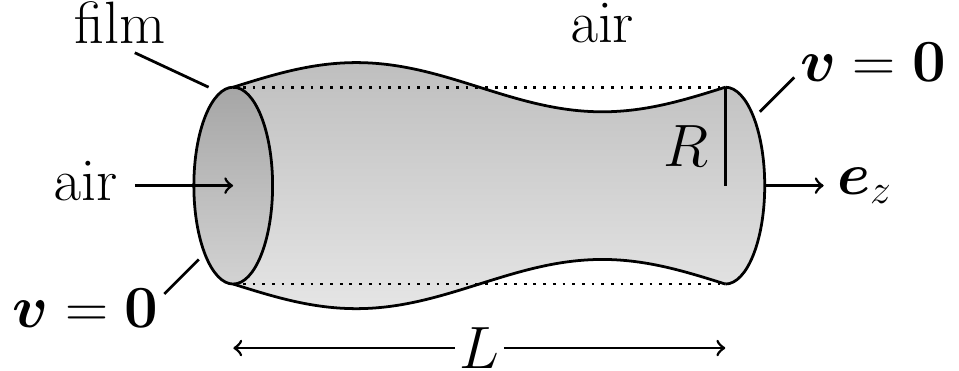}
		\vspace{-0pt}
		\caption{initial shape schematic}
		\label{fig:fig_ale_initial_cylinder_schematic}
	\end{subfigure}
	\hspace{4pt}
	\begin{subfigure}[b]{0.28\columnwidth}
		\centering
		\includegraphics[width=0.97\textwidth]{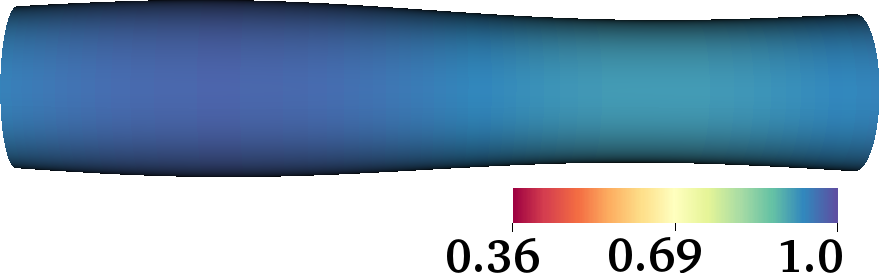}
		\vspace{14pt}
		\caption{fluid film at $t = 15$}
		\label{fig:fig_ale_cylinder_shape_early}
	\end{subfigure}
	\hspace{4pt}
	\begin{subfigure}[b]{0.28\columnwidth}
		\centering
		\includegraphics[width=0.97\textwidth]{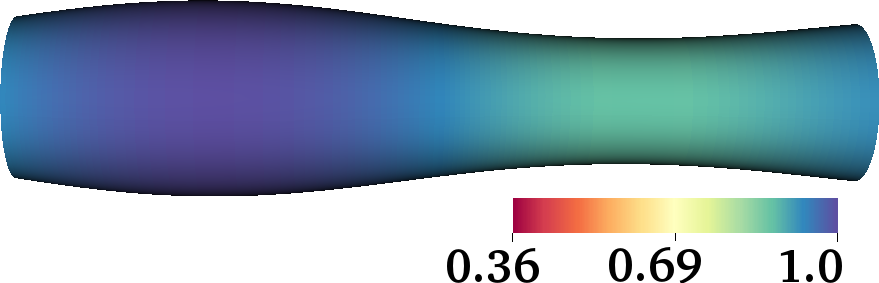}
		\vspace{13pt}
		\caption{fluid film at $t = 20$}
		\label{fig:fig_ale_cylinder_shape_20}
	\end{subfigure}
	\\[4pt]
	~\hspace{4pt}
	\begin{subfigure}[b]{0.28\columnwidth}
		\centering
		\includegraphics[width=0.97\textwidth]{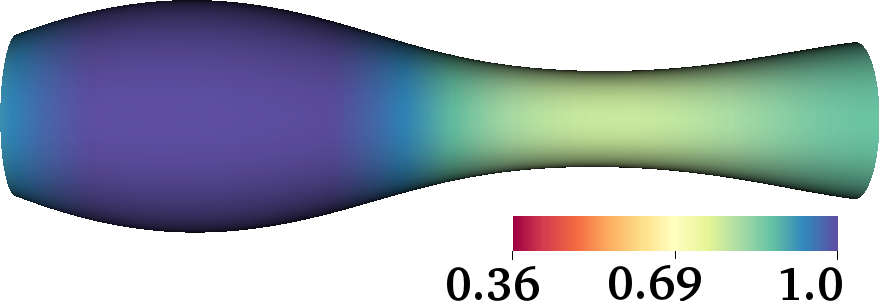}
		\vspace{7pt}
		\caption{fluid film at $t = 25$}
		\label{fig:fig_ale_cylinder_shape_25}
	\end{subfigure}
	\hspace{4pt}
	\begin{subfigure}[b]{0.28\columnwidth}
		\centering
		\includegraphics[width=0.97\textwidth]{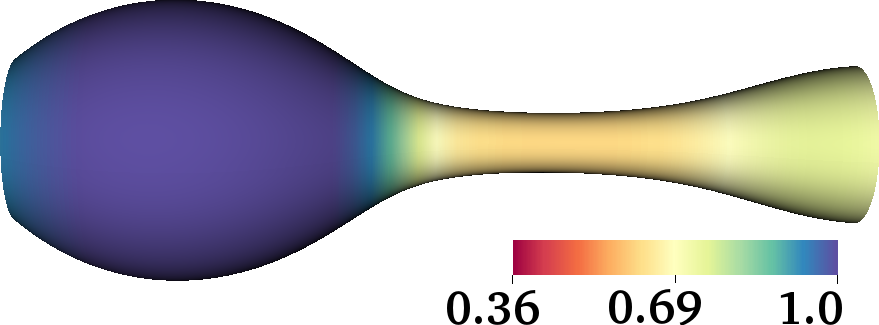}
		\vspace{7pt}
		\caption{fluid film at $t = 30$}
		\label{fig:fig_ale_cylinder_shape_30}
	\end{subfigure}
	\hspace{4pt}
	\begin{subfigure}[b]{0.28\columnwidth}
		\centering
		\includegraphics[width=0.97\textwidth]{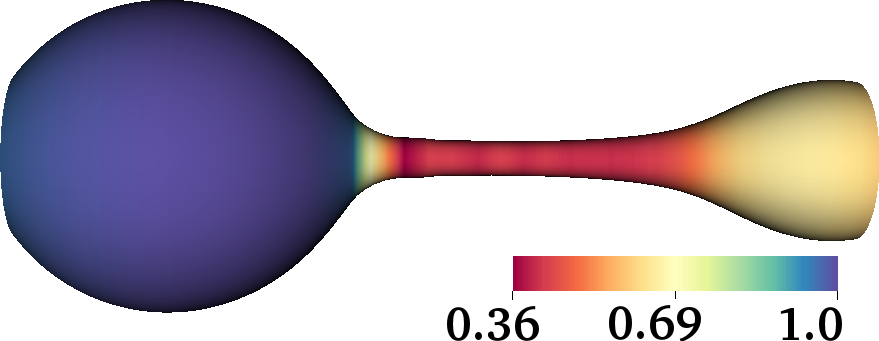}
		\vspace{5pt}
		\caption{fluid film at $t = 35$}
		\label{fig:fig_ale_cylinder_shape_late}
	\end{subfigure}
	\caption{
		Curved and deforming fluid film.
		(a) Schematic of the initial film shape, according to Eq.~\eqref{eq:ale_cylinder_initial_position}, where the maximum radius perturbation is 1\%.
		(b)--(f) Snapshots of an unstable and deforming film at $t = 15$ (b), $t = 20$ (c), $t = 25$ (d), $t = 30$ (e), and $t = 35$ (f), where the color indicates the surface tension.
		At early times (b), the shape perturbation results in surface tension gradients which drive a destabilizing flow.
		At late times (f), spherical bulbs appear on the ends of the cylinder.
		At all times, we enforce $\bmv = \bm{0}$ on both edges.
		A video of the instability can be found at 
		\href{https://youtu.be/FUx8fGXuzqY}{\texttt{youtu.be/FUx8fGXuzqY}}
		or supplemental movie \hyperref[sec:sec_appendix_movie_1]{E.1}.
		Simulation parameters are $R = 1$, $L = 10$, $\lambda_0 = 1$, $\zeta = 1$, $\Delta t = 0.1$, $h_\theta = \, ^1 \!/_{10}$, and $h_z = \, ^1 \!/_{40}$.
	}
	\label{fig:fig_ale_cylinder}
\end{figure}

At time $t = 0$, we perturb the radius of the stationary cylinder such that the initial position is given by
\begin{equation} \label{eq:ale_cylinder_initial_position}
	\bmx (\theta, z; \, t=0)
	= R \, \Big[
		1
		+ \epsilon \, \sin \Big(
			\dfrac{2 \pi z}{L}
		\Big)
	\Big] \, \bm{e}_r
	+ z \, \bm{e}_z
	~,
\end{equation}
where the small dimensionless parameter $\epsilon$ is set to $0.01$ in our simulations.
A schematic of the initial surface shape is shown in Fig.~\ref{fig:fig_ale_initial_cylinder_schematic}.
We evolve the fluid film from its initial state using our LE method and observe if it is stable or unstable with respect to the initial perturbation.
Over the course of our simulation, we maintain a constant pressure drop across the fluid surface.
This pressure drop enters as a body force
$\rho \bm{b} = p \bm{n}$, and its contribution to the tangent matrix and residual vector is provided in Appendix~\ref{sec:sec_appendix_normal_body_force}.
We find the cylinder is stable to the initial perturbation when its length $L < 2 \pi R$ and unstable when $L > 2 \pi R$.
In the latter case, the initial perturbation continues to grow and eventually reaches a configuration that has spherical bulbs on the two ends (Figs.~\ref{fig:fig_ale_cylinder_shape_early}--\ref{fig:fig_ale_cylinder_shape_late}, supplemental movie \hyperref[sec:sec_appendix_movie_1]{E.1}, video provided at
\href{https://youtu.be/FUx8fGXuzqY}{\texttt{youtu.be/FUx8fGXuzqY}}).
These spherical shapes are believed to form because they are the stable surfaces compatible with our boundary conditions and incompressibility constraint.
We proceed to use simple physical arguments to understand why the initial perturbation is unstable.

%
% *** Physical Explanation of the Instability
%

\subsubsection{Physical Explanation of the Instability} \label{sec:sec_physical_explanation_instability}

The instability arises because our initial shape perturbation changes the mean curvature of the surface, which in turn changes the surface tension through the Young--Laplace equation.
The resultant surface tension gradients then drive in-plane fluid flow, as can be seen from the in-plane equations~\eqref{eq:eom_in_plane}.
When
$L > 2 \pi R$,
fluid flows from the narrow region of the cylinder to the wide region (see Fig.~\ref{fig:fig_ale_initial_cylinder_schematic}), resulting in an unstable film.
However, when
$L < 2 \pi R$,
the surface flow is directed from the wide region to the narrow region, which causes the initial bulge to shrink over time such that the surface returns to its cylindrical configuration.

To understand this general idea in more detail, we begin with the Young--Laplace equation, written as
$\lambda = - p / (2 H)$.
For an unperturbed cylinder,
$H = -1 / (2 R)$.
The initial perturbation \eqref{eq:ale_cylinder_initial_position} alters the mean curvature of the film by modifying both radii of curvature: it changes the radius of a circular cross-section of the cylinder, and the sinusoidal shape along $z$ introduces a nonzero radius of curvature in the axial direction.
Analytically, the mean curvature $H$ of the initially perturbed shape \eqref{eq:ale_cylinder_initial_position} is calculated according to Sec.~\ref{sec:sec_surface_geometry} as
\begin{equation} \label{eq:ale_cylinder_mean_curvature}
	H
	= \dfrac{-1}{2R}
	+ \dfrac{\epsilon}{2 R} \bigg[
		1 - \Big(
			\dfrac{2 \pi R}{L}
		\Big)^{\! 2}
	\bigg] \, \sin \Big(
		\dfrac{2 \pi z}{L}
	\Big)
	~.
\end{equation}
Consider the quantity in square brackets in Eq.~\eqref{eq:ale_cylinder_mean_curvature}, which consists of two terms.
The first term comes from the change in the circular cross-section, while the second term, which contains a factor of $L^{-2}$, comes from the change in radius along the $z$-direction.
According to Eq.~\eqref{eq:ale_cylinder_mean_curvature}, when
$L > 2 \pi R$,
the mean curvature $H$ becomes less negative where the cylinder bulges outwards and more negative where it bulges inwards.
The Young--Laplace equation then requires $\lambda$ to become larger at the outward bulge and smaller at the inward bulge (see Fig.~\ref{fig:fig_ale_cylinder_shape_early}), and the in-plane equations \eqref{eq:eom_in_plane} indicate fluid flows from regions of low $\lambda$ to high $\lambda$.
As a result, fluid flows along the surface tension gradient which in this case causes the instability to grow (see Fig.~\ref{fig:fig_ale_cylinder_shape_late}).
When
$L < 2 \pi R$,
on the other hand, the effect on $\lambda$ is reversed and the in-plane flow serves to stabilize the perturbed cylindrical shape.

% *** start figure *** %
\begin{figure}[!b]
	\centering
	\includegraphics[width=0.40\textwidth]{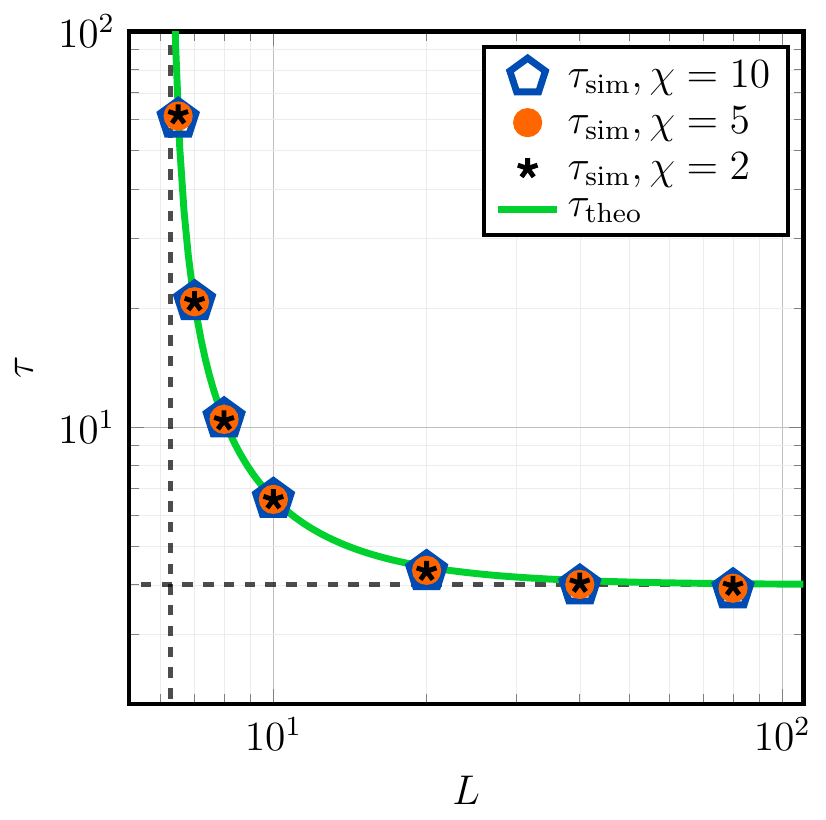}
	\caption{
		Curved and deforming fluid film: instability time scale.
		Plot of the time scale $\tau$ as a function of length $L$, where $R = 1$, $\lambda_0 = 1$, $\zeta = 1$, $\Delta t = 0.1$, $h_\theta = \, ^1/_{30}$, $h_z = \, ^1/_{60}$, and the initial film shape is given by Eq.~\eqref{eq:ale_cylinder_initial_position}.
		There is excellent agreement between the theoretical prediction (green curve, see Eq.~\ref{eq:ale_cylinder_perturbed_time_scale}) and the simulation results \eqref{eq:ale_cylinder_time_scale_result} for $\chi = 2$ (black stars), $\chi = 5$ (orange circles), and $\chi = 10$ (blue pentagons).
		The vertical dashed line shows the critical length $L = 2 \pi R$ above which cylindrical fluid films are unstable, and the horizontal dashed line indicates $\tau = 4 \zeta / \lambda_0$ when $L \rightarrow \infty$.
	}
	\label{fig:fig_ale_cylinder_deformation_time_study}
\end{figure}
% *** end figure *** %

%
% *** Instability Time Scale
%

\subsubsection{Instability Time Scale} \label{sec:sec_ale_instability_time_scale}

In addition to describing the instability with simple physical arguments, we performed a linear stability analysis and found the fluid film equations are indeed unstable to the initial perturbation \eqref{eq:ale_cylinder_initial_position} when
$L > 2 \pi R$
(see Appendix~\ref{sec:sec_appendix_cylinder_stability}).
Our analysis also revealed a theoretical time scale $\tau_{\mathrm{theo}}$ for the instability which, when
$L > 2 \pi R$,
is given by
\begin{equation} \label{eq:ale_cylinder_perturbed_time_scale}
	\tau_{\mathrm{theo}}
	=
	\Big(
		\dfrac{4 \, \zeta}{\lambda_0}
	\Big)
	\,
	\bigg[
		1
		- \Big(
			\dfrac{2 \pi R}{L}
		\Big)^{\! 2}
	\bigg]^{-1}
	~.
\end{equation}
Dimensionally, the ratio $\zeta / \lambda_0$ is expected to set the time scale of the deforming fluid film, as found in a study of topological transitions in two-dimensional dry foams \cite{durand-stone-prl-2006}.
Equation \eqref{eq:ale_cylinder_perturbed_time_scale} indicates
$\tau_{\mathrm{theo}} \rightarrow 4 \zeta / \lambda_0$
as
$L \rightarrow \infty$
and
$\tau_{\mathrm{theo}} \rightarrow \infty$
as
$L \rightarrow 2 \pi R^+$,
as shown by the green curve in Fig.~\ref{fig:fig_ale_cylinder_deformation_time_study}.

We also calculate the time scale $\tau_{\mathrm{sim}}$ from our full LE simulations as a function of length $L$ and compare it to the analytical prediction \eqref{eq:ale_cylinder_perturbed_time_scale}.
Assuming the unstable perturbation \eqref{eq:ale_cylinder_initial_position} initially grows exponentially in time, at small times the surface shape is given by
\begin{equation} \label{eq:ale_cylinder_early_position}
	\bmx (\theta, z, \, t)
	= R \, \big( \,
		1
		+ \epsilon \, \tilde{r} \, \mathrm{e}^{t / \tau_{\mathrm{sim}}} \,
	\big) \, \bm{e}_r
	+ z \, \bm{e}_z
	~,
\end{equation}
where
$\tilde{r} := \sin (2 \pi z / L)$
is defined for notational convenience.
We seek the time $t^*$ for which the initial instability has grown by the chosen multiplicative factor $\chi$, such that the surface position is given by
\begin{equation} \label{eq:ale_cylinder_early_position_chi}
	\bmx (\theta, z, \, t)
	= R \, \big( \,
		1
		+ \epsilon \, \chi \, \tilde{r} \,
	\big) \, \bm{e}_r
	+ z \, \bm{e}_z
	~.
\end{equation}
If $\chi = 2$, for example, we numerically measure the time $t^*$ when the initial perturbation doubles in size.
By comparing Eqs.~\eqref{eq:ale_cylinder_early_position} and \eqref{eq:ale_cylinder_early_position_chi}, the time scale $\tau_{\mathrm{sim}}$ can be numerically calculated as
\begin{equation} \label{eq:ale_cylinder_time_scale_result}
	\tau_{\mathrm{sim}}
	= \dfrac{t^*}{\ln \chi}
	~.
\end{equation}
We emphasize that in Eq.~\eqref{eq:ale_cylinder_time_scale_result}, $\chi$ is a chosen marker for the growth of the instability and $t^*$ is numerically measured to calculate $\tau_{\mathrm{sim}}$.
Figure \ref{fig:fig_ale_cylinder_deformation_time_study} shows the excellent agreement between numerically calculated \eqref{eq:ale_cylinder_time_scale_result} and analytical \eqref{eq:ale_cylinder_perturbed_time_scale} time scales, as a function of cylinder length, and demonstrates the simulations correctly predict the limiting time scale
$\tau = 4 \zeta / \lambda_0$
as
$L \rightarrow \infty$.

%
% *** Stability of Non-axisymmetric Perturbations
%

\subsubsection{Stability of Non-axisymmetric Perturbations} \label{sec:sec_non_axi_perturbations}

\begin{figure}[!b]
	\centering
	\begin{subfigure}[b]{0.29\columnwidth}
		\centering
		\includegraphics[width=0.97\textwidth]{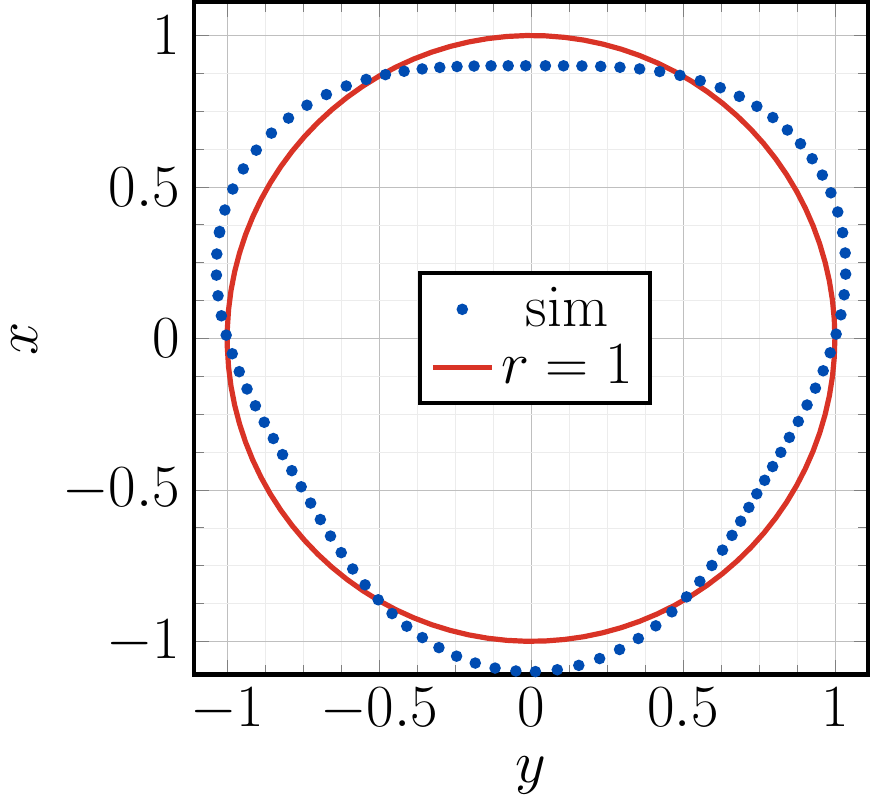}
		\caption{$t = 0$, $z=2.5$}
		\label{fig:fig_ale_cylinder_non_axi_a}
	\end{subfigure}
	\hspace{4pt}
	\begin{subfigure}[b]{0.29\columnwidth}
		\centering
		\includegraphics[width=0.97\textwidth]{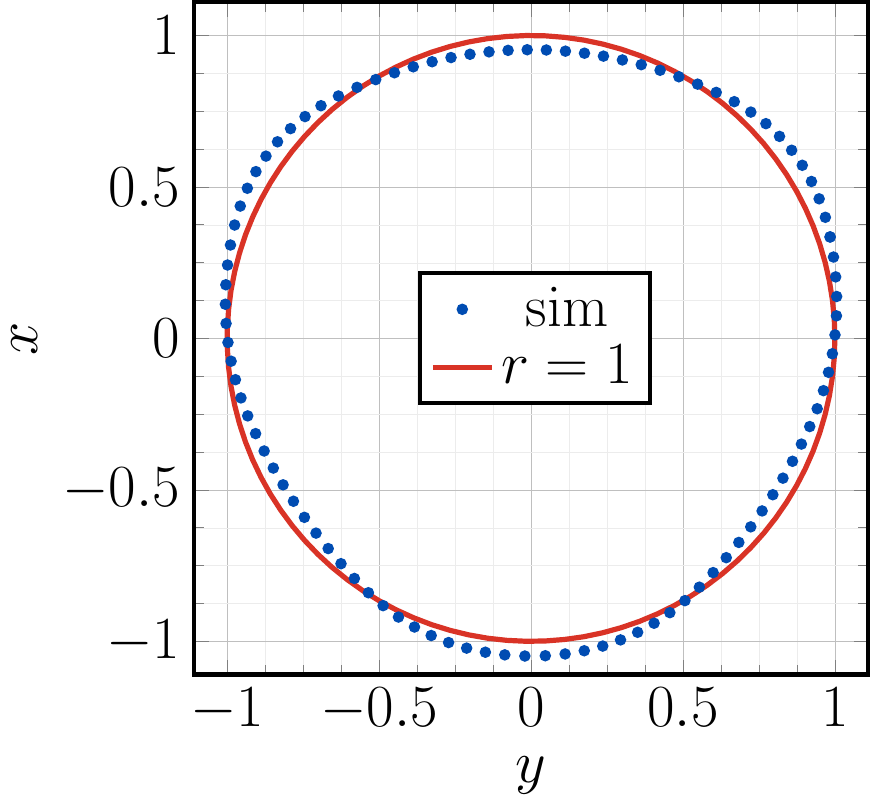}
		\caption{$t = 0.01$, $z=2.5$}
		\label{fig:fig_ale_cylinder_non_axi_b}
	\end{subfigure}
	\hspace{4pt}
	\begin{subfigure}[b]{0.29\columnwidth}
		\centering
		\includegraphics[width=0.97\textwidth]{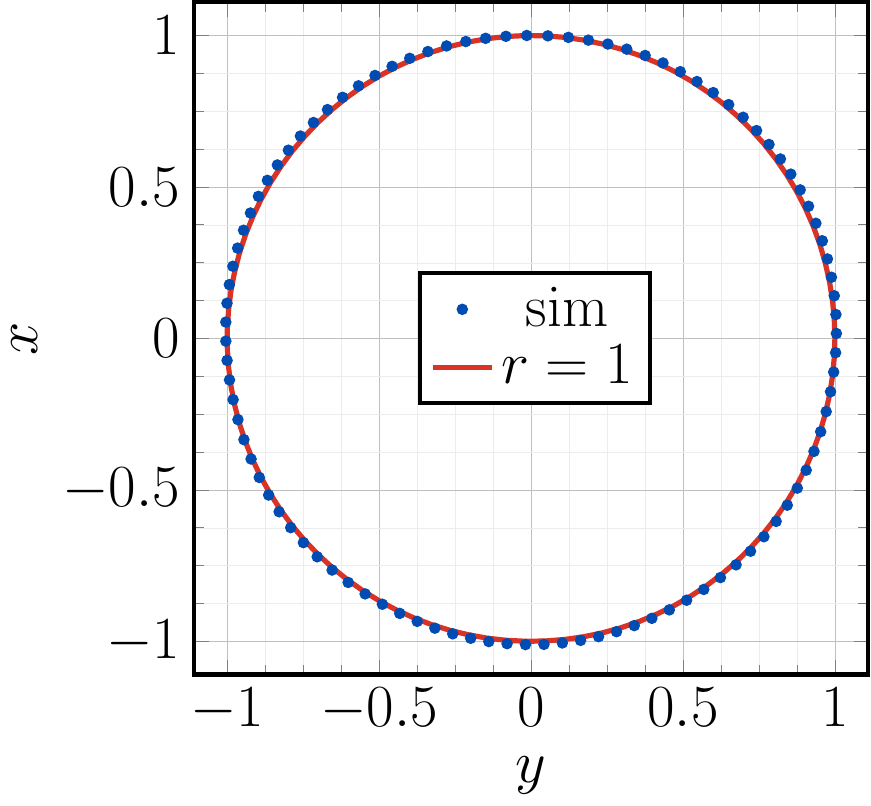}
		\caption{$t = 0.2$, $z=2.5$}
		\label{fig:fig_ale_cylinder_non_axi_c}
	\end{subfigure}
	\caption{
		Curved and deforming fluid film.
		The initial non-axisymmetric film shape is given by Eq.~\eqref{eq:ale_cylinder_initial_position_non_axi}, and the film evolves in time until it is once again axisymmetric.
		(a)--(c) cross-sections of the fluid film at $z = L/4$ show how the initial perturbation relaxes in time.
		At all times, we enforce $\bmv = \bm{0}$ on both ends.
		Simulation parameters are $R = 1$, $L = 10$, $\lambda_0 = 1$, $\zeta = 1$, $\Delta t = 0.01$, $h_\theta = \, ^1 \!/_{10}$, and $h_z = \, ^1 \!/_{40}$.
		We choose $\epsilon = 0.1$, rather than 0.01 as in all other cases, for ease in the visualization.
	}
	\label{fig:fig_ale_cylinder_non_axi}
\end{figure}

The linear stability analysis (Appendix~\ref{sec:sec_appendix_cylinder_stability}) also considered non-axisymmetric modes, and found the resultant surface tension gradients always stabilized the cylindrical configuration.
Thus, all non-axisymmetric modes are stable.
Here we consider the non-axisymmetric perturbation
\begin{equation} \label{eq:ale_cylinder_initial_position_non_axi}
	\begin{split}
		\bmx (\theta, z; \, t=0)
		\, &= \, R \, \Big[
			1
			+ \epsilon \, \sin \big(
				3 \theta
			\big) \, \sin \Big(
				\dfrac{2 \pi z}{L}
			\Big)
		\Big] \, \bm{e}_r
		+ z \, \bm{e}_z
		\\[3pt]
		\, &= \, R \, \Big(
			1
			+ \epsilon \, \tilde{r}
		\Big) \, \bm{e}_r
		+ z \, \bm{e}_z
		~,
	\end{split}
\end{equation}
where we redefine $\tilde{r}$ as
$\tilde{r} := \sin (3 \theta) \, \sin(2 \pi z / L)$
for notational convenience.
The corresponding mean curvature is calculated as
\begin{equation} \label{eq:ale_cylinder_mean_curvature_non_axi}
	H
	= \dfrac{-1}{2R}
	- \dfrac{\epsilon \, \tilde{r}}{2 R} \bigg[
		8
		+ \Big(
			\dfrac{2 \pi R}{L}
		\Big)^{\! 2}
	\bigg]
	~.
\end{equation}
According to Eq.~\eqref{eq:ale_cylinder_mean_curvature_non_axi}, if the cylinder bulges outwards
($\tilde{r} > 0$)
then $H$ is more negative and the surface tension decreases.
On the other hand, if the cylinder bulges inwards
($\tilde{r} < 0$)
then $H$ is less negative and the surface tension increases.
The resulting in-plane flow goes from the outward bulges to the inward bulges, and always stabilizes the film shape.
We confirm the theoretical result with our full non-axisymmetric simulations, as shown in Fig.~\ref{fig:fig_ale_cylinder_non_axi}, and note all non-axisymmetric perturbations are stable (see Appendix~\ref{sec:sec_appendix_cylinder_stability}).

% *** start figure *** %
\begin{figure}[p]
	\centering
	\begin{subfigure}[b]{0.27\columnwidth}
		\centering
		\includegraphics[width=0.99\textwidth]{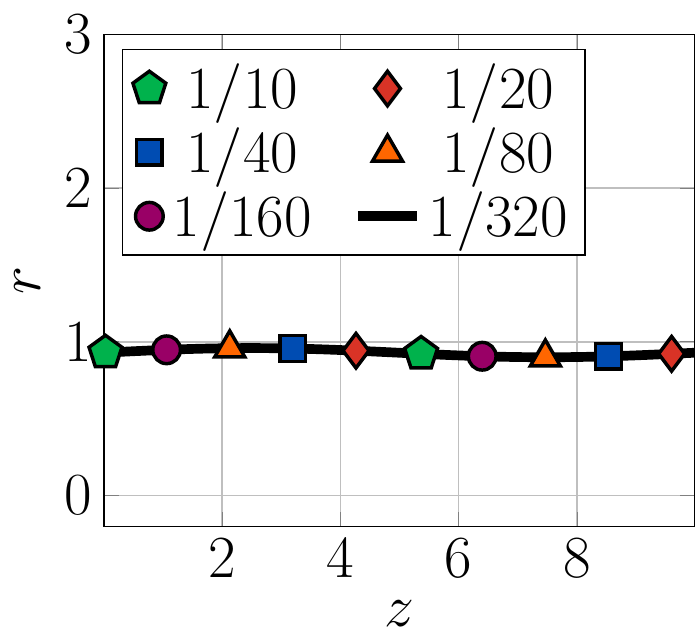}
		\label{fig:fig_ale_ts_r_1}
	\end{subfigure}
	\hspace{-20pt}
	\begin{subfigure}[b]{0.27\columnwidth}
		\centering
		\includegraphics[width=0.99\textwidth]{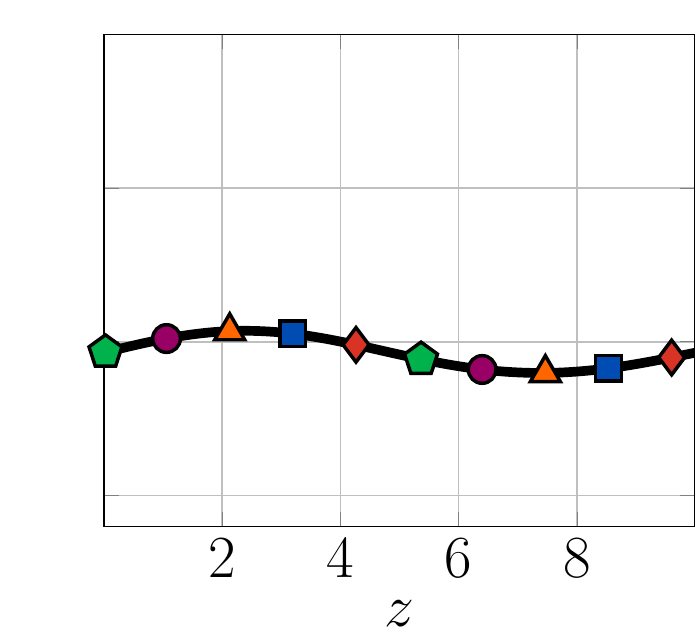}
		\label{fig:fig_ale_ts_r_3}
	\end{subfigure}
	\hspace{-20pt}
	\begin{subfigure}[b]{0.27\columnwidth}
		\centering
		\includegraphics[width=0.99\textwidth]{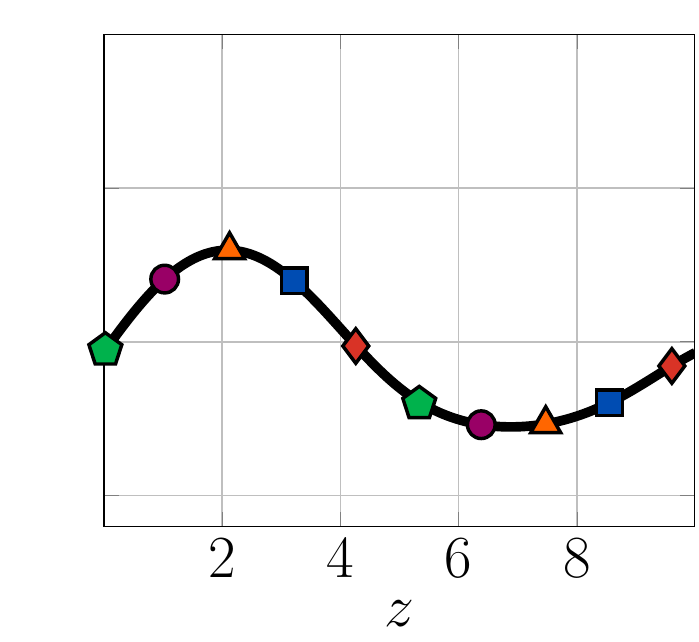}
		\label{fig:fig_ale_ts_r_5}
	\end{subfigure}
	\hspace{-20pt}
	\begin{subfigure}[b]{0.27\columnwidth}
		\centering
		\includegraphics[width=0.99\textwidth]{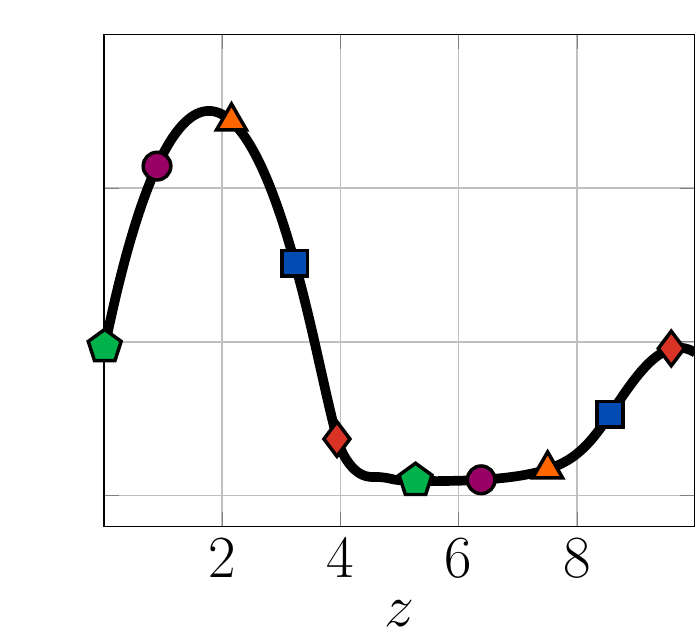}
		\label{fig:fig_ale_ts_r_7}
	\end{subfigure}
	\\
	\begin{subfigure}[b]{0.295\columnwidth}
		\centering
		\includegraphics[width=0.99\textwidth]{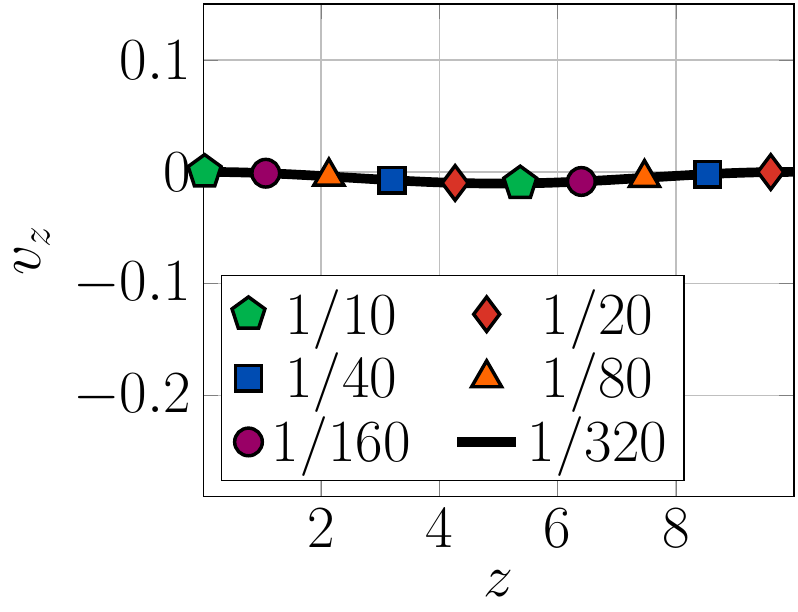}
		\label{fig:fig_ale_ts_vz_1}
	\end{subfigure}
	\hspace{-35pt}
	\begin{subfigure}[b]{0.295\columnwidth}
		\centering
		\includegraphics[width=0.99\textwidth]{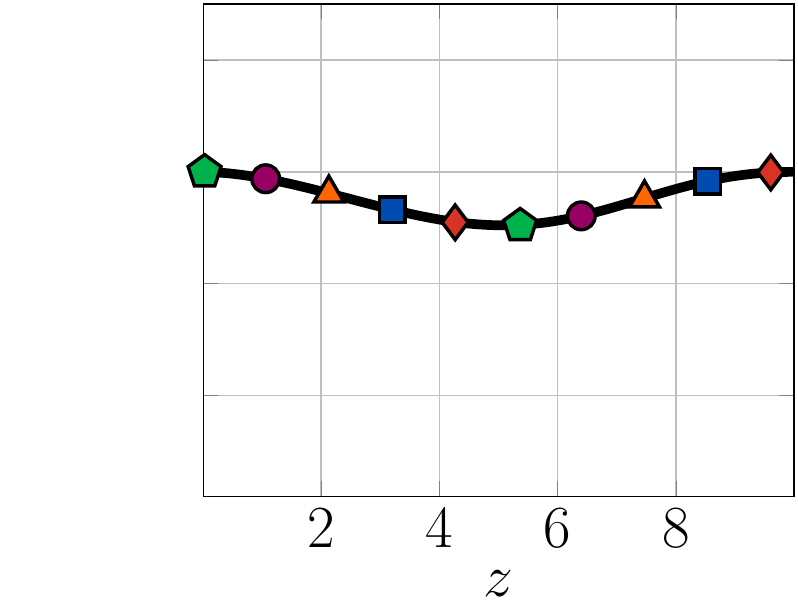}
		\label{fig:fig_ale_ts_vz_3}
	\end{subfigure}
	\hspace{-35pt}
	\begin{subfigure}[b]{0.295\columnwidth}
		\centering
		\includegraphics[width=0.99\textwidth]{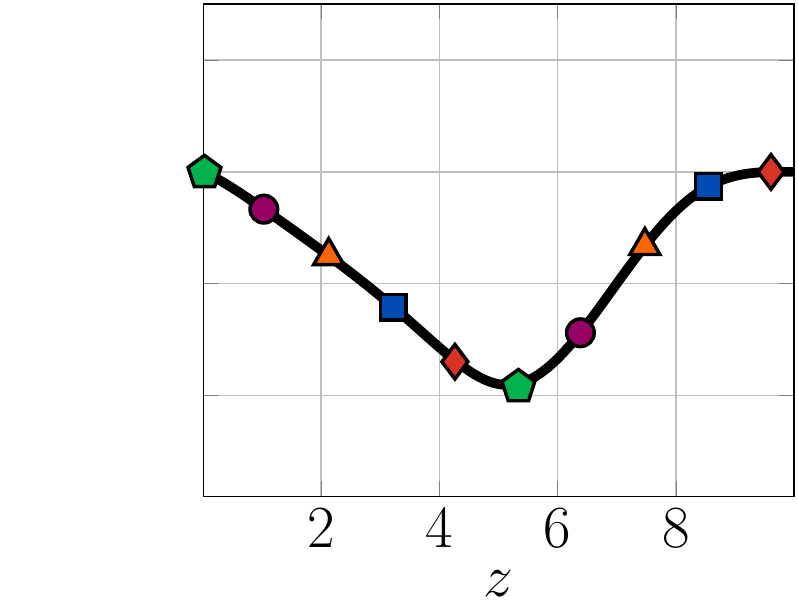}
		\label{fig:fig_ale_ts_vz_5}
	\end{subfigure}
	\hspace{-35pt}
	\begin{subfigure}[b]{0.295\columnwidth}
		\centering
		\includegraphics[width=0.99\textwidth]{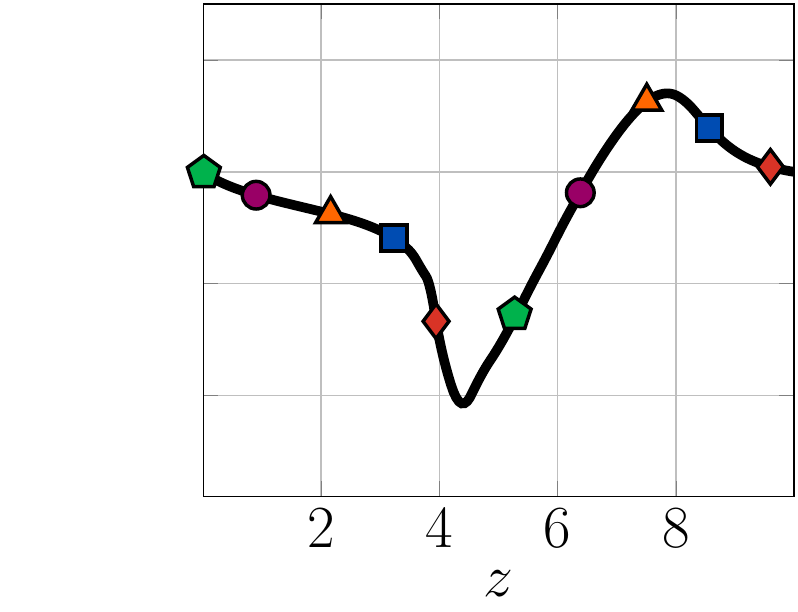}
		\label{fig:fig_ale_ts_vz_7}
	\end{subfigure}
	%\vspace{-13pt}
	\caption{
		Curved and deforming fluid film: results of time step refinement.
		We perturb a cylinder of length 10 and radius one, on a $30 \times 60$ mesh, where the time step $\Delta t$ ranges from $1/10$ to $1/320$ in powers of two (indicated in the legend).
		In all cases, the maximum initial radial perturbation is $\epsilon = 0.01$.
		Going from left to right, we plot the position (top row) and $z$-velocity (bottom row) at times $t$ = 5, 15, 25, and 35.
		In all cases, the different runs are visually indistinguishable.
	}
	\label{fig:fig_ale_ts}
\end{figure}
% *** end figure *** %
% *** start figure *** %
\begin{figure}[p]
	\centering
	\begin{subfigure}[b]{0.30\columnwidth}
		\centering
		\includegraphics[width=0.99\textwidth]{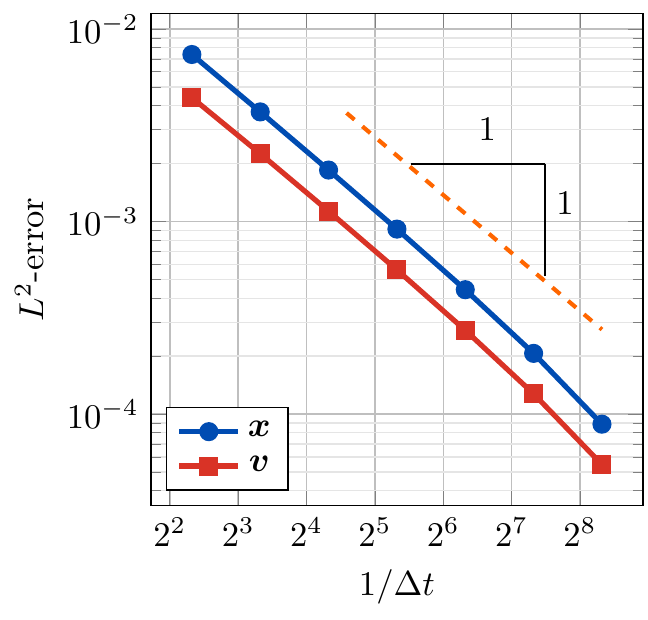}
		\caption{refinement in $\Delta t$}
		\label{fig:fig_ale_time_study_error}
	\end{subfigure}
	\hspace{4pt}
	\begin{subfigure}[b]{0.30\columnwidth}
		\centering
		\includegraphics[width=0.99\textwidth]{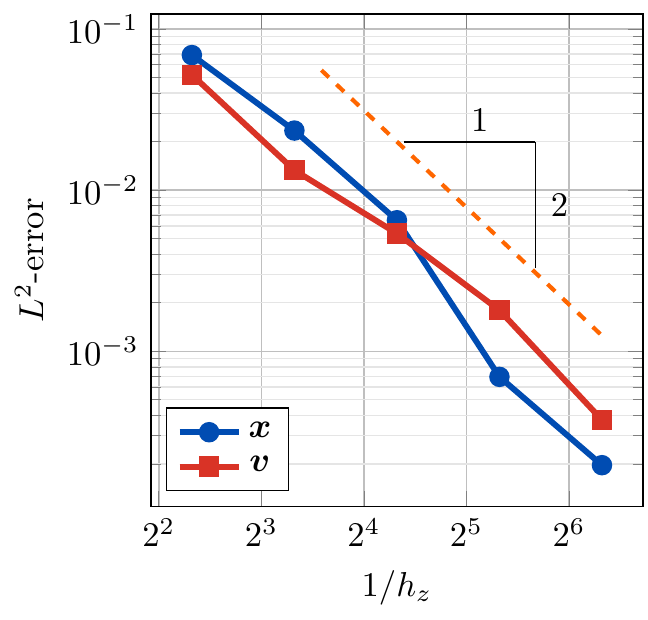}
		\caption{refinement in $z$}
		\label{fig:fig_ale_mesh_study_error_z}
	\end{subfigure}
	\hspace{4pt}
	\begin{subfigure}[b]{0.30\columnwidth}
		\centering
		\includegraphics[width=0.99\textwidth]{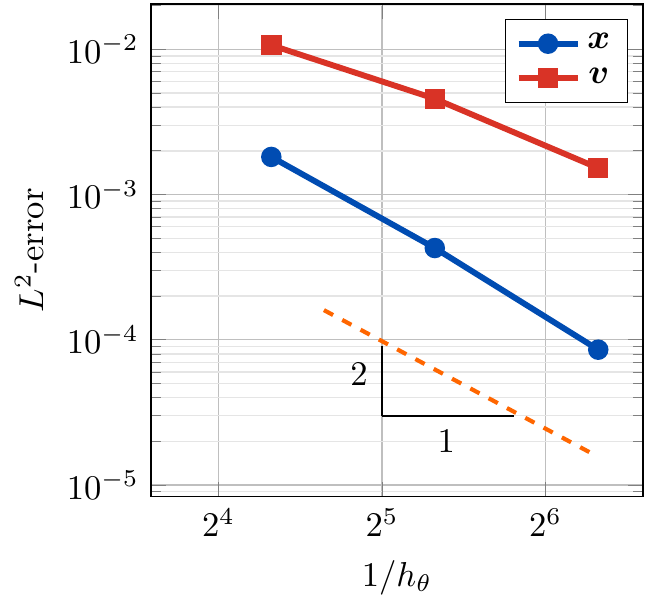}
		\caption{refinement in $\theta$}
		\label{fig:fig_ale_mesh_study_error_theta}
	\end{subfigure}
	\vspace{4pt}
	\caption{
		Curved and deforming fluid film: time step and mesh refinement.
		The $L^2$-error of the surface position (blue circles) and material velocity (red squares) of the fluid film at time $t = 35$, with simulation parameters $\epsilon = 0.01$, $R = 1$, $L = 10$, $\lambda_0 = 1$, and $\zeta = 1$.
		(a) Refining the time step $\Delta t$, on a $30 \times 60$ mesh, with true solution approximated as the $\Delta t = \, ^1/_{1280}$ simulation.
		(b) Refining the mesh width in the $z$-direction, $h_z$, with $\Delta t = 0.1$ and $h_\theta = \, ^1/_{20}$.
		The numerical result with $h_z = \, ^1/_{160}$ is treated as the true solution.
		(c) Refining the mesh width in the $\theta$-direction, $h_\theta$, with $\Delta t = 0.1$ and $h_z = \, ^1/_{40}$.
		The true solution is approximated as the simulation result with $h_\theta = \, ^1/_{160}$.
		In all cases, the position and velocity converge at the same rate: linearly with $\Delta t$ refinement, and quadratically on both $z$- and $\theta$-refinement.
	}
	\label{fig:fig_ale_error}
\end{figure}
% *** end figure *** %

%
% *** Time Step and Mesh Refinement
%

\subsubsection{Time Step and Mesh Refinement} \label{sec:sec_ale_refinement}

Finally, we study the convergence of our axisymmetric numerical results in three cases: refining
$\Delta t$, refining $h_z$, and refining $h_\theta$.
In each convergence study, all simulations are run until time $t = 35$, at which point the initially cylindrical film has undergone significant deformation.
Figure~\ref{fig:fig_ale_cylinder_shape_late} shows the shape of the fluid film at $t = 35$; Fig.~\ref{fig:fig_ale_ts} shows axial profiles of the radius and fluid $z$-velocity at different snapshots in time.
In the latter, the fluid film shape and velocity are visually indistinguishable for different choices of the time step, as in each case $\Delta t$ is considerably less than the time scale $\tau \approx 6.6$ seconds (Eq.~\eqref{eq:ale_cylinder_perturbed_time_scale} with $L = 10$).

The $L^2$-error is calculated for the fluid film position $\bmx$ and material velocity $\bmv$ at time $t = 35$, with the true solution being approximated as the finest simulation run (further details are provided in Fig.~\ref{fig:fig_ale_error}).
Refining the time step $\Delta t$ shows linear scaling in the position and material velocity, as shown in Fig.~\ref{fig:fig_ale_time_study_error}.
Both the position and velocity are expected to scale linearly because we used a backward Euler temporal discretization, in which the position is not a fundamental unknown but rather calculated from the mesh velocity according to Eq.~\eqref{eq:ale_mesh_position}.
The $L^2$-error scales quadratically for both $z$-refinement (Fig.~\ref{fig:fig_ale_mesh_study_error_z}) and $\theta$-refinement (Fig.~\ref{fig:fig_ale_mesh_study_error_theta}).
Given our implementation of $C^1$-continuous bi-quadratic velocities \eqref{eq:velocity_solution_space} and $C^0$-continuous bi-linear surface tensions \eqref{eq:lambda_solution_space} with uniform B-splines, the quadratic scaling on mesh refinement is also expected \cite{cottrell-iga}.
Thus, our LE simulation results demonstrate the anticipated convergence behavior upon both time step and mesh refinement.

%
% *** Lagrangian Implementation
%

\section{Lagrangian Implementation} \label{sec:sec_lagrangian}

\begin{figure}[!t]
	\vspace{-3pt}
	\centering
	\begin{subfigure}[b]{0.30\columnwidth}
		\centering
		\includegraphics[width=0.99\textwidth]{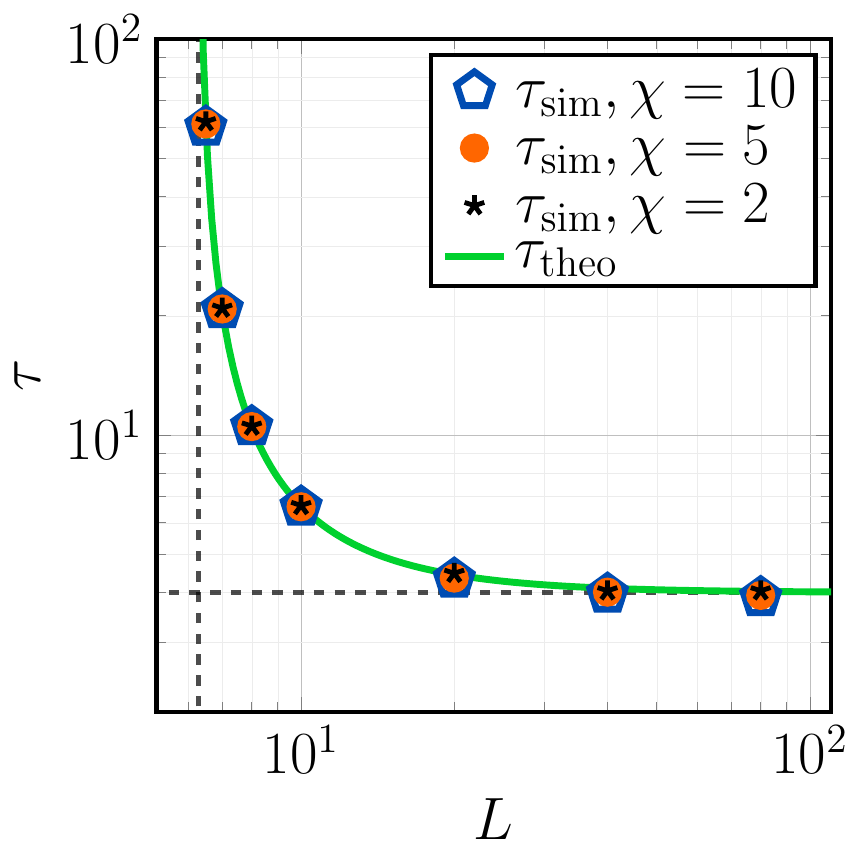}
		\caption{instability time scale}
		\label{fig:fig_lag_cylinder_deformation_time_study}
	\end{subfigure}
	~
	\begin{subfigure}[b]{0.33\columnwidth}
		\centering
		\includegraphics[width=0.99\textwidth]{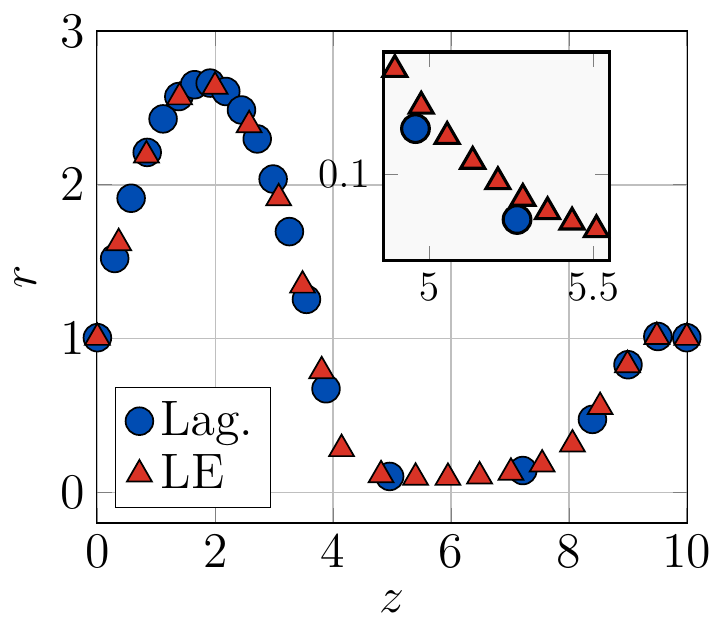}
		\caption{mesh spacing, $t = 35$}
		\label{fig:fig_lagrangian_eulerian_nodal_spacing}
	\end{subfigure}
	~
	\begin{subfigure}[b]{0.30\columnwidth}
		\centering
		\includegraphics[width=0.99\textwidth]{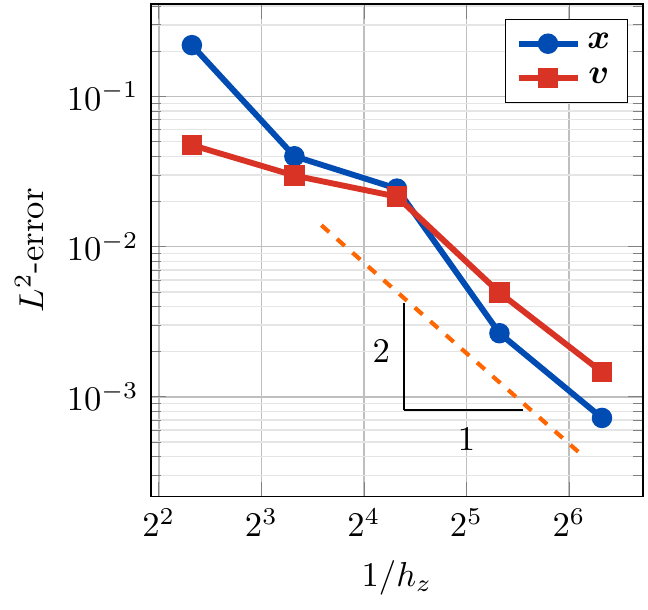}
		\caption{error comparison, $t = 35$}
		\label{fig:fig_lagrangian_eulerian_error}
	\end{subfigure}
	\\
	\vspace{6pt}
	\centering
	\begin{subfigure}[b]{0.31\columnwidth}
		\centering
		\includegraphics[width=0.99\textwidth]{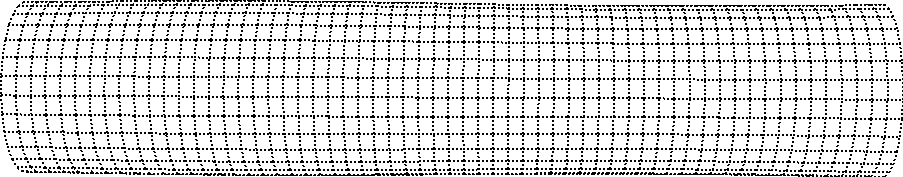}\\[28pt]
		\includegraphics[width=0.99\textwidth]{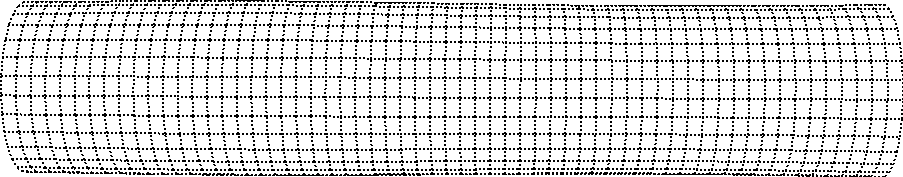}
		\vspace{-1pt}
		\caption{$t = 5$}
		\label{fig:fig_lag_ale_t_05}
	\end{subfigure}
	~
	\begin{subfigure}[b]{0.31\columnwidth}
		\centering
		\includegraphics[width=0.99\textwidth]{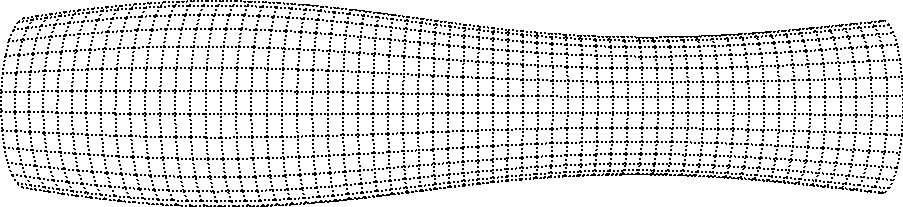}\\[23pt]
		\includegraphics[width=0.99\textwidth]{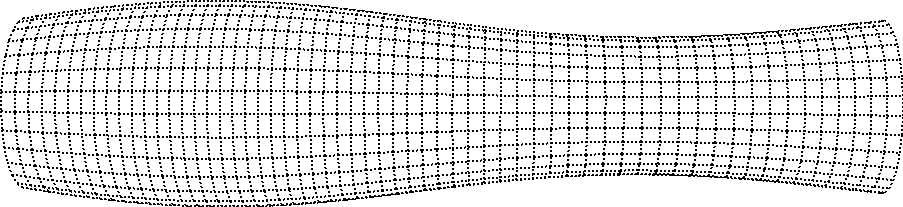}
		\vspace{-3pt}
		\caption{$t = 20$}
		\label{fig:fig_lag_ale_t_20}
	\end{subfigure}
	~
	\begin{subfigure}[b]{0.31\columnwidth}
		\centering
		\includegraphics[width=0.99\textwidth]{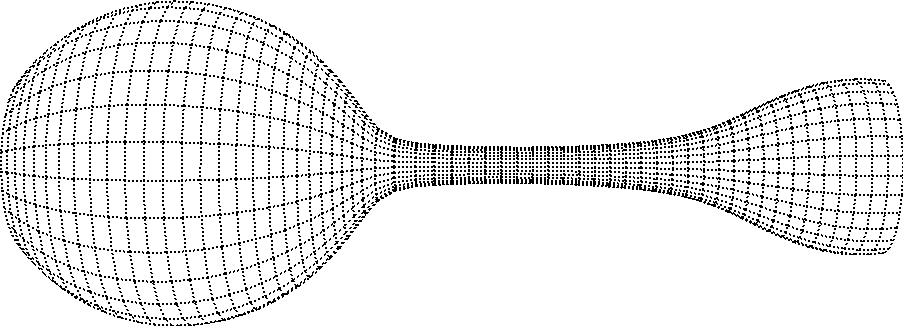}\\[5pt]
		\includegraphics[width=0.99\textwidth]{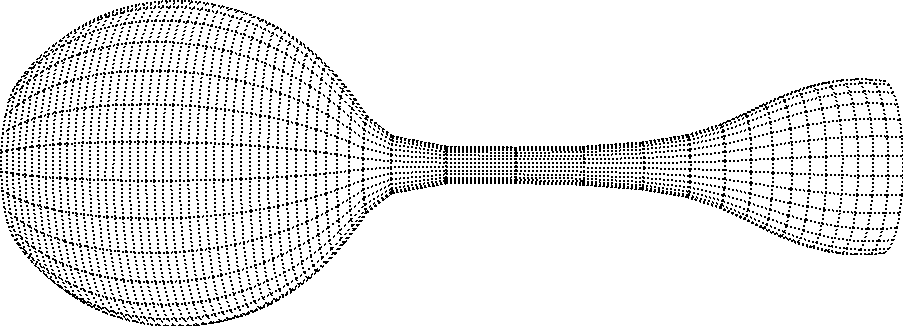}
		\caption{$t = 35$}
		\label{fig:fig_lag_ale_t_35}
	\end{subfigure}
	\caption{
		Numerical results from the Lagrangian implementation of an initially cylindrical, perturbed fluid film.
		(a) Instability time scale, reproducing Fig.~\ref{fig:fig_ale_cylinder_deformation_time_study} with a Lagrangian scheme (all simulation parameters are unchanged).
		Once again, the theoretical time scale calculation of Eq.~\eqref{eq:ale_cylinder_perturbed_time_scale}, depicted by the green curve, agrees quantitatively with the simulation results---calculated according to Eq.~\eqref{eq:ale_cylinder_time_scale_result}, and shown for
		$\chi = 2$ (black stars),
		$\chi = 5$ (orange circles), and
		$\chi = 10$ (blue pentagons).
		(b) Lagrangian and LE mesh locations as a function of axial length $z$ at time
		$t = 35$,
		depicted respectively by blue circles and red triangles.
		As the solution is axisymmetric, only radial mesh positions are provided.
		In both cases, a $20 \times 160$ mesh is used and only one in every eight mesh points is shown.
		(inset) Every mesh point in the range
		$4.9 \le z \le 5.5$,
		highlighting the lack of Lagrangian mesh points in this region.
		(c) Lagrangian $L^2$-error of the surface position (blue circles) and material velocity (red squares) of the fluid film at time
		$t = 35$,
		upon $z$-refinement.
		The error converges quadratically, as in the identical LE simulations (Fig.~\ref{fig:fig_ale_mesh_study_error_z}), however Lagrangian errors are $\sim 5$ times larger than their LE counterparts.
		(d)--(f) LE (top) and Lagrangian (bottom) simulations for a $30 \times 60$ mesh at times $t = 5$ (d), $t = 20$ (e), and $t = 35$ (f), with only the mesh shown to further demonstrate the nodal spacing.
		The Lagrangian mesh flows in-plane as the surface deforms, such that the spherical bulbs are more spatially resolved than the central tube.
		As a result, the LE mesh more accurately resolves the fluid film shape throughout the simulation (supplemental movie \hyperref[sec:sec_appendix_movie_3]{E.3}, see video at
		\href{https://youtu.be/wnXuK6d3WTQ}{\texttt{youtu.be/wnXuK6d3WTQ}}).
	}
	\label{fig:fig_lag_results}
\end{figure}

As previously mentioned, our ALE formulation accommodates different mesh velocity equations.
In this section, a Lagrangian scheme is implemented to demonstrate the generality of our formulation.
First, the lid-driven cavity problem is reexamined, which demonstrates how Lagrangian methods are unsuitable for problems involving shear flows.
The Lagrangian implementation is then used to model the unstable cylindrical fluid film described in Sec.~\ref{sec:sec_curved_deforming_films}, and is shown to correctly capture the time scale associated with the instability.
However, in the Lagrangian simulations, mesh nodes move due to in-plane fluid flow.
As a consequence, at later times certain regions of the surface contain only a few nodes and are spatially unresolved.
Thus, Lagrangian simulations have errors larger than those of their LE counterparts.

Our finite element framework easily accommodates a Lagrangian formulation, as a Lagrangian scheme is recovered when
\begin{equation} \label{eq:lagrangian_mesh_velocity_main}
	\bmvm
	\, - \, \bmv
	\, = \, \bm{0}
	~,
\end{equation}
which replaces Eq.~\eqref{eq:ale_mesh_velocity_eulerian_equation} as the strong form of the mesh equation.
Rather than satisfying Eq.~\eqref{eq:lagrangian_mesh_velocity_main} weakly, we simply enforce the velocity and mesh velocity degrees of freedom to be equal in our discretized system of equations at every Newton--Raphson step.
Our numerical implementation is described in Appendix~\ref{sec:sec_lagrangian_finite_element_formulation}.

A Lagrangian formulation is unsuitable for even simple situations with in-plane velocity gradients.
For example, a Lagrangian scheme fails to reproduce the lid-driven cavity result of Sec.~\ref{sec:sec_flat_cavity} (supplemental movie \hyperref[sec:sec_appendix_movie_2]{E.2}, video at 
\href{https://youtu.be/FCoShaa_FhM}{\texttt{youtu.be/FCoShaa\_FhM}}).
We thus analyze our Lagrangian implementation by simulating a perturbed cylindrical fluid film, and comparing the results to those of the LE simulations, found in Sec.~\ref{sec:sec_curved_deforming_films}.
We first repeat the calculation of the deformation time scale (see Sec.~\ref{sec:sec_ale_instability_time_scale}), and find excellent agreement between theory and Lagrangian simulations (Fig.~\ref{fig:fig_lag_cylinder_deformation_time_study}).
Next, the distribution of nodes at time
$t = 35$
are compared for both the Lagrangian and LE implementations.
While the LE nodes maintain roughly equal spacing, the Lagrangian nodes move with the in-plane flow and become more concentrated in the spherical bulbs (Fig.~\ref{fig:fig_lagrangian_eulerian_nodal_spacing}).
Relative to the LE results, in Lagrangian simulations the central tubular region is spatially underresolved at later times
(Figs.~\ref{fig:fig_lag_ale_t_05}--\ref{fig:fig_lag_ale_t_35},
supplemental movie \hyperref[sec:sec_appendix_movie_3]{E.3},
video at \href{https://youtu.be/wnXuK6d3WTQ}{\texttt{youtu.be/wnXuK6d3WTQ}}).
Accordingly, Lagrangian simulations have comparatively larger errors upon $z$-refinement (Fig.~\ref{fig:fig_lagrangian_eulerian_error}, c.f. Fig.~\ref{fig:fig_ale_mesh_study_error_z}).

%
% *** Conclusions
%

\section{Conclusions} \label{sec:sec_conclusions}

In this paper, we developed an ALE theory and formulation for two-dimensional materials which are arbitrarily curved, may be deforming over time, and exhibit in-plane fluidity.
Within the setting of differential geometry, we introduced a new surface parametrization to endow the surface with a mesh whose in-plane motion can be specified arbitrarily.
In particular, the mesh need not convect with the in-plane material velocity, and thus our framework does not suffer from the limitations of a Lagrangian formulation.
With the ALE theory, and the results of a previous irreversible thermodynamic study~\cite{sahu-mandadapu-pre-2017}, we used the standard tools of finite element analysis to devise an isoparametric, fully implicit finite element method for two-dimensional curved and deforming fluid films.
In particular, we
(i) developed a weak formulation of the problem,
(ii) spatially discretized it with a standard Bubnov--Galerkin approximation,
(iii) temporally discretized the residual equations with the backward Euler method, and
(iv) linearized the resulting system of equations using the Newton--Raphson method.
We also highlighted our use and implementation of the Dohrmann--Bochev method, which overcame the LBB instability.
Finally, we provided several numerical examples to showcase the merits of our method.
We showed how our formulation is easily adapted to solve for fixed-surface flows, with the added benefit of determining the pressure drop required to constrain the fluid to the surface.
We also used both LE and pure Lagrangian numerical implementations to model cylindrical films which can deform over time.
We found such films are unstable above a critical length, and demonstrated that our numerical results are in agreement with the predictions of a linear stability analysis.

Our main motivation behind this work is to develop a robust, theoretically sound, and sufficiently general method for modeling lipid membranes.
To this end, we intend to implement a mesh velocity scheme which is not simply in-plane Eulerian \eqref{eq:ale_mesh_velocity_eulerian_equation} or Lagrangian \eqref{eq:lagrangian_mesh_velocity_main} in our future work---which, in principle, should overcome the squeezing of the surface-fixed coordinate nodes depicted in Fig.~\ref{fig:fig_theta_velocity}.
Furthermore, as fluid films are a computational precursor to lipid membranes, we plan to extend our implementation to model single-component lipid membranes by adding bending contributions to the weak formulation.
This analysis will be presented as Part II of the current work \cite{sahu-mandadapu-ale-ii}.
With a numerical method to simulate single-component lipid membranes, our framework can be extended to describe various phenomena governing biological membranes---such as the coupling between the bulk fluid surrounding the membrane and intramembrane lipid flows~\cite{gross-atzberger-jcp-2018} and in-plane species diffusion~\cite{sigurdsson-atzberger-sm-2016}.
Moreover, one could analyze multi-component systems, which consist of different lipid and transmembrane proteins.
Recent microscopic studies demonstrate how protein--lipid and protein--protein interactions play an important role in membrane bending \cite{brannigan-bpj-2007, west-bpj-2009, blood-voth-pnas-2006, atzberger-jcp-2013} and lipid domain formation \cite{katira-elife-2016}; experiments also demonstrate the coupling between lipid phase separation and membrane bending \cite{baumgart-ncomm-2015}.
Modeling such phenomena~\cite{wang-du-jmb-2008} would be a natural extension of our work.

%
% *** Acknowledgements
%

%\vfill
\section*{Acknowledgements} \label{sec:sec_acknowledgements}

We are indebted to Prof.~Panos Papadopoulos for discussions on LBB conditions, and for bringing the Dohrmann--Bochev method to our attention.
We thank Michelle Liu for assistance in visualizing our numerical results, Jo\"{e}l Tchoufag for many useful discussions, and Prof.~Howard Stone for helpful comments and suggestions.
A.S.~would also like to thank Prof.~Phil Colella for his \texttt{C++} makefile and suggested coding best practices.

A.S.~is supported by the Computational Science Graduate Fellowship from the U.S. Department of Energy, as well as U.C. Berkeley.
Y.A.D.O.~would like to thank the support of U.C. Berkeley.
R.A.S.~acknowledges the support of the German Research Foundation (DFG) through Grant No. GSC 111, as well as the Aachen--California Network of Academic Exchange.
K.K.M.~acknowledges the support of U.C. Berkeley and U.S.\ Department of Energy Contract No.\ DE-AC02-05CH11231, FWP no.\ CHPHYS02.

	%
	% *** APPENDIX
	%

	\begin{appendices}
		\numberwithin{equation}{section}

%
% *** Analytical Fluid Film Calculations
%

\section{Analytical Fluid Film Calculations} \label{sec:sec_appendix_calculations}

In this appendix, we provide several theoretical calculations for two-dimensional curved and deforming fluid films.
First, we determine the material stresses using an irreversible thermodynamic framework~\cite{sahu-mandadapu-pre-2017}.
The Helmholtz free energy density for a two-dimensional fluid film is calculated by using a Lagrange multiplier $\lambda$ to enforce areal incompressibility.
The in-plane and out-of-plane equations of motion of such a film are then determined, and simplified in three cases: a completely flat film, a cylindrical film, and a cylindrical film with a bulge in the center.
Finally, a linear stability analysis is performed on an initially cylindrical fluid film that is allowed to deform, which is found to be unstable when its length exceeds its circumference.
We end by calculating the time scale associated with the instability.

%
% *** Helmholtz Free Energy Density
%

\subsection{Irreversible Thermodynamics and Material Stresses} \label{sec:sec_appendix_irrev_thermo}

We begin by briefly overviewing how the stresses of a general two-dimensional material are obtained via the balance law formulation and irreversible thermodynamic framework of Ref.~\cite{sahu-mandadapu-pre-2017}.
As discussed in Sec.~\ref{sec:sec_strong_form_fluid}, the stress vectors $\bm{T}^\calpha$ are the boundary tractions across curves of constant $\zeta^\alpha$.
Without loss of generality, the stress vectors can be decomposed in the $\{ \bm{a}_\calpha, \bm{n} \}$ basis as
\begin{equation} \label{eq:appendix_stress_vector}
	\bm{T}^\calpha
	= \, N^{\calpha \cbeta} \bm{a}_\cbeta
	\, + \, S^\calpha \bm{n}
	~,
\end{equation}
where $N^{\calpha \cbeta}$ and $S^\calpha$ are the in-plane and out-of-plane (shear) traction components, respectively.
An angular momentum balance reveals there are two conditions to be satisfied in relating $N^{\calpha \cbeta}$ and $S^\calpha$ to the couple-stress components $M^{\calpha \cbeta}$---the latter of which are related to the couples acting on the edge of a surface patch (see Ref.~\cite{sahu-mandadapu-pre-2017} for more details).
The first condition requires
\begin{equation} \label{eq:appendix_angular_momentum_sigma}
	\sigma^{\calpha \cbeta}
	:= N^{\calpha \cbeta}
	- b^\cbeta_\cmu \, M^{\cmu \calpha}
\end{equation}
to be symmetric, an analogous condition to the symmetry of the stress tensor for bulk systems.
Physically, $\sigma^{\calpha \cbeta}$ contains the couple-free components of the in-plane stresses.
The second condition requires
\begin{equation} \label{eq:appendix_angular_momentum_s}
	S^\calpha
	= - M^{\cbeta \calpha}_{; \cbeta}
	~,
\end{equation}
which indicates gradients of the moments lead to shear forces, in analogy to well-known beam bending examples in which the shear force is proportional to the spatial derivative of the moment.
For two-dimensional surfaces which cannot sustain moments, such as fluid films, the couple-stress components
$M^{\calpha \cbeta} = 0$,
such that $N^{\calpha \cbeta}$ is symmetric and the shear forces
$S^\calpha = 0$.

To determine the stresses in a general two-dimensional material, an irreversible thermodynamic framework is employed.
We introduce a Helmholtz free energy per unit mass, $\psi$, for the material, which as a thermodynamic state function captures only the reversible behavior of the material.
For the two-dimensional materials of interest, we assume $\psi$ depends on only the covariant metric $a_{\calpha \cbeta}$, the curvature components $b_{\calpha \cbeta}$, and the temperature
$T = T(\zeta^\alpha, \, t)$
with the functional dependence
\begin{equation} \label{eq:appendix_helmholtz_general}
	\psi
	= \psi (a_{\calpha \cbeta}, \, b_{\calpha \cbeta}, \, T)
	~.
\end{equation}
With a general form of the Helmholtz free energy density satisfying Eq.~\eqref{eq:appendix_helmholtz_general}, the irreversible thermodynamic framework developed in Ref.~\cite{sahu-mandadapu-pre-2017} shows that in the linear irreversible regime, the couple-free stress components $\sigma^{\calpha \cbeta}$ and the couple-stress components $M^{\calpha \cbeta}$ are given by
\begin{equation} \label{eq:appendix_stress_general}
	\sigma^{\calpha \cbeta}
	= \rho \Big(
		\dfrac{\partial \psi}{\partial a_{\calpha \cbeta}}
		+ \dfrac{\partial \psi}{\partial a_{\cbeta \calpha}}
	\Big)
	+ \pi^{\calpha \cbeta}
	\qquad \quad
	\text{and}
	\qquad \quad
	M^{\calpha \cbeta}
	= \dfrac{\rho}{2} \Big(
		\dfrac{\partial \psi}{\partial b_{\calpha \cbeta}}
		+ \dfrac{\partial \psi}{\partial b_{\cbeta \calpha}}
	\Big)
	+ \omega^{\calpha \cbeta}
	~.
\end{equation}
In Eq.~\eqref{eq:appendix_stress_general}, the first terms on the right-hand side provide the elastic contribution to the stresses and couple-stresses from the free energy $\psi$, while $\pi^{\calpha \cbeta}$ and $\omega^{\calpha \cbeta}$ describe the stresses due to in-plane and out-of-plane dissipative phenomena, respectively.
In all cases considered, we assume there is no dissipation from rates of change of curvature, such that
$\omega^{\calpha \cbeta} = 0$.
Furthermore, for isotropic and incompressible materials with in-plane fluidity, the viscous stresses are given by Eq.~\eqref{eq:viscous_stresses}.
Note that studies based on the Euler--Lagrange equations for lipid membranes do not contain forces resulting from the viscous contribution $\pi^{\calpha \cbeta}$ to the stress components $\sigma^{\calpha \cbeta}$.

At this point, we have a general method for determining the governing equations for an incompressible, two-dimensional material with in-plane fluidity.
By choosing the form of the Helmholtz free energy density $\psi$, we calculate the stress and couple-stress components according to Eq.~\eqref{eq:appendix_stress_general}, determine the components of the tractions~\eqref{eq:appendix_stress_vector} through Eqs.~\eqref{eq:appendix_angular_momentum_sigma} and \eqref{eq:appendix_angular_momentum_s}, and substitute the tractions into the linear momentum balance \eqref{eq:strong_equations_of_motion}.
In what follows, we determine the Helmholtz free energy density for a fluid film, and then carry out such a procedure.

%
% *** Helmholtz Free Energy Density
%

\subsubsection{Helmholtz Free Energy Density} \label{sec:sec_appendix_helmholtz_fluid}

In this section, we calculate the Helmholtz free energy density of an incompressible fluid film, which is a constraint enforcing areal incompressibility through the Lagrange multiplier $\lambda$.
To this end, we introduce a reference configuration of the surface, write the total surface energy such that there is no area change between the current and reference configurations, and then determine the form of the Helmholtz free energy density $\psi$.

Consider an arbitrarily curved patch of surface $\scp$ which is deforming over time, and whose position is given by $\check{\bmx} (\zeta^\alpha, t)$.
At some time $t_0$, the reference patch $\scpo$ is specified by its position $\check{\bmx}_0$, defined to be
\begin{equation} \label{eq:appendix_reference_surface_position}
	\check{\bmx}_0
	:= \check{\bmx} (\zeta^\alpha, t_0)
	~.
\end{equation}
The reference patch $\scpo$ is static and its properties by definition do not change over time.
The areal mass density of the reference patch is denoted $\rho_0$, and the Jacobian determinant $J$ between the reference and current patch is given by
$J := \sqrt{ \rule{0mm}{1.30ex} } \overline{ \det \, a_{\calpha \cbeta} } $\,.
A differential areal element $\mathrm{d}A$ on the reference patch is related to the corresponding areal element $\mathrm{d}a$ on the current patch according to
$\mathrm{d}a = J \, \mathrm{d}A$.
The global form of the conservation of mass can be written as
\begin{equation} \label{eq:appendix_global_mass_balance}
	\int_{\scpo}
		\rho_0
	~\mathrm{d}A
	= \int_\scp
		\rho
	~\mathrm{d}a
	~,
	\hspace{20pt}
	\text{which implies}
	\hspace{20pt}
	\int_{\scpo}
		\rho_0
	~\mathrm{d}A
	= \int_{\scpo}
	\rho \, J
	~\mathrm{d}A
	~.
\end{equation}
As the reference patch $\scpo$ is stationary, Eq.~\eqref{eq:appendix_global_mass_balance} shows
\begin{equation} \label{eq:appendix_local_mass_balance}
	\rho_0
	= \rho \, J
	~.
\end{equation}

Areal incompressibility is enforced through the equation $J = 1$, which indicates areas do not change~\eqref{eq:appendix_local_mass_balance}.
We introduce the Helmholtz free energy of the patch, $W$, as
\begin{equation} \label{eq:appendix_total_helmholtz}
	W
	= \int_{\scpo}
		\lambda \, \big(
			J
			- 1
		\big)
	~\mathrm{d}A
	~,
\end{equation}
where $\lambda$ is a Lagrange multiplier enforcing the incompressibility constraint
$J = 1$.
As $\psi$ is the Helmholtz free energy per unit mass, the total Helmholtz free energy of the patch is also given by
\begin{equation} \label{eq:appendix_total_helmholtz_area}
	W
	= \int_\scp
		\rho \, \psi
	~\mathrm{d}a
	= \int_{\scpo}
		\rho \, \psi \, J
	~\mathrm{d}A
	= \int_{\scpo}
		\rho_0 \, \psi
	~\mathrm{d}A
	~.
\end{equation}
In Eq.~\eqref{eq:appendix_total_helmholtz_area}, we used the relation
$\mathrm{d}a = J \, \mathrm{d}A$
to map the integral to the reference patch, and then substituted Eq.~\eqref{eq:appendix_local_mass_balance}.
Due to the arbitrariness of $\scpo$, the integrands of Eq.~\eqref{eq:appendix_total_helmholtz} and the right-hand side of Eq.~\eqref{eq:appendix_total_helmholtz_area} are equal, and thus
\begin{equation} \label{eq:appendix_helmholtz_calculation}
	\psi
	~=~ \dfrac{\lambda}{\rho_0} \Big(
		J
		- 1
	\Big)
	~=~ \lambda \Big(
		\dfrac{J}{\rho_0}
		- \dfrac{1}{\rho_0}
	\Big)
	~=~ \lambda \, \Big(
		\dfrac{1}{\rho}
		- \dfrac{1}{\rho_0}
	\Big)
	~.
\end{equation}
With the result of Eq.~\eqref{eq:appendix_helmholtz_calculation}, as well as the identities
$\partial \rho / \partial a_{\calpha \cbeta} = - \tfrac{1}{2} \, \rho \, a^{\calpha \cbeta}$
and
$\partial \rho / \partial b_{\calpha \cbeta} = 0$,
we find $\sigma^{\calpha \cbeta}$ to be given by Eq.~\eqref{eq:fluid_stresses} and
$M^{\calpha \cbeta} = 0$,
such that
$\bm{T}^\calpha = \sigma^{\calpha \cbeta} \bm{a}_\cbeta$.

%
% *** In-Plane and Out-of-Plane Equations of Motion
%

\subsection{In-Plane and Out-of-Plane Equations of Motion} \label{sec:sec_appendix_component_eqns}

As described in Appendix~\ref{sec:sec_appendix_irrev_thermo}, given the in-plane stresses $\sigma^{\calpha \cbeta}$ \eqref{eq:fluid_stresses}, the stress vectors of a fluid film are given by
$\bm{T}^\calpha = \sigma^{\calpha \cbeta} \bm{a}_\cbeta$.
By substituting the stress vectors into Eq.~\eqref{eq:strong_equations_of_motion}, applying the product rule, and using the identity
$\bm{a}_{\cbeta; \calpha} = b_{\calpha \cbeta} \bm{n}$,
we obtain the equations of motion as
\begin{equation} \label{eq:appendix_strong_eqns_of_motion}
	\rho \dot{\bmv}
	\, = \, \rho \bm{b}
	\, + \, \sigma^{\calpha \cbeta}_{; \calpha} \bm{a}_\cbeta
	\, + \, \sigma^{\calpha \cbeta} b_{\calpha \cbeta} \bm{n}
	~.
\end{equation}
We proceed to show how Eqs.~\eqref{eq:eom_in_plane} and \eqref{eq:shape_eqn} can be found by taking the dot product of Eq.~\eqref{eq:appendix_strong_eqns_of_motion} with the unit normal $\bm{n}$ and the in-plane basis vectors $\bm{a}_\calpha$, respectively.

Using Eq.~\eqref{eq:fluid_stresses}
for $\sigma^{\calpha \cbeta}$, realizing
$H = \tfrac{1}{2} a^{\calpha \cbeta} b_{\calpha \cbeta}$, and taking the dot product of Eq.~\eqref{eq:appendix_strong_eqns_of_motion} with the unit normal $\bm{n}$ yields
\begin{equation} \label{eq:appendix_shape_general}
	\begin{split}
		\rho \dot{\bmv} \cdot \bm{n}
		\, &= \, p
		\, + \, \sigma^{\calpha \cbeta} b_{\calpha \cbeta}
		\\[3pt]
		&= \, p
		\, + \, 2 \lambda H
		\, + \, \pi^{\calpha \cbeta} b_{\calpha \cbeta}
		~,
	\end{split}
\end{equation}
where the pressure drop $p$ across the surface is given by
$p = \rho \bm{b} \cdot \bm{n}$.
The component form of $\pi^{\calpha \cbeta}$ can be more conveniently written as \cite{sahu-mandadapu-pre-2017}
\begin{equation} \label{eq:appendix_viscous_stresses}
	\pi^{\calpha \cbeta}
	= \zeta \Big(
		v^\calpha_{; \cmu} \, a^{\cbeta \cmu}
		+ v^\cbeta_{; \cmu} \, a^{\calpha \cmu}
		- 2 \, v \, b^{\calpha \cbeta}
	\Big)
	~.
\end{equation}
Substituting Eq.~\eqref{eq:appendix_viscous_stresses} into Eq.~\eqref{eq:appendix_shape_general} and using the identity
$b^{\calpha \cbeta} b_{\calpha \cbeta} = 4 H^2 - 2 K$
leads to the out-of-plane shape equation
\begin{equation} \label{eq:appendix_shape}
	\begin{split}
		\rho \dot{\bmv} \cdot \bm{n}
		\, &= \, p
		\, + \, 2 \lambda H
		\, + \, \zeta \Big(
			v^\calpha_{; \cmu} \, a^{\cbeta \cmu}
			+ v^\cbeta_{; \cmu} \, a^{\calpha \cmu}
			- 2 \, v \, b^{\calpha \cbeta}
		\Big) b_{\calpha \cbeta}
		\\[3pt]
		&= \, p
		\, + \, 2 \lambda H
		\, + \, \zeta \, \Big(
			2 \, b^{\calpha \cbeta} \, v_{\calpha; \cbeta}
			- 4 \, v \, \big( 2 H^2 - K\big)
		\Big)
		~.
	\end{split}
\end{equation}
The result of Eq.~\eqref{eq:appendix_shape} is presented in the main text as Eq.~\eqref{eq:shape_eqn}.
When there are no flows, Eq.~\eqref{eq:appendix_shape} simplifies to the well-known Young--Laplace equation for fluid films,
$p + 2 \lambda H = 0$.

To find the in-plane equations of motion, Eq.~\eqref{eq:appendix_strong_eqns_of_motion} is contracted with the in-plane basis vector $\bm{a}_\calpha$ and Eq.~\eqref{eq:fluid_stresses}
is substituted into the result to yield
\begin{equation} \label{eq:appendix_in_plane_general}
	\begin{split}
		\rho \dot{\bmv} \cdot \bm{a}_\calpha
		\, &= \, \rho b_\calpha
		\, + \, a_{\calpha \cbeta} \, \sigma^{\cgamma \cbeta}_{; \cgamma}
		\\[3pt]
		&= \, \rho b_\calpha
		\, + \, \lambda_{, \calpha}
		\, + \, a_{\calpha \cbeta} \, \pi^{\cgamma \cbeta}_{; \cgamma}
		~.
	\end{split}
\end{equation}
The last term in Eq.~\eqref{eq:appendix_in_plane_general} is simplified by first taking the covariant derivative of the viscous stresses provided in Eq.~\eqref{eq:appendix_viscous_stresses} to find
\begin{equation} \label{eq:appendix_viscous_stress_gradient_first}
	\begin{split}
		\pi^{\cgamma \cbeta}_{; \cgamma}
		&= \zeta \Big(
			v^\cgamma_{; \cmu} \, a^{\cbeta \cmu}
			+ v^\cbeta_{; \cmu} \, a^{\cgamma \cmu}
			- 2 \, v \, b^{\cgamma \cbeta}
		\Big)_{; \cgamma}
		\\[3pt]
		&= \zeta \Big(
			v^\cgamma_{; \cmu \cgamma} \, a^{\cbeta \cmu}
			+ v^\cbeta_{; \cmu \cgamma} \, a^{\cmu \cgamma}
			- 2 \, v_{, \cgamma} \, b^{\cgamma \cbeta}
			- 2 \, v \, b^{\cgamma \cbeta}_{; \cgamma}
		\Big)
		~.
	\end{split}
\end{equation}
Using the identities
$b^{\cgamma \cbeta}_{; \cgamma} = 2 \, H_{, \cgamma} \, a^{\cgamma \cbeta}$
and
$v^\cgamma_{; \cmu \cgamma} = v^\cgamma_{; \cgamma \cmu} + K \, v^\cgamma \, a_{\cgamma \cmu}$,
along with Eq.~\eqref{eq:strong_incompressibility}, Eq.~\eqref{eq:appendix_viscous_stress_gradient_first} can be written as
\begin{equation} \label{eq:appendix_viscous_stress_gradient}
	\begin{split}
		\pi^{\cgamma \cbeta}_{; \cgamma}
		&= \zeta \Big(
			v^\cbeta_{; \cmu \cgamma} \, a^{\cmu \cgamma}
			+ K \, v^\cbeta
			+ a^{\cbeta \cmu} \, v^\cgamma_{; \cgamma \cmu}
			- 2 \, v_{, \cgamma} \, b^{\cgamma \cbeta}
			- 4 v \, H_{, \cgamma} \, a^{\cgamma \cbeta}
		\Big)
		\\[3pt]
		&= \zeta \Big(
			v^\cbeta_{; \cmu \cgamma} \, a^{\cmu \cgamma}
			+ K \, v^\cbeta
			+ a^{\cbeta \cmu} \, \big(
				2 \, v_{, \cmu} \, H
				+ 2 \, v \, H_{, \cmu}
			\big)
			- 2 \, v_{, \cgamma} \, b^{\cgamma \cbeta}
			- 4 v \, H_{, \cgamma} \, a^{\cgamma \cbeta}
		\Big)
		\\[3pt]
		&= \zeta \Big(
			v^\cbeta_{; \cmu \cgamma} \, a^{\cmu \cgamma}
			+ K \, v^\cbeta
			+ 2 \, v_{, \cmu} \, H \, a^{\cbeta \cmu}
			- 2 \, v_{, \cgamma} \, b^{\cgamma \cbeta}
			- 2 \, v \, H_{, \cgamma} \, a^{\cgamma \cbeta}
		\Big)
		~.
	\end{split}
\end{equation}
Substituting Eq.~\eqref{eq:appendix_viscous_stress_gradient} into Eq.~\eqref{eq:appendix_in_plane_general}, one obtains the in-plane equations
\begin{equation} \label{eq:appendix_eom_in_plane}
	\rho \dot{\bmv} \cdot \bm{a}_\calpha
	\, = \, \rho b_\calpha
	\, + \, \lambda_{, \calpha}
	\, + \, \zeta \Big(
		a^{\cmu \cgamma} \, v_{\calpha; \cmu \cgamma}
		+ K \, v_\calpha
		+ 2 \, v_{, \calpha} \, H
		- 2 \, v_{, \cgamma} \, b^{\cgamma}_{\calpha}
		- 2 \, v \, H_{, \calpha}
	\Big)
	~,
\end{equation}
which, upon exchange of dummy indices, is identical to Eq.~\eqref{eq:eom_in_plane}.
Equation \eqref{eq:appendix_eom_in_plane} is the extension of the Navier--Stokes equations to curved and deforming fluid films.
We note that Eqs.~\eqref{eq:appendix_shape} and \eqref{eq:appendix_eom_in_plane} were first presented in Refs.~\cite{scriven-1960, aris}.

%
% *** Flat Plane Equations
%

\subsubsection{Flat Plane Equations} \label{sec:sec_appendix_flat_eqns}

A flat surface is parametrized by
$\zeta^1 := x$
and
$\zeta^2 := y$,
such that the position $\bmx$ is given by
\begin{equation} \label{eq:appendix_flat_position}
	\bm{x} ( x, \, y )
	= x \, \bm{e}_x
	+ y \, \bm{e}_y
	~.
\end{equation}
Employing the results of Sec.~\ref{sec:sec_surface_geometry} yields
$\bm{a}_\alpha = \bm{e}_\alpha$,
$\bm{n} = \bm{e}_z$,
$a_{\alpha \beta} = \delta_{\alpha \beta}$,
$b_{\alpha \beta} = 0$,
$H = 0$,
$K = 0$,
and
$\Gamma^\alpha_{\lambda \mu} = 0$,
where $\delta_{\alpha \beta}$ is the standard Kronecker delta.
The velocity $\bm{v}$ is written as
\begin{equation} \label{eq:appendix_flat_velocity}
	\bmv
	= v_x \, \bm{e}_x
	+ v_y \, \bm{e}_y
	~.
\end{equation}
There is no normal component of the velocity in the normal direction,
$v = \bmv \cdot \bm{n} = 0$, 
because the fluid is constrained to lie in the plane.
We substitute the geometric quantities and velocity equations into the continuity equation~\eqref{eq:strong_incompressibility}, shape equation~\eqref{eq:appendix_shape}, and in-plane equations~\eqref{eq:appendix_eom_in_plane} to obtain
\begin{equation} \label{eq:appendix_flat_equations}
	\begin{split}
		&\hspace{25pt}
		v_{x, x}
		\, + \, v_{y, y}
		\, = \, 0
		~,
		\hspace{35pt}
		p
		= 0
		~,
		\hspace{38pt}
		\text{and}
		\\[3pt]
		&\rho \Big(
			v_{\alpha, t}
			\, + \, v_x \, v_{\alpha, x}
			\, + \, v_y \, v_{\alpha, y}
		\Big)
		\, = \, \rho b_{\alpha}
		\, + \, \lambda_{, \alpha}
		\, + \, \zeta \Big(
			v_{\alpha, x x}
			\, + \, v_{\alpha, y y}
		\Big)
		~.
	\end{split}
\end{equation}
The first equation in \eqref{eq:appendix_flat_equations} is the familiar continuity equation, the second equation says there is no pressure drop across a fluid flowing on a flat plane, and the last equations are the two-dimensional Navier--Stokes equations in which the pressure has been replaced by the negative surface tension.

%
% *** Fixed Cylinder Equations
%

\subsubsection{Fixed Cylinder Equations} \label{sec:sec_appendix_cylinder_eqns}

A cylindrical surface is parametrized with
$\zeta^1 := \theta$
and
$\zeta^2 := z$,
which are the standard polar angle and axial position of a cylindrical coordinate system.
The surface position $\bmx$ on the cylinder is given by
\begin{equation} \label{eq:appendix_cylinder_position}
	\bmx (\theta, \, z)
	= R \, \bm{e}_r (\theta)
	+ z \, \bm{e}_z
	~,
\end{equation}
which yields
$\bm{a}_1 = R \bm{e}_\theta$,
$\bm{a}_2 = \bm{e}_z$,
$\bm{n}   = \bm{e}_r$,
$a_{\alpha \beta} = \mathrm{diag} \, ( R^2, \, 1 )$,
$a^{\alpha \beta} = \mathrm{diag} \, ( R^{-2}, \, 1 )$,
$b_{\alpha \beta} = \mathrm{diag} \, (-R, \, 0 )$,
$H = -1 / (2R)$,
$K = 0$,
and
$\Gamma^\alpha_{\lambda \mu} = 0$.
The velocity $\bmv$ is written as
\begin{equation} \label{eq:appendix_cylinder_velocity}
	\bmv
	= v^\theta \, \bm{a}_1
	+ v^z \, \bm{a}_2
	~.
\end{equation}
Due to the different units of $\bm{a}_1$ and $\bm{a}_2$, $v^\theta$ has units of angular velocity and $v^z$ has units of velocity.
As is standard for a fixed surface, the normal velocity
$v = \bmv \cdot \bm{n} = 0$.
Substituting the geometric equations into the continuity equation~\eqref{eq:strong_incompressibility}, shape equation~\eqref{eq:appendix_shape}, and in-plane equations~\eqref{eq:appendix_eom_in_plane}, and ignoring inertial terms, one obtains, respectively,
\begin{gather}
	v^\theta_{, \theta}
	+ v^z_{, z}
	\, = \, 0
	~,
	\label{eq:appendix_cylinder_continuity}
	\\[4pt]
	R \, p
	\, = \, 2 \zeta \, v^\theta_{, \theta}
	\, + \, \lambda
	~,
	\label{eq:appendix_cylinder_shape}
	\\[4pt]
	\zeta \big(
		v^\theta_{, \theta \theta}
		+ R^2 v^\theta_{, z z}
	\big)
	\, + \, \lambda_{, \theta}
	\, = \, 0
	~,
	\label{eq:appendix_cylinder_theta_eqn}
	\\
	\shortintertext{and}
	\zeta \big(
		v^z_{, \theta \theta}
		+ R^2 v^z_{, z z}
	\big)
	\, + \, R^2 \lambda_{, z}
	\, = \, 0
	~.
	\label{eq:appendix_cylinder_z_eqn}
\end{gather}
In the absence of flows, the shape equation \eqref{eq:appendix_cylinder_shape} again reduces to the familiar Young--Laplace equation for cylinders:
$p = \lambda / R$.

%
% *** Equations for a Cylinder with a Bulge
%

\subsubsection{Equations for a Cylinder with a Bulge} \label{sec:sec_appendix_cylinder_bulge_eqns}

In this section, we analytically calculate the velocity, surface tension, and normal pressure solution of the bulged cylinder shown in Fig.~\ref{fig:fig_cylinder_schematic}.
We assume the solution is axisymmetric, such that all unknowns depend only on the axial position $z$, and also neglect inertial terms.
Our analytical solution is used to validate the numerical solution, as described in the main text.

The bulged cylinder is parametrized with the axial distance along the cylinder $z$ and polar angle $\theta$.
The cylinder radius, which is now a function of axial position only, is denoted $r(z)$ and given by
\begin{equation} \label{eq:appendix_cylinder_bulge_radius}
	r(z)
	=
	\begin{cases}
		1
		&\hspace{5pt} 0 \le z^* \le \, ^{3 \!} /_{\! 11} \\[5pt]
		1 \, + \, \tfrac{2}{25} \big(11 \, z^* - \, 3 \big)^2
		&\hspace{5pt} ^{3 \!} /_{\! 11} \le z^* \le \, ^{7 \!} /_{\! 22}\\[5pt]
		1 \, + \, \tfrac{1}{25} \big( 1 - 2 \, (11 \, z^* - \, 4)^2 \big)
		&\hspace{5pt} ^{7 \!} /_{\! 22} \le z^* \le \, ^{4 \!} /_{\! 11}\\[5pt]
		1.04
		&\hspace{5pt} ^{4 \!} /_{\! 11} \le z^* \le \, ^{1 \!} /_{\! 2}\\[5pt]
		r \, \big(1 - L z^*\big)
		&\hspace{5pt} ^{1 \!} /_{\! 2} \le z^* \le \, 1 \hspace{20pt} ,
	\end{cases}
\end{equation}
where $z^* := z / L$.
Defining the quantities
$\zeta^1 := \theta$
and
$\zeta^2 := z$,
the surface position is given by
\begin{equation} \label{eq:appendix_cylinder_bulge_position}
	\bmx (\theta, \, z)
	= r(z) \, \bm{e}_r (\theta)
	+ z \, \bm{e}_z
	~.
\end{equation}
From the surface position we calculate
$\bm{a}_1 = r \bm{e}_\theta$,
$\bm{a}_2 = r' \bm{e}_r + \bm{e}_z$,
$\bm{n} = (\bm{e}_r - r' \bm{e}_z)/(1 + (r')^2)$,
$a_{\alpha \beta} = \mathrm{diag} \, ( r^2, \, 1 + (r')^2 )$,
$a^{\alpha \beta} = \mathrm{diag} \, ( r^{-2}, \, 1 / ( 1 + (r')^2 ) )$,
and
$b_{\alpha \beta} = 1/(1 + (r')^2) \cdot \mathrm{diag} \, (-r, \, r'')$,
where $( \, \cdot \, )'$ denotes differentiation with respect to $z$.
The mean and Gaussian curvatures are
\begin{equation} \label{eq:appendix_cylinder_bulge_curvatures}
	H = \dfrac{
		r \, r''
		- \big(
			1 + (r')^2
		\big)
	} {
		2 r \big(
			1 + (r')^2
		\big)
	}
	\hspace{20pt}
	\mathrm{and}
	\hspace{20pt}
	K
	= \dfrac{
		- r''
	} {
		r \big(
			1 + (r')^2
		\big)^3
	}
	~,
\end{equation}
respectively.
The Christoffel symbols are also calculated, and those which are nonzero are given by
\begin{equation} \label{eq:appendix_cylinder_bulge_christoffel}
	\Gamma^1_{1 2}
	= \Gamma^1_{2 1}
	= \dfrac{r'}{r}
	~,
	\hspace{20pt}
	\Gamma^2_{1 1}
	= \dfrac{- r' \, r}{1 + (r')^2}
	~,
	\hspace{20pt}
	\mathrm{and}
	\hspace{24pt}
	\Gamma^2_{2 2}
	= \dfrac{1}{2} \,
	\dfrac{\mathrm{d}}{\mathrm{d} z} \Big[
		\ln \big(
			1
			+ (r')^2
		\big)
	\Big]
	~.
\end{equation}

With the known geometric quantities, we solve for the velocity and surface tension on the fixed surface.
The velocity $\bmv$ is first decomposed in the $\{ \bm{a}_1, \, \bm{a}_2 \}$ basis as
\begin{equation} \label{eq:appendix_cylinder_bulge_velocity}
	\bmv
	= v^\theta \, \bm{a}_1
	+ v^z \, \bm{a}_2
	~,
\end{equation}
where again the normal velocity
$\bmv \cdot \bm{n} = 0$
because the fluid is constrained to flow on the fixed surface.
Substituting the velocity decomposition \eqref{eq:appendix_cylinder_bulge_velocity} and Christoffel symbols \eqref{eq:appendix_cylinder_bulge_christoffel} into the continuity equation \eqref{eq:strong_incompressibility} and setting $v^\theta = 0$, due to our axisymmetric assumption, yields
\begin{equation} \label{eq:appendix_continuity_cylinder_bulge}
	(v^z)'
	\, + \, v^z \cdot \dfrac{\mathrm{d}}{\mathrm{d}z} \Big[
		\ln \big(
			r \sqrt{1 + (r')^2} \,
		\big)
	\Big]
	= 0
	~.
\end{equation}
Eq.~\eqref{eq:appendix_continuity_cylinder_bulge} is separable as the cylinder shape $r(z)$ is known, and by specifying the inlet velocity $v^z (z = 0)$ to be a given value $V$, we find the $z$-velocity solution to be
\begin{equation} \label{eq:appendix_cylinder_bulge_z_velocity}
	v^z (z)
	= V \cdot \dfrac{r(0)}{r(z)} \cdot \Bigg(
		\dfrac{
			1
			+ \big(
				r'(0)
			\big)^2
		} {
			1
			+ \big(
				r'(z)
			\big)^2
		}
	\Bigg)^{1/2}
	~.
\end{equation}

Substituting the known $z$-velocity \eqref{eq:appendix_cylinder_bulge_z_velocity} and geometric relations into the in-plane $v^z$ equation \eqref{eq:appendix_eom_in_plane} and neglecting inertia, one obtains
\begin{equation} \label{eq:appendix_cylinder_bulge_z_in_plane_eom}
	\dfrac{\mathrm{d} \lambda}{\mathrm{d} z}
	= \zeta \, v^z \cdot \dfrac{
		r'' \big(
			2 + (r')^2
		\big)
	} {
		r \big(
			1 + (r')^2
		\big)
	}
	~.
\end{equation}
Since $v^z$ is known from the continuity equation \eqref{eq:appendix_continuity_cylinder_bulge}, Eq.~\eqref{eq:appendix_cylinder_bulge_z_in_plane_eom} determines the surface tension $\lambda = \lambda(z)$.
By setting $\lambda (z = 0)$ to be a constant, $\lambda_0$, and integrating Eq.~\eqref{eq:appendix_cylinder_bulge_z_in_plane_eom}, we find the surface tension is given by
\begin{equation} \label{eq:appendix_cylinder_bulge_lambda}
	\lambda (z)
	= \lambda_0
	+ \int_0^z
		\zeta \, v^z (s) \cdot \dfrac{
			r''(s) \, \big[
				2 + \big( r'(s) \big)^2
			\big]
		} {
			r(s) \, \big[
				1 + \big( r'(s) \big)^2
			\big]
		}
	~\mathrm{d}s
	~.
\end{equation}
Finally, by substituting the $z$-velocity, surface tension, and geometric quantities into the shape equation \eqref{eq:appendix_shape}, the pressure $p = p(z)$ is found to be
\begin{equation} \label{eq:appendix_cylinder_bulge_pressure}
	p (z)
	\, = \, - 2 \lambda H
	\, + \, 2 \zeta \, \bigg[
		\dfrac{r'}{r} \bigg(
			\dfrac{r''}{ \big( 1 + (r')^2 \big)^2 }
			\, + \, \dfrac{1}{r \, \big( r + (r')^2 \big) }
		\bigg)
	\bigg]
	\, v^z
	~.
\end{equation}
Eqs.~\eqref{eq:appendix_cylinder_bulge_velocity}, \eqref{eq:appendix_cylinder_bulge_lambda}, \eqref{eq:appendix_cylinder_bulge_pressure}, and our assumption
$v^\theta = 0$
constitute a solution to the problem, which is used to analyze the numerical solutions in Sec.~\ref{sec:sec_fixed_cylinder_flow} in the main text.

%
% *** Choice of Cylinder Length
%

\paragraph{Choice of Cylinder Length} \label{sec:sec_choice_cylinder_length}

In our numerical solution of the bulged cylinder, $C^1$-continuous piecewise quadratic basis functions are used to represent the surface \eqref{eq:fluid_velocity_subspace}.
In this basis, $r''$ is discontinuous, and as a result, the mean curvature $H$ and the pressure drop $p$ are discontinuous as well.
Consequently, they are poorly represented in our finite-dimensional function space.
To avoid such discontinuities in Sec.~\ref{sec:sec_fixed_cylinder_flow}, a large cylinder length
$L = 4,\!000$
is chosen such that $r'$ and $r''$ are approximately zero.
In this limit, $H$ is given approximately by
$H(z) \approx -1/(2 \, r(z))$,
which is continuous and, as it is a function of $z$, leads to a nontrivial solution.

%
% *** Stability Analysis of a Deforming Cylinder
%

\subsection{Stability Analysis of a Deforming Cylinder} \label{sec:sec_appendix_cylinder_stability}

In this section, we analyze the linear stability of an initially cylindrical fluid film, which is able to deform over time.
The film is acted on by a constant pressure drop $p$ across its surface, and is prescribed to have zero velocity everywhere on its boundary.
A base state solution to the cylindrical fluid equations of Sec.~\ref{sec:sec_appendix_cylinder_eqns} is provided, and then the first-order perturbed equations are presented.
The perturbed quantities are expanded in a Fourier basis, and the dispersion relation is solved for.
It is found that fluid films are unstable when their length $L$ is greater than their circumference, $2 \pi R$.
The instability analysis also reveals the time scale of the growing, unstable modes.
We end by showing how an axisymmetric assumption, and the manipulation of the perturbed equations, leads to the same theoretical predictions.

A solution to the unperturbed cylindrical equations \eqref{eq:appendix_cylinder_continuity}--\eqref{eq:appendix_cylinder_z_eqn} is given by
\begin{equation} \label{eq:appendix_cylinder_unperturbed_solution}
	v^z = V
	~,
	\hspace{30pt}
	v^\theta = 0
	~,
	\hspace{30pt}
	\mathrm{and}
	\hspace{34pt}
	\lambda = \lambda_0
	~.
\end{equation}
In Eq.~\eqref{eq:appendix_cylinder_unperturbed_solution}, $V$ is a constant speed and $\lambda_0$ is related to the constant pressure drop $p$ according to
$\lambda_0 = R \, p$.
A perturbation of the film surface position is introduced as
\begin{equation} \label{eq:appendix_cylinder_perturbation}
	\bmx (\theta, z, t)
	= \big(
		R
		+ \epsilon \, \tilde{r} (\theta, z, t)
	\big) \, \bm{e}_r
	\, + \, z \, \bm{e}_z
	~,
\end{equation}
where $\epsilon$ is a small parameter and $\tilde{r} (\theta, z, t)$ is an $\mathcal{O}(R)$ radial perturbation.
Importantly, the perturbation introduced in Eq.~\eqref{eq:appendix_cylinder_perturbation} both changes the geometry of the surface, and also allows for a normal velocity $v = \bmv \cdot \bm{n}$ given by
$v = \epsilon \, \tilde{r}_{, t}$.
In this section, a `tilde' $( \tilde{\, \cdot \,} )$ is used over all perturbed variables, such that the $\theta$-velocity, $z$-velocity, and surface tension are written as
\begin{equation} \label{eq:appendix_cylinder_perturbed_unknowns}
	v^\theta
	= \epsilon \, \tilde{v}^\theta
	~,
	\hspace{10pt}
	v^z
	= V
	+ \epsilon \, \tilde{v}^z
	~,
	\hspace{10pt}
	\mathrm{and}
	\hspace{12pt}
	\lambda
	= \lambda_0
	+ \epsilon \, \tilde{\lambda}
	~.
\end{equation}
Setting the perturbed quantities to zero in Eq.~\eqref{eq:appendix_cylinder_perturbed_unknowns} recovers the base state solution~\eqref{eq:appendix_cylinder_unperturbed_solution}.

By calculating the perturbed geometric quantities describing the surface, substituting them along with Eq.~\eqref{eq:appendix_cylinder_perturbed_unknowns} into the governing equations, and keeping only terms of order $\epsilon$, we obtain the perturbed equations for an initially cylindrical fluid film.
The details of this calculation are omitted, however, the results are presented as
\begin{gather}
	0 \, = \,
	R \, \big(
		\tilde{v}^\theta_{, \theta}
		+ \tilde{v}^z_{, z}
	\big)
	\, + \, \tilde{r}_{, t}
	\, + \, V \, \tilde{r}_{, z}
	~,
	\label{eq:appendix_cylinder_perturbed_continuity}
	\\[6pt]
	0 \, = \,
	\zeta \, \big(
		\tilde{r}_{, \theta t}
		+ R \, \tilde{v}^\theta_{, \theta \theta}
		+ R^3 \, \tilde{v}^\theta_{, z z}
		+ V \, \tilde{r}_{, \theta z}
	\big)
	\, + \, R \, \tilde{\lambda}_{, \theta}
	~,
	\label{eq:appendix_cylinder_perturbed_theta}
	\\[6pt]
	0 \, = \,
	\zeta \, \big(
		-R^2 \, \tilde{r}_{, z t}
		+ R \, \tilde{v}^z_{, \theta \theta}
		+ R^3 \, \tilde{v}^z_{, z z}
		- V \, R^2 \, \tilde{r}_{, z z}
	\big)
	\, + \, R^3 \, \tilde{\lambda}_{, z}
	~,
	\label{eq:appendix_cylinder_perturbed_z}
	\\
	\shortintertext{and}
	0 \, = \,
	- 2 \zeta \, \big(
		R \, \tilde{v}^\theta_{, \theta}
		+ \tilde{r}_{, t}
		+ V \, \tilde{r}_{, z}
	\big)
	\, + \, \lambda_0 \, \big(
		\tilde{r}
		+ \tilde{r}_{, \theta \theta}
		+ R^2 \, \tilde{r}_{, z z}
	\big)
	\, - \, R \, \tilde{\lambda}
	~.
	\label{eq:appendix_cylinder_perturbed_shape}
\end{gather}
Eqs.~\eqref{eq:appendix_cylinder_perturbed_continuity}--\eqref{eq:appendix_cylinder_perturbed_shape} are the perturbed continuity, in-plane $\theta$, in-plane $z$, and shape equations, respectively.

The four equations \eqref{eq:appendix_cylinder_perturbed_continuity}--\eqref{eq:appendix_cylinder_perturbed_shape} contain four unknowns:
$\tilde{r}$, $\tilde{v}^\theta$, $\tilde{v}^z$, and $\tilde{\lambda}$.
To perform a linear stability analysis, we follow the canonical treatment of Ref.~\cite{chandrasekhar}, and expand all unknowns in a Fourier basis as
\begin{align*}
	&
	\tilde{r} (\theta, z, t)
	= \sum_{m, q} \hat{r} (m, q) ~ \mathrm{e}^{i ( m\theta + qz - \omega t)}
	~,
	\hspace{60pt}
	\tilde{v}^\theta (\theta, z, t)
	= \sum_{m, q} \hat{v}^\theta (m, q) ~ \mathrm{e}^{i ( m\theta + qz - \omega t)}
	~,
	\stepcounter{equation}
	\tag{\theequation}\label{eq:appendix_cylinder_fourier_expansion}
	\\[4pt]
	&\tilde{v}^z (\theta, z, t)
	= \sum_{m, q} \hat{v}^z (m, q) ~ \mathrm{e}^{i ( m\theta + qz - \omega t)}
	~,
	\hspace{20pt}
	\mathrm{and}
	\hspace{20pt}
	\tilde{\lambda} (\theta, z, t)
	= \sum_{m, q} \hat{\lambda} (m, q) ~ \mathrm{e}^{i ( m\theta + qz - \omega t)}
	~,
	\hspace{20pt}
\end{align*}
where
$m \in \mathbb{Z}$ such that all quantities are periodic in $\theta$ and $q$ are the allowed wavenumbers given by $q = n \pi / L$ for $n \in \mathbb{Z}$.
The coefficients
$\hat{r}$, $\hat{v}^\theta$, $\hat{v}^z$, and $\hat{\lambda}$ are the Fourier coefficients.
Substituting Eq.~\eqref{eq:appendix_cylinder_fourier_expansion} into the perturbed equations \eqref{eq:appendix_cylinder_perturbed_continuity}--\eqref{eq:appendix_cylinder_perturbed_shape} and recognizing modes of different $m$ and $q$ are independent leads to the equations
\begin{gather}
	0 \, = \, 
	R \, \big(
		m \, \hat{v}^\theta
		+ q \, \hat{v}^z
	\big)
	\, + \, \hat{r} \, \big(
		V \, q
		- \omega
	\big)
	~,
	\label{eq:appendix_cylinder_perturbed_continuity_fourier}
	\\[4pt]
	0 \, = \,
	\hat{r} \big(
		m \, \omega
		- m \, q \, V
	\big)
	+ \hat{v}^\theta \, \big(
		- m^2 R
		- q^2 R^3
	\big)
	\, + \, i \, \dfrac{m \, R}{\zeta} \hat{\lambda}
	~,
	\label{eq:appendix_cylinder_perturbed_theta_fourier}
	\\[4pt]
	0 \, = \,
	\hat{r} \big(
		- \omega \, q \, R
		+ V \, q^2 R
	\big)
	\, + \, \hat{v}^z \big(
		- m^2
		- q^2 R^2
	\big)
	\, + \, i \, \dfrac{R^2 q}{\zeta} \, \hat{\lambda}
	~,
	\label{eq:appendix_cylinder_perturbed_z_fourier}
	\\
	\shortintertext{and}
	0 \, = \,
	\hat{r} \, \Big(
		2 \, i \, \zeta (\omega - q \, V)
		+ \lambda_0 \, \big(
			1 - m^2 - R^2 q^2
		\big)
	\Big)
	- 2 \, i \, m \, \zeta \, R \, \hat{v}^\theta
	- R \, \hat{\lambda}
	~.
	\label{eq:appendix_cylinder_perturbed_shape_fourier}
\end{gather}
Eqs.~\eqref{eq:appendix_cylinder_perturbed_continuity_fourier}--\eqref{eq:appendix_cylinder_perturbed_shape_fourier} are linear in the unknowns and can be written as the matrix equation
$\mathbf{A} \, \mathbf{y} = \bm{0}$,
for
$\mathbf{y} = (\hat{r}, \, \hat{v}^\theta, \, \hat{v}^z, \, \hat{\lambda})^{\mathrm{T}}$
and coefficient matrix $\mathbf{A}$.
For a nontrivial solution to exist, the matrix $\mathbf{A}$ cannot be invertible, which is guaranteed by setting $(\det \mathbf{A})$ equal to zero.
Doing so yields the dispersion relation
\begin{equation} \label{eq:appendix_cylinder_perturbed_dispersion}
	\omega
	= q \, V
	\, + \, i \, \dfrac{\lambda_0}{4 \, \zeta \, R^4 q^4}
	\cdot \big(
		m^2
		+ R^2 q^2
	\big)^2 \, \big(
		1
		- m^2
		- R^2 q^2
	\big)
	~.
\end{equation}
Given the form of our ansatz is
$\mathrm{e}^{i ( m\theta + qz - \omega t)}$
for all terms \eqref{eq:appendix_cylinder_fourier_expansion}, our solution is unstable when
$\mathrm{Im} \, \{\omega\} > 0$,
which only occurs when
$m^2 + R^2 q^2 < 1$.
Accordingly, unstable modes are axisymmetric
$(m = 0)$
and only occur when
$ q < 1/R$,
or equivalently
$L > n \pi R$.

Note that our linear stability analysis predicts the
$n = 1$
mode is unstable when
$L > \pi R$.
However, for such an instability to grow, material is required to be drawn in to increase the area of the fluid film, a phenomena which is incompatible with our zero-velocity boundary conditions.
Accordingly, given our choice of boundary conditions, instabilities only occur when
$L > 2 \pi R$.
For the
$n = 2$ mode, the time scale $\tau$ is given by
$1 / \mathrm{Im} \{ \omega \}$,
found to be
\begin{equation} \label{eq:appendix_cylinder_full_time_scale}
	\tau
	=
	\Big(
		\dfrac{4 \, \zeta}{\lambda_0}
	\Big)
	\,
	\bigg[
		1
		- \Big(
			\dfrac{2 \pi R}{L}
		\Big)^{\! 2}
	\bigg]^{-1}
	~.
\end{equation}
The theoretical time scale $\tau$ in Eq.~\eqref{eq:appendix_cylinder_full_time_scale} is presented in Eq.~\eqref{eq:ale_cylinder_perturbed_time_scale} of the main text.

With the understanding that axisymmetric, sinusoidal perturbations lead to unstable solutions, a simpler stability analysis can be performed.
Assuming all perturbed quantities are independent of $\theta$, no base state $z$-velocity
$(V = 0)$,
and no perturbed $\theta$-velocity
$(\tilde{v}^\theta = 0)$,
Eqs.~\eqref{eq:appendix_cylinder_perturbed_continuity}--\eqref{eq:appendix_cylinder_perturbed_shape} simplify to
\begin{gather}
	0 \, = \,
	R \, \tilde{v}^z_{, z}
	\, + \, \tilde{r}_{, t}
	~,
	\label{eq:appendix_cylinder_perturbed_continuity_simple}
	\\[6pt]
	0 \, = \,
	- \zeta \, \tilde{r}_{, z t}
	+ \, \zeta \, R \, \tilde{v}^z_{, z z}
	\, + \, R \, \tilde{\lambda}_{, z}
	~,
	\label{eq:appendix_cylinder_perturbed_z_simple}
	\\
	\shortintertext{and}
	0 \, = \,
	- 2 \, \zeta \, \tilde{r}_{, t}
	\, + \, \lambda_0 \, \tilde{r}
	\, + \, \lambda_0 \, R^2 \, \tilde{r}_{, z z}
	\, - \, R \, \tilde{\lambda}
	~.
	\label{eq:appendix_cylinder_perturbed_shape_simple}
\end{gather}
We consider an initial perturbation of the form
\begin{equation} \label{eq:appendix_cylinder_perturbed_simple}
	\tilde{r} (z, t = 0)
	= R \, \sin \big(
		q \, z
	\big)
	~,
\end{equation}
where as before
$q = n \pi / L$,
and seek to determine how this perturbation evolves in time.

To understand the time evolution of the perturbation, captured in $\tilde{r}_{, t}$, a series of algebraic manipulations are carried out.
The initial perturbation \eqref{eq:appendix_cylinder_perturbed_simple} is substituted into the perturbed shape equation \eqref{eq:appendix_cylinder_perturbed_shape_simple}, leading to
\begin{equation} \label{eq:appendix_cylinder_perturbed_simple_step_1}
	-2 \, \zeta \, \tilde{r}_{, t}
	\, + \, \big(
		1 - q^2 R^2
	\big) \, \lambda_0 \, \tilde{r}
	\, - \, R \, \tilde{\lambda}
	\, = \, 0
	~.
\end{equation}
Taking the partial derivative of Eq.~\eqref{eq:appendix_cylinder_perturbed_continuity_simple} with respect to $z$ and subtracting Eq.~\eqref{eq:appendix_cylinder_perturbed_z_simple} yields
\begin{equation} \label{eq:appendix_cylinder_perturbed_simple_step_2}
	2 \, \zeta \, \tilde{r}_{, t z}
	\, - \, R \, \tilde{\lambda}_{, z}
	\, = \, 0
	~.
\end{equation}
To Eq.~\eqref{eq:appendix_cylinder_perturbed_simple_step_2} we add the partial derivative of Eq.~\eqref{eq:appendix_cylinder_perturbed_simple_step_1} with respect to $z$ and obtain
\begin{equation} \label{eq:appendix_cylinder_perturbed_simple_step_3}
	\big(
		1 - q^2 R^2
	\big) \, \lambda_0 \, \tilde{r}_{, z}
	\, - \, 2 \, R \, \tilde{\lambda}_{, z}
	\, = \, 0
	~.
\end{equation}
Integrating Eq.~\eqref{eq:appendix_cylinder_perturbed_simple_step_3} with respect to $z$, recognizing the integration constant is zero, and substituting the result into Eq.~\eqref{eq:appendix_cylinder_perturbed_simple_step_1} leads to
\begin{equation} \label{eq:appendix_cylinder_perturbed_simple_result}
	\big(
		1 - q^2 R^2
	\big) \, \lambda_0 \, \tilde{r}
	\, = \, 4 \, \zeta \, \tilde{r}_{, t}
	~.
\end{equation}
Eq.~\eqref{eq:appendix_cylinder_perturbed_simple_result} describes the time evolution of $\tilde{r}$ only in terms of $\tilde{r}$ itself, and thus describes when the perturbation will grow or decay in time.
When $q^2 R^2 < 1$, or equivalently when $L > n \pi R$, the initial perturbation grows in time and the cylinder is unstable.
As before, the boundary conditions exclude the
$n = 1$
mode, and cylindrical fluid films are expected to be unstable when
$L > 2 \pi R$.

Eq.~\eqref{eq:appendix_cylinder_perturbed_simple_result} reveals the time scale $\tau$ for the evolution of the initial perturbation.
For the $n = 2$ mode, the time scale is given by
\begin{equation} \label{eq:appendix_cylinder_perturbed_simple_tau}
	\tau
	=
	\Big(
		\dfrac{4 \, \zeta}{\lambda_0}
	\Big)
	\,
	\bigg[
		1
		- \Big(
			\dfrac{2 \pi R}{L}
		\Big)^{\! 2}
	\bigg]^{-1}
	~,
\end{equation}
in agreement with the full perturbation analysis time scale \eqref{eq:appendix_cylinder_full_time_scale}.

%
% *** Numerical Solution Method
%

\section{Numerical Solution Method} \label{sec:sec_appendix_numerical_solution_method}

In this section, we discretize the fluid surface
and define a finite number of basis functions on the discretized domain.
The fundamental unknowns, as well as their arbitrary variations, are expressed in terms of these basis functions according to the Bubnov--Galerkin approximation.
The resulting residual equations are temporally discretized using the backward Euler method, and then solved using Newton--Raphson iteration.
We end by presenting the details of our numerical implementation of the Dohrmann--Bochev method~\cite{dohrmann-bochev-ijnmf-2004}, which removes numerical inf--sup instabilities arising from the incompressibility of the fluid film.

%
% *** Discretization and Bubnov--Galerkin Approximation
%

\subsection{Discretization and Bubnov--Galerkin Approximation} \label{sec:sec_fluid_bubnov_galerkin_approx}

For the space $\mcvh$ \eqref{eq:fluid_velocity_subspace}, we introduce the $\nn$ (\underline{n}umber of \underline{n}odes) basis functions $\{ N_I (\zeta^\alpha) \}$ such that
$\mcvh = (\textrm{span} \, \{ N_I (\zeta^\alpha) \} )^3$,
and an arbitrary velocity
$\bmvh \in \mcvh$ can be expressed as
\begin{equation} \label{eq:velocity_basis_expansion}
	\bmv (\zeta^\alpha, t)
	= \sum_{J = 1}^{\nn} N_J (\zeta^\alpha) \, \bmv_J (t)
	\, = \, \mN \, \mvt
	~.
\end{equation}
In Eq.~\eqref{eq:velocity_basis_expansion} and from now on, the subscript `$h$' is dropped for notational convenience.
All variables refer to the approximate solution unless otherwise noted.
We introduced the $3 \times (3 \cdot \nn)$ matrix of shape function values, $\mN$, and the $(3 \cdot \nn) \times 1$ vector of velocity degrees of freedom at time $t$, $\mvt$, given respectively by
\begin{equation} \label{eq:matrix_shape_fn_array}
	\setlength\arraycolsep{4pt}
	\mN
	:= \Big[ \hspace{5pt}
		N_{1} \, \mone \hspace{9pt}
		N_{2} \, \mone \hspace{5pt}
		\ldots \hspace{5pt}
		N_{\nn} \, \mone \hspace{5pt}
	\Big]
	\hspace{30pt}
	\mathrm{and}
	\hspace{30pt}
	\mvt
	:=
	\begin{bmatrix}
		~ \\[-7pt]
		\, \bm{v}_1 (t) \, \\[-2pt]
		\, \vdots \, \\[2pt]
		\, \bm{v}_{\nn} (t) \, \\[3pt]
	\end{bmatrix}
	~,
\end{equation}
where $\mone$ is the $3 \times 3$ identity matrix.
In practice, the entire matrix $\mN$ is never computed.
Rather, as discussed in Sec.~\ref{sec:sec_fluid_local_calculations}, the shape functions over a single element are calculated and stored in a local shape function matrix.

As the mesh velocity $\bmvm \in \mcvh$, we similarly expand the mesh velocity as
\begin{equation} \label{eq:mesh_velocity_basis_expansion}
	\bmvm (\zeta^\alpha, t)
	\, = \, \mN \, \mvmt
	~,
\end{equation}
where the mesh velocity degree of freedom vector $\mvmt$ is defined analogously to $\mvt$ in Eq.~\eqref{eq:matrix_shape_fn_array}.
The identity matrix $\mone$, which appears in Eqs.~\eqref{eq:velocity_basis_expansion} and \eqref{eq:mesh_velocity_basis_expansion} via the shape function matrix $\mN$~\eqref{eq:matrix_shape_fn_array}$_1$,
is chosen to be the Cartesian identity matrix.
Accordingly, the velocity and mesh velocity degrees of freedom contained in $\mv$ and $\mvm$ are Cartesian components as well.
Thus, when $\bmv$ and $\bmvm$ are calculated through Eqs.~\eqref{eq:velocity_basis_expansion} and \eqref{eq:mesh_velocity_basis_expansion}, we find their representation in the canonical Cartesian basis.

The space of surface tensions $\Lambdah$ is spanned by the $\nln$ (\underline{n}umber of \underline{L}agrange multiplier \underline{n}odes) Lagrange multiplier basis functions $\{ \bar{N}_J (\zeta^\alpha) \}$, written as
$\Lambdah = \textrm{span} \, \{ \bar{N}_J (\zeta^\alpha) \}$.
Here and from now on, discretized quantities with a `bar' accent correspond to the Lagrange multiplier $\lambda$.
%This is not to be confused with the `hat' accent over Greek indices in the Lagrangian parametrization.
We introduce $\mhN$ as the $1 \times \nln$ vector of Lagrange multiplier basis functions and $\mlambdat$ as the corresponding $\nln \times 1$ vector of Lagrange multiplier degrees of freedom, given by
\begin{equation} \label{eq:matrix_shape_fn_array_lagrange_multiplier}
	\setlength\arraycolsep{4pt}
	\mhN
	:= \Big[ \hspace{5pt}
		\bar{N}_1 \hspace{11pt}
		\bar{N}_2 \hspace{6pt}
		\ldots \hspace{6pt}
		\bar{N}_{\nln} \hspace{5pt}
	\Big]
	\hspace{30pt}
	\mathrm{and}
	\hspace{30pt}
	\mlambdat
	:=
	\begin{bmatrix}
		~ \\[-7pt]
		\, \bar{\lambda}_1 (t) \, \\[-2pt]
		\, \vdots \, \\[2pt]
		\, \bar{\lambda}_{\nln} (t) \, \\[3pt]
	\end{bmatrix}
	~,
\end{equation}
such that the surface tension can be expanded as
\begin{equation} \label{eq:lambda_basis_expansion}
	\lambda (\zeta^\alpha, t)
	= \mhN \, \mlambdat
	~.
\end{equation}
Note the basis functions contained in $\mN$ and $\mhN$ are determined by the discretization of the domain $\Omega$, so the only remaining unknowns are contained in $\mvt$, $\mvmt$, and $\mlambdat$, which are together called the degrees of freedom of the system.

The arbitrary variations $\delta \bmv (\zeta^\alpha)$, $\delta \bmvm (\zeta^\alpha)$, and $\delta \lambda (\zeta^\alpha)$ are discretized with the same basis functions as their unknown counterparts, according to the Bubnov--Galerkin approximation, and are expanded as
\begin{equation} \label{eq:variation_element_basis_expansion}
	\delta \bmv (\zeta^\alpha)
	= \mN \, \mdeltav
	~,
	\hspace{20pt}
	\delta \bmvm (\zeta^\alpha)
	= \mN \, \mdeltavm
	~,
	\hspace{20pt}
	\mathrm{and}
	\hspace{23pt}
	\delta \lambda (\zeta^\alpha)
	= \mhN \, \mdeltalambda
	~.
\end{equation}
The variation vectors $\mdeltav$, $\mdeltavm$, and $\mdeltalambda$ are given by prefixing a `$\delta$' to every entry of their respective counterparts, $\mv$, $\mvm$, and $\mlambda$.

%
% *** Residual Vector Equations
%

\subsection{Residual Vector Equations} \label{sec:sec_residual_vector_equations}

As shown in Appendix~\ref{sec:sec_appendix_residual_vector}, by substituting the discretized unknown variations \eqref{eq:variation_element_basis_expansion} into Eq.~\eqref{eq:fluid_weak_form_compact} and introducing the shorthand
$
	\tilde{\mcg} (t) := \mcg (
		\bm{u} (\zeta^\alpha, \, t), \,
		\delta \bm{u} (\zeta^\alpha)
	)
$,
where $\bmu$ represents all unknowns and $\delta \bmu$ is its arbitrary variation, the weak form can be written as
\begin{equation} \label{eq:weak_expanded_variations}
	\begin{split}
		\tilde{\mcg} (t)
		\, = \, \mdeltav^{\mathrm{T}} \mrvt
		\, &+ \,\, \mdeltavm^{\mathrm{T}} \mrmt
		\, + \,\, \mdeltalambda^{\mathrm{T}} \mrlambdat
		\hspace{3pt} = \hspace{3pt} 0
		\\[5pt]
		&\hspace{4pt}
		\forall
		\hspace{3pt}
		\mdeltav \in \mathbb{R}^{3 \cdot \nn}, ~
		\mdeltavm \in \mathbb{R}^{3 \cdot \nn}, ~
		\mdeltalambda \in \mathbb{R}^{\nln}
		~,
	\end{split}
\end{equation}
for any time $t$.
Since the weak form is linear in the unknown variations, the global velocity, mesh velocity, and surface tension residual vectors are respectively given by
\begin{equation} \label{eq:weak_residuals}
	\mrvt
	:= \dfrac{\partial \tilde{\mcg} (t)}{\partial \mdeltav}
	~,
	\hspace{15pt}
	\mrmt
	:= \dfrac{\partial \tilde{\mcg} (t)}{\partial \mdeltavm}
	~,
	\hspace{15pt}
	\mathrm{and}
	\hspace{18pt}
	\mrlambdat
	:= \dfrac{\partial \tilde{\mcg} (t)}{\partial \mdeltalambda}
	~.
\end{equation}
The calculation of the residual vectors, according to Eqs.~\eqref{eq:weak_expanded_variations} and \eqref{eq:weak_residuals}, is provided in Appendix~\ref{sec:sec_appendix_residual_vector}.
Since the discretized unknown variations are arbitrary, Eq.~\eqref{eq:weak_expanded_variations} is equivalent to simultaneously requiring the global residual vectors to be zero, written as
\begin{equation} \label{eq:residuals_zero}
	\mrvt
	= \mzero
	~,
	\hspace{15pt}
	\mrmt
	= \mzero
	~,
	\hspace{15pt}
	\mathrm{and}
	\hspace{20pt}
	\mrlambdat
	= \mzero
	~.
\end{equation}

%
% *** Time Integration
%

\subsection{Time Integration} \label{sec:sec_time_integration}

The relations provided in Eq.~\eqref{eq:residuals_zero} are true at any time $t$, however, they are only solved numerically at a set of $N$ discrete times
$\{ t_1, \, t_2, \ldots , t_{N} \}$.
Thus, assuming a known solution $\bmu(\zeta^\alpha, \, t_n)$ satisfying Eq.~\eqref{eq:residuals_zero} at time $\tn$, we seek the unknown solution $\bmu(\zeta^\alpha, \, \tno)$ satisfying Eq.~\eqref{eq:residuals_zero} at time $\tno$.
To this end, the fundamental unknowns are expressed as
\begin{align}
	\bmv (\zeta^\alpha, \, \tno)
	\, &= \, \bmv (\zeta^\alpha, \, \tn)
	\, + \, \Delta \bmv (\zeta^\alpha, \, \tno)
	~,
	\label{eq:velocity_time_expansion}
	\\[8pt]
	\bmvm (\zeta^\alpha, \, \tno)
	\, &= \, \bmvm (\zeta^\alpha, \, \tn)
	\, + \, \Delta \bmvm (\zeta^\alpha, \, \tno)
	~,
	\label{eq:mesh_velocity_time_expansion}
	\\[-4pt]
	\shortintertext{and}
	\lambda (\zeta^\alpha, \, \tno)
	\, &= \, \lambda (\zeta^\alpha, \, \tn)
	\, + \, \Delta \lambda (\zeta^\alpha, \, \tno)
	~.
	\label{eq:lambda_time_expansion}
	\\
	\intertext{Furthermore, Eq.~\eqref{eq:ale_mesh_position} is discretized with the backward Euler method to yield}
	\check{\bmx} (\zeta^\alpha, \, \tno)
	\, &= \, \check{\bmx} (\zeta^\alpha, \, \tn)
	\, + \, \Delta t \, \Delta \bmvm (\zeta^\alpha, \, \tno)
	~,
	\label{eq:position_time_expansion}
\end{align}
where
$\Delta t := \tno - \tn$.
According to Eq.~\eqref{eq:position_time_expansion}, changes in the mesh velocity affect the surface position, which in turn affects the various geometric terms found in the residual vector equations~\eqref{eq:residuals_zero}.

Discretizing the fundamental unknowns (\ref{eq:velocity_basis_expansion}, \ref{eq:mesh_velocity_basis_expansion}, \ref{eq:lambda_basis_expansion}), and introducing the notation
$\mvn = [ \mathbf{v} (\tn) ]$,
for example, Eqs.~\eqref{eq:velocity_time_expansion}--\eqref{eq:position_time_expansion} are equivalently expressed as
\begin{align}
	\bmv (\zeta^\alpha, \, \tno)
	\, &= \, \mN \, \mvno
	\, = \, \mN \, \mvn
	\, + \, \mN \, \mDeltavno
	~,
	\label{eq:velocity_time_expansion_discretized}
	\\[8pt]
	\bmvm (\zeta^\alpha, \, \tno)
	\, &= \, \mN \, \mvmno
	\, = \, \mN \, \mvmn
	\, + \, \mN \, \mDeltavmno
	~,
	\label{eq:mesh_velocity_time_expansion_discretized}
	\\[8pt]
	\lambda (\zeta^\alpha, \, \tno)
	\, &= \, \mhN \, \mlambdano
	\, = \, \mhN \, \mlambdan
	\, + \, \mhN \, \mDeltalambdano
	~,
	\label{eq:lambda_time_expansion_discretized}
	\\[-4pt]
	\shortintertext{and}
	\check{\bmx} (\zeta^\alpha, \, \tno)
	\, &= \, \mN \, \mxno
	\, = \, \mN \, \mxn
	\, + \, \Delta t \, \mN \, \mDeltavmno
	~.
	\label{eq:position_time_expansion_discretized}
\end{align}
Note that while the position $\check{\bmx}$ is not a fundamental unknown, it is discretized as in Eq.~\eqref{eq:position_time_expansion_discretized} such that the change in mesh velocity, $\mDeltavmno$, is properly accounted for when solving for $\bmu(\zeta^\alpha, \, \tno)$.
Substituting Eqs.~\eqref{eq:velocity_time_expansion_discretized}--\eqref{eq:position_time_expansion_discretized} into the residual equations \eqref{eq:residuals_zero} at time $\tno$ leads to a set of nonlinear algebraic equations in the unknowns
$\mDeltavno$,
$\mDeltavmno$,
and
$\mDeltalambdano$,
which are collectively gathered in the vector $\mDeltauno$.
These equations are solved via the Newton--Raphson method.

%
% *** Newton--Raphson Iteration
%

\subsection{Newton--Raphson Iteration} \label{sec:sec_newton_raphson_iteration}

We begin by initially guessing the solution at time $\tno$ is equal to the known solution at time $\tn$, written as
$\mbuno_{0} = \mbun$,
for which
$\mDeltauno_{0} = \mzero$.
We then generate a sequence of iterative solutions $\mbuno_\so$, where
\begin{equation} \label{eq:newton_raphson_delta_u}
	\mbuno_\so
	\, = \, \mbuno_\sz
	+ \mDeltauno_\so
	~.
\end{equation}
Substituting Eqs.~\eqref{eq:velocity_time_expansion_discretized}--\eqref{eq:newton_raphson_delta_u} into Eq.~\eqref{eq:residuals_zero}, evaluated at time $\tno$, and keeping terms to first order in $\mDeltauno_\so$ yields
\begin{align}
	\begin{split}
		\\[10pt]
		\hspace{-30pt}
	\end{split}
	\begin{split}
		\mzero
		\, &= \, \mrvno_\so
		\\[6pt]
		\mzero
		\, &= \, \mrmno_\so
		\\[6pt]
		\mzero
		\, &= \, \mrlambdano_\so
	\end{split}
	~
	\begin{split}
		&\dot{=} ~ \mrvno_\sz
		\\[6pt]
		&\dot{=} ~ \mrmno_\sz
		\\[6pt]
		&\dot{=} ~ \mrlambdano_\sz
	\end{split}
	\hspace{-35pt}
	\begin{split}
		&+ \, \mKvv \, \mDeltav
		\\[6pt]
		&+ \, \mKmv \, \mDeltav
		\\[6pt]
		&+ \, \mKlv \, \mDeltav
	\end{split}
	\begin{split}
		&+ \, \mKvm \, \mDeltavm
		\\[6pt]
		&+ \, \mKmm \, \mDeltavm
		\\[6pt]
		&+ \, \mKlm \, \mDeltavm
	\end{split}
	\hspace{-36pt}
	\begin{split}
		&+ \, \mKvl \, \mDeltalambda
		~,
		\\[6pt]
		&+ \, \mKml \, \mDeltalambda
		~,
		\\[6pt]
		&+ \, \mKll \, \mDeltalambda
		~.
	\end{split}
	\label{eq:element_tangent}
\end{align}
In Eq.~\eqref{eq:element_tangent}, the nine components of the tangent matrix $\mKno_\sz$ are the partial derivatives of the residual vector equations \eqref{eq:weak_residuals} with respect to the unknown vectors.
The stiffness matrix components for an LE implementation are provided in Appendix~\ref{sec:sec_appendix_tangent_calcs}.
For a different mesh velocity equation, only $\mKmv$, $\mKmm$, $\mKml$, and $\mrm$ are modified, however, other changes may be required to avoid well-known ALE issues associated with arbitrary mesh motions---such as the violation of the geometric conservation law \cite{thomas-aiaa-1979,farhat-cmame-1996}.

Equation~\eqref{eq:element_tangent} may be compactly written as
\begin{equation} \label{eq:finite_element_matrix_equation}
	\mKno_\sz ~ \mDeltauno_\so
	\, = \, - \mrno_\sz
	~,
\end{equation}
where the tangent matrix $\mKno_\sz$ and residual vector $\mrno_\sz$ are both calculated from $\mbuno_\sz$, which is known.
The degrees of freedom are updated according to Eq.~\eqref{eq:newton_raphson_delta_u}, and the process is repeated until the 2-norm of $\mDeltauno_\so$ falls below a specified iteration threshold $\epsilon_{\texttt{iter}}$.
At this point, the solution at time $\tno$, 
$\bmu (\zeta^\alpha, \, \tno)$,
is assumed to be specified by $\mbuno_\so$ through Eqs.~\eqref{eq:velocity_basis_expansion}, \eqref{eq:mesh_velocity_basis_expansion}, and \eqref{eq:lambda_basis_expansion}.

%
% *** Local Element Calculations
%

\subsection{Local Element Calculations} \label{sec:sec_fluid_local_calculations}

The residual vector and tangent matrix in Eq.~\eqref{eq:finite_element_matrix_equation} consists of integrals over the parametric domain $\Omega$.
However, the basis functions $N_I$ and $\bar{N}_I$ have compact support and are only nonzero over a fixed number of elements.
In practice, integrals are calculated locally over a single element $\Omegae$ to generate the elemental (or local) residual vector $\mre$ and tangent matrix $\mKe$, which are then assembled to form their global counterparts $\mr$ and $\mK$, respectively \cite{zienkiewicz-taylor-fem}.

Over any element, the fixed number of nonzero basis functions is denoted $\nen$ (\underline{n}umber of \underline{e}lement \underline{n}odes) and the fixed number of nonzero Lagrange multiplier basis functions is denoted $\neln$ (\underline{n}umber of \underline{e}lement \underline{L}agrange multiplier \underline{n}odes).
The corresponding elemental basis function matrices are given by
\begin{align}
	\mNe
	&:= \Big[ \hspace{5pt}
		N_{1}^e ~ \mone \hspace{9pt}
		N_{2}^e ~ \mone \hspace{5pt}
		\ldots \hspace{5pt}
		N_{\nen}^e ~ \mone \hspace{5pt}
	\Big]
	\label{eq:matrix_local_shape_fn_array}
	\\
	\shortintertext{and}
	\mhNe
	&:= \Big[ \hspace{5pt}
		\bar{N}_1^e \hspace{11pt}
		\bar{N}_2^e \hspace{6pt}
		\ldots \hspace{6pt}
		\bar{N}_{\neln}^e \hspace{5pt}
	\Big]
	~.
	\label{eq:matrix_local_LM_shape_fn_array}
\end{align}
Over a single element, at time $\tn$, the fundamental unknowns are expanded as
\begin{equation} \label{eq:fundamental_unknown_expanion_local}
	\bmv (\zeta^\alpha, \tn)
	\, = \, \mNe \, \mven
	~,
	\hspace{10pt}
	\bmvm (\zeta^\alpha, \tn)
	\, = \, \mNe \, \mvmen
	~,
	\hspace{10pt}
	\mathrm{and}
	\hspace{13pt}
	\lambda (\zeta^\alpha, \tn)
	\, = \, \mhNe \, \mlambdaen
	~,
\end{equation}
where $\mven$, $\mvmen$, and $\mlambdaen$ are the local degrees of freedom at time $\tn$.
Eq.~\eqref{eq:fundamental_unknown_expanion_local} is the local analog of Eqs.~\eqref{eq:velocity_basis_expansion}, \eqref{eq:mesh_velocity_basis_expansion}, and \eqref{eq:lambda_basis_expansion}.
Furthermore, as the shape function matrices in Eqs.~\eqref{eq:matrix_local_shape_fn_array} and \eqref{eq:matrix_local_LM_shape_fn_array} correspond to all nonzero degrees of freedom over a single element, integrals over $\Omega$ are calculated as
\begin{equation}
	\int_\Omega \Big(
		\ldots \mN \ldots \mhN \ldots
	\Big) ~\Jm ~\mathrm{d}\Omega
	\,
	=
	\sum_{e = 1}^{\nume} \, \int_{\Omegae} \! \Big(
		\ldots \mNe \ldots \mhNe \ldots
	\Big) ~\Jm ~\mathrm{d}\Omega
	~,
	\label{eq:fluid_element_sum}
\end{equation}
where the matrices $\mN$ and $\mhN$ in the integrand on the left-hand side are replaced by $\mNe$ and $\mhNe$, respectively, on the right-hand side.
Standard finite element techniques are used to assumble $\mr$ and $\mK$ from their local counterparts, $\mre$ and $\mKe$, according to Eq.~\eqref{eq:fluid_element_sum} \cite{zienkiewicz-taylor-fem}.
From now on, only local residual vector and tangent matrix calculations are provided.
The details of the numerical integration are left to Appendix~\ref{sec:sec_appendix_numerical_integration}.

%
% *** Dohrmann--Bochev Implementation
%

\subsection{Dohrmann--Bochev Implementation} \label{sec:sec_bochev_finite_element_implementation}

In this section, we demonstrate how the Dohrmann--Bochev contribution to the weak form \eqref{eq:D_B_weak} is incorporated into our finite element framework.
Since the global projection criterion \eqref{eq:D_B_projection_global} can be simplified to a condition over individual elements \eqref{eq:D_B_projection}, the Dohrmann--Bochev method naturally takes advantage of local element calculations.
In particular, Eq.~\eqref{eq:D_B_projection} allows one to express $\blambda$ in terms of $\lambda$ over any element.
With this result, the Dohrmann--Bochev direct Galerkin expression \eqref{eq:D_B_weak} can be rewritten such that its contributions to the residual vector and tangent matrix are easily calculated.

We begin by considering Eq.~\eqref{eq:D_B_projection} over a single element $\Omegae$.
Local basis functions $\breve{N}_i^e$ are chosen for the space $\mathbb{P}_1 (\Omegae)$~\eqref{eq:D_B_breve_lambda_subspace}.
Note that in this manuscript, local basis functions are indexed by lower-case Latin letters, while global basis functions are indexed by upper-case Latin letters.
During numerical integration, the element $\Omegae$ is mapped onto the unit square $\Omega_\square$, which is parametrized by $(\xi, \eta) \in [-1, 1] \times [-1, 1]$ (see Appendix~\ref{sec:sec_appendix_numerical_integration}).
On the unit square, the basis functions of $\mathbb{P}_1 (\Omegae)$, $\breve{N}^e_i$, are defined to span the space of all planes over the domain.
As any plane can be expressed as
$a + b \xi + c \eta = 0$
for constants
$a, \, b, \, c \in \mathbb{R}$,
the basis functions of $\mathbb{P}_1 (\Omegae)$ are given by
$\breve{N}^e_1 = 1$,
$\breve{N}^e_2 = \xi$,
and
$\breve{N}^e_3 = \eta$.
We define the row vector of basis functions, the column vector of projected unknowns at time $\tn$, and the column vector of projected unknown variations, respectively, as
\begin{equation} \label{eq:D_B_matrix_shape_fn_array_lagrange_multiplier}
	\setlength\arraycolsep{4pt}
	\mbNe
	:= \Big[ \hspace{5pt}
		\breve{N}_1^e \hspace{11pt}
		\breve{N}_2^e \hspace{11pt}
		\breve{N}_3^e \hspace{5pt}
	\Big]
	~,
	\hspace{20pt}
	\mblambdaen
	:=
	\begin{bmatrix}
		~ \\[-9pt]
		\, \breve{\lambda}_1 (\tn) \, \\[3pt]
		\, \breve{\lambda}_2 (\tn) \, \\[3pt]
		\, \breve{\lambda}_3 (\tn) \, \\[3pt]
	\end{bmatrix}
	~,
	\hspace{20pt}
	\mathrm{and}
	\hspace{30pt}
	\mdeltablambdae
	:=
	\begin{bmatrix}
		~ \\[-9pt]
		\, \delta \breve{\lambda}_1 \, \\[3pt]
		\, \delta \breve{\lambda}_2 \, \\[3pt]
		\, \delta \breve{\lambda}_3 \, \\[3pt]
	\end{bmatrix}
	~.
\end{equation}
On a single element $\Omegae$ at time $\tno$, the projected quantities $\blambda$ and $\delta \blambda$ can be written as
\begin{equation} \label{eq:D_B_expansion_basis_fns}
	\blambda (\zeta^\alpha, \, \tno)
	= \mbNe \, \mblambdaeno
	\hspace{25pt}
	\textrm{and}
	\hspace{30pt}
	\delta \blambda (\zeta^\alpha)
	= \mbNe \, \mdeltablambdae
	~.
\end{equation}

With Eq.~\eqref{eq:D_B_expansion_basis_fns}, and the separation of the projection criterion over individual elements \eqref{eq:D_B_projection}, $\blambda$ can be expressed in terms of $\lambda$ alone.
Substituting Eqs.~\eqref{eq:lambda_basis_expansion}, \eqref{eq:variation_element_basis_expansion}$_3$, and \eqref{eq:D_B_expansion_basis_fns} into Eq.~\eqref{eq:D_B_projection} and rearranging terms yields
\begin{equation} \label{eq:D_B_discretized_projection_general}
	\mdeltablambdae^{\mathrm{T}}
	\, \bigg\{
		\int_{\Omegae}
			\mbNe^{\mathrm{T}} \, \mhNe
		~\mathrm{d}\Omega
		~
	\bigg\}
	\, \mlambdaeno
	\, = \,
	\mdeltablambdae^{\mathrm{T}}
	\, \bigg\{
		\int_{\Omegae}
			\mbNe^{\mathrm{T}} \, \mbNe
		~\mathrm{d}\Omega
		~
	\bigg\}
	\, \mblambdaeno
	\hspace{15pt}
	\forall \,\, \mdeltablambdae
	~.
\end{equation}
Defining the matrices $\mGdb$ and $\mHdb$ according to
\begin{equation}
	\mGdb
	:= \int_{\Omegae}
		\mbNe^{\mathrm{T}} \, \mhNe
	~\mathrm{d}\Omega
	\hspace{25pt}
	\mathrm{and}
	\hspace{25pt}
	\mHdb
	:= \int_{\Omegae}
		\mbNe^{\mathrm{T}} \, \mbNe
	~\mathrm{d}\Omega
	~,
	\label{eq:D_B_matrices}
\end{equation}
and owing to the arbitrariness of $\mdeltablambdae$, Eq.~\eqref{eq:D_B_discretized_projection_general} simplifies to
\begin{equation} \label{eq:D_B_discretized_projection}
	\mGdb \, \mlambdaeno
	\, = \, \mHdb \, \mblambdaeno
	~.
\end{equation}
The mass matrix $\mHdb$ in Eq.~\eqref{eq:D_B_discretized_projection} is invertible, such that the projected surface tension coefficient matrix $\mblambdaeno$ is given in terms of the original surface tension coefficient matrix $\mlambdaeno$ as
\begin{equation} \label{eq:D_B_lambda_coeff_projection_relation}
	\mblambdaeno
	\, = \, \mHdb^{-1} \, \mGdb \, \mlambdaeno
	~.
\end{equation}
Eqs.~\eqref{eq:D_B_expansion_basis_fns}$_1$ and \eqref{eq:D_B_lambda_coeff_projection_relation} provide a way to calculate $\blambda$ given $\lambda$, on any element $\Omegae$ and at any time $\tno$, in accordance with the projection definition~\eqref{eq:D_B_projection}.

Our final step is to substitute our results into Eq.~\eqref{eq:D_B_weak} and calculate the Dohrmann--Bochev contribution to the tangent matrix and residual vector.
To this end, the Dohrmann--Bochev contribution to the weak formulation \eqref{eq:D_B_weak} is rewritten in a more convenient form.
The integrand of Eq.~\eqref{eq:D_B_weak} is expressed as
$
	(
		\delta \lambda
		- \delta \blambda
	) \,
	(
		\lambda
		- \blambda
	)
	= (
		\lambda \, \delta \lambda
		- \blambda \, \delta \blambda
	)
	- \blambda (
		\delta \lambda
		- \delta \blambda
	)
	- \delta \blambda (
		\lambda
		- \blambda
	)
$,
such that with the projection definition \eqref{eq:D_B_projection}, the Dohrmann--Bochev contribution to the weak form \eqref{eq:D_B_weak} is given by
\begin{equation} \label{eq:D_B_weak_simplified}
	\mcgDB
	\, = \,
	\sum_{e = 1}^{\nume} \,
		\dfrac{\alphaDB}{\zeta} \int_\Omegae
		\big(
			\lambda \, \delta \lambda
			- \blambda \, \delta \blambda
		\big)
	~\mathrm{d}\Omega
	~,
\end{equation}
where $\mcgDB$ is expressed as a sum over elements to take advantage of Eq.~\eqref{eq:D_B_lambda_coeff_projection_relation}.
Substituting Eqs.~\eqref{eq:lambda_basis_expansion}, \eqref{eq:variation_element_basis_expansion}$_3$, \eqref{eq:D_B_expansion_basis_fns}, and \eqref{eq:D_B_lambda_coeff_projection_relation} into Eq.~\eqref{eq:D_B_weak_simplified} yields
\begin{align*}
	\mcgDB
	\, &= \sum_{e = 1}^{\nume} \,
	\mdeltalambdae^\mT \,
	\dfrac{\alphaDB}{\zeta} \bigg\{
		\int_\Omegae \!\!
			\mhNe^{\mathrm{T}} \mhNe
		\, \mathrm{d}\Omega
		\, - \,
		\mGdb^{\mathrm{T}} \mHdb^{-1}
		\bigg(
			\int_\Omegae \!\!
				\mbNe^{\mathrm{T}} \mbNe
			\,\mathrm{d}\Omega \!
		\bigg)
		\, \mHdb^{-1} \mGdb
	\bigg\} \, \mlambdaeno
	\\[5pt]
	&= \, \sum_{e = 1}^{\nume} \,\,
	\mdeltalambdae^\mT
	\, \bigg\{
		\dfrac{\alphaDB}{\zeta}
		\int_\Omegae
			\mhNe^{\mathrm{T}} \mhNe
		~\mathrm{d}\Omega
		~
		- ~
		\dfrac{\alphaDB}{\zeta} \, \mGdb^{\mathrm{T}} \mHdb^{-1} \mGdb
	\bigg\} \, \mlambdaeno
	~,
	\stepcounter{equation}
	\tag{\theequation}\label{eq:D_B_weak_discretized_step_1}
\end{align*}
where the second line is simplified by recognizing the integral in parenthesis is $\mHdb$ \eqref{eq:D_B_matrices}.
As the result of Eq.~\eqref{eq:D_B_weak_discretized_step_1} is linear in $\mdeltalambdae$ and $\mlambdaeno$, its contributions to the elemental tangent matrix $\mKlleno$ and residual vector $\mrlambdaeno$ are easily calculated, as described in Appendix~\ref{sec:sec_appendix_fe_calculations}.
The local tangent matrices and residual vectors are then assembled to form their global counterparts using standard procedures.

%
% *** Dohrmann--Bochev Example
%

\subsubsection{Dohrmann--Bochev Example} \label{sec:sec_flat_dohrmann_bochev_example}

\begin{figure}[!t]
	\centering
	\begin{subfigure}[b]{0.24\columnwidth}
		\centering
		\includegraphics[width=\textwidth]{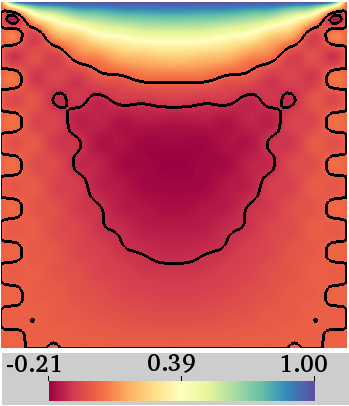}
		\caption{$v_x$, without DB}
		\label{fig:fig_cavity_no_DB_vx}
	\end{subfigure}
	\begin{subfigure}[b]{0.24\columnwidth}
		\centering
		\includegraphics[width=\textwidth]{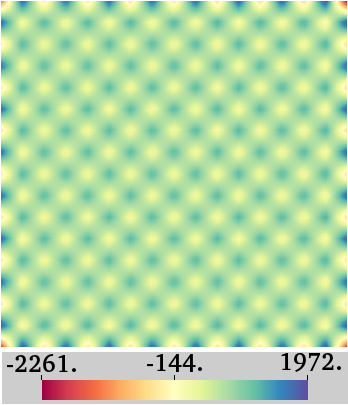}
		\caption{$\lambda$, without DB}
		\label{fig:fig_cavity_no_DB_lambda}
	\end{subfigure}
	\begin{subfigure}[b]{0.24\columnwidth}
		\centering
		\includegraphics[width=\textwidth]{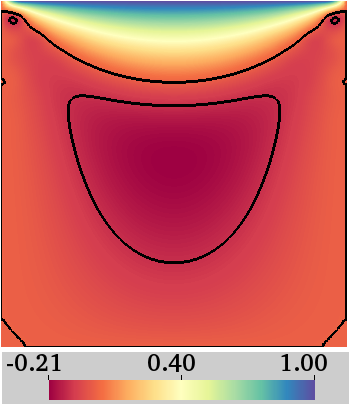}
		\caption{$v_x$, with DB}
		\label{fig:fig_cavity_with_DB_vx}
	\end{subfigure}
	\begin{subfigure}[b]{0.24\columnwidth}
		\centering
		\includegraphics[width=\textwidth]{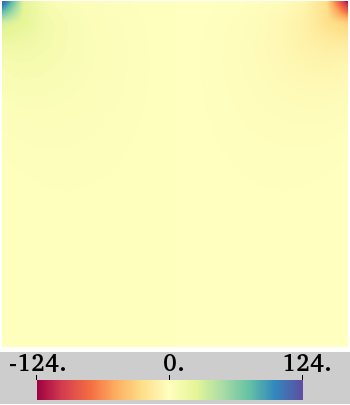}
		\caption{$\lambda$, with DB}
		\label{fig:fig_cavity_with_DB_lambda}
	\end{subfigure}
	\caption{
		Lid-driven cavity problem: the importance of Dohrmann--Bochev stabilization.
		(a),(b) Plots of the $x$-velocity (a) and surface tension (b) without Dohrmann--Bochev stabilization.
		Checkerboard patterns appear, with a characteristic length of $^{1 \!}/_{16}$ on a $16 \times 16$ mesh---a characteristic feature of the LBB instability.
		(c),(d) The same simulation is run with Dohrmann--Bochev stabilization terms, with the calculated $x$-velocity $v_x$ (c) and surface tension $\lambda$ (d) not exhibiting oscillations.
		In (a) and (c), contours are provided at $v_x = 0$ and $v_x = -0.12$.
	}
	\label{fig:fig_dohrmann_bochev}
\end{figure}
The lid-driven cavity problem demonstrates the need for inf--sup stabilization.
When the Dohrmann--Bochev terms are turned off and the Stokes flow problem is solved on a coarse $16 \times 16$ mesh, checkerboard patterns in the $x$-velocity (Fig.~\ref{fig:fig_cavity_no_DB_vx}) and surface tension (Fig.~\ref{fig:fig_cavity_no_DB_lambda}) are observed relative to the corresponding solution when the stabilization is included (Figs.~\ref{fig:fig_cavity_with_DB_vx} and \ref{fig:fig_cavity_with_DB_lambda}, respectively).
Moreover, a single square in the checkerboard pattern has length and width $^{1 \!}/_{16}$, and as the mesh size is changed, the checkerboard pattern changes in tandem.
Such oscillations indicate the violation of the inf--sup condition \cite{larson-bengzon-2013}.
Fig.~\ref{fig:fig_cavity_with_DB_lambda} is also useful in highlighting the sharp surface tension spikes at the top two corners of the domain, which prevent us from having a meaningful error analysis of the surface tension in this problem---as discussed in Sec.~\ref{sec:sec_flat_cavity}.

%
% *** Lagrangian Finite Element Formulation
%

\subsection{Lagrangian Finite Element Formulation} \label{sec:sec_lagrangian_finite_element_formulation}

As described in Sec.~\ref{sec:sec_ale_mesh_velocity_equations}, a Lagrangian scheme is recovered when the mesh velocity and material velocity are equal, i.e.
\begin{equation} \label{eq:lagrangian_mesh_velocity}
	\bmvm
	\, - \, \bmv
	\, = \, \bm{0}
	~.
\end{equation}
The velocities $\bmvm$ and $\bmv$ belong to the same space of functions (see Sec.~\ref{sec:sec_strong_solution_spaces}), as do their approximate counterparts $\bmvmh$ and $\bmvh$~\eqref{eq:fluid_velocity_subspace}.
We thus attain a Lagrangian scheme by setting all the nodal mesh and material velocities to be equal at all times $t$, written as
$\mvmt - \mvt = \mzero$.
Assuming
$\mvmtzero - \mvtzero = \mzero$
at the initial time $t_0$, the discretized Lagrangian mesh equation is given by
\begin{equation} \label{eq:lagrangian_strong_form}
	\mDeltavmno_\so
	\, - \, \mDeltavno_\so
	\, = \, \mzero
	~,
\end{equation}
for all times $t_{n+1}$ and iterations $i+1$ (see Secs.~\ref{sec:sec_time_integration} and \ref{sec:sec_newton_raphson_iteration}).

To implement a Lagrangian scheme, Eq.~\eqref{eq:lagrangian_strong_form} replaces the linearized and discretized mesh residual equation \eqref{eq:element_tangent}$_2$ of the ALE formulation.
Accordingly, the Lagrangian analog to Eqs.~\eqref{eq:element_tangent} and \eqref{eq:finite_element_matrix_equation} is given by
\begin{equation} \label{eq:lagrangian_matrix_large}
	\setlength\arraycolsep{8pt}
	\begin{bmatrix}
		~		& ~		& ~			\\[-8pt]
		\mKvv	& \mKvm	& \mKvl		\\[3pt]
		-\mone	& \mone	& \mzero	\\[3pt]
		\mKlv	& \mKlm	& \mKll		\\[3pt]
	\end{bmatrix}
	\,
	\begin{bmatrix}
		~				\\[-7pt]
		\mDeltav		\\[3pt]
		\mDeltavm		\\[3pt]
		\mDeltalambda	\\[3pt]
	\end{bmatrix}
	\, = \, -
	\begin{bmatrix}
		~			\\[-7pt]
		\mrv		\\[4pt]
		\mzero		\\[3pt]
		\mrlambda	\\[3pt]
	\end{bmatrix}
	~,
\end{equation}
which is equivalently expressed as
\begin{equation} \label{eq:lagrangian_matrix}
	\setlength\arraycolsep{6pt}
	\begin{bmatrix}
		~				& ~			\\[-9pt]
		\bigl(\,
			\mKvv + \mKvm
		\,\bigr)		& \mKvl		\\[7pt]
		\bigl(\,
			\mKlv + \mKlm
		\,\bigr)		& \mKll		\\[3pt]
	\end{bmatrix}
	\,
	\begin{bmatrix}
		~				\\[-8pt]
		\mDeltav		\\[8pt]
		\mDeltalambda	\\[3pt]
	\end{bmatrix}
	\, = \, -
	\begin{bmatrix}
		~			\\[-8pt]
		\mrv		\\[8pt]
		\mrlambda	\\[3pt]
	\end{bmatrix}
	~.
\end{equation}
With Eq.~\eqref{eq:lagrangian_matrix}, we implement a Lagrangian scheme within our ALE framework.
The tangent matrix and residual vector components in Eq.~\eqref{eq:lagrangian_matrix} are calculated as before (see Appendix~\ref{sec:sec_appendix_fe_calculations}), with the velocity $\bmv$ replacing the mesh velocity $\bmvm$.
Note that in a Lagrangian implementation, nonzero velocity boundary conditions lead to a moving boundary, according to Eqs.~\eqref{eq:ale_mesh_position} and \eqref{eq:lagrangian_mesh_velocity}.

%
% *** Finite Element Method Calculations
%

\section{Tangent Matrix \& Residual Vector Calculations} \label{sec:sec_appendix_fe_calculations}

In this section, the residual vectors and tangent matrices required for our LE finite element implementation are calculated.
To do so, several geometric quantities are linearized.
We then detail how the residual vectors and tangent matrices are modified when the surface is constrained to not deform (relevant to Secs.~\ref{sec:sec_flat_numerics} and \ref{sec:sec_flow_curved_surface}) and when the body force is a pressure drop acting everywhere normal to the surface (used in Sec.~\ref{sec:sec_curved_deforming_films}).
In the former case, inertial terms are included in the calculation as well.
However, these terms are not included when the fluid film can deform.
We end with a brief description of how integrals are numerically calculated, and then present an overview of our code, which shows the structure of the residual vector and tangent matrix calculations.

%
% *** Residual Vector
%

\subsection{Residual Vector} \label{sec:sec_appendix_residual_vector}

In this section, the residual vectors $\mrvt$, $\mrmt$, and $\mrlambdat$ are calculated such that they satisfy Eqs.~\eqref{eq:weak_expanded_variations} and \eqref{eq:weak_residuals}.
We adopt a convention where if $\mdeltave$ is a column vector, then
$\partial \mcg / \partial \mdeltave $
is a column vector as well.
Also, the matrix $\mN_{, \calpha}$ contains the partial derivatives of the basis functions with respect to $\zeta^\alpha$, and will be frequently used in our implementation.

We substitute the discretized arbitrary variations~\eqref{eq:variation_element_basis_expansion} and simplified form of the Dohrmann--Bochev weak form \eqref{eq:D_B_weak_discretized_step_1} into the direct Galerkin expression (\ref{eq:fluid_weak_form_with_DB}, \ref{eq:direct_galerkin_expression_including_DB}), remove inertial terms as they are assumed to be negligible, and rearrange the remaining terms to obtain
\begin{align*}
	\tilde{\mcg} (t)
	\, = \,
	&- \int_\Omega
		\, \mdeltav^\mT \mN^\mT \, \rho \bm{b}
	~\Jm
	~\mathrm{d} \Omega
	+ \int_\Omega 
		\, \mdeltav^\mT \mN^\mT_{, \calpha} \,\, \pi^{\calpha \cbeta} \, \bm{a}_\cbeta
		~\Jm
	~\mathrm{d} \Omega
	\,
	+ \int_\Omega
		\, \mdeltav^\mT \mN^\mT_{, \calpha} \,\, \bm{a}^\calpha \, \lambda
		~\Jm
	~\mathrm{d} \Omega
	\\[7pt]
	&
	+ \int_\Omega
		\, \mdeltalambda^\mT \mhN^\mT  \, \big(
			\bm{a}^\calpha \cdot \bm{v}_{, \calpha}
		\big)
		~\Jm
	~\mathrm{d}\Omega
	\,
	+ \, \alpham \! \int_\Omega
		\, \mdeltavm^\mT \mN^\mT \, \Big(
			\bmvm
			- \big(
				\bm{n} \otimes \bm{n}
			\big) \bmv
		\Big)
		~\Jm
	~\mathrm{d} \Omega
	\stepcounter{equation}
	\tag{\theequation}\label{eq:appendix_fluid_weak_form_linear_variation}
	\\[5pt]
	&
	- \, \sum_{e = 1}^{\nume} \,\,
	\mdeltalambdae^\mT
	\, \bigg\{
		\dfrac{\alphaDB}{\zeta}
		\int_\Omegae \!
			\mhNe^{\mathrm{T}} \mhNe
		~\mathrm{d}\Omega
		\, - \,
		\dfrac{\alphaDB}{\zeta} \, \mGdb^{\mathrm{T}} \mHdb^{-1} \mGdb
	\bigg\} \, \mlambdaet
	~,
\end{align*}
where the first negative sign in the third line arises because $\mcgDB$ is subtracted in Eq.~\eqref{eq:direct_galerkin_expression_including_DB}.
In Eq.~\eqref{eq:appendix_fluid_weak_form_linear_variation}, we used the relation (for example) 
$ \delta \bmv \cdot \rho \bm{b} = \mdeltav^{\mT} \mN^{\mT} \, \rho \bm{b}$, where $\mN$ is defined in Eq.~\eqref{eq:matrix_shape_fn_array}$_1$.
The variation coefficient vectors in Eq.~\eqref{eq:appendix_fluid_weak_form_linear_variation} are independent of $\zeta^\alpha$, and are moved outside the integrals to yield
\begingroup
\allowdisplaybreaks
\begin{align*}
	\tilde{\mcg} (t)
	\, = \,
	&- \mdeltav^\mT \! \int_\Omega
		\, \mN^\mT \, \rho \bm{b}
	~\Jm
	~\mathrm{d} \Omega
	+ \mdeltav^\mT \! \int_\Omega 
		\, \mN^\mT_{, \calpha} \,\, \pi^{\calpha \cbeta} \, \bm{a}_\cbeta
		~\Jm
	~\mathrm{d} \Omega
	\,
	+ \mdeltav^\mT \! \int_\Omega
		\, \mN^\mT_{, \calpha} \,\, \bm{a}^\calpha \, \lambda
		~\Jm
	~\mathrm{d} \Omega
	\\[7pt]
	&
	+ \mdeltalambda^\mT \! \int_\Omega
		\, \mhN^\mT  \, \big(
			\bm{a}^\calpha \cdot \bm{v}_{, \calpha}
		\big)
		~\Jm
	~\mathrm{d}\Omega
	\,
	+ \mdeltavm^\mT \, \alpham \! \int_\Omega
		\, \mN^\mT \, \Big(
			\bmvm
			- \big(
				\bm{n} \otimes \bm{n}
			\big) \bmv
		\Big)
		~\Jm
	~\mathrm{d} \Omega
	\stepcounter{equation}
	\tag{\theequation}\label{eq:appendix_fluid_weak_form_linear_variation_step_two}
	\\[5pt]
	&
	- \, \sum_{e = 1}^{\nume} \,\,
	\mdeltalambdae^\mT
	\, \bigg\{
		\dfrac{\alphaDB}{\zeta}
		\int_\Omegae \!
			\mhNe^{\mathrm{T}} \mhNe
		~\mathrm{d}\Omega
		\, - \,
		\dfrac{\alphaDB}{\zeta} \, \mGdb^{\mathrm{T}} \mHdb^{-1} \mGdb
	\bigg\} \, \mlambdaet
	~.
\end{align*}
\endgroup
Defining the velocity, mesh velocity, and surface tension residual vectors as
\begin{align}
	\mrvt \,
	&= \,
	- \int_\Omega
		\, \mN^\mT \, \rho \bm{b}
		~\Jm
	~\mathrm{d} \Omega
	\,
	+ \int_\Omega 
		\, \mN^\mT_{, \calpha} \,\, \Big(
			\pi^{\calpha \cbeta} \, \bm{a}_\cbeta
			+ \bm{a}^\calpha \, \lambda
		\Big)
		~\Jm
	~\mathrm{d} \Omega
	~,
	\label{eq:appendix_mrve_ale}
	\\[8pt]
	\mrmt \,
	&= \,
	\alpham \! \int_\Omega
		\, \mN^\mT \, \Big(
			\bmvm
			- \big(
				\bm{n} \otimes \bm{n}
			\big) \bmv
		\Big)
		~\Jm
	~\mathrm{d} \Omega
	~,
	\label{eq:appendix_mrme_ale}
	\\
	\shortintertext{and}
	\mrlambdat \,
	&= \,
	\int_\Omega
		\, \mhN^\mT  \, \big(
			\bm{a}^\calpha \cdot \bm{v}_{, \calpha}
		\big)
		~\Jm
	~\mathrm{d}\Omega
	\label{eq:appendix_mrlambdae_ale}
	\\[-2pt]
	\nonumber
	\, &\hspace{30pt}
	- \, \sum_{e = 1}^{\nume} \,\,
	\bigg\{
		\dfrac{\alphaDB}{\zeta}
		\int_\Omegae \!
			\mhNe^{\mathrm{T}} \mhNe
		~\mathrm{d}\Omega
		\, - \,
		\dfrac{\alphaDB}{\zeta} \, \mGdb^{\mathrm{T}} \mHdb^{-1} \mGdb
	\bigg\} \, \mlambdaet
	~,
\end{align}
Eq.~\eqref{eq:appendix_fluid_weak_form_linear_variation_step_two} can be written as in Eq.~\eqref{eq:weak_expanded_variations}.

%
% *** Tangent Matrix
%

\subsection{Tangent Matrix} \label{sec:sec_appendix_tangent_calcs}

The tangent matrix components resulting from the linearization given in Eq.~\eqref{eq:element_tangent} are defined as
\begingroup
\allowdisplaybreaks
\begin{equation} \label{eq:appendix_tangent_component_defs}
	\begin{split}
		\mKvv
		&:= \dfrac{\partial (\mrvno_\sz)}{\partial (\mvno_\sz)^\mT}
		~,
		\\[8pt]
		\mKmve
		&:= \dfrac{\partial (\mrmno_\sz)}{\partial (\mvno_\sz)^\mT}
		~,
		\\[8pt]
		\mKlve
		&:= \dfrac{\partial (\mrlambdano_\sz)}{\partial (\mvno_\sz)^\mT}
		~,
	\end{split}
	\hspace{30pt}
	\begin{split}
		\mKvm
		&:= \dfrac{\partial (\mrvno_\sz)}{\partial (\mvmno_\sz)^\mT}
		~,
		\\[8pt]
		\mKmm
		&:= \dfrac{\partial (\mrmno_\sz)}{\partial (\mvmno_\sz)^\mT}
		~,
		\\[8pt]
		\mKlm
		&:= \dfrac{\partial (\mrlambdano_\sz)}{\partial (\mvmno_\sz)^\mT}
		~,
	\end{split}
	\hspace{30pt}
	\begin{split}
		\mKvl
		&:= \dfrac{\partial (\mrvno_\sz)}{\partial (\mlambdano_\sz)^\mT}
		~,
		\\[8pt]
		\mKml
		&:= \dfrac{\partial (\mrmno_\sz)}{\partial (\mlambdano_\sz)^\mT}
		~,
		\\[8pt]
		\mKll
		&:= \dfrac{\partial (\mrlambdano_\sz)}{\partial (\mlambdano_\sz)^\mT}
		~,
	\end{split}
\end{equation}
\endgroup
with all components calculated from the known degree of freedom vector at iteration $i$, $\mbuno_\sz$, alone.
Before calculating the tangent matrix components, several geometric quantities are linearized for their subsequent use in the tangent matrix calculation.

%
% *** Unknown Linearization
%

\subsubsection{Unknown Linearization} \label{sec:sec_appendix_linearization}

To linearize the residual vectors $\mrvno_\so$, $\mrmno_\so$, and $\mlambdano_\so$ about the state $\bmu(\tno)_\sz$, we rewrite Eq.~\eqref{eq:newton_raphson_delta_u} as
\begin{equation} \label{eq:appendix_unknown_steps}
	\bmv_\so
	= \bmv_\sz
	+ \Delta \bmv_\so
	~,
	\hspace{10pt}
	\bmvm_\so
	= \bmvm_\sz
	+ \Delta \bmvm_\so
	~,
	\hspace{10pt}
	\mathrm{and}
	\hspace{13pt}
	\lambda_\so
	= \lambda_\sz
	+ \Delta \lambda_\so
	~,
\end{equation}
where the time dependence is not written for notational simplicity.
From now on, we also omit the subscript $(i + 1)$ from the change in fundamental unknowns.
In this section, we calculate how various geometric quantities, namely the position $\bmx$, basis vectors $\bm{a}_\calpha$ and $\bm{a}^\calpha$, metric components $a_{\calpha \cbeta}$ and $a^{\calpha \cbeta}$, normal $\bm{n}$, and Jacobian $\Jm$, differ between iterations $i$ and $i + 1$ given the relations in Eq.~\eqref{eq:appendix_unknown_steps}.

With the backward Euler time discretization \eqref{eq:position_time_expansion}, the surface position is iterated according to
\begin{equation} \label{eq:appendix_position_discretization}
	\check{\bmx}_\so
	= \check{\bmx}_\sz
	+ \Delta t \, \Delta \bmvm
	~,
\end{equation}
from which we define
$\Delta \check{\bmx} := \Delta t \, \Delta \bmvm$.
With an equation for the position, and the definitions
$\bm{a}_\calpha = \check{\bmx}_{, \calpha}$
and
$a_{\calpha \cbeta} = \bm{a}_\calpha \cdot \bm{a}_\cbeta$,
it is straightforward to calculate
\begin{align}
	\bm{a}_{\calpha}^\so
	\, &= \, \bm{a}_{\calpha}^\sz
	\, + \, \Delta t \, \Delta \bmvm_{, \calpha}
	\\
	\shortintertext{and}
	a_{\calpha \cbeta}^\so
	\, &= \, a_{\calpha \cbeta}^\sz
	\, + \, \Delta t \, \Delta \bmvm_{, \calpha} \cdot \bm{a}_\cbeta^\sz
	\, + \, \Delta t \, \Delta \bmvm_{, \cbeta} \cdot \bm{a}_\calpha^\sz
	~.
\end{align}
The shorthand
$\Delta \bm{a}_\calpha := \Delta t \, \Delta \bmvm_{, \calpha}$
and
$
\Delta a_{\calpha \cbeta}
:= \Delta t \, \Delta \bmvm_{, \calpha} \cdot \bm{a}_\cbeta^\sz
+ \Delta t \, \Delta \bmvm_{, \cbeta} \cdot \bm{a}_\calpha^\sz
$
will often be used in our linearization calculations.

To determine $a^{\calpha \cbeta}_\so$, the identity
$
\partial a^{\calpha \cbeta} / \partial a_{\cmu \cnu}
= - \tfrac{1}{2} ( a^{\calpha \cmu} \, a^{\cbeta \cnu}
+ a^{\calpha \cnu} \, a^{\cbeta \cmu} )
$
is required, such that
$\Delta a^{\calpha \cbeta}$ is calculated according to
$\Delta a^{\calpha \cbeta} = (\partial a^{\calpha \cbeta} / \partial a_{\cmu \cnu}) \, \Delta a_{\cmu \cnu}$.
After rearranging terms and using the symmetry of the metric components, one obtains
\begin{equation}
	a^{\calpha \cbeta}_\so
	\, = \, a^{\calpha \cbeta}_\sz
	\, - \, a^{\calpha \cmu}_\sz \, a^{\cbeta \cnu}_\sz \, \Big[
		\Delta t \, \Delta \bmvm_{, \cmu} \cdot \bm{a}_\cnu^\sz
		\, + \, \Delta t \, \Delta \bmvm_{, \cnu} \cdot \bm{a}_\cmu^\sz
	\Big]
	~.
\end{equation}
The calculation of $\bm{a}^\calpha_\so$ is straightforward, as
$\bm{a}^\calpha = a^{\calpha \cbeta} \, \bm{a}_\cbeta$;
however terms are rearranged to simplify our numerical implementation.
Introducing the identity tensor $\bm{1}$ in curvilinear coordinates as
$\bm{1} = \bm{a}_\clambda \otimes \bm{a}^\clambda + \bm{n} \otimes \bm{n}$,
we find
\begin{equation} \label{eq:appendix_Delta_bm_a_ualpha_calc}
	\begin{split}
		\Delta \bm{a}^\calpha
		\, &= \, \Delta (a^{\calpha \cbeta} \, \bm{a}_\beta)
		\, = \, a^{\calpha \cbeta} \, \Delta \bm{a}_\cbeta
		\, + \, \dfrac{\partial a^{\calpha \cbeta}}{\partial a_{\cmu \cnu}} \, \Delta a_{\cmu \cnu} \, \bm{a}_\cbeta
		\\[0pt]
		&= \, a^{\calpha \cbeta} \, \Delta \bm{a}_\cbeta
		\, - \, \bm{a}_\cbeta \, \big(
			\Delta \bm{a}_\cmu \cdot \bm{a}_\cnu
			+ \Delta \bm{a}_\cnu \cdot \bm{a}_\cmu
		\big) \, a^{\calpha \cmu} \, a^{\cbeta \cnu}
		\\[4pt]
		&= \, a^{\calpha \cbeta} \, \Delta \bm{a}_\cbeta
		\, - \, \big(
			\Delta \bm{a}_\cmu \cdot \bm{a}_\cnu
		\big) \, \bm{a}_\cbeta \, a^{\calpha \cmu} \, a^{\cbeta \cnu}
		\, - \, \big(
			\Delta \bm{a}_\cnu \cdot \bm{a}_\cmu
		\big) \, \bm{a}_\cbeta \, a^{\calpha \cmu} \, a^{\cbeta \cnu}
		\\[4pt]
		&= \, a^{\calpha \cbeta} \,\, \bm{1} \,\, \Delta \bm{a}_\cbeta
		\, - \, \big(
			\Delta \bm{a}_\cmu \cdot \bm{a}^\cbeta
		\big) \, \bm{a}_\cbeta \, a^{\calpha \cmu}
		\, - \, \big(
			\Delta \bm{a}_\cnu \cdot \bm{a}^\calpha
		\big) \, \bm{a}^\cnu
		\\[4pt]
		&= \, a^{\calpha \cbeta} \Big[
			\bm{a}_\clambda \otimes \bm{a}^\clambda
			+ \bm{n} \otimes \bm{n}
		\Big] \, \Delta \bm{a}_\cbeta
		\, - \, \big(
			\bm{a}_\cbeta \otimes \bm{a}^\cbeta
		\big) \, \Delta \bm{a}_\cmu \, a^{\calpha \cmu}
		\, - \, \big(
			\bm{a}^\cnu \otimes \bm{a}^\calpha
		\big) \, \Delta \bm{a}_\cnu
		\\[4pt]
		&= \, \Big[
			a^{\calpha \cbeta} \, \big(
				\bm{n} \otimes \bm{n}
			\big)
			- \bm{a}^\cbeta \otimes \bm{a}^\calpha
		\Big] \, \Delta \bm{a}_\cbeta
		~.
	\end{split}
\end{equation}
In the third line of Eq.~\eqref{eq:appendix_Delta_bm_a_ualpha_calc} the metric components were used to raise indices, in the fourth line the identity tensor was introduced and the last two terms were rewritten using dyadic products, and in the last line dummy indices were rearranged.
With the result of Eq.~\eqref{eq:appendix_Delta_bm_a_ualpha_calc}, one obtains
\begin{equation} \label{eq:appendix_linearization_bm_a_ualpha}
	\bm{a}^\calpha_\so
	\, = \, \bm{a}^\calpha_\sz
	\, + \, \Big[
		a^{\calpha \cbeta}_\sz \, \bm{n}_\sz \otimes \bm{n}_\sz
		\, - \, \bm{a}^\cbeta_\sz \otimes \bm{a}^\calpha_\sz
	\Big] \, \Delta t \, \Delta \bmvm_{, \cbeta}
	~.
\end{equation}

To calculate $\Delta \bm{n}$, note
$\bm{n} \cdot \bm{n} = 1$,
which implies
$\Delta \bm{n} \cdot \bm{n} = 0$,
so $\Delta \bm{n}$ is completely specified by its in-plane components:
$\Delta \bm{n} = \bm{a}^\calpha ( \Delta \bm{n} \cdot \bm{a}_\calpha )$.
Next, observe
$
\Delta \bm{n} \cdot \bm{a}_\calpha
= \Delta (\bm{n} \cdot \bm{a}_\calpha)
- \bm{n} \cdot \Delta \bm{a}_\calpha
= - \bm{n} \cdot \Delta \bm{a}_\calpha
$,
as $\bm{n}$ is orthogonal to $\bm{a}_\calpha$.
Writing $\Delta \bm{n}$ as
$\Delta \bm{n} = - (\bm{a}^\calpha \otimes \bm{n}) \Delta \bm{a}_\calpha$,
we find $\bm{n}_\so$ to be given by
\begin{equation} \label{eq:appendix_linearization_bm_n}
	\bm{n}_\so
	\, = \, \bm{n}_\sz
	\, - \, \big(
		\bm{a}^\alpha_\sz \otimes \bm{n}_\sz
	\big) \, \Delta t \, \Delta \bmvm_{, \calpha}
	~.
\end{equation}
Finally, to calculate $\Delta \Jm$, we begin with the definition of the normal vector $\bm{n}$ as
$\bm{n} = ( \bm{a}_\cone \times \bm{a}_\ctwo ) / \Jm$,
such that
$
\Delta \bm{n}
= ( \Delta \bm{a}_\cone \times \bm{a}_\ctwo ) / \Jm
+ ( \bm{a}_\cone \times \Delta \bm{a}_\ctwo ) / \Jm
+ ( \bm{a}_\cone \times \bm{a}_\ctwo ) \, \Delta (1 / \Jm)
$.
Substituting
$
( \bm{a}_\cone \times \bm{a}_\ctwo ) \, \Delta (1 / \Jm)
= - \bm{n} \, ( \Delta \Jm / \Jm )
$
into the expression for $\Delta \bm{n}$, contracting both sides with $\bm{n}$, and remembering
$\Delta \bm{n} \cdot \bm{n} = 0$
leads to
\begin{equation}
	\Delta \Jm
	= \Delta \bm{a}_\cone \cdot \big( \bm{a}_\ctwo \times \bm{n} \big)
	+ \Delta \bm{a}_\ctwo \cdot \big( \bm{n} \times \bm{a}_\ctwo \big)
	~.
\end{equation}
Substituting
$\bm{n} = ( \bm{a}_\cone \times \bm{a}_\ctwo ) / \Jm$
into the above equation, and using the vector triple product rule, yields
\begin{equation}
	\begin{split}
		\Delta \Jm
		&= \dfrac{1}{\Jm} \Big[
			\Delta \bm{a}_\cone \cdot \big(
				a_{\ctwo \ctwo} \, \bm{a}_\cone
				- a_{\ctwo \cone} \, \bm{a}_\ctwo
			\big)
			+ \Delta \bm{a}_\ctwo \cdot \big(
				a_{\cone \cone} \, \bm{a}_\ctwo
				- a_{\cone \ctwo} \, \bm{a}_\cone
			\big)
		\Big]
		\\[4pt]
		&= \Jm \, \Big[
			\Delta \bm{a}_\cone \cdot \big(
				a^{\cone \cone} \, \bm{a}_\cone
				+ a^{\cone \ctwo} \, \bm{a}_\ctwo
			\big)
			+ \Delta \bm{a}_\ctwo \cdot \big(
				a^{\ctwo \ctwo} \, \bm{a}_\ctwo
				+ a_{\ctwo \cone} \, \bm{a}_\cone
			\big)
		\Big]
		\\[4pt]
		&= \Jm \, \Big[
			\Delta \bm{a}_\cone \cdot \bm{a}^\cone
			+ \Delta \bm{a}_\ctwo \cdot \bm{a}^\ctwo
		\Big]
		~=~ \Jm \, \Big[\,
			\bm{a}^\calpha \cdot \Delta \bm{a}_\calpha
		\,\Big]
		~,
	\end{split}
\end{equation}
where in the second line we substituted
$a_{\cone \cone} = (\Jm)^2 \, a^{\ctwo \ctwo}$,
$a_{\cone \ctwo} = a_{\ctwo \cone} = - (\Jm)^2 \, a^{\ctwo \cone}$,
and
$a_{\ctwo \ctwo} = (\Jm)^2 \, a^{\cone \cone}$ because $a^{\calpha \cbeta}$ is the matrix inverse of $a_{\calpha \cbeta}$, and in the third line simplified with
$\bm{a}^\calpha = a^{\calpha \cbeta} \, \bm{a}_\cbeta$.
Accordingly, $\Jm_\so$ is found to be
\begin{equation} \label{eq:appendix_linearization_jacobian}
	\Jm_\so
	\, = \, \Jm_\sz
	\, + \, \Jm_\sz \, \big(
		\bm{a}^\calpha_\sz
		\cdot \Delta t \, \Delta \bmvm_{, \calpha}
	\big)
	~.
\end{equation}

%
% *** Local Component Calculations
%

\subsubsection{Local Component Calculations} \label{sec:sec_appendix_tangent_component_calcs}

The tangent matrix components are calculated by substituting Eq.~\eqref{eq:appendix_unknown_steps}, along with the corresponding changes in geometric quantities described above, into the residual equations \eqref{eq:appendix_mrve_ale}--\eqref{eq:appendix_mrlambdae_ale} and rearranging terms to match the structure shown in Eq.~\eqref{eq:element_tangent}.
As described in Sec.~\ref{sec:sec_fluid_local_calculations}, in practice only local residual vectors and tangent matrices are numerically evaluated.
We accordingly provide local results here, where elemental vector and matrix quantities are denoted with a superscript $e$.

We present the calculation of $\mKlve$, $\mKlme$, and $\mKlle$ to demonstrate our procedure, and the remaining components of the tangent matrix will be provided without calculation.
Substituting Eq.~\eqref{eq:appendix_unknown_steps} into Eq.~\eqref{eq:appendix_mrlambdae_ale} and keeping only terms up to first order in $\Delta \bmu$ over the element $\Omegae$, one obtains
\begin{align*}
	\mrlambdaeno_\so
	\, &= \,
	\int_\Omegae
		\, \mhNe^\mT  \, \big(
			\bm{a}^\calpha_\sz \cdot \bm{v}_{, \calpha}^\sz
		\big)
		~\Jm_\sz
	~\mathrm{d}\Omega
	\, + \,
	\int_\Omegae
		\, \mhNe^\mT  \, \big(
			\Delta \bm{a}^\calpha \cdot \bm{v}_{, \calpha}^\sz
		\big)
		~\Jm_\sz
	~\mathrm{d}\Omega
	\\[4pt]
	& \hspace{24pt}
	+ \,
	\int_\Omegae
		\, \mhNe^\mT  \, \big(
			\bm{a}^\calpha_\sz \cdot \Delta \bm{v}_{, \calpha}
		\big)
		~\Jm_\sz
	~\mathrm{d}\Omega
	\, + \,
	\int_\Omegae
		\, \mhNe^\mT  \, \big(
			\bm{a}^\calpha_\sz \cdot \bm{v}_{, \calpha}^\sz
		\big)
		~\Delta J
	~\mathrm{d}\Omega
	\stepcounter{equation}
	\tag{\theequation}\label{eq:appendix_tangent_example_first}
	\\[5pt]
	&\hspace{48pt}
	\, - \,
	\bigg\{
		\dfrac{\alphaDB}{\zeta}
		\int_\Omegae \!
			\mhNe^{\mathrm{T}} \mhNe
		~\mathrm{d}\Omega
		\, - \,
		\dfrac{\alphaDB}{\zeta} \, \mGdb^{\mathrm{T}} \mHdb^{-1} \mGdb
	\bigg\} \, \Big(
		\mlambdaen
		\, + \, \mDeltalambdae
	\Big)
	~.
\end{align*}
Substituting $\Delta \bm{a}^\calpha$ from Eq.~\eqref{eq:appendix_linearization_bm_a_ualpha} and $\Delta J$ from Eq.~\eqref{eq:appendix_linearization_jacobian} into Eq.~\eqref{eq:appendix_tangent_example_first}, recognizing the first term on the right-hand side and the Dohrmann--Bochev term on the last line involving $\mlambdaen$ combine to be the residual at iteration $i$, and rearranging terms, we find
\begin{equation} \label{eq:appendix_tangent_example_second}
	\begin{split}
		\mrlambdaeno_\so
		\, &= \,
		\mrlambdaeno_\sz
		\, + \,
		\int_\Omegae
			\! \mhNe^\mT  \, \bm{v}_{, \calpha}^\sz \cdot \Big[
				a^{\calpha \cbeta}_\sz \, \bm{n}_\sz \otimes \bm{n}_\sz
				\, - \, \bm{a}^\cbeta_\sz \otimes \bm{a}^\calpha_\sz
			\Big] \, \Delta t \, \Delta \bmvm_{, \cbeta}
			~\Jm_\sz
		~\mathrm{d}\Omega
		\\[4pt]
		&\hspace{10pt}
		+ \,
		\int_\Omegae
			\! \mhNe^\mT  \, \big(
				\bm{a}^\calpha_\sz \cdot \Delta \bm{v}_{, \calpha}
			\big)
			~\Jm_\sz
		~\mathrm{d}\Omega
		\, + \,
		\int_\Omegae
			\! \mhNe^\mT  \, \big(
				\bm{a}^\calpha_\sz \cdot \bm{v}_{, \calpha}^\sz
			\big)
			\,
			\big(
				\bm{a}^\clambda_\sz
				\cdot \Delta t \, \Delta \bmvm_{, \clambda}
			\big)
			~\Jm
		~\mathrm{d}\Omega
		\\[7pt]
		&\hspace{20pt}
		\, - \,
		\bigg\{
			\dfrac{\alphaDB}{\zeta}
			\int_\Omegae \!
				\mhNe^{\mathrm{T}} \mhNe
			~\mathrm{d}\Omega
			\, - \,
			\dfrac{\alphaDB}{\zeta} \, \mGdb^{\mathrm{T}} \mHdb^{-1} \mGdb
		\bigg\} \, \mDeltalambdae
		~.
	\end{split}
\end{equation}
Equation \eqref{eq:appendix_tangent_example_second} is now expressed entirely in terms of known quantities at iteration $i$ and the changes to the fundamental unknowns.
The changes in the fundamental unknowns are discretized as in Eqs.~\eqref{eq:velocity_time_expansion_discretized}--\eqref{eq:lambda_time_expansion_discretized}, substituted into Eq.~\eqref{eq:appendix_tangent_example_second}, and the unknown vectors are moved outside the resultant integrals to give
\begin{align*}
	\mrlambdaeno_\so
	\, = \,
	\mrlambdaeno_\sz
	\, &+ \,
	\bigg\{
		\int_\Omegae
			\! \mhNe^\mT  \, \bm{v}_{, \calpha}^\sz \cdot \Big[
				a^{\calpha \cbeta}_\sz \, \bm{n}_\sz \otimes \bm{n}_\sz
				\, - \, \bm{a}^\cbeta_\sz \otimes \bm{a}^\calpha_\sz
			\Big] \, \mNe_{, \cbeta} \, \Delta t
			~\Jm_\sz
		~\mathrm{d}\Omega
	\bigg\} \, \mDeltavme
	\\[4pt]
	&
	+ \,
	\bigg\{
		\int_\Omegae
			\! \mhNe^\mT  \, \big(
				\bm{a}^\calpha_\sz \cdot \bm{v}_{, \calpha}^\sz
			\big)
			\, \bm{a}^\clambda_\sz \, \mNe_{, \clambda} \, \Delta t
			~\Jm
		~\mathrm{d}\Omega
	\bigg\} \, \mDeltavme
	\\[6pt]
	&
	+ \,
	\bigg\{
		\int_\Omegae
			\! \mhNe^\mT  \, \bm{a}^\calpha_\sz \, \mNe_{, \calpha}
			~\Jm_\sz
		~\mathrm{d}\Omega
	\bigg\} \, \mDeltave
	\\[5pt]
	&
	\, - \,
	\bigg\{
		\dfrac{\alphaDB}{\zeta}
		\int_\Omegae \!
			\mhNe^{\mathrm{T}} \mhNe
		~\mathrm{d}\Omega
		\, - \,
		\dfrac{\alphaDB}{\zeta} \, \mGdb^{\mathrm{T}} \mHdb^{-1} \mGdb
	\bigg\} \, \mDeltalambdae
	~.
	\stepcounter{equation}
	\tag{\theequation}\label{eq:appendix_tangent_example_third}
\end{align*}
In comparing Eq.~\eqref{eq:appendix_tangent_example_third} with Eq.~\eqref{eq:element_tangent}$_3$, the tangent matrices for this equation are given by
\begin{align}
	\mKlve
	\, &= \,
	\int_\Omegae
		\! \mhNe^\mT  \, \bm{a}^\calpha_\sz \, \mNe_{, \calpha}
		~\Jm_\sz
	~\mathrm{d}\Omega
	~,
	\label{eq:appendix_mKlve}
	\\[4pt]
	\begin{split}
		\mKlme
		\, &= \,
		\int_\Omegae
			\! \mhNe^\mT  \, \bm{v}_{, \calpha}^\sz \cdot \Big[
				a^{\calpha \cbeta}_\sz \, \bm{n}_\sz \otimes \bm{n}_\sz
				\, - \, \bm{a}^\cbeta_\sz \otimes \bm{a}^\calpha_\sz
			\Big] \, \mNe_{, \cbeta} \, \Delta t
			~\Jm_\sz
		~\mathrm{d}\Omega
		\\[4pt]
		\, &\hspace{25pt}
		+ \,
		\int_\Omegae
			\! \mhNe^\mT  \, \big(
				\bm{a}^\calpha_\sz \cdot \bm{v}_{, \calpha}^\sz
			\big)
			\, \bm{a}^\clambda_\sz \, \mNe_{, \clambda} \, \Delta t
			~\Jm
		~\mathrm{d}\Omega
		~,
	\end{split}
	\label{eq:appendix_mKlme}
	\\
	\shortintertext{and}
	\mKlle \, &=
	\, - \, \dfrac{\alphaDB}{\zeta}
	\int_\Omegae \!
		\mhNe^{\mathrm{T}} \mhNe
	~\mathrm{d}\Omega
	\, + \,
	\dfrac{\alphaDB}{\zeta} \, \mGdb^{\mathrm{T}} \mHdb^{-1} \mGdb
	~.
	\label{eq:appendix_mKlle}
\end{align}
The presence of the matrices $\mGdb$ and $\mHdb$ in Eqs.~\eqref{eq:appendix_tangent_example_third} and \eqref{eq:appendix_mKlle} require a minor modification to the structure of standard finite element codes.
For each element, $\mGdb$ and $\mHdb$ are calculated as integrals over the element \eqref{eq:D_B_matrices}; they are themselves assembled during the iteration loop over Gauss quadrature points.
Once the loop is complete, the contributions of the last
terms in Eq.~\eqref{eq:appendix_mrlambdae_ale},
as well as the terms in Eq.~\eqref{eq:appendix_mKlle}, are calculated and assembled.
The organization of this calculation is highlighted in Appendix~\ref{sec:sec_appendix_code_structure}.

An identical procedure is followed to determine the remaining local tangent matrix components.
For Eq.~\eqref{eq:element_tangent}$_1$ such an analysis yields
\begingroup
\allowdisplaybreaks
\begin{align*}
	\mKvve
	\, &= \,
	\zeta \int_\Omegae \!
		\mNe^\mT_{, \calpha} \, \big(
			\bm{a}^\cgamma_\sz \otimes \bm{a}^\calpha_\sz
		\big) \, \mNe_{, \cgamma}
		~\Jm_\sz
	~\mathrm{d}\Omega
	\, + \, \zeta \int_\Omegae \!
		\mNe^\mT_{, \calpha} \, \big(
			\bm{a}_\cbeta^\sz \otimes \bm{a}^\cbeta_\sz
		\big) \, a^{\cgamma \calpha}_\sz \, \mNe_{, \cgamma}
		~\Jm_\sz
	~\mathrm{d}\Omega
	~,
	\stepcounter{equation}
	\tag{\theequation}\label{eq:appendix_mKvve}
	\\[4pt]
	%\begin{split}
		\mKvme
		\, &= \,
		- \int_\Omegae
			\! \mNe^\mT \, \rho \bm{b} \, \big(
				\bm{a}^\calpha_\sz \cdot \mNe_{, \calpha}
			\big) \, \Delta t
			~\Jm_\sz
		~\mathrm{d}\Omega
		\\[4pt]
		&\hspace{-00pt}
		+ \zeta \int_\Omegae
			\! \mNe^\mT_{, \calpha} \, \bm{a}^\cgamma_\sz \, \big(
				\bmv_{, \cgamma}^\sz \cdot \bm{n}_\sz
			\big) \, a^{\calpha \cbeta}_\sz \, \big(
				\bm{n}_\sz \cdot \mNe_{, \cbeta}
			\big) \, \Delta t
			~\Jm_\sz
		~\mathrm{d}\Omega
		\\[4pt]
		&\hspace{-00pt}
		- \, \zeta \int_\Omegae
			\! \mNe^\mT_{, \calpha} \, \bm{a}^\cgamma_\sz \, \big(
				\bmv_{, \cgamma}^\sz \cdot \bm{a}^\cbeta_\sz
			\big) \, \big(
				\bm{a}^\calpha_\sz \cdot \mNe_{, \cbeta}
			\big) \, \Delta t
			~\Jm_\sz
		~\mathrm{d}\Omega
		\\[4pt]
		&\hspace{-00pt}
		+ \zeta \int_\Omegae
			\! \mNe^\mT_{, \calpha} \, \bm{n}_\sz \, \big(
				\bm{a}^\calpha_\sz \cdot \bm{v}_{, \cgamma}^\sz
			\big) \, a^{\cgamma \cbeta}_\sz \, \big(
				\bm{n}_\sz \cdot \mNe_{, \cbeta}
			\big) \, \Delta t
			~\Jm_\sz
		~\mathrm{d}\Omega
		\\[4pt]
		&\hspace{-00pt}
		- \, \zeta \int_\Omegae
			\! \mNe^\mT_{, \calpha} \, \big(
				\bm{a}^\calpha_\sz \cdot \bmv_{, \cgamma}^\sz
			\big) \, \big(
				\bm{a}^\cbeta_\sz \otimes \bm{a}^\cgamma_\sz
			\big) \, \mNe_{, \cbeta} \, \Delta t
			~\Jm_\sz
		~\mathrm{d}\Omega
		\\[4pt]
		&\hspace{-00pt}
		+ \zeta \int_\Omegae
			\! \mNe^\mT_{, \calpha} \, \big(
				\bmv_{, \cgamma}^\sz \cdot \bm{a}^\calpha_\sz
			\big) \, \bm{a}^\cgamma_\sz \, \big(
				\bm{a}^\cbeta_\sz \cdot \mNe_{, \cbeta}
			\big) \, \Delta t
			~\Jm_\sz
		~\mathrm{d}\Omega
		\\[4pt]
		&\hspace{-00pt}
		+ \zeta \int_\Omegae
			\! \mNe^\mT_{, \calpha} \, \bm{a}^\clambda_\sz \, a^{\cgamma \calpha} \, \big(
				\bm{n}_\sz \cdot \bmv_{, \cgamma}^\sz
			\big) \, \big(
				\bm{n}_\sz \cdot \mNe_{, \clambda}
			\big) \, \Delta t
			~\Jm_\sz
		~\mathrm{d}\Omega
		\\[4pt]
		&\hspace{-00pt}
		- \zeta \int_\Omegae
			\! \mNe^\mT_{, \calpha} \, \bm{a}_\cbeta^\sz \, a^{\calpha \cgamma}_\sz \, \big(
				\bmv_{, \cgamma}^\sz \cdot \bm{a}^\clambda_\sz
			\big) \, \big(
				\bm{a}^\cbeta_\sz \cdot \mNe_{, \clambda}
			\big) \, \Delta t
			~\Jm_\sz
		~\mathrm{d}\Omega
		\\[4pt]
		&\hspace{-00pt}
		- \zeta \int_\Omegae
			\! \mNe^\mT_{, \calpha} \, \bm{a}_\cbeta^\sz \, \big(
				\bm{a}^\cbeta_\sz \cdot \bmv_{, \cgamma}^\sz
			\big) \, a^{\cgamma \cmu}_\sz \, a^{\calpha \cnu}_\sz \, \big(
				\bm{a}_\cnu^\sz \cdot \mNe_{, \cmu}
				+ \bm{a}_\cmu^\sz \cdot \mNe_{, \cnu}
			\big) \, \Delta t
			~\Jm_\sz
		~\mathrm{d}\Omega
		\\[4pt]
		&\hspace{-00pt}
		+ \zeta \int_\Omegae
			\! \mNe^\mT_{, \calpha} \, a^{\calpha \cgamma}_\sz \, \big(
				\bmv_{, \cgamma}^\sz \cdot \bm{a}^\cbeta_\sz
			\big) \, \mNe_{, \cbeta} \, \Delta t
			~\Jm_\sz
		~\mathrm{d}\Omega
		\\[4pt]
		&\hspace{-00pt}
		+ \zeta \int_\Omegae
			\! \mNe^\mT_{, \calpha} \, a^{\calpha \cgamma}_\sz \, \bm{a}_\cbeta^\sz \, \big(
				\bm{a}^\cbeta_\sz \cdot \bmv_{, \cgamma}^\sz
			\big) \, \big(
				\bm{a}^\clambda_\sz \cdot \mNe_{, \clambda}
			\big) \, \Delta t
			~\Jm_\sz
		~\mathrm{d}\Omega
		\\[4pt]
		&\hspace{-00pt}
		+ \int_\Omegae
			\! \mNe^\mT_{, \calpha} \, a^{\calpha \cbeta}_\sz \, \bm{n}_\sz \otimes \bm{n}_\sz \, \mNe_{, \cbeta} \, \lambda_\sz \, \Delta t
			~\Jm_\sz
		~\mathrm{d}\Omega
		\\[4pt]
		&\hspace{-00pt}
		- \int_\Omegae
			\! \mNe^\mT_{, \calpha} \, \bm{a}^\cbeta_\sz \otimes \bm{a}^\calpha_\sz \, \mNe_{, \cbeta} \, \lambda_\sz \, \Delta t
			~\Jm_\sz
		~\mathrm{d}\Omega
		\\[4pt]
		&\hspace{-00pt}
		+ \int_\Omegae
			\! \mNe^\mT_{, \calpha} \, \bm{a}^\calpha_\sz \, \lambda_\sz \, \big(
				\bm{a}^\cbeta_\sz \cdot \mNe_{, \cbeta}
			\big) \, \Delta t
			~\Jm_\sz
		~\mathrm{d}\Omega
		~,
	\stepcounter{equation}
	\tag{\theequation}\label{eq:appendix_mKvme}
	%\end{split}
	\\
	\shortintertext{and}
	\mKvle
	\, &= \,
	\int_\Omegae
		\! \mNe^\mT_{, \calpha} \, \bm{a}^\calpha_\sz \, \mhNe
		~\Jm_\sz
	~\mathrm{d}\Omega
	~,
	\stepcounter{equation}
	\tag{\theequation}\label{eq:appendix_mKvle}
\end{align*}
\endgroup
where body forces $\rho \bm{b}$ are assumed to be independent of the surface geometry.
For Eq.~\eqref{eq:element_tangent}$_2$ we find
\begingroup
\allowdisplaybreaks
\begin{align*}
	\mKmve
	\, &= \,
	- \alpham \!\! \int_\Omegae
		\! \mNe^\mT \, \big(
			\bm{n}_\sz \otimes \bm{n}_\sz
		\big) \, \mNe
		~\Jm_\sz
	~\mathrm{d} \Omega
	~,
	\stepcounter{equation}
	\tag{\theequation}\label{eq:appendix_mKmve}
	\\[6pt]
	%\begin{split}
		\mKmme
		\, &= \,
		\alpham \!\! %\bigg\{
			\int_\Omegae
				\! \mNe^\mT \, \bmvm \big(
					\bm{a}^\calpha_\sz \cdot \mNe_{, \calpha}
				\big) \Delta t
				~\Jm_\sz
			~\mathrm{d}\Omega
			\, + \alpham \!\! \int_\Omegae
				\! \mNe^\mT \, \bm{n}_\sz \big(
					\bmv \cdot \bm{a}^\calpha_\sz
				\big) \, \big(
					\bm{n}_\sz \cdot \mNe_{, \calpha}
				\big)\, \Delta t
				~\Jm_\sz
			~\mathrm{d}\Omega
			\\[3pt]
			&\hspace{25pt}
			+ \,
			\alpham \!\! \int_\Omegae
				\! \mNe^\mT \, \big(
					\bmv_\sz \cdot \bm{n}_\sz
				\big)
				\,
				\big(
					\bm{a}^\calpha_\sz \otimes \bm{n}_\sz
				\big) \, \mNe_{, \calpha} \, \Delta t
				~\Jm_\sz
			~\mathrm{d}\Omega
			\, + \alpham \!\! \int_\Omegae
				\! \mNe^\mT \, \mNe
				~\Jm_\sz
			~\mathrm{d}\Omega
			\\[3pt]
			&\hspace{50pt}
			- \,
			\alpham \!\! \int_\Omegae
				\! \mNe^\mT \, \big(
					\bm{n}_\sz \otimes \bm{n}_\sz
				\big) \, \bmv_\sz \, \big(
					\bm{a}^\calpha_\sz \cdot \mNe_{, \calpha}
				\big) \, \Delta t
				~\Jm_\sz
			~\mathrm{d}\Omega
		%\, \bigg\}
		~,
	%\end{split}
	\stepcounter{equation}
	\tag{\theequation}\label{eq:appendix_mKmme}
	\\
	\intertext{and}
	\mKmle \, &= \, \mzero
	~.
	\stepcounter{equation}
	\tag{\theequation}\label{eq:appendix_mKmle}
\end{align*}
\endgroup

%
% *** Modification for a Fixed Surface
%

\subsection{Modification for a Fixed Surface} \label{sec:sec_appendix_fixed_surface}

In this section, we continue to work only with the local residual vector and tangent matrix.
Constraining the fluid surface to be fixed introduces a new unknown variable $p$, and corresponding variation $\delta p$, which are locally expanded in the $\{ N_i (\zeta^\alpha) \}$ basis as
\begin{equation} \label{eq:appendix_pressure_expansion}
	p (\zeta^\alpha, \, \tno)
	= \mNeo \, \mpeno
	\hspace{30pt}
	\mathrm{and}
	\hspace{30pt}
	\delta p (\zeta^\alpha)
	= \mNeo \, \mdeltape
	~,
\end{equation}
where $\mpeno$ are the local pressure degrees of freedom at time $\tno$.
As $p$ is a scalar quantity, $\mpeno$ and $\mdeltape$ are $\nen \times 1$ vectors and are multiplied by the $1 \times \nen$ vector of basis functions, which is denoted by $\mNeo$ in contrast to the $3 \times (3 \cdot \nen)$ matrix $\mNe$ introduced in Eq.~\eqref{eq:matrix_shape_fn_array}.

As the mesh is constrained to remain stationary, $\bmvm = \bm{0}$, all geometric quantities are fixed, and there are no mesh velocity degrees of freedom.
The weak form provided in Eq.~\eqref{eq:fluid_weak_form_fixed} is rewritten at time $\tno$ as in Eq.~\eqref{eq:weak_expanded_variations}, yielding
\begin{equation} \label{eq:appendix_weak_expanded_variations_fixed}
	\begin{split}
		\tilde{\mcg} (\tno)
		\, = \, \sum_{e = 1}^{\nume} \Big(
			\mdeltave^{\mathrm{T}} \mrveno
			\, + \, \mdeltalambdae^{\mathrm{T}} \mrlambdaeno
			\, + \, \mdeltape^{\mathrm{T}} \mrpeno
		\Big)
		\hspace{5pt} = \hspace{5pt} 0
		~,
	\end{split}
\end{equation}
for all variation vectors $\mdeltave$, $\mdeltalambdae$, and $\mdeltape$.
The residual vector $\mrpeno$ is defined by
\begin{equation} \label{eq:appendix_weak_pressure_residual_def}
	\mrpeno
	:= \dfrac{\partial \tilde{\mcg} (\tno)}{\partial \mdeltape}
	~,
\end{equation}
for the direct Galerkin expression provided in Eqs.~\eqref{eq:fluid_weak_form_fixed} and \eqref{eq:appendix_weak_expanded_variations_fixed}.
Following the same procedure as in Appendix~\ref{sec:sec_appendix_residual_vector} leads to
\begin{equation} \label{eq:appendix_weak_pressure_residual}
	\mrpeno
	\, = \, \int_\Omega \,
		\mNeo^\mT \, \bm{n} \cdot \bmv
		~\Jm
	~\mathrm{d}\Omega
	~.
\end{equation}
Furthermore, the first term in Eq.~\eqref{eq:appendix_mrve_ale} is modified by replacing $\rho \bm{b}$ with $p \bm{n}$, which gives
\begin{equation} \label{eq:appendix_mrve_fixed}
	\mrveno \,
	= \,
	- \int_\Omega
		\, \mNe^\mT \, p \, \bm{n}
		~\Jm
	~\mathrm{d} \Omega
	\,
	+ \int_\Omega 
		\, \mNe^\mT_{, \calpha} \,\, \Big(
			\pi^{\calpha \cbeta} \, \bm{a}_\cbeta
			+ \bm{a}^\calpha \, \lambda
		\Big)
		~\Jm
	~\mathrm{d} \Omega
	~.
\end{equation}
The form of $\mrlambdaeno$ is unchanged from Eq.~\eqref{eq:appendix_mrlambdae_ale}.

In removing the mesh velocity degrees of freedom and introducing the pressure degrees of freedom, there are once more nine tangent matrix components.
The matrices $\mKvveno$, $\mKvleno$, $\mKlveno$, and $\mKlleno$ are unchanged, and the weak form \eqref{eq:fluid_weak_form_fixed} will modify only $\mKpveno$ and $\mKvpeno$.
Accordingly, $\mKpleno$, $\mKlpeno$, and $\mKppeno$ are all zero.
As Eq.~\eqref{eq:appendix_weak_pressure_residual} is linear in $\bmv$, the tangent matrix component $\mKpveno$ is given by
\begin{align}
	\mKpveno
	\, &= \, \int_\Omega \,
		\mNeo^\mT \, \big(
			\bm{n} \cdot \mNe
		\big)
		~\Jm
	~\mathrm{d}\Omega
	~.
	\label{eq:appendix_mKpve}
	\\
	\intertext{The first term in Eq.~\eqref{eq:appendix_mrve_fixed} is linear in $p$, and we similarly calculate}
	\mKvpeno
	\, &= \, \int_\Omega \,
		\mNe^\mT \, \bm{n} \,\, \mNeo
		~\Jm
	~\mathrm{d}\Omega
	~.
	\label{eq:appendix_mKvpe}
\end{align}
With the residual vectors (\ref{eq:appendix_weak_pressure_residual}, \ref{eq:appendix_mrve_fixed}) and tangent matrix components (\ref{eq:appendix_mKpve}, \ref{eq:appendix_mKvpe}), our general ALE finite element framework has been adapted to solve for fixed-surface fluid flows.

%
% *** Inertial Contributions
%

\subsubsection{Inertial Contributions} \label{sec:sec_appendix_inertia_fixed_surface}

For simplicity, the subscript $\rho$ is used to denote a quantity which arises only due to inertia.
In this section, the inertial contributions to the local residual vector and tangent matrix are calculated.
The inertial contribution to the weak form component $\mcg_{\mathrm{v}}$ \eqref{eq:fluid_weak_eqn_motion}---denoted $\mcg_{\mathrm{v}, \rho}$---can be written as
\begin{equation} \label{eq:appendix_weak_inertia}
	\begin{split}
		\mcg_{\mathrm{v}, \rho}
		\, = \int_\Omega
			\delta \bmv \cdot \rho \dot{\bmv}
			~\Jm
		~\mathrm{d}\Omega
		\, &= \sum_{e = 1}^{\nume} \, \int_\Omegae
			\! \mdeltave^\mT \, \mNe^\mT \, \rho \dot{\bmv}
			~\Jm
		~\mathrm{d}\Omega
		\\[3pt]
		&\hspace{-35pt}
		= \, \sum_{e = 1}^{\nume} \,\, \mdeltave^\mT \int_\Omegae
			\! \mNe^\mT \, \rho \dot{\bmv}
			~\Jm
		~\mathrm{d}\Omega
		~.
	\end{split}
\end{equation}
Substituting Eq.~\eqref{eq:ale_acceleration}, with
$\bmvm = \bm{0}$
by construction, into Eq.~\eqref{eq:appendix_weak_inertia} leads to the inertial contribution to the local residual vector at time $\tno$, which is expressed as
\begin{equation} \label{eq:appendix_residual_inertia_step_1}
	\mrvrhoeno
	\, = \int_\Omegae
		\! \mNe^\mT \, \rho \bmv'
		~\Jm
	~\mathrm{d}\Omega
	\, + \int_\Omegae
		\! \mNe^\mT \, \rho \big(
			\bmv_{, \calpha} \otimes \bm{a}^\calpha
		\big) \, \bmv
		~\Jm
	~\mathrm{d}\Omega
	~.
\end{equation}
The time derivative $\bmv'$ is approximated with an implicit backwards-Euler discretization, given by
\begin{equation} \label{eq:appendix_velocity_backwards_euler}
	\bmv' (\zeta^\alpha, \tno)
	= \dfrac{1}{\Delta t} \Big(
		\bmv (\zeta^\alpha, \tno)
		- \bmv (\zeta^\alpha, \tn)
	\Big)
	~.
\end{equation}
Substituting Eq.~\eqref{eq:appendix_velocity_backwards_euler} into the residual vector in Eq.~\eqref{eq:appendix_residual_inertia_step_1}, evaluated at time $\tno$, approximating $\bmv (\zeta^\alpha, \tno)$ with
$\bmv_\so = \bmv_\sz + \Delta \bmv$,
and keeping terms to first order in $\Delta \bmv$, we obtain
\begin{equation} \label{eq:appendix_residual_inertia_step_2}
	\begin{split}
		\mrvrhoeno_\so
		\, &= \, \mrvrhoeno_\sz
		\, + \int_\Omegae
			\! \mNe^\mT \, \dfrac{\rho}{\Delta t} \, \Delta \bmv
			~\Jm
		~\mathrm{d}\Omega
		\\[3pt]
		&
		\hspace{20pt}
		+ \int_\Omegae
			\! \mNe^\mT \, \rho \big(
				\Delta \bmv_{, \calpha} \otimes \bm{a}^\calpha
			\big) \, \bmv_\sz
			~\Jm
		~\mathrm{d}\Omega
		\, + \int_\Omegae
			\! \mNe^\mT \, \rho \big(
				\bmv_{, \calpha}^\sz \otimes \bm{a}^\calpha
			\big) \, \Delta \bmv
			~\Jm
		~\mathrm{d}\Omega
		~,
	\end{split}
\end{equation}
where $\mrvrhoeno_\sz$ is given by Eq.~\eqref{eq:appendix_residual_inertia_step_1} evaluated at $\bmv(t + \Delta t)_\sz$, written as
\begin{equation} \label{eq:appendix_residual_inertia}
	\mrvrhoeno_\sz
	\, = \int_\Omegae
		\! \mNe^\mT \, \dfrac{\rho}{\Delta t} \, \Big(
			\bmv_\sz - \bmv(t)
		\Big)
		~\Jm
	~\mathrm{d}\Omega
	\, + \int_\Omegae
		\! \mNe^\mT \, \rho \big(
			\bmv_{, \calpha}^\sz \otimes \bm{a}^\calpha
		\big) \, \bmv_\sz
		~\Jm
	~\mathrm{d}\Omega
	~.
\end{equation}
From Eq.~\eqref{eq:appendix_residual_inertia_step_2}, the tangent matrix contribution from the inertial terms is found to be given by
\begin{equation} \label{eq:appendix_tangent_inertia}
	\begin{split}
		\mKvvrhoeno
		\, &=
		\int_\Omegae
			\! \mNe^\mT \, \dfrac{\rho}{\Delta t} \, \mNe
			~\Jm
		~\mathrm{d}\Omega
		\, + \int_\Omegae
			\! \mNe^\mT \, \rho \big(
				\bmv_\sz \cdot \bm{a}^\calpha
			\big) \, \mNe_{, \calpha}
			~\Jm
		~\mathrm{d}\Omega
		\\[3pt]
		&\hspace{35pt}
		+ \int_\Omegae
			\! \mNe^\mT \, \rho \, \big(
				\bmv_{, \calpha}^\sz \otimes \bm{a}^\calpha
			\big) \, \mNe
			~\Jm
		~\mathrm{d}\Omega
		~.
	\end{split}
\end{equation}

%
% *** Modification for a Normal Body Force
%

\subsection{Modification for a Normal Body Force} \label{sec:sec_appendix_normal_body_force}

In this section, we discuss how the finite element formulation for a curved and deforming fluid film is modified when the constant body force $\rho \bm{b}$ is replaced by $p \bm{n}$, a pressure normal to the surface.
The body force term in Eq.~\eqref{eq:appendix_mrve_ale} is modified, and the local residual vector at time $\tno$ is rewritten as
\begin{equation} \label{eq:appendix_mrve_ale_normal_body}
	\mrveno \,
	= \,
	- \int_\Omegae
		\! \mNe^\mT \, p \bm{n}
		~\Jm
	~\mathrm{d} \Omega
	\,
	+ \int_\Omegae
		\! \mNe^\mT_{, \calpha} \,\, \Big(
			\pi^{\calpha \cbeta} \, \bm{a}_\cbeta
			+ \bm{a}^\calpha \, \lambda
		\Big)
		~\Jm
	~\mathrm{d} \Omega
	~.
\end{equation}
The linearization of the first term yields an additional contribution to the tangent matrix $\mKvmeno$, given by
\begin{equation} \label{eq:appendix_normal_body_residual_contribution}
	- \int_\Omegae
		\! \mNe^\mT \, p \, \Delta \bm{n}
		~\Jm_\sz
	~\mathrm{d} \Omega
	~.
\end{equation}
Denoting the additional tangent matrix contribution as $\mKvmpeno$, and calculate it by substituting $\Delta \bm{n}$ \eqref{eq:appendix_linearization_bm_n} into Eq.~\eqref{eq:appendix_normal_body_residual_contribution} and moving the mesh velocity degrees of freedom outside the integral, leads to
\begin{equation} \label{eq:appendix_mKvmpe_normal_body}
	\mKvmpeno \, = \,
	\int_\Omegae \!
		\mNe^\mT \, \big(
			\bm{a}^\calpha_\sz \otimes \bm{n}_\sz
		\big) \, \mNe_{, \calpha} \,\, p \, \Delta t
		~\Jm
	~\mathrm{d}\Omega
	~.
\end{equation}
By modifying our residual vector and tangent matrix as in Eqs.~\eqref{eq:appendix_normal_body_residual_contribution} and \eqref{eq:appendix_mKvmpe_normal_body}, we simulate body forces normal to the surface as it deforms over time.

%
% *** Numerical Integration
%

\subsection{Numerical Integration} \label{sec:sec_appendix_numerical_integration}

To numerically evaluate integrals over a single element, we first consider a one-dimensional integral over the non-empty portion of the element boundary $\Gammane$ for which $\zeta^2$ is fixed and $\zeta^1$ ranges from $\zeta^1_{\, a}$ to $\zeta^1_{\, b}$.
We map the parametric domain
$\zeta^1 \in [\zeta^1_{\, a} , \, \zeta^1_{\, b}]$
to the interval
$[-1, \, 1]$,
which is parametrized by the scaled variable $\xi$, and on this rescaled domain we sum the values of the integrand at
$\xi = \{ -\sqrt{^3\!/_{\!5}}, \, 0, \, \sqrt{^3\!/_{\!5}} \}$
with corresponding weights
$\{ ^5\!/_{\!9}, \, ^8\!\!/_{\!9}, \, ^5\!\!/_{\!9} \}$, as is standard for numerical integration using Gaussian quadrature \cite{abramowitz-stegun}.

To calculate two-dimensional areal integrals over an element of the parametric domain $\Omegae$, we perform an analogous procedure.
The element $\Omegae$ is parametrized by
$( \zeta^1, \, \zeta^2) = [\zeta^1_{\, a}, \, \zeta^1_{\, b}] \times [\zeta^2_{\, a}, \, \zeta^2_{\, b}]$,
which we map to the unit square $\Omega_\square$, parametrized by
$(\xi, \eta) \in [-1, \, 1] \times [-1, \, 1]$.
The integrand is then calculated at the nine quadrature points with their corresponding weights.

%
% *** Code Structure
%

\subsection{Code Structure} \label{sec:sec_appendix_code_structure}

We have now presented the computational details relevant to our ALE finite element formulation.
A high-level overview of how our code is structured is shown in Algorithm~\ref{algo}.

%\vspace{-13pt}
%// code structure, written in C++ pseudocode
\begin{lstlisting}[caption={\texttt{C++} pseudocode of the ALE finite element method},label=algo,float=p]
 // mesh and basis function calculations
 generate_mesh(); generate_basis_functions();

 for (time_index = 0; time_index < num_time_steps; ++time_index) {

	initialize_u_vector(); initialize_delta_u();

	while (norm(delta_u) > newton_tolerance) {
		// initialize global residual vector and tangent matrix
		initialize_r_vector(); initialize_K_matrix();

		for (element_index = 0; element_index < num_elements; ++element_index) {
			// Dohrmann--Bochev H and G matrices
			initialize_DB_H_matrix(); initialize_DB_G_matrix();

			// local residual vector and tangent matrix
			initialize_element_r_vector(); initialize_element_K_matrix();

			for (gauss_pt_idx = 0; gauss_pt_idx < num_gauss_pts; ++gauss_pt_idx) {
				// residual vector calculations: Appendix C.1
				increment_element_r_vector(u_vector);
				// tangent matrix calculations: Appendix C.2
				increment_element_K_matrix(u_vector);

				// Dohrmann--Bochev H and G matrix assembly: Appendix C.2
				increment_DB_H_matrix(u_vector);
				increment_DB_G_matrix(u_vector);
			}

			// add Dohrmann--Bochev terms to tangent matrix and residual vector:
			// Appendix C.2
			add_DB_terms(DB_H_matrix, DB_G_matrix,
					element_K_matrix, element_r_vector);

			// assemble global counterparts
			assemble_K_matrix(K_matrix, element_K_matrix);
			assemble_r_vector(r_vector, element_r_vector);
		}

		// apply boundary conditions
		apply_boundary_conditions(K_matrix, r_vector);

		solve_delta_u(delta_u, K_matrix, r_vector);
		u_vector += delta_u;
	}

	output_u_vector();
 }
\end{lstlisting}

%
% *** Numerical Benchmarks
%

\section{Numerical Benchmarks} \label{sec:sec_appendix_numerical_benchmarks}

In this section, we supplement the validation of the LE finite element method in Sec.~\ref{sec:sec_numerical_simulations} by considering four simple benchmark problems for which analytical solutions are known.
In each case, the numerical calculation of the $x$-velocity and surface tension are plotted against their analytical results.
The $L^2$-errors of the velocity and surface tension upon mesh refinement are also provided.
All simulations are run on a stationary, flat mesh corresponding to the spatial domain
$(x, \,y) \in [0, 1] \times [0, 1]$,
and in three of the four cases the analytical solution lies in our finite-dimensional solution space $\mcuh$.
The remaining case demonstrates the convergence of the surface tension as the mesh is refined.
Dohrmann--Bochev stabilization is used in all cases.

%
% *** Hydrostatic Fluid
%

\subsection{Hydrostatic Fluid} \label{sec:sec_appendix_hydrostatic}

\begin{figure}[!t]
	\centering
	%\vspace{-10pt}
	\begin{subfigure}[b]{0.30\columnwidth}
		\centering
		\includegraphics[height=0.95\textwidth]{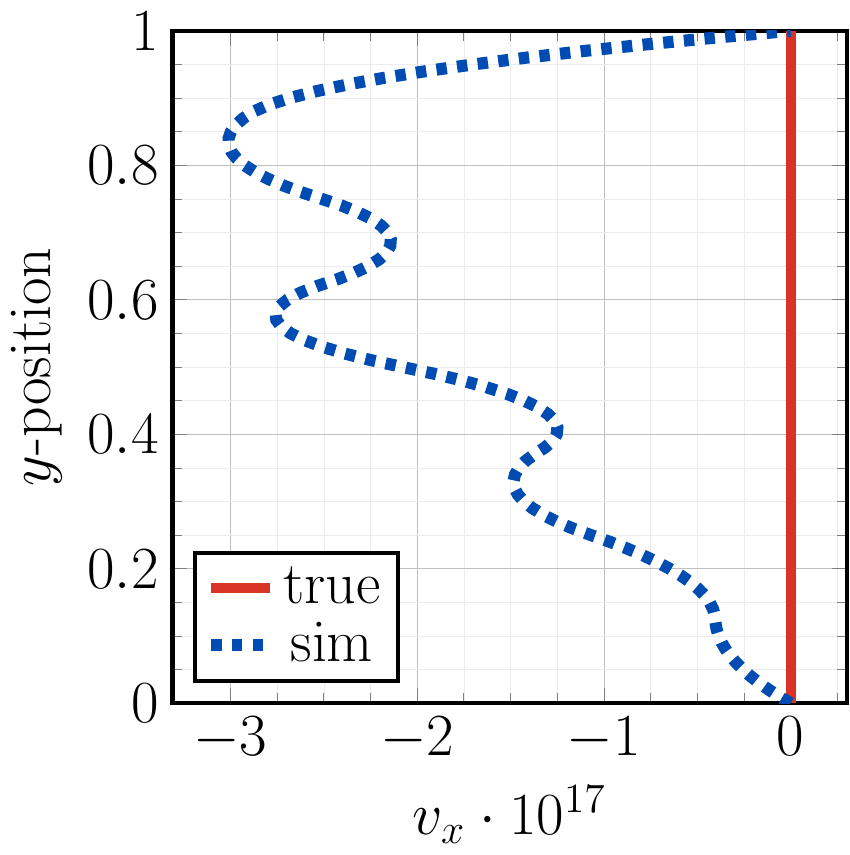}
		\caption{$x$-velocity at $x = 0.5$}
		\label{fig:fig_hydro_vx}
	\end{subfigure}
	\begin{subfigure}[b]{0.30\columnwidth}
		\centering
		\hfill
		\includegraphics[height=0.94\textwidth]{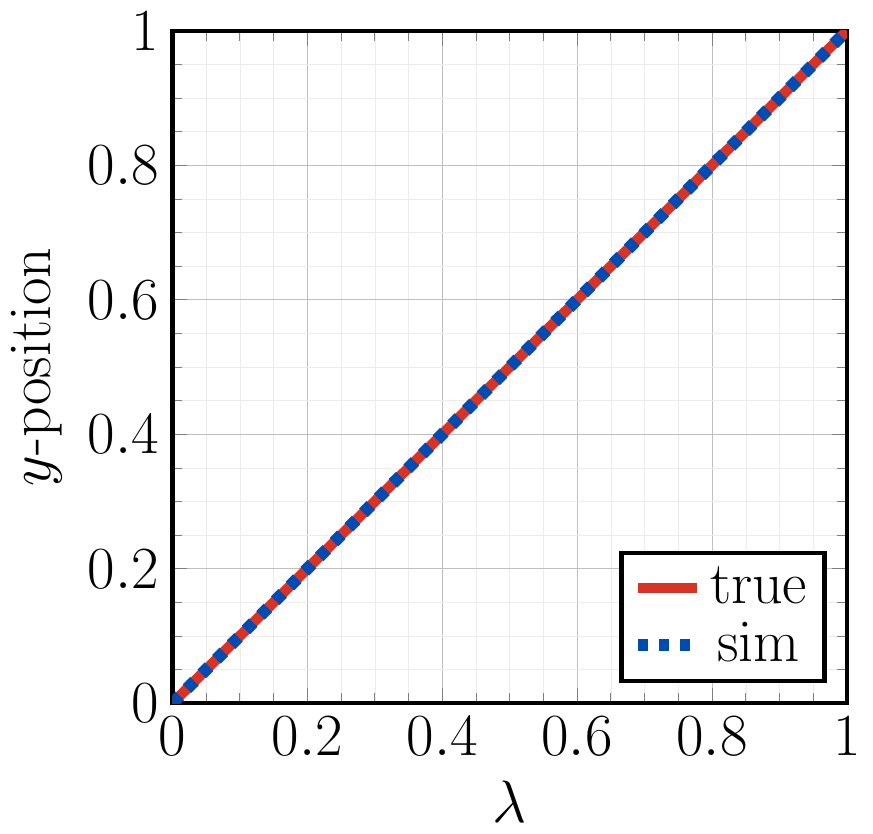}
		\caption{surface tension at $x = 0.5$}
		\label{fig:fig_hydro_lambda}
	\end{subfigure}
	\begin{subfigure}[b]{0.30\columnwidth}
		\centering
		\hfill
		\includegraphics[height=0.91\textwidth]{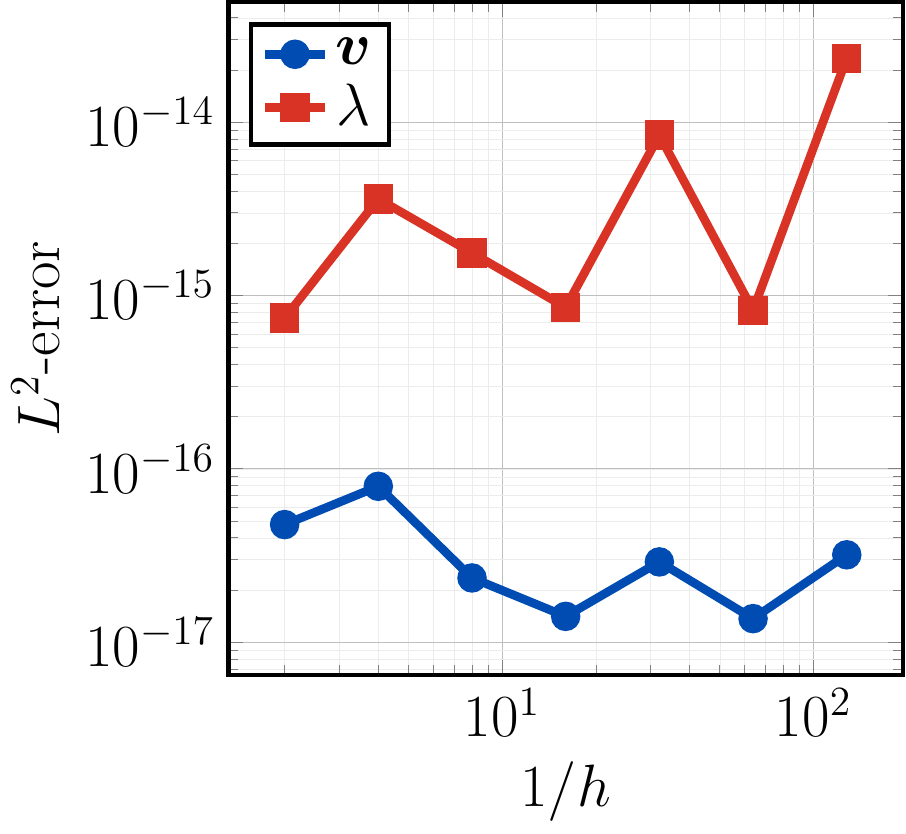}
		\caption{error, refining $x$ and $y$}
		\label{fig:fig_hydro_error}
	\end{subfigure}
	%\vspace{-1pt}
	\caption{
		Numerical simulation of the hydrostatic problem.
		The true solution (solid red line) is compared to the numerical solution (dashed blue line) for the $x$-velocity (a) and surface tension (b).
		Note the $x$-velocity error in (a) is multiplied by $10^{17}$.
		The $L^2$-error in the velocity (blue circles) and surface tension (red squares) is plotted in (c), as a function of the element height and width $h$, which ranges from $^{1 \!}/_2$ to $^{\! 1}/_{128}$.
		Errors in the $x$-velocity remain below the machine precision $\epsilon_{\texttt{mp}} \approx 1.1 \cdot 10^{-16}$, while errors in the surface tension are $\mathcal{O} (10^{-14})$.
		Errors in general are expected to increase as the number of elements, and therefore the number of degrees of freedom, increase.
	}
	\label{fig:fig_hydrostatic}
\end{figure}

We first consider a fluid at rest which is acted upon by gravity, such that the body force is given by
$\rho \bm{b} = - \rho g \, \bm{e}_y$.
Units are chosen such that
$\rho g = 1$,
and the surface tension is specified to be zero at the bottom of the domain.
The analytical solution is given by
$\bmv = \bm{0}$
and
$\lambda = y$.
We solve the problem numerically by specifying
$\bmv = \bm{0}$
on all four boundaries and
$\lambda = 0$
on the bottom edge.
Results from our numerical solution are shown in Fig.~\ref{fig:fig_hydrostatic}; note the scale of the $x$-axis in part (a).

%
% *** Couette Flow
%

\subsection{Couette Flow} \label{sec:sec_appendix_couette}

We next simulate Couette flow, in which fluid fills the space between two parallel plates; the bottom plate is stationary while the top plate moves tangentially at constant speed $V$.
The bottom plate is located at
$y = 0$,
the top plate at
$y = 1$,
and furthermore units are chosen such that
$V = 1$.
In this case, the analytical solution is given by
$\bmv = y \, \bm{e}_x$
and
$\lambda = 0$.

\begin{figure}[t!]
	\centering
	\begin{subfigure}[b]{0.30\columnwidth}
		\centering
		\includegraphics[width=0.95\textwidth]{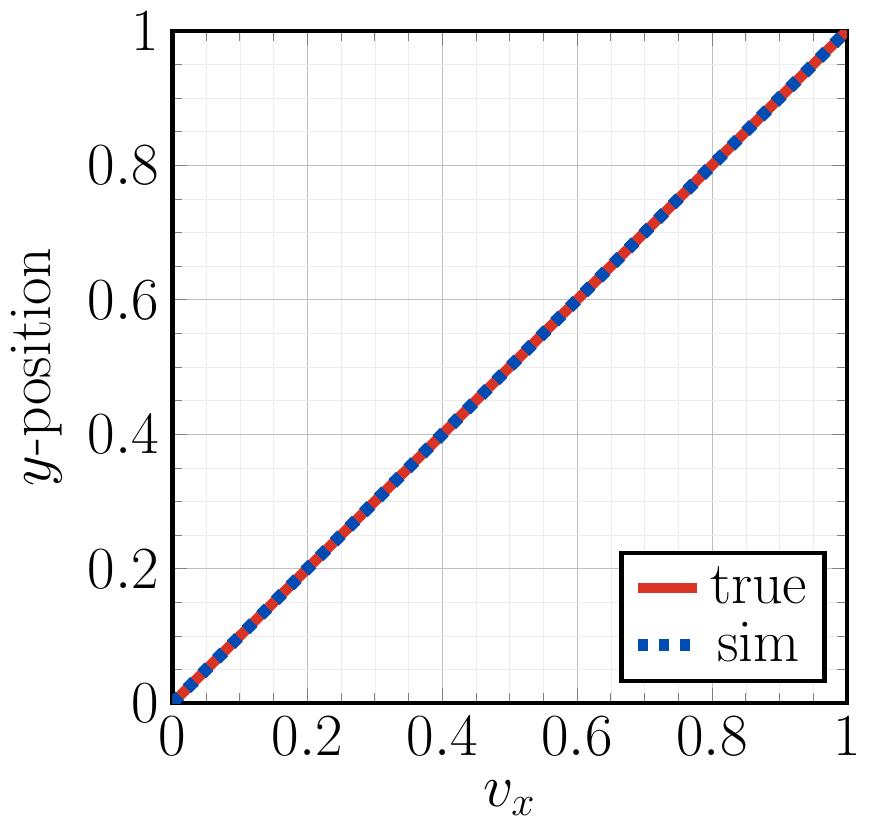}
		\caption{$x$-velocity at $x = 0.5$}
		\label{fig:fig_couette_vx}
	\end{subfigure}
	\begin{subfigure}[b]{0.30\columnwidth}
		\centering
		\hfill
		\includegraphics[width=0.95\textwidth]{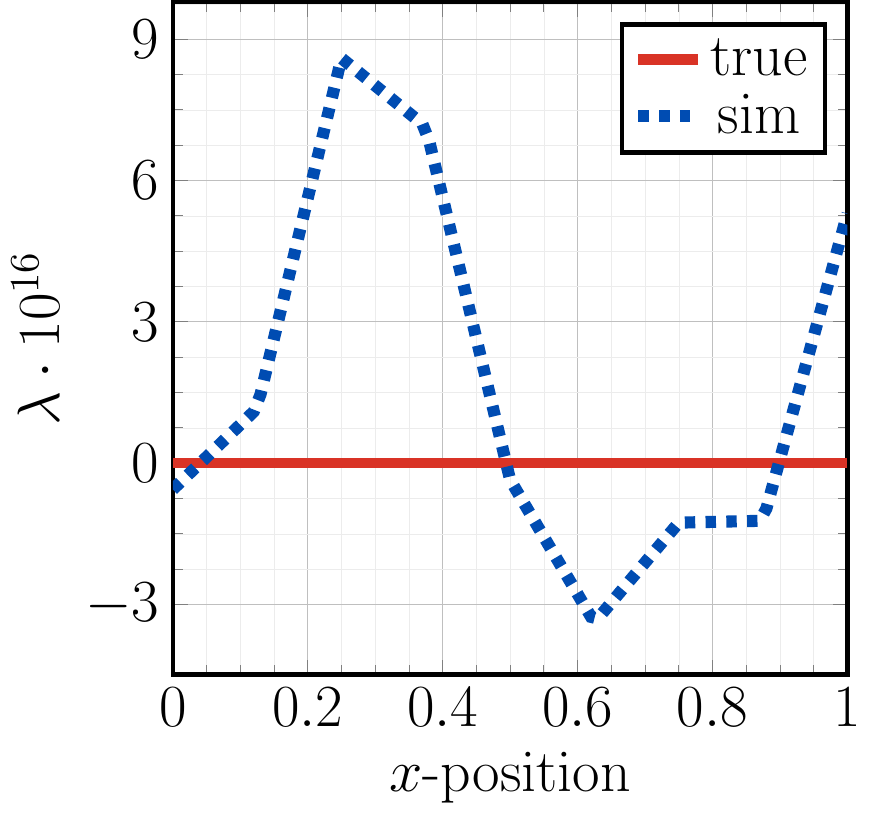}
		\caption{surface tension at $y = 0.5$}
		\label{fig:fig_couette_lambda}
	\end{subfigure}
	\begin{subfigure}[b]{0.30\columnwidth}
		\centering
		\hfill
		\includegraphics[width=0.95\textwidth]{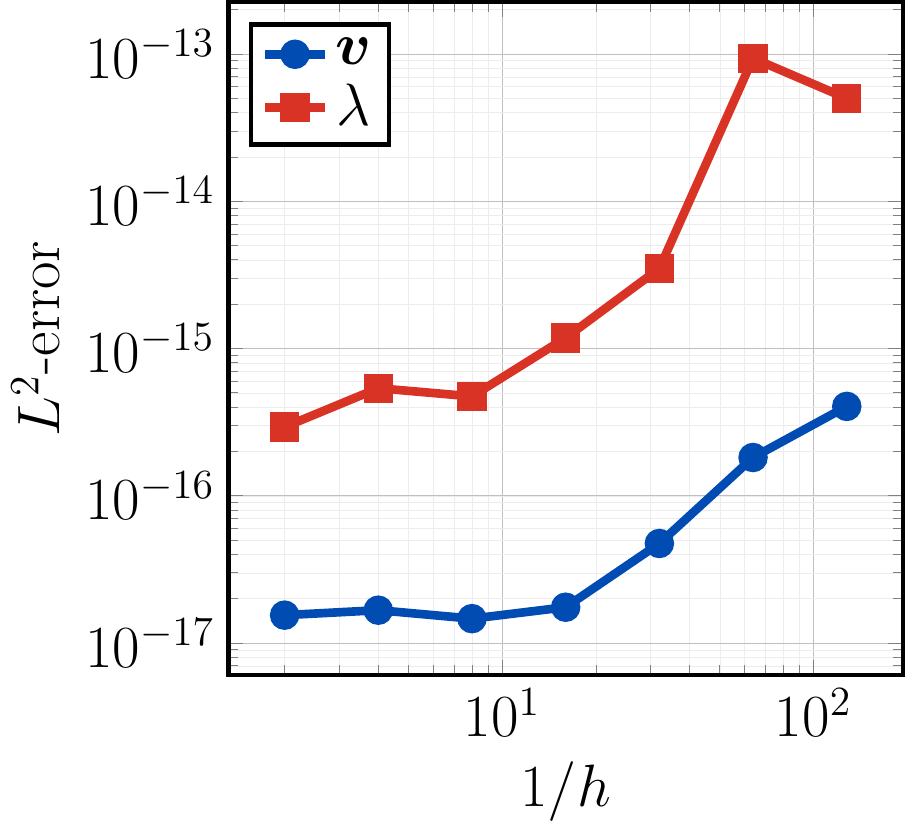}
		\caption{error, refining $x$ and $y$}
		\label{fig:fig_couette_error}
	\end{subfigure}
	%\vspace{3pt}
	\caption{
		Numerical solution of Couette flow.
		The $x$-velocity as a function of $y$-position and surface tension as a function of $x$-position are provided in (a) and (b), respectively, and show excellent agreement between numerical and analytical solutions (dashed blue lines and solid red lines, respectively).
		Note the $y$-axis in (b) is multiplied by $10^{16}$.
		The $L^2$-errors in the velocity (blue circles) and surface tension (red squares) is shown in (c) as a function of the mesh size $h$, as the mesh is refined in both the $x$ and $y$ directions.
		The increase in the error may again be due to the addition of small errors in each degree of freedom, as the number of degrees freedom increases.
	}
	\label{fig:fig_couette}
\end{figure}

In our numerical solution, we specify Dirichlet boundary conditions rather than formulating the degrees of freedom to be periodic on the left and right edges of the domain.
Thus
$\bmv = \bm{e}_x$ on the top edge,
$\bmv = \bm{0}$ on the bottom edge, and
$\bmv = y \, \bm{e}_x$ on the left and right edges.
Furthermore, $\lambda$ is specified to be zero on the left edge.
The results of our simulation are provided in Fig.~\ref{fig:fig_couette}; note the $y$-axis in (b) is multiplied by $10^{16}$.

\begin{figure}[t!]
	\centering
	\begin{subfigure}[b]{0.30\columnwidth}
		\centering
		\includegraphics[width=0.95\textwidth]{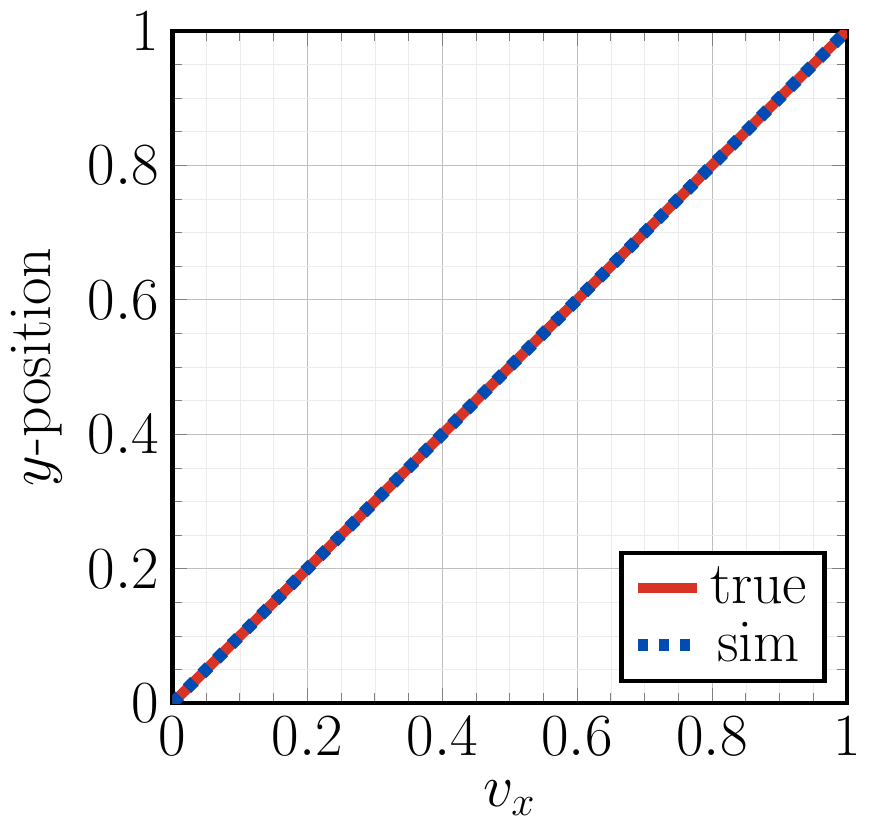}
		\caption{$x$-velocity at $x = 0.5$}
		\label{fig:fig_couette_body_vx}
	\end{subfigure}
	\begin{subfigure}[b]{0.30\columnwidth}
		\centering
		\hfill
		\includegraphics[width=0.95\textwidth]{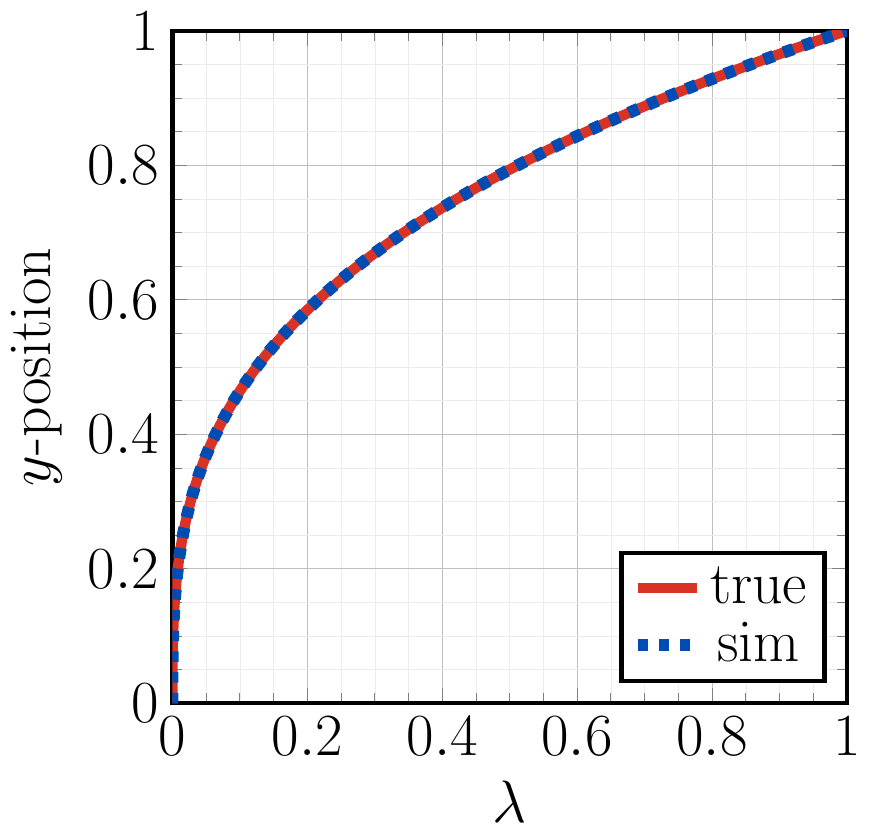}
		\caption{surface tension at $x = 0.5$}
		\label{fig:fig_couette_body_lambda}
	\end{subfigure}
	\begin{subfigure}[b]{0.30\columnwidth}
		\centering
		\hfill
		\includegraphics[width=0.95\textwidth]{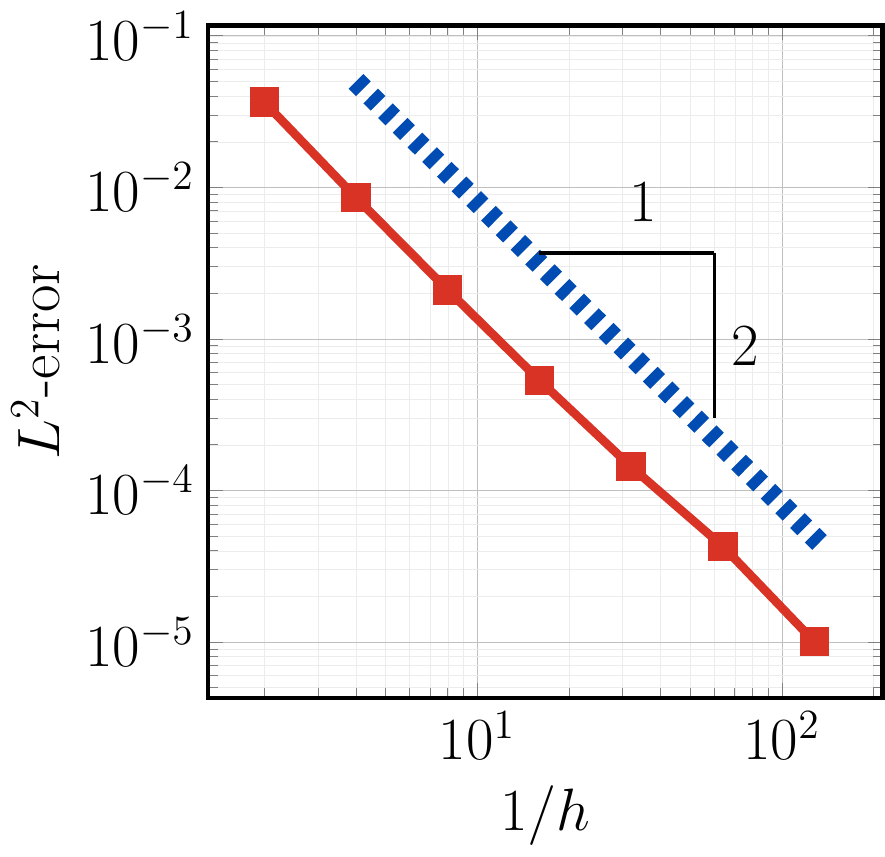}
		\caption{error, refining $x$ and $y$}
		\label{fig:fig_couette_body_error}
	\end{subfigure}
	\caption{
		Numerical solution of Couette flow, with a body force $\rho \bm{b} = - 3 y^2 \, \bm{e}_y$.
		The $x$-velocity (a) and surface tension (b) as a function of $y$-position show excellent agreement between numerical and analytical solutions (dashed blue lines and solid red lines, respectively).
		The $L^2$-errors in the surface tension (red squares) is shown in (c) as a function of the mesh size $h$, as the mesh is refined in both the $x$ and $y$ directions, and converges quadratically as expected.
		As the analytical solution for the velocity lies in $\mcvh$, the velocity errors (not shown) are $\mathcal{O}(10^{-16})$ as in Fig.~\ref{fig:fig_couette_error}.
	}
	\label{fig:fig_couette_body}
	%\vspace{-5pt}
\end{figure}

\begin{figure}[t!]
	\centering
	\begin{subfigure}[b]{0.30\columnwidth}
		\centering
		\includegraphics[height=0.95\textwidth]{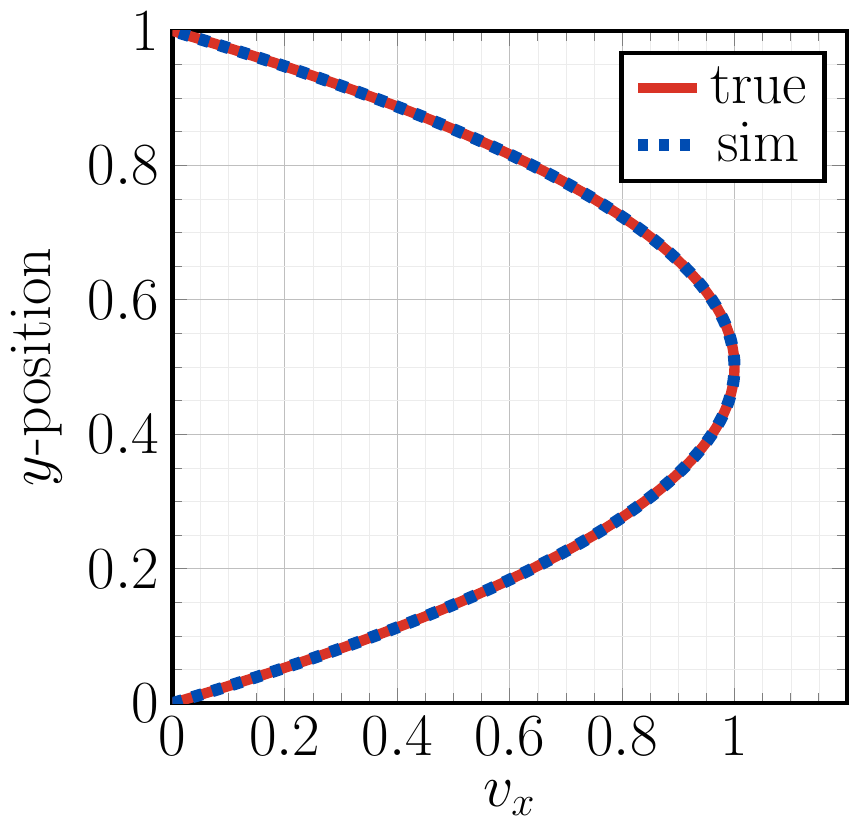}
		\caption{$x$-velocity at $x = 0.5$}
		\label{fig:fig_poise_vx}
	\end{subfigure}
	\begin{subfigure}[b]{0.30\columnwidth}
		\centering
		\hfill
		\includegraphics[height=0.95\textwidth]{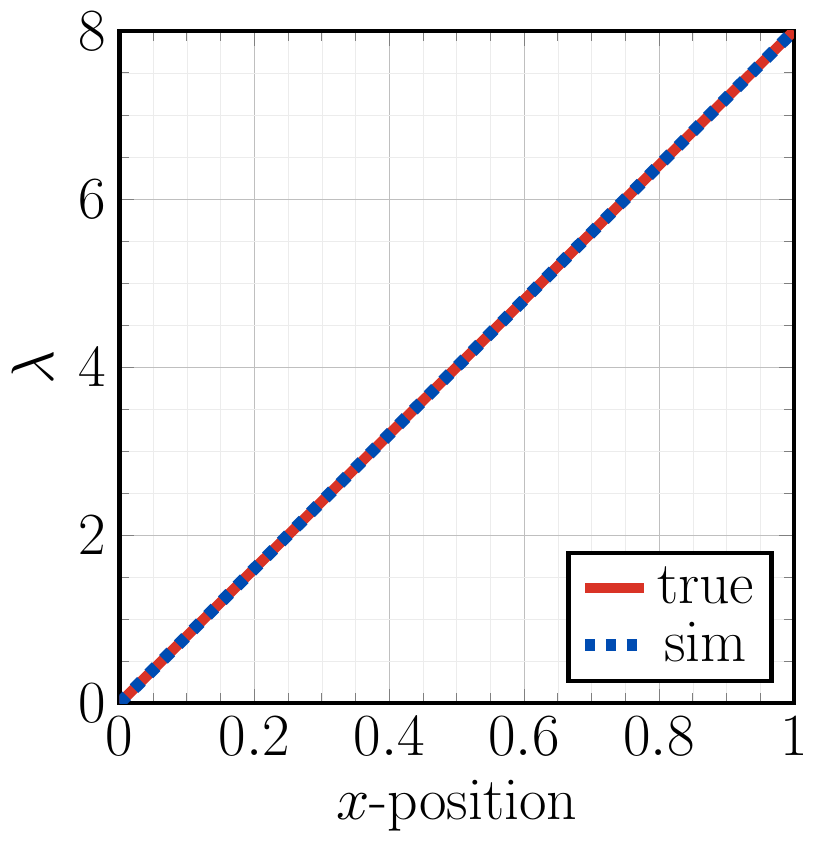}
		\caption{surface tension at $y = 0.5$}
		\label{fig:fig_poise_lambda}
	\end{subfigure}
	\begin{subfigure}[b]{0.30\columnwidth}
		\centering
		\hfill
		\includegraphics[height=0.91\textwidth]{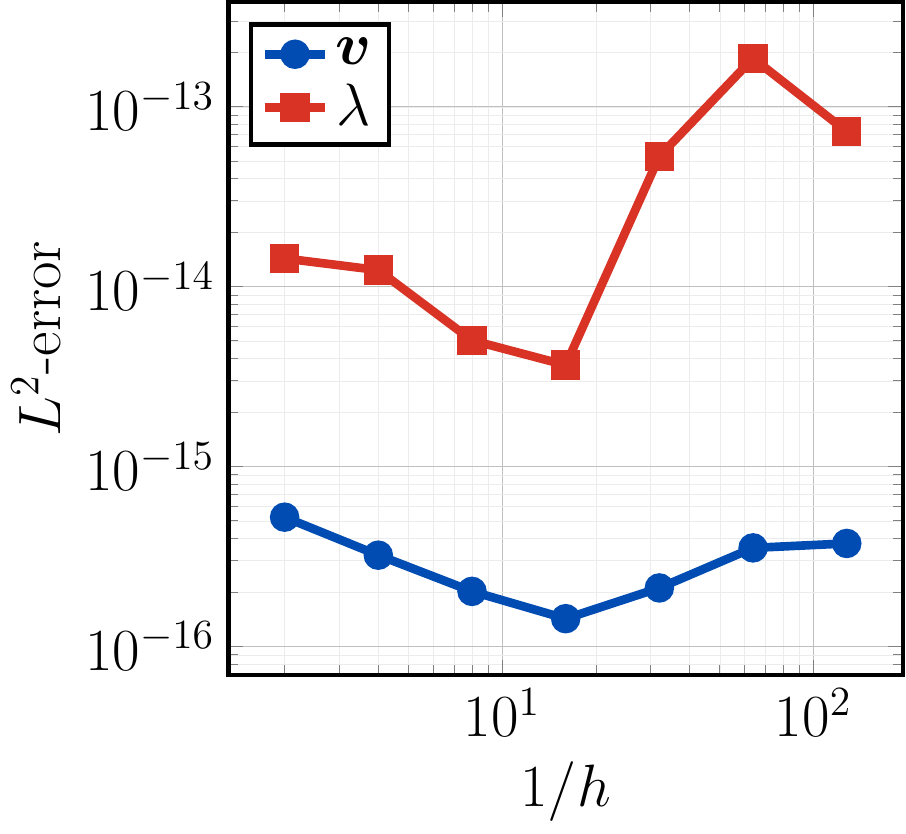}
		\caption{error, refining $x$ and $y$}
		\label{fig:fig_poise_error}
	\end{subfigure}
	\caption{
		Numerical solution of Hagen--Poiseuille flow.
		The $x$-velocity is parabolic in $y$ (a) and the surface tension is linear in $x$ (b); in both figures the numerical result is shown with a dashed blue line and the analytical result with a solid red line.
		The $L^2$-error in velocity (blue circles) and surface tension (red squares) is shown in (c).
	}
	\label{fig:fig_poise}
\end{figure}

%
% *** Couette Flow with a Body Force
%

\subsection{Couette Flow with a Body Force} \label{sec:sec_appendix_couette_body}

To demonstrate the convergence of the surface tension in our numerical method, we again consider the case of Couette flow, yet now with a body force
$\rho \bm{b} = -3y^2 \, \bm{e}_y$.
In this case, the analytical solution is given by
$\bmv = y \, \bm{e}_x$
and
$\lambda = y^3$.
The numerical solutions and convergence of the surface tension are shown in Fig.~\ref{fig:fig_couette_body}.

%
% *** Hagen--Poiseuille Flow
%

\subsection{Hagen--Poiseuille Flow} \label{sec:sec_appendix_hagen_poiseuille}

The final benchmark problem considered is Hagen--Poiseuille flow, for which a surface tension change across the length of a stationary channel drives flow.
No-slip boundary conditions on the top and bottom walls of the channel, located at
$y = 1$
and
$y = 0$,
respectively, are assumed.
Units are chosen such that the surface tension change per length
$\Delta \lambda / L = 8$.
Furthermore setting the viscosity to be unity and arbitrarily choosing
$\lambda = 0$
at
$x = 0$,
the analytical solution is given by
$\bmv = 4 \, y \, (1 - y) \, \bm{e}_x$
and
$\lambda = 8 x$.

We specify Dirichlet boundary conditions on all four edges of the square domain, rather than incorporating the surface tension change across the domain into the weak form through the boundary traction.
We set
$\bmv = 0$
on the top and bottom edges,
$\bmv = 4 \, y \, (1 - y) \, \bm{e}_x$
on the left and right edges, and specify
$\lambda = 0$
on the left edge.
Our simulation finds a parabolic velocity profile throughout the domain and correctly calculates the surface tension change across the domain.
Numerical results are shown in Fig.~\ref{fig:fig_poise}.
Once again, there is excellent agreement with the analytical solution.

%
% *** Movies
%

\section{Movies} \label{sec:sec_appendix_movies}

% *** Curved and deforming fluid film *** %
\paragraph{1. Curved and deforming fluid film} \label{sec:sec_appendix_movie_1}

The movie provided at
\href{https://youtu.be/FUx8fGXuzqY}{\texttt{youtu.be/FUx8fGXuzqY}}
shows an initially cylindrical fluid film which is perturbed and then allowed to dynamically evolve over time.
The color and color bar in the movie show the surface tension, whose variations mediate an instability via in-plane fluid flow.
The LE simulation is run on a $10 \times 40$ mesh, from time $t=0$ to $t=35$.
Snapshots of the simulation are found in Fig.~\ref{fig:fig_ale_cylinder} in the main text.

% *** Lagrangian scheme fails lid-driven cavity benchmark *** %
\paragraph{2. Lagrangian scheme fails lid-driven cavity benchmark} \label{sec:sec_appendix_movie_2}

\begin{figure}[h!]
	\centering
	\vspace{10pt}
	\begin{subfigure}[b]{0.31\columnwidth}
		\centering
		\includegraphics[width=0.99\textwidth]{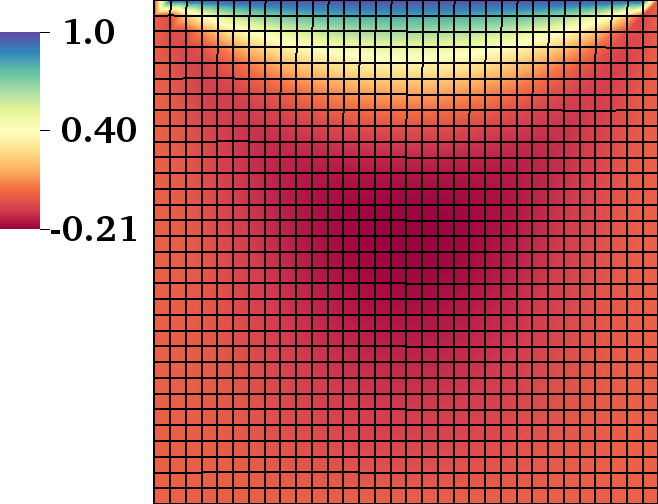}
		\caption{$t = 1$}
		\label{fig:fig_lag_cavity_t_05}
	\end{subfigure}
	~
	\begin{subfigure}[b]{0.31\columnwidth}
		\centering
		\includegraphics[width=0.99\textwidth]{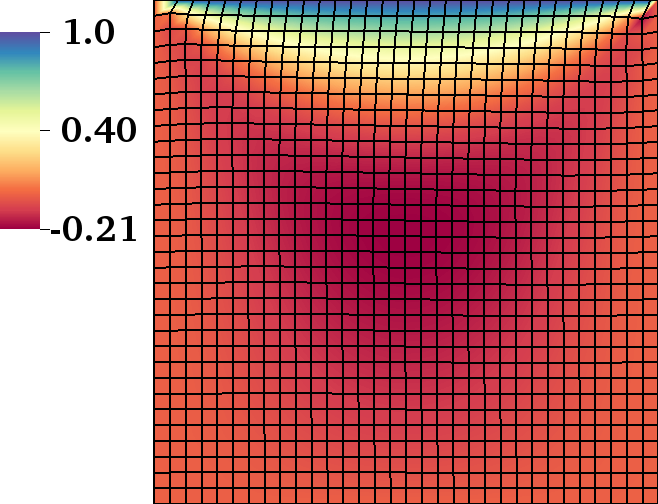}
		\caption{$t = 10$}
		\label{fig:fig_lag_cavity_t_10}
	\end{subfigure}
	~
	\begin{subfigure}[b]{0.31\columnwidth}
		\centering
		\includegraphics[width=0.99\textwidth]{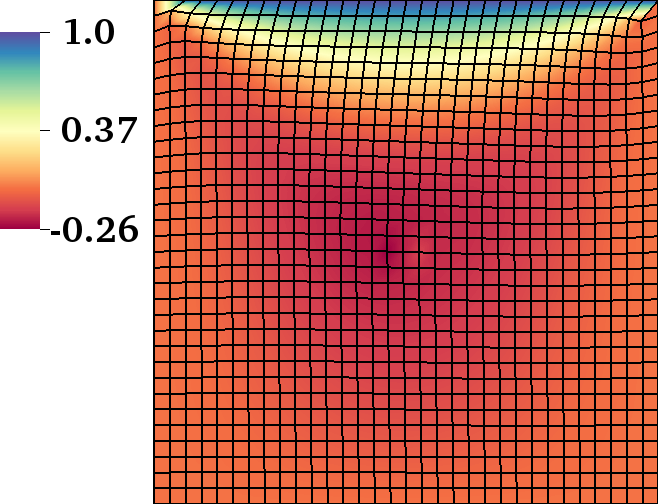}
		\caption{$t = 20$}
		\label{fig:fig_lag_cavity_t_15}
	\end{subfigure}
	\vspace{-3pt}
	\caption{
		Lagrangian simulation of the lid-driven cavity problem at times $t = 1$ (a), $t = 10$ (b), and $t = 20$ (c).
		The video is provided at
		\href{https://youtu.be/FCoShaa_FhM}{\texttt{youtu.be/FCoShaa\_FhM}}; a description is given in Sec.~\hyperref[sec:sec_appendix_movie_2]{E.2}.
	}
	\label{fig:fig_lag_cavity_results}
	\vspace{-15pt}
\end{figure}

This movie, which can be found at
\href{https://youtu.be/FCoShaa_FhM}{\texttt{youtu.be/FCoShaa\_FhM}}
with snapshots presented in Fig.~\ref{fig:fig_lag_cavity_results}, shows a Lagrangian finite element scheme modeling the well-known lid-driven cavity problem.
Due to in-plane velocity gradients, elements deform over time.
The color and color bar in the movie indicate the $x$-component of the velocity, which is incorrect at late times.
The cavity is represented by a unit square, the mesh is $32 \times 32$ elements, and the time step $\Delta t = 1/640$.
After 20 time steps, nodes on the top edge move by a mesh spacing due to the velocity boundary condition at the top edge.
Note in particular the top right corner, where after 20 time steps two nodes coincide, causing the simulation to fail.

% *** Comparison of Lagrangian and LE schemes in modeling a deforming fluid film *** %
\paragraph{3. Comparison of Lagrangian and LE schemes in modeling a fluid film} \label{sec:sec_appendix_movie_3}

This movie, provided at
\href{https://youtu.be/wnXuK6d3WTQ}{\texttt{youtu.be/wnXuK6d3WTQ}},
compares an LE scheme (top) and Lagrangian scheme (bottom) in modeling an unstable cylindrical fluid film.
The movie runs from time $t=0$ to $t=35$, the color and color bars indicate the surface tension, and both simulations use a $30 \times 60$ mesh.
While both schemes capture the same solution, the Lagrangian nodes move in-plane due to the fluid flow.
As a result, at later times there are comparatively more nodes in the left spherical bulb, and fewer nodes in the central tubular region.
The underresolution of the tube in the Lagrangian scheme leads to larger errors.
Snapshots from this movie are shown in Figs.~\ref{fig:fig_lag_ale_t_05}--\ref{fig:fig_lag_ale_t_35} in the main text.

%
% *** List of important symbols
%

\section{List of Important Symbols} \label{sec:sec_symbol_list}

\vspace{-16pt}
\setlength{\columnsep}{15pt}
\begin{multicols}{2}
\noindent
\setlength{\tabcolsep}{3pt}
\begin{tabular}{ @{}l p{0.38\textwidth} }
	$\bm{1}$
	& identity tensor in $\mathbb{R}^3$ \\
	$\bm{a}_\alpha$ 
	& in-plane covariant basis vectors \\ 
	$\bm{a}_\calpha$ 
	& in-plane covariant basis vectors, \\&induced by $\zeta^\alpha$ parametrization \\ 
	$\bm{a}^\alpha$ 
	& in-plane contravariant basis vectors \\ 
	$\bm{a}_{\alpha , \beta}$
	& partial derivative of $\bm{a}_\alpha$ w.r.t.~$\theta^\beta$ \\
	$\bm{a}_{\alpha ; \beta}$
	& covariant derivative of $\bm{a}_\alpha$ w.r.t.~$\theta^\beta$ \\
	$a_{\alpha \beta}$
	& covariant metric \\ 
	$a^{\alpha \beta}$
	& contravariant metric \\
	$\alpham$
	& parameter for mesh weak form \\&unit consistency \\
	$\alphaDB$
	& Dohrmann--Bochev parameter \\
	$\bm{b}$
	& body force per unit mass \\
	$b_{\alpha \beta}$
	& covariant curvature components \\ 
\end{tabular}
\begin{tabular}{ l p{0.38\textwidth} }
	$b^{\alpha \beta}$
	& contravariant curvature components \\
	$\bm{c}$
	& material velocity minus mesh velocity\\
	%& relative velocity between material and mesh \\
	$c^\calpha$
	& component of $\bm{c}$, expressed in the \\& $\zeta^\alpha$ parametrization \\
	$C^m(\Omega)$
	& space of scalar functions on $\Omega$ with \\& $m$ continuous derivatives \\
	$\chi$
	& multiplicative factor of cylinder \\&perturbation \\
	$ \delta (\, \cdot \,)$
	& arbitrary variation of $(\, \cdot \,)$ \\
	$ [ \delta (\, \cdot \,)^{e} ]$
	& discretized local variation of $(\, \cdot \,)$ \\
	$\mcg$
	& direct Galerkin expression \\
	$\Gamma$
	& boundary of parametric domain $\Omega$ \\
	$\Gamman$
	& Neumann portion of $\Gamma$ \\
	$\Gammad$
	& Dirichlet portion of $\Gamma$ \\
\end{tabular}
\begin{tabular}{ l p{0.38\textwidth} }
	$\Gamma^\alpha_{\lambda \mu}$
	& Christoffel symbols of the second kind \\
	$h$
	& finite element size \\
	$H$
	& mean curvature \\ 
	$H^2(\Omega)$
	& Sobolev space of order two on $\Omega$ \\
	$J$
	& areal surface expansion relative to \\&reference configuration \\ 
	$\Jm$
	& Jacobian determinant from $\Omega$ to $\scp$ \\ 
	$\JGammam$
	& Jacobian determinant from $\Gamma$ to $\bscp$ \\
	$K$
	& Gaussian curvature \\ 
	$\mK$
	& global tangent matrix \\
	$\mKe$
	& local tangent matrix of element $e$ \\
	$L^2(\Omega)$
	& space of square-integrable \\&functions on $\Omega$ \\
	$\lambda$
	& surface tension, enforces areal \\&incompressibility \\
	$\blambda$
	& projection of $\lambda$ onto $\bLambda$ \\
	$\Lambda$
	& functional space of surface tensions $\lambda$ \\
	$\bLambda$
	& functional space of projected \\&surface tensions $\blambda$ \\
	$\Lambdah$
	& finite-dimensional subspace of $\Lambda$ \\
	$\mlambda$
	& global surface tension degrees of \\&freedom (DOFs) \\
	$\mlambdae$
	& local surface tension DOFs \\
	$\mblambdae$
	& local projected surface tension DOFs \\
	%$\bm{m}$
	%& bending moment per unit length at the patch boundary \\
	%$\bm{M}$
	%& director traction at the patch boundary \\
	%$\bm{M}^\alpha$
	%& couple-stress vector along a curve \\&of constant $\theta^\alpha$ \\
	%$M^{\alpha \beta}$
	%& contravariant bending moment \\&components \\
	%$\bm{\mu}$
	%& couple-stress tensor \\
	$\bm{n}$
	& normal vector to the surface \\ 
	$\nume$
	& number of elements \\
	$\nen$
	& number of element nodes \\
	$\neln$
	& number of Lagrange multiplier \\&element nodes \\
	$\nln$
	& number of Lagrange multiplier nodes \\
	$\nn$
	& number of nodes \\
	$\mN$
	& global shape function matrix \\
	$\mNe$
	& local shape function matrix \\
	$\mhN$
	& global Lagrange multiplier shape \\&function matrix \\
	$\mhNe$
	& local Lagrange multiplier shape \\&function matrix \\
	$\mbNe$
	& local projected Lagrange multiplier shape function matrix \\
	%$N^{\alpha \beta}$
	%& in-plane contravariant stress components \\
	$\bm{\nu}$
	& in-plane unit normal to the \\&patch boundary \\
	$\Omega$
	& parametric domain of $\zeta^\alpha$ \\
	$\Omegae$
	& discretized element of $\Omega$ \\
	%$\omega^{\alpha \beta}$
	%& contravariant bending dissipation \\&components \\
	$p$
	& pressure normal to the surface \\
	$\mathcal{P}$
	& functional space of normal pressures \\
	$\mathcal{P}^h$
	& finite-dimensional subspace of $\mathcal{P}$ \\
	$\scp$
	& patch of the surface under consideration \\
\end{tabular}
\begin{tabular}{ l p{0.38\textwidth} }
	%$\scpo$
	%& reference surface patch \\
	$\bscp$
	& boundary of the patch $\scp$ \\
	$\bnscp$
	& Neumann portion of $\bscp$ \\
	$\bdscp$
	& Dirichlet portion of $\bscp$ \\
	$\mathbb{P}_n(\Omegae)$
	& space of polynomial functions \\&of order $n$ on $\Omegae$ \\
	$\pi^{\alpha \beta}$
	& in-plane contravariant viscous \\&stress components \\
	%$\psi$
	%& Helmholtz free energy density \\&per unit mass \\
	$\mathbb{Q}_n(\Omegae)$
	& space of bi-polynomial functions \\&of order $n$ on $\Omegae$ \\
	$\mr$
	& global residual vector \\
	$\mre$
	& local residual vector of element $e$ \\
	$\rho$
	& areal mass density of the current surface \\
	%$\rho_0$
	%& areal mass density of the reference surface \\
	%$S^\alpha$
	%& out-of-plane contravariant \\&stress components \\
	%$\bm{\sigma}$
	%& Cauchy stress tensor \\
	$\sigma^{\alpha \beta}$
	& in-plane contravariant couple-free \\&components of the surface stress \\
	$\bm{T}$
	& traction at the patch boundary \\
	$\bm{T}^\alpha$
	& stress vector along a curve of constant $\theta^\alpha$ \\
	%$\bm{\tau}$
	%& in-plane unit tangent at the \\&patch boundary \\
	$\tau_{\mathrm{theo}}$
	& theoretical cylindrical fluid film \\&time scale \\
	$\tau_{\mathrm{sim}}$
	& numerical cylindrical fluid film time scale \\
	$\theta^\alpha$
	& surface-fixed parametrization \\&of the surface \\
	$\bmu$
	& general vector of all unknowns \\ 
	$\mbu$
	& global DOF vector for all unknowns \\ 
	$\mbue$
	& local DOF vector for all unknowns \\ 
	$\bmv$
	& material velocity \\ 
	$\mve$
	& local velocity DOFs \\
	$\dot{\bmv}$
	& material acceleration \\ 
	$\mcv$
	& functional space of velocities and \\&mesh velocities \\
	$\mcvh$
	& finite-dimensional subspace of $\mcv$ \\&resulting from discretization \\
	$\mcv_0$
	& subset of $\mcv$ which vanishes on $\Gammad$ \\
	$\mcvho$
	& finite-dimensional subspace of $\mcv_0$ \\
	$\bmvm$
	& mesh velocity \\ 
	$\mvme$
	& local mesh velocity DOF vector \\
	%$W$
	%& total patch Helmholtz free energy \\
	$\bmx$ 
	& surface position in $\mathbb{R}^3$ \\
	$\hat{\bmx}$ 
	& surface position, expressed in $\xi^\alpha$ \\&coordinates \\
	$\check{\bmx}$ 
	& surface position, expressed in $\zeta^\alpha$ \\&coordinates \\
	$\bmx_{\mathrm{b}}$
	& position of a point on the patch boundary \\
	%$\bmx_0$
	%& position of reference patch $\scpo$ \\
	$\xi^\alpha$
	& convected coordinate parametrization of the surface \\
	$\zeta$
	& shear viscosity coefficient for \\&in-plane flow \\
	$\zeta^\alpha$
	& mesh parametrization of the surface \\
	$ \lvert \lvert \, \cdot \, \rvert \rvert_0 $
	& $L^2$-norm of $(\, \cdot \,)$ \\
	$\otimes$
	& dyadic or outer product between \\&two vectors \\
\end{tabular}
\setlength{\tabcolsep}{6pt}
\end{multicols}

	\end{appendices}

	%
	% *** REFERENCES
	%

	\addcontentsline{toc}{section}{References}
	\bibliographystyle{content/bibStyle}
	\small
	\bibliography{content/refs}

\end{document}